%&tex
%% Master TeX file for the Orbifold2  paper

%auto-ignore
% macros.tex
%% TeX macros for Orbifold2 paper

\input phyzzx

\input pstricks
\psset{unit=1mm,linewidth=0.5pt,linecolor=black,arrowscale=1.5}
\newrgbcolor{darkgreen}{0 0.6 0}
\newrgbcolor{paleyellow}{1 1 0.5}
\newrgbcolor{purple}{0.5 0 0.5}
\newrgbcolor{darkblue}{0 0 0.7}
\newrgbcolor{darkred}{0.7 0 0}
\newgray{palegray}{0.9}
\newrgbcolor{aquamarine}{0 0.4 0.5}

%%%%%%%%%%%%%%%%%%%%%%%%%%%%%%%%%%%%%%%%%%%%%%%%%%%%%%%%%%%%
% Adding a new font family --- what a pain in the ****
%
\newfam\msymfam
\font\seventeenmsym =msbm10 scaled\magstep3    \skewchar\seventeenmsym='60
\font\fourteenmsym  =msbm10 scaled\magstep2     \skewchar\fourteenmsym='60
\font\twelvemsym    =msbm10 scaled\magstep1       \skewchar\twelvemsym='60
\font\tenmsym       =msbm10                          \skewchar\tenmsym='60
\font\ninemsym      =msbm9                          \skewchar\ninemsym='60
\font\sevenmsym     =msbm7                         \skewchar\sevenmsym='60
\font\sixmsym       =msbm6                           \skewchar\sixmsym='60
\font\fivemsym      =msbm5                          \skewchar\fivemsym='60
\begingroup
\catcode`\@=11
\xdef\msym{\n@expand\f@m\msymfam}
\let\fourteenf@nts@ld=\fourteenf@nts
\xdef\fourteenf@nts{\fourteenf@nts@ld \textfont\msymfam=\fourteenmsym 
    \scriptfont\msymfam=\tenmsym \scriptscriptfont\msymfam=\sevenmsym }
\let\twelvef@nts@ld=\twelvef@nts
\xdef\twelvef@nts{\twelvef@nts@ld \textfont\msymfam=\twelvemsym 
    \scriptfont\msymfam=\ninemsym \scriptscriptfont\msymfam=\sixmsym }
\let\tenf@nts@ld=\tenf@nts
\xdef\tenf@nts{\tenf@nts@ld \textfont\msymfam=\tenmsym 
    \scriptfont\msymfam=\sevenmsym \scriptscriptfont\msymfam=\fivemsym }
\endgroup
\twelvepoint
%%%%%%%%%%%%%%%%%%%%%%%%%%%%%%%%%%%%%%%%%%%%%%%%%%%%%%%%%%
\def\IZ{{\msym Z}}
\def\IC{{\msym C}}
\def\IR{{\msym R}}
\def\IS{{\msym S}}

\def\M{{\aquamarine\it M}}
\def\FP{{\cal O}}
\def\I{{\cal I}5}
\def\Ori{{\bf O8}}

\def\bx{{\bf x}}
\def\tI{{$\rm type~I'$}}
\def\TN{\mathord{\rm TN}}

\def\dualrel#1{\buildrel #1\over\longleftrightarrow}
\def\diag{\mathop{\rm diag}\nolimits }
\def\dim{\mathop{\rm dim}\nolimits }
\def\symrep{{\vcenter{\hrule height 0.2ex \hbox to 2.2ex{%
        \vrule height 1ex width 0.2ex\hfil
        \vrule width 0.2ex\hfil\vrule width 0.2ex}%
    \hrule  height 0.2ex}}}
\def\asymrep{{\vcenter{\hbox{\vrule width 0.2ex\vbox to 2.2ex{%
        \hrule width 1ex height 0.2ex\vfil
        \hrule height 0.2ex\vfil\hrule height 0.2ex}%
    \vrule  width 0.2ex}}}}
\def\thrindex{{\vcenter{\hbox{\vrule width 0.2ex\vbox to 3.2ex{%
        \hrule width 1ex height 0.2ex\vfil
        \hrule width 1ex height 0.2ex\vfil
        \hrule width 1ex height 0.2ex\vfil
        \hrule width 1ex height 0.2ex}%
    \vrule  width 0.2ex}}}}

\def\Re{\mathop{\rm Re}\nolimits}

%%%%%%%%%%%%%%%%%%%

%%%%%%%%%%%%%%%%%%%%%%%%%%%%%%%%%%%%%%%%%%%%%%%%
\def\papersize{\hsize=35pc \vsize=50pc 
   \pagebottomfiller=0pc plus 10pc
   \skip\footins=\bigskipamount \normalspace }

%auto-ignore
% biblio.tex
%% References for Orbifold2 paper
%% Compiled by Stefan Theisen and Jacob Sonnenschein

\REF\HWI{P.~Ho\v rava and E.~Witten,
``Heterotic and type I string dynamics from eleven dimensions'',
Nucl.\ Phys.\ B {\bf 460} (1996) 506
[hep-th/9510209].}

\REF\HWII{P.~Ho\v rava and E. Witten, ``Heterotic and type I string
dynamics from eleven dimensions'', Nucl.\ Phys.\ B {\bf 506} (1996) 506
[hep-th/9510209].}

\REF\KSTY{V. S.~Kaplunovsky, J. Sonnenschein, S. Theisen
and S. Yankielowicz,
``On the Duality between Perturbative Heterotic Orbifolds and M-Theory
on $T^4/Z_N$'',
Nucl.\ Phys.\  B {\bf 590} (2000) 123 [hep-th/9912144].}

\REF\FLOI{M.~Faux, D.~L\"ust and B.~A.~Ovrut,
``Intersecting orbifold planes and local anomaly cancellation in  M-theory'',
Nucl.\ Phys.\ B {\bf 554} (1999) 437
[hep-th/9903028].}

\REF\FLOII{M.~Faux, D.~L\"ust and B.~A.~Ovrut,
``Local anomaly cancellation, M-theory orbifolds and phase-transitions'',
Nucl.\ Phys.\ B {\bf 589} (2000) 269
[hep-th/0005251];
``An M-theory perspective on heterotic K3 orbifold compactifications'',
hep-th/0010087;
``Twisted sectors and Chern-Simons terms in M-theory orbifolds'',
hep-th/0011031.}

\REF\DasMuk{K.~Dasgupta and S.~Mukhi,
``Orbifolds of {\it M}-theory'',
Nucl.\ Phys.\ B {\bf 465} (1996) 399
[hep-th/9512196].}

\REF\WittenMfive{E.~Witten,
``Five-branes and {\it M}-Theory on an Orbifold'',
Nucl.\ Phys.\  B {\bf 463} (1996) 383,
[hep-th/9512219].}

\REF\AFIUV{G.~Aldazabal, A.~Font, L.~E.~Ibanez, A.~M.~Uranga and 
G.~Violero,
``Non-Perturbative Heterotic D=6, N=1 Orbifold Vacua'',
Nucl.\ Phys.\  B {\bf 519} (1998) 239,
[hep-th/9706158].}

\REF\BDS{A.~Bilal, J.-P.~Derendinger and R.~Sauser,
``M-Theory on $S^1/Z_2$: New Facts from a Careful Analysis'',
Nucl.\ Phys.\ B {\bf 576} (2000) 347
[hep-th/9912150].}

\REF\Townsend{P. Townsend, ``The eleven-dimensional supermembrane
revisited'',  Phys.\ Lett.\  B {\bf 350} (1995) 184 [hep-th/9501068].}

\REF\HanZaf{A. Hanany and A. Zaffaroni, ``Branes and six-dimensional
supersymmetric theories'',  Nucl.\ Phys.\ B {\bf 529} (1998) 180 
[hep-th/9712145];
``Monopoles in string theory",  JHEP {\bf 9912} (1999) 014 [hep-th/9911113].}

\REF\Seiberg{N. Seiberg,
``Five dimensional SUSY field theories, non-trivial fixed points 
and string dynamics", 
Phys.\ Lett.\ B {\bf 388} (1996) 753 [hep-th/9608111].}

\REF\SM{D.~R.~Morrison and N.~Seiberg,
``Extremal transitions and five-dimensional supersymmetric field  theories'',
Nucl.\ Phys.\ B {\bf 483} (1997) 229\hfill\break [hep-th/9609070].}

\REF\AFIQ{G.~Aldazabal, A.~Font, L.~E.~Ibanez and F.~Quevedo,
``Chains of N=2, D=4 heterotic/type II duals'',
Nucl.\ Phys.\ B {\bf 461} (1996) 85 [hep-th/9510093].}

\REF\DMW{M. Duff, R. Minasian and E. Witten,
``Evidence for Heterotic/Heterotic Duality'', 
Nucl.\ Phys.\  B {\bf 465} (1996) 413
[hep-th/9601036].}

\REF\PolII{J.~Polchinski, ``String Theory, Vol. II'', 
Cambridge University Press, 1998.}

\REF\DHS{M.~Dine, P.~Huet and N.~Seiberg,
``Large And Small Radius In String Theory'',
Nucl.\ Phys.\ B {\bf 322} (1989) 301.}

\REF\DLP{J.~Dai, R.~G.~Leigh and J.~Polchinski,
``New Connections Between String Theories'',
Mod.\ Phys.\ Lett.\ A {\bf 4} (1989) 2073.}

\REF\PW{J.~Polchinski and E.~Witten,
``Evidence for Heterotic - Type I String Duality'',
Nucl.\ Phys.\ B {\bf 460} (1996) 525
[hep-th/9510169].}

\REF\KS{S.~Kachru and E.~Silverstein,
``On gauge bosons in the matrix model approach to M theory'',
Phys.\ Lett.\ B {\bf 396} (1997) 70
[hep-th/9612162].}

\REF\DL{D.~A.~Lowe,
``Bound states of type I' D-particles and enhanced gauge symmetry'',
Nucl.\ Phys.\ B {\bf 501} (1997) 134
[hep-th/9702006], 
``Heterotic matrix string theory'',
Phys.\ Lett.\ B {\bf 403} (1997) 243
[hep-th/9704041].}

\REF\MNT{D.~Matalliotakis, H.~Nilles and S.~Theisen,
``Matching the BPS spectra of heterotic --  type-I -- type-I' strings'',
Phys.\ Lett.\ B {\bf 421} (1998) 169
[hep-th/9710247].}

\REF\BGL{O.~Bergman, M.~R.~Gaberdiel and G.~Lifschytz,
``String creation and heterotic-type I' duality'',
Nucl.\ Phys.\ B {\bf 524} (1998) 524 [hep-th/9711098].}

\REF\BGS{C.~P.~Bachas, M.~B.~Green and A.~Schwimmer,
``(8,0) quantum mechanics and symmetry enhancement in type I' superstrings'',
JHEP {\bf 9801} (1998) 006
[hep-th/9712086].}

\REF\CV{F.~A.~Cachazo and C.~Vafa,
``Type I' and real algebraic geometry'',
[hep-th/0001029].}

\REF\DKV{M.~R.~Douglas, S.~Katz and C.~Vafa,
``Small instantons, del Pezzo surfaces and type I' theory'',
Nucl.\ Phys.\ B {\bf 497} (1997) 155
[hep-th/9609071].}

\REF\WittenTN{E. Witten, ``New ``Gauge'' Theories In Six Dimension'',
JHEP {\bf 9801} (1998) 001; Adv. Theor. Math.Phys. {\bf 2} (1998) 61
[hep-th/9710065].} 

\REFS\SOSP{ S.~Elitzur, A. Giveon, D. Kutasov, E. Rabinovici
and A. Schwimmer ``Brane dynamics and N=1 supersymmetric gauge
theories", Nucl.\ Phys.\  B {\bf 505} (1997) 202 [hep 9704104];
K. Landsteiner and E. Lopez, ``New curves from branes"
Nucl.\ Phys.\  B {\bf 516} (1998) 273 [hep-th 9708118]; 
N. Evans, C.V.~Johnson and A.D. Sharpere, ``Orientifolds,
branes, and duality of 4D gauge theories", 
Nucl.\ Phys.\  B {\bf 505} (1997) 251 [hep-th 9703210];
A. Brandhuber, J. Sonnenschein, S. Theisen and S. Yankielowicz,
``M theory and Seiberg-Witten curves: orthogonal and symplectic
groups",  Nucl.\ Phys.\ B {\bf 504} (1997) 175 [hep-th 9705232];
for a review, \cf\
A.~Giveon and D.~Kutasov,
``Brane dynamics and gauge theory'',
Rev.\ Mod.\ Phys.\  {\bf 71} (1999) 983
[hep-th/9802067].}

\REF\Wittenbend{E. Witten ``Solutions of 4d field theories via M-theory",
Nucl./ Phys./ B {\bf 500} (1997) 3 [hep-th 9703166].}

\REF\BruKar{I. Brunner and A. Karch, ``Branes at orbifolds versus 
Hanany Witten in six dimensions,"  JHEP {\bf 03} (1998) 003
[hepth/9712143].}  

\REF\Romans{J.L.~Romans, ``Massive N=2a Supergravity in 10 Dimensions'',
Phys.\ Lett.\ B {\bf 169} (1986) 147.}

\REF\HanWit{A.~Hanany and E.~Witten, 
``Type IIB superstrings, BPS monopoles, and three-dimensional 
gauge dynamics'',
Nucl.\ Phys.\ B {\bf 492} (1997) 152
[hep-th/9611230].}

%auto-ignore
% title.tex
%% The title page for Orbifold2 paper
%% Written by Vadim Kaplunovsky

\Pubnum={UTTG--11--01\cr TAUP--2677--01\cr 
AEI-2001-104\cr
hep-th/yymmnnn}
\date={August 2001}

\titlepage

\title{%
    On Heterotic Orbifolds, \M~Theory and 
    Type I$'$ Brane Engineering
    \foot{%
        Research supported in part by
	the US-Israeli Binational Science Foundation, the US National Science
	Foundation (grant \#PHY--00--71512), the Robert A.~Welsh Foundation,
	the German--Israeli Foundation for Scientific Research (GIF),
	by the EEC (contract \#HPRN--CT--2000--00122), 
	by the Israel Science Foundation, and by DFG--SFB--375.
	}
    \foot{We advice printing this paper in color}  
    }

\author{%
    Elie~Gorbatov,\attach{a}
    Vadim~S.~Kaplunovsky,\attach{a}\break
    Jacob~Sonnenschein,\attach{b}
    Stefan~Theisen\attach{c}
    {\twelverm and} Shimon~Yankielowicz\attach{b}
    }

\medskip
\begingroup
\Tenpoint\sl
\noindent
\leftskip=1.5em
\item a Theory Group, Physics Dept., University of Texas, Austin, TX78712, USA.
\item b School of Physics and Astronomy, Beverly and Raimond Sackler Faculty
of Exact Sciences,
Tel Aviv University, Tel-Aviv 69978, Israel.
\item c Planck--Institut f\"ur Gravitationsphysik, Albert--Einstein--Institut,\brk
	Am M\"uhlenberg~1, D--14476 Golm, Germany.
\par
\endgroup
\endpage

\abstract
Ho\v rava--Witten \M~$\rm theory\leftrightarrow heterotic$ string duality
poses special problems for the twisted sectors of heterotic orbifolds.
In our previous paper~[\KSTY] we explained how in \M~theory
the twisted states couple to gauge fields apparently living on  M9 branes at
{\sl both} ends of the eleventh dimension {\sl at the same time}.
The resolution involves 7D gauge fields which live on fixed planes of the
$(T^4/\IZ_N)\times (\IS^1/\IZ_2)\times\IR^{5,1}$ orbifold and lock 
onto the 10D gauge fields along the intersection planes.
The physics of such intersection planes does not follow directly from the
\M~theory but there are stringent kinematic constraints
due to duality and local consistency, which allowed us to deduce the local
fields and the boundary conditions at each intersection.
\subpar\noindent
In this paper we explain various phenomena at the intersection planes
in terms of duality between Ho\v rava--Witten and \tI~superstring theories.
The orbifold fixed planes are dual to stacks of D6 branes, the M9 planes
are dual to \Ori\ orientifold planes accompanied by D8 branes,
and the intersections are dual to brane junctions.
We engineer several junction types which lead to distinct patterns of
7D/10D gauge field locking, 7D symmetry breaking and/or local 6D fields.
Another aspect of brane engineering is putting the junctions together;
sometimes, the combined effect is rather spectacular from the HW point of view
and the quantum numbers of some twisted states have to `bounce' off both
ends of the eleventh dimension before their heterotic identity becomes clear.
\subpar\noindent
Some models involve D6/\Ori\ junctions where the string coupling diverges
towards the orientifold plane.
We use the $\rm heterotic\leftrightarrow HW\leftrightarrow I'$ duality
to predict what {\sl should} happen at such junctions.
For example, pinning down an NS5 half-brane to a definite location on a
$\lambda=\infty$~\Ori\ plane requires precisely four D6 branes.

\endpage

%%%%%%%%%%%%%%%%%%%%%%%%%%%%%%%%%%%%%%%%%%
%auto-ignore
% chapter1.tex
%% Chapter 1 of Orbifold2 paper
%% Written by Stefan Theisen

\chapter{Introduction}

It is by now well established that duality symmetries
relate all five ten-dimensio\-nal 
perturbative string theories but that
they do not constitute a closed set. Rather
eleven-dimensional supergravity has to be included as one of the 
possible effective low-energy descriptions. This implies that the 
underlying fundamental theory --- called \M~theory --- is not simply 
a theory of strings,  but its true nature remains rather 
\M ysterious. 

Of particular interest is the Ho\v rava--Witten duality between the 
heterotic $E_8\times E_8$ string and the 11D \M~theory compactified on 
a finite interval $I=\IS^1/\IZ_2$, where the gauge degrees of freedom 
are localized at the two end-of-the-word boundary branes 
${\rm M9}_1$ and ${\rm M9}_2$,\foot{Our notations in this paper follow
the $D$-braned convention in which extended objects --- branes or fixed
planes --- are labelled by their space rather than space-time 
dimensionalities. Thus, an M9 brane has nine space dimensions 
plus one time, hence it carries an 10D SYM theory on its world-volume;
likewise, an \FP6\ plane has six space dimensions and carries 
a 7D SYM on its word-volume, \etc.}  
one $E_8$ factor on each side [\HWI,\HWII]. 
This duality was derived in ten flat 
Minkowski dimensions and should hold in lower 
dimensions after compactification.
In our previous paper [\KSTY] we studied the $T^4/\IZ_N$ orbifold  
compactifications of this duality to $d=6$ in which both 
$E_8$ gauge symmetries are broken by the orbifold action, 
$E_8^{(1)}\times E_8^{(2)}\to G^{(1)}\times G^{(2)}$
(\cf\ also [\FLOI,\FLOII] and [\DasMuk,\WittenMfive,\AFIUV,\BDS]).
In the twisted sectors of such orbifolds, the particles are usually charged 
under {\it both} $G^{(1)}$  and $G^{(2)}$, 
which raises a paradox in the dual 11D
\M~theory description; with the $G^{(1)}$ confined to one end of the world
and the $G^{(2)}$ confined to the other end, where in the eleventh dimension 
do we put the massless twisted states?\foot{We focus on the massless states
because their exact masslessness is protected by their chirality and their 
origin in the dual \M~theory must therefore be local. The massive states
are neither chiral nor BPS (in $d=6, {\cal N}=1$ SUSY) which leaves a wider 
choice  for their \M~theory origins. For example, they could become 
extended objects stretched between ${\rm M9}_1$ and 
${\rm M9}_2$, hence $G^{(1)}\times G^{(2)}$ charges without a paradox.} 

We found that the local charges of the troublesome twisted states 
do not directly belong to 
$G^{(1)}\times G^{(2)}\subset E_8^{(1)}\times E_8^{(2)}$
but rather to $G_7\times G^{(2)}$ where $G_7$ is a non-perturbative 
7D gauge symmetry localized on a fixed plane $\FP6=\IR^{5,1}\times 
\IS^1/\IZ_2\times\rm a$ fixed point of the orbifold action. 
The $G_7$ symmetry mixes with a similar factor of $G^{(1)}$
along the $\I_1={\rm M9}_1\cap\FP6$ intersection plane. 
Consequently, the diagonal symmetry {\it appears} to be a subgroup 
of $G^{(1)}$ but geometrically it extends beyond the 
${\rm M9}_1$ brane along the \FP6 
fixed-planes towards the other end of the eleventh dimension. 
On $\I_2={\rm M9}_2\cap \FP6$ we have both $G_7$ and $G^{(2)}$ gauge fields 
and the twisted fields living there acquire both charges in a local fashion.
The apparent paradox thus arises from a mis-identification of $G_7$
as a subgroup of $G^{(1)}$. This is natural in the perturbative 
heterotic theory but one has to more careful in \M~theory.

As an example, 
consider the $T^4/\IZ_2$ orbifold with $G^{(1)}=
E_7\times SU(2)^{\rm p}$, 
$G^{(2)}=SO(16)$ and $G_7=SU(2)^{\rm np}$. 
According to ref.~[\KSTY] we have the following 
picture: 
$$
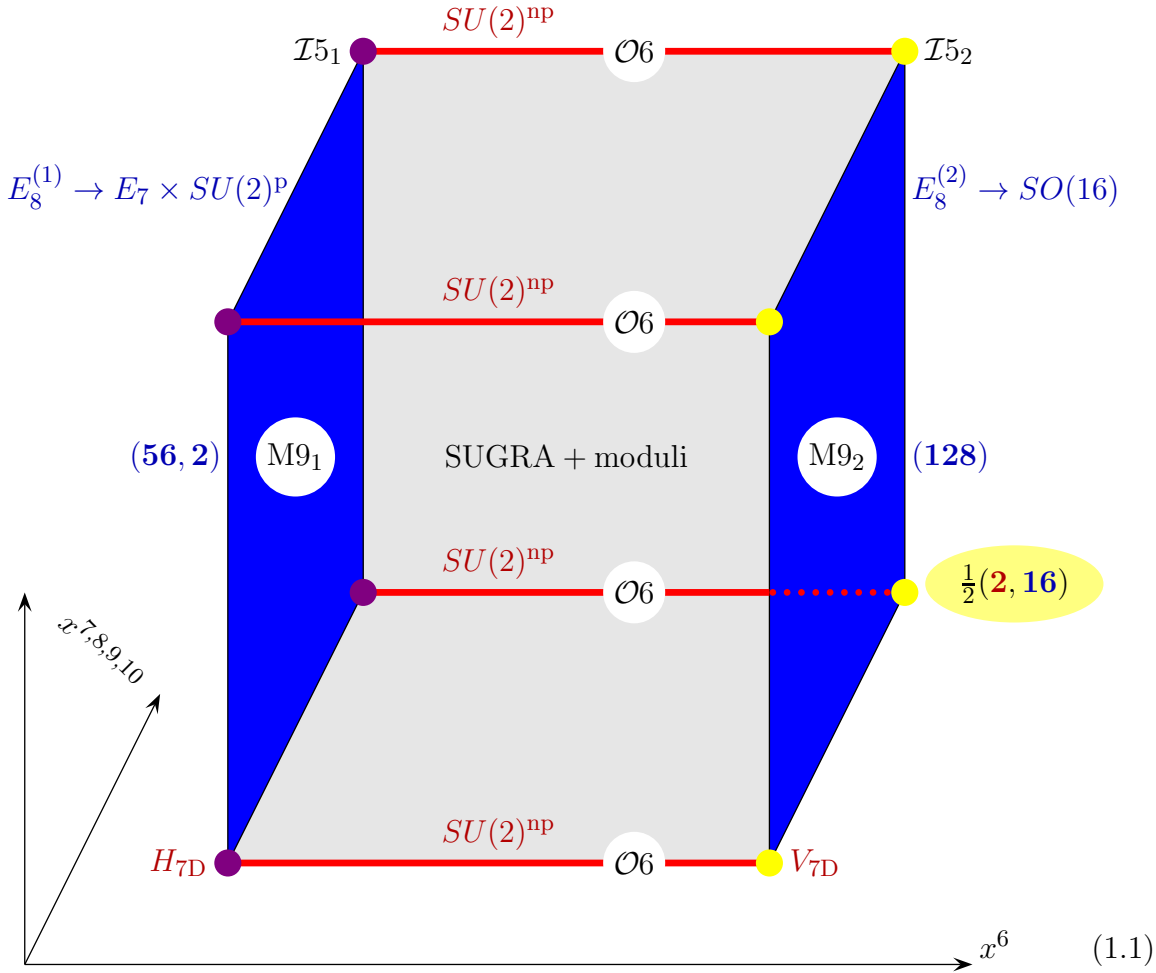
\psset{unit=0.9mm}
\pspicture[](-70,-70)(+80,+70)
\pspolygon*[linecolor=palegray](-50,-60)(-50,+20)(-30,+60)(+50,+60)(+50,-20)(+30,-60)
\pspolygon[fillstyle=solid,fillcolor=blue](-50,-60)(-50,+20)(-30,+60)(-30,-20)
\pspolygon[fillstyle=solid,fillcolor=blue](+30,-60)(+30,+20)(+50,+60)(+50,-20)
\rput(0,0){$\rm SUGRA+moduli$}
\cput*(-40,0){$\rm M9_1$}
\cput*(+40,0){$\rm M9_2$}
\rput[r](-41,+40){$\darkblue E_8^{(1)}\to E_7\times SU(2)^{\rm p}$}
\rput[r](-51,0){$\darkblue ({\bf56},{\bf2})$}
\rput[l](+51,+40){$\darkblue E_8^{(2)}\to SO(16)$}
\rput[l](+51,0){$\darkblue ({\bf128})$}
\psline[linecolor=red,linewidth=1](-50,-60)(+30,-60)
\psline[linecolor=red,linewidth=1](-50,+20)(+30,+20)
\psline[linecolor=red,linewidth=1](-30,-20)(+30,-20)
\psline[linecolor=red,linewidth=1,linestyle=dotted](+30,-20)(+50,-20)
\psline[linecolor=red,linewidth=1](-30,+60)(+50,+60)
\cput*(+10,+60){\FP6}
\cput*(+10,+20){\FP6}
\cput*(+10,-20){\FP6}
\cput*(+10,-60){\FP6}
\rput[b](-10,+62){$\darkred SU(2)^{\rm np}$}
\rput[b](-10,+22){$\darkred SU(2)^{\rm np}$}
\rput[b](-10,-18){$\darkred SU(2)^{\rm np}$}
\rput[b](-10,-58){$\darkred SU(2)^{\rm np}$}
\pscircle*[linecolor=purple](-50,-60){2}
\pscircle*[linecolor=purple](-50,+20){2}
\pscircle*[linecolor=purple](-30,-20){2}
\pscircle*[linecolor=purple](-30,+60){2}
\pscircle*[linecolor=yellow](+30,-60){2}
\pscircle*[linecolor=yellow](+30,+20){2}
\pscircle*[linecolor=yellow](+50,-20){2}
\pscircle*[linecolor=yellow](+50,+60){2}
\rput[r](-33,+60){$\I_1$}
\rput[l](+53,+60){$\I_2$}
\rput[r](-53,-60){$\darkred H_{\rm 7D}$}
\rput[l](+33,-60){$\darkred V_{\rm7D}$}
\rput[l](+53,-20){\psovalbox*[fillcolor=paleyellow]
{$\half({\bf\darkred2},{\bf\darkblue16})$}}
\psline{->}(-80,-75)(+60,-75)
\rput[bl](+61,-74){$x^6$}
\psline{->}(-80,-75)(-80,-20)
\psline{->}(-80,-75)(-60,-35)
\rput[b]{315}(-70,-30){$x^{7,8,9,10}$}
\endpspicture
\eqn\ZTWO
$$
(For simplicity we have depicted only four of the sixteen \FP6\ planes.) 
At the $\I_1$ intersections (denoted by purple dots) 7D $SU(2)^{\rm np}$ 
gauge fields lock onto $SU(2)^{\rm p}$ gauge fields, 
$$
A_\mu(x^6=0)=A_\mu^{\rm 10D}(x^{7,8,9,10}=0)
~~\hbox{for same}~x^{0,\dots,5},\,\mu=0,\dots,5\,.
\eqn\HWLockingBC
$$
By supersymmetry (there are eight unbroken supercharges at \I\ intersections)
similar Dirichlet-like boundary conditions apply for the fermionic
partners of the vector fields. Between the boundaries on \FP6\ there exist 
16 supercharges and a 16--SUSY vector multiplet comprises both,
a 8--SUSY vector multiplet $V_{\rm 7D}$ and a 8--SUSY hypermultiplet
$H_{\rm 7D}$. The $H_{\rm 7D}$ components have Neumann boundary 
conditions at $\I_1$.
At the other end of the $11^{\rm th}$ dimension, at $\I_2$ (yellow dot) 
it is $H_{\rm 7D}$ which suffers Dirichlet boundary conditions while 
$V_{\rm 7D}$ enjoys Neumann boundary conditions.  
Consequently the net gauge symmetry at $\I_2$ is $SU(2)\times SO(16)$ 
which allows for local half-hypermultiplets in the 
$({\bf 2},{\bf 16})$ representation.

In [\KSTY] we gave three lines of evidence for the 
mixing of M9 and $\FP6$ gauge groups: 
(i) It is the only way  
to reconcile the massless spectra 
of heterotic orbifolds with locality in the dual \M~theory description.
(ii) The heterotic gauge coupling, which is known exactly in six 
dimensions, shows that some gauge groups cannot be of purely M9
origin but must mix with the non-perturbative factors. 
(iii) Each $\I$ intersection plane carries a chiral 
field theory which suffers from local anomalies involving 
massless particles living on the $\I$ itself, 
on M9, on $\FP6$ and in the 11D
bulk as well as inflow and intersection anomalies 
due to Chern--Simons terms in \M~theory.
For the local fields and boundary conditions proposed in ref.[\KSTY]
the anomalies cancel out.

We inferred the boundary conditions for various 7D SYM fields 
(living on the \FP6 fixed planes) from kinematic considerations but did not 
say a word about their dynamical origins, much as Ho\v rava and Witten 
argued that M9 boundary branes of \M~theory {\sl must} carry $E_8$ SYM fields
but did not explain {\sl how} such fields actually arise in \M~theory 
compactified on $\IS^1/\IZ_2$. 
In particular, we did not explain how the two \I\ ends of the same 
\FP6\ fixed plane give rise to different boundary conditions and 
why only $\I_2$ has local 6D hypermultiplets. 

In this paper we give a dynamical explanation of all the boundary
conditions and local fields proposed in [\KSTY].
Our main idea is to map the \FP6 orbifold planes in \M~theory 
onto coincident KK magnetic monopoles [\Townsend]
and hence to coincident 
D6 branes in the type IIA superstring.
Consequently, the Ho\v rava--Witten (HW) theory maps onto the \tI\
superstring theory and each M9 boundary brane becomes an \Ori\ orientifold
plane accompanied by eight D8 branes. The \I\ intersection planes 
of HW \M~theory become brane junctions of several distinct types,
hence diverse boundary conditions and local 6D fields at different 
junctions. For example, $N$ D6 branes ending on an \Ori\ plane 
give rise to local half-hypermultiplets in the bi-fundamental 
representation of $G^{(2)}=SO(2k)$ and $G_7=SU(N)$ broken to  
$Sp(N/2)$. On the other hand, D6 branes ending on D8 branes 
in a one-on-one fashion give rise to locking boundary conditions for the 
gauge fields and consequently mixing of the relevant symmetries. 
All of this is explained in detail in section~4. 

The rest of this paper is organized as follows:
Section~2 is a summary of our previous work [\KSTY]. 
We explain the kinematics of HW duals of heterotic orbifolds 
and provide rules and formulae for checking local anomaly cancellation
and correctness of 6D gauge couplings. We also summarize the specific
models used later in this paper.

Section~3 is a review of the $\rm HW\leftrightarrow I'$ duality in $d=9$,
which is also relevant to the untwisted sectors of the heterotic orbifolds. 
We learn how to build \tI\ duals of $G^{(1)}\times G^{(2)}$ 
gauge groups of various 
orbifold models, including the $E_n$ group factors which take us 
beyond the perturbative \tI\ regime and recall the r\^ole played by 
half D0 branes stuck to an orientifold plane. 
We also discuss the special case of $E_0$ factors.  

We return to the twisted sectors in section~4 where we introduce 
the D6 branes and explain the fundamentals of brane engineering 
the \tI\ duals of HW orbifolds. As an example, we engineer the duals 
of the $\IZ_2$ orbifold depicted in fig.~\ZTWO\ and show how the 
boundary conditions and local fields proposed in [\KSTY] arise dynamically 
from the \tI\ superstring theory. 

In section~5 we consider $\IZ_3$  and $\IZ_4$  orbifolds where the 
7d/10D gauge symmetry mixing involves a proper subgroup of the 
non-perturbative 7D symmetry $G_7$.
We use brane engineering to derive rather complicated boundary 
conditions for various 8--SUSY hyper and vector multiplet components 
$H_{\rm 7D}$ and $V_{\rm 7D}$ of all the 7D, 16--SUSY $SU(N)$ 
vector multiplets. 
We also find localized 6D massless states (at both intersection planes) 
whose local quantum numbers eventually map onto those of the heterotic twisted 
states, --- but the mapping is way too complicated to find without the benefit
of a dual \tI/D6 brane model.

Section~6 adds NS5 half-branes stuck at \Ori\ planes to our brane 
engineering tool kit. $N$ D6 branes ending on such an NS5 half-brane
give rise to 6D hypermultiplets in a $\asymrep$ tensor representation 
of the $SU(N)$ gauge symmetry [\HanZaf]. 
Many heterotic orbifolds have twisted 
states with such quantum numbers and we give a simple $\IZ_6$ example. 

In section~7 we reverse the flow of the $\rm heterotic\leftrightarrow
HW\leftrightarrow I'$ duality and use HW orbifolds to infer the 
physics of strongly coupled brane junctions. We find that it 
takes precisely four D6 branes ending on a $\lambda=\infty$ 
\Ori\ plane (carrying an extended $E_1$ symmetry) to somehow 
pin down a zero-tension NS5 half-brane to the junction; consequently:
the local symmetry at the junction is $SU(4)\times(E_1=SU(2))$ and the 
local 6D hypermultiplets comprise $\half({\bf 6},{\bf 2})$.
We also consider $N$ D6 branes ending on ($\lambda=\infty$, $\rm charge =-9$)
$\bf O8^*$ [\Seiberg,\SM]
planes and find that somehow such junctions require 
$N\equiv 0 \bmod 3$. 

Section~8 gives a brief summary of our results and open problems. 

Finally, in the Appendix we consider the 6D gauge couplings 
and the local anomaly cancellation in the new models discussed in 
sections~ 6 and 7.

%auto-ignore
% chapter2.tex
%% Chapter 2 of Orbifold2 paper
%% Written by Jacob Sonnenschein and Stefan Theisen

\chapter{Heterotic vs. the  Ho\v{r}ava--Witten \M~theory -- a review}

We summarize the main points of ref.[\KSTY] where we 
have studied the duality between ${\cal N}=1$ supersymmetric 
heterotic string compactifications on 
$\IR^{5,1}\otimes(T^4/\IZ_N)$ and the Ho\v{r}ava--Witten  \M~theory on
$\IR^{5,1}\otimes (T^4/\IZ_N)\otimes(\IS^1/\IZ_2)$. The condition 
of having a supersymmetric compactification restricts 
$N\in\lbrace2,3,4,6\rbrace$.
Some of the models 
treated in [\KSTY] will be reexamined in later sections and we provide 
the necessary data here. Also, we collect those results of [\KSTY]
which are needed to check consistency of the construction: 
the correct heterotic gauge couplings and local anomaly cancellation.

The construction of six-dimensional heterotic orbifolds was reviewed 
\eg\ in [\AFIQ]. It involves the specification of a shift vector 
$\delta$ which realizes the embedding of the $\IZ_N$ twist 
on the gauge degrees of freedom. 
The compactification breaks the gauge group to $G^{(1,2)}\subset E_8^{(1,2)}$
where $G^{(1,2)}$ depends on the choice of $\delta$. 
All massless 
states in the untwisted sector are charged either under $G^{(1)}$ or 
under $G^{(2)}$ and they live on the corresponding M9. Generically, in the 
twisted sectors there are states which are charged under both $G^{(1)}$ 
and $G^{(2)}$. 
As a specific example consider the $\IZ_2$ orbifold 
with shift vector 
$\delta=(\delta_1;\delta_2)=({1\over2},{1\over2},0,0,0,0,0,0;1,0,0,0,0,0,0,0)$
and gauge group $G^{(1)}\times G^{(2)}=(E_7\times SU(2))\times SO(16)$.
The massless matter in the untwisted sector consists of hypermultiplets
transforming as $({\bf 56},{\bf 2};{\bf 1})$ and $({\bf 1},{\bf 1};{\bf 128})$
and four neutral moduli hypermultiplets. In the twisted sector there 
are sixteen half-hypermultiplets --- one at each $\IZ_2$ fixed point ---
transforming as $({\bf 1},{\bf 2};{\bf 16})$.

The basic set-up of the dual \M~theory includes the following ingredients.  
We denote by $x^{0,\dots,5}$ the coordinates of $\IR^{5,1}$,
$x^6\in[0,\pi R_{11}]$ is the coordinate along the 
\M~theory interval $\IS^1/\IZ_2$
and $x^{7,\dots,{10}}$ are the coordinates on $T^4/\IZ_N$.
There is one M9 brane at each end of the interval and an orbifold
fixed plane, denoted by $\FP6$, for each of the fixed points of the 
$\IZ_N$ action on $T^4$. 
\foot{Recall that 
a $\IZ_2 (\IZ_3)$ orbifold has 16 (9) $\IZ_2 (\IZ_3)$ fixed points. 
A $\IZ_4$ orbifold has four $\IZ_4$ fixed points and six $\IZ_2$
points. Finally, a $\IZ_6$ orbifold has (1,4,5) ($\IZ_6,\IZ_3,\IZ_2$)
fixed points.}   
They intersect each M9 in an $\I$, \ie\ $\I={\rm M9}\cap\FP6$.
\foot{If we want to distinguish the 
two ends of the interval at $x^6=0$ and $x^6=\pi R_{11}$ 
we use sub- or superscripts `1' and `2', \eg\ M9$_1$, \etc.}
Compactification on $\IS^1/\IZ_2$ leads to the Ho\v rava--Witten theory 
with an $E_8$ factor on each end-of-the-world M9. Compactifying further 
on $T^4/\IZ_N$ breaks the gauge group, as in the heterotic case, with  
$G^{(1)}$ confined to ${\rm M9}_{1}$ and $G^{(2)}$ to ${\rm M9}_2$. The charged
matter corresponding to the untwisted sector of the dual heterotic theory 
is localized either on ${\rm M9}_1$ or on ${\rm M9}_2$, 
depending on whether they 
are charged under $G^{(1)}$ or $G^{(2)}$. The twisted matter states 
are localized on the fixed planes $\FP6$ and, since they carry charge, 
on the end-on-the-world nine-branes, i.e. on the \I's.  
There seems to be no way to localize those states which are charged
under both $G^{(1)}$ and $G^{(2)}$. 

The solution of the puzzle involves non-perturbative gauge fields 
which are localized on the \FP6\ planes.  
On a $\FP6$ plane corresponding to a $A_{n-1}$ singularity 
one has the gauge group $SU(n)^{\rm np}$. The states corresponding 
to the Cartan generators originate from the 
\M~theory three-form $C$ and those corresponding to the roots from 
M2 branes wrapping vanishing cycles. 
Supersymmetry requires that these states are components of  
7D vector multiplets. The presence of boundaries complicates 
the situation. In particular the boundary conditions of the 
7D fields at the ends of the interval have to be specified.  
Under $\IZ_2: \IS^1\to \IS^1/\IZ_2$ eight supercharges are even and 
eight are odd and hence supersymmetry is broken 16--SUSY$\to$8--SUSY. 
The 7D 16--SUSY vector multiplets decompose into a 6D 8--SUSY 
vector+hypermultiplet ($V_{\rm 7D}$ and $H_{\rm 7D}$)
with opposite --- free {\it vs.} fixed --- boundary conditions. 
Another constraint is that in the heterotic picture, i.e. when we 
collapse the interval to a point, there should be only the perturbative 
heterotic states. This means \eg\ for the $\IZ_2$ model that the
7D fields do not have massless zero-mods in 6D, i.e. neither 
$V_{\rm 7D}$ nor $H_{\rm 7D}$ has Neumann boundary conditions on both ends.
\foot{In section~5 we consider models where some  
$H_{\rm 7D}$ components have zero modes. Nevertheless, the condition that in 
the heterotic limit there are no additional states must and will be satisfied.}

Clearly, in the $\IZ_2$ 
example, since there are no non-perturbative $SO(16)$ gauge fields
at our disposal, we have to place one half-hypermultiplet 
of the twisted sector on each $\I_2$. The $SU(2)$ charge it carries 
must be that of $SU(2)^{\rm np}$ and it transforms as 
$({\bf16},{\bf2})$ of 
$SO(16)^{(\rm p)}|_{{\rm M9}_2}\times SU(2)^{(\rm np)}|_{\FP6}$. 
$V_{\rm 7D}$ must have Neumann boundary conditions at $\I_2$
while at $\I_1$ 
the fields in $V_{\rm 7D}$ `lock' to the $SU(2)$ fields on ${\rm M9}_1$, \ie\
they satisfy
$$
A_\mu(x^6=0)=A_\mu^{\rm 10D}(x^{7,8,9,10}=0)
~~\hbox{for same}~x^{0,\dots,5},\,\mu=0,\dots,5
\eqno\HWLockingBC
$$
and likewise for the gauginos in the 6D vector multiplet. 
The  $SU(2)$ visible in the heterotic description is the
diagonal subgroup $SU(2)^{\rm het}={\rm diag}[SU(2)^{\rm p}
\times\left(SU(2)^{\rm np}\right){}^{\!\!16}]$.
The boundary conditions of $H_{\rm 7D}$ are opposite to those of $V_{\rm 7D}$:
Neumann on $\I_1$
and Dirichlet on $\I_2$. At any given \I\ only those 7D fields 
contribute to the massless spectrum which satisfy Neumann 
boundary conditions there.  
The main burden of the analysis of any given model is 
to determine the correct massless spectrum at the \I s. We will see
that this is a highly non-trivial problem in all but the simplest models. 
The situation for the $\IZ_2$ model is summarized in fig. \ZTWO. 

Let us now give the evidence for this proposal which we have amassed
in [\KSTY] and which we had verified for several other models. In addition
to reproducing the correct perturbative heterotic spectrum, we showed 
that the correct heterotic gauge coupling, which can be computed exactly, 
could be derived from the dual \M~theory and also that the anomalies on 
each \I\ cancel locally. Both checks rely heavily on the set-up 
and will now be summarized in turn. 

We start with the $\underline{\hbox{6d gauge couplings.}}$
The gauge kinetic energy of the six-dimensional
low-energy effective ${\cal N}=1$
SYM theory is, in string frame, [\DMW]
$$
{\cal L}\sim{1\over\alpha'}\sum_{\alpha}(v_\alpha e^{-\phi}
+\tilde v_\alpha)\tr F_\alpha^2\,.
\eqn\gaugekin 
$$
Here $\phi$ is the heterotic dilaton, 
$e^{-\phi}={{\rm Vol(K3)}\over \lambda_{\rm het}^2\alpha'^2}$,
and $\lambda_{\rm het}$ the heterotic string coupling constant.
The sum is over all gauge group factors.
$v$ and $\tilde v$ are 
dimensionless constants.
{}For perturbative non-abelian gauge groups, $v=1$ --- it is, in fact, the
level of the Kac-Moody algebra --- and $\tilde v$ arises at one loop.
For non-perturbative gauge groups, on the other hand, $v=0$
and $\tilde v$ is fixed at tree level.  

The coefficients $v$ and $\tilde v$ are related, via supersymmetry,
to the coefficients of the anomaly polynomial which must factorize
to allow a Green--Schwarz mechanism to cancel the anomaly.
Factorizability of the anomaly polynomial in the form\foot{Here we 
have already used the necessary condition 
$n_H-n_v=244$ for the GS mechanism to work.}
$$
\eqalign{
{\cal A}\ &
\equiv\ \coeff23\Tr_{H-V}({\cal F}^4)\ 
-\ \coeff16\tr( R^2)\times\Tr_{H-V}({\cal F}^2)\
+\ \left(\tr (R^2)\right)^2\cr
&=\ \biggl(\sum_i v_i\tr(F_i^2)\,-\,\tr(R^2)\biggr)\times
\biggl(\sum_i \tilde v_i\tr(F_i^2)\,-\,\tr(R^2)\biggr).\cr
}\eqn\NetAnomalyFactorization
$$
imposes the constraint
$$
b_\alpha=6(v_\alpha+\tilde v_\alpha),\eqn\bv
$$
where $b_\alpha$ is the coefficient of the one-loop beta-function of the
$d=4,~{\cal N}=2$ SYM theory that one obtains upon further compactification
on $T^2$.

The M9 branes carry magnetic charges under $k_{1,2}=n_{1,2}-12$
where $n_{1,2}$ are the instanton numbers which satisfy 
$n_1+n_2=24$. In the orbifold limit the instantons are located at 
the fixed points and can have fractional instanton number. 
The relation between $k$ and $n$ follows from the 
Bianchi identity of the field strength of $C$.  
If one integrates the anomaly polynomial of the heterotic theory 
over a {\it smooth} K3 one derives $\tilde v_{1,2}={1\over2}k_{1,2}$, 
\ie\ $\tilde v_1+\tilde v_2=0$. In these compactifications there are 
no non-perturbative gauge fields and we thus conclude
that $\tilde v^{\rm p}={k\over 2}$.

This does, however, not hold for 
the compactification on K3 {\it orbifolds}. In fact, for the $\IZ_2$ 
model one finds $k_1=-4$ and $\tilde v(E_7)=-2,\,
\tilde v(SO(16))=2$ but $\tilde v(SU(2))=-2+16$. 
The result for the $SU(2)$ factor indicates that 
it is indeed the diagonal subgroup of the perturbative $SU(2)$ and 
16 non-perturbative $SU(2)$'s on the \FP6's. 
Generally, for those group factors
which have a perturbative and a nonperturbative component, 
$\tilde v=\tilde v^{\rm p}+\tilde v^{\rm np}$.

Even though the non-perturbative fields do not contribute 
additional degrees of freedom in the heterotic limit, 
they effect the heterotic gauge coupling via
$$
{1\over g^2_{\rm het}}={1\over g^2_{\rm M9}}+\sum_i {1\over g^2_{\FP6}}\,.
\eqn\gaugecoup
$$
The sum is over all those non-perturbative gauge groups which mix 
with the perturbative gauge group on M9. 
Combining this with 
${1\over g^2_{\rm M9}}={1\over\alpha'}
\left({{\rm vol(K3)}\over\lambda_{\rm het}^2\alpha'^2}v
+\tilde v^{\rm p}\right)$ 
and $\tilde v^{\rm p}={k\over2}$ we find for any factor 
$G\subset E_8$ of the heterotic gauge group which mixes with a 
non-perturbative gauge group $G\subset SU(n)$ located at a 
$\IZ_n$ fixed plane, 
$$
{1\over g^2_{G}}={v\over g^2_{E_8}}
+{\tilde v^{\rm np}\over g^2_{SU(n)}}+({\rm 1-loop}),
\eqn\GaugeCouplings
$$
where the one-loop contribution is 
${\tilde v^{\rm p}\over\alpha'}={k\over2\alpha'}$.
Later we will compute ${1\over g_G^2}$ and 
use eq.\GaugeCouplings\ to determine $\tilde v^{\rm np}$ for 
various models. The result depends on the details of the 
mixing of perturbative and non-perturbative gauge groups
which will turn out to be highly non-trivial in the presence 
of $U(1)$ factors. These values have to agree with those 
derived from \NetAnomalyFactorization.

The second consistency check is 
$\underline{\hbox{local anomaly cancellation}}$ 
on each \FP6. 
In the \M~theory description of the heterotic orbifold we have allocated all
massless fields (perturbative and non-perturbative)
to the bulk (gravity and moduli) and the
various types of planes (M9, \FP6\ or \I) which are present. 
The anomalies have to cancel locally, {\it i.e.}
on any such plane separately. In the bulk and on the \FP6\ this is
automatic, they are odd-dimensional. On each of the two M9 branes,
away from the intersection planes \I,
there are 16 supercharges and an entire $E_8$ gauge group.
Anomaly cancellation works in exactly the
same way as in the Ho\v{r}ava--Witten theory.
The situation on the intersection planes $\I$, however, involves
new features: here
supersymmetry is broken further to eight super-charges and the gauge group
is broken to a subgroup. 
The anomaly on the intersection
planes  gets contributions from three sources: the quantum, 
inflow and intersection contributions. 
The total anomaly polynomial is 
$$
{\cal A}={\cal A}({\rm Quantum})+{\cal A}({\rm inflow})
+{\cal A}({\rm intersection})\,.
\eqn\TotalAnomaly
$$

\noindent
{\it Quantum contributions}: they arise from the massless states which are 
charged under the gauge group $G^{\rm 6D}_{\rm local}$ 
operating at a particular 
$\I$. Fields residing in the bulk, on the M9 planes,  
on the $\FP6$ plane which is bounded by the $\I$ plane and
the fields confined to $\I$ contribute.
We will denote the multiplet content of the charged M9 fields which 
contribute to the anomaly by $Q_{10}$. This splits in hypermultiplets and 
vector multiplets which contribute with opposite signs. Likewise 
we introduce the notation $Q_7$ and $Q_6$ for the charged fields 
on \FP6 and \I\ which contribute to the anomaly. We also use
$Q=Q_{10}+Q_7+Q_6$. 
The net number of fields is denoted by ${\rm dim}(Q)$.
$Q_7$ gets contributions from $H_{\rm 7D}$ and $V_{\rm 7D}$.
Since the boundary conditions are local and are not communicated  
across the interval, the contributions of the 7D fields 
to the anomaly have to be distributed
\`a priori over the two \I\ boundaries of \FP6.
However, at each end only 
the components with Neumann boundary conditions do actually 
contribute (but with a factor $\half$).
$Q_6$ consists of all the fields which are 
localized on the \I\ plane. To determine $Q_{10}$ we have 
to distribute the M9 fields over all \I s on the same side
of the interval. For non-prime orbifolds one has to be careful.
{\it E.g.} for a $\IZ_4$ orbifold we first have to subtract the 
contribution form the 6 $\IZ_2$ fixed points and then distribute the 
rest over the four $\IZ_4$ fixed points. $Q_{10}$ can be succinctly written 
as follows [\KSTY]: denote by $\alpha$ the $\IZ_N$ generator whose action 
on $E_8$ is realized by the shift vector $\delta$. Then for an $\IZ_N$ 
plane $Q_{10}=-T(\alpha)({\bf 248})$ 
where 
$$
T({x})=\cases{{x\over16},& $N=2\,,$\cr
                  {x\over9}, & $N=3\,,$\cr
                  {x\over8}+{x^2\over32}, & $N=4\,,$\cr
                  {x\over6}+{x^2\over18}+{x^3\over 48}, & $N=6\,.$\cr}
\eqn\talpha
$$
For the $\IZ_2$ model of fig.\ZTWO\ we have 

\noindent
$G^{\rm local}_1=E_7^{\rm 10D}\times SU(2)^{\rm diag},\,
G^{\rm local}_2=SO(16)^{\rm 10D}\times SU(2)^{\rm 7D},$

\noindent
$\alpha_1({\bf 248})=\bf{(133,1)+(1,3)-(56,2)}$ ~and~ 
$\alpha_2({\bf 248})=\bf{(120)-(128)}$. 

\noindent
The local charged spectra are
\noindent
\settabs 4\columns
\+ $Q_6^{(1)}=\emptyset,$&
$Q_7^{(2)}=-{1\over2}\bf{(1,3)},$&
$Q_{10}^{(1)}={1\over16}[\bf{(56,2)-(133,1)-(1,3)}],$&&\cr
\+$Q_6^{(2)}={1\over2}\bf{(16,2)},$&
$Q_7^{(1)}={1\over2}\bf{(1,3)},$&
$Q_{10}^{(2)}={1\over16}[\bf{(128,1)-(120,1)}]$&&\cr
\noindent
from which  
${\rm dim}(Q)_{1}=0$ and ${\rm dim}(Q)_2=15$ follows.

The last contribution to the quantum anomaly comes from the bulk fields. 
They have to be distributed over all fixed planes at both ends of the 
interval. One has again to be careful for non-prime orbifolds. In 
particular for the moduli contribution one has to remember that 
$\IZ_N$ orbifolds have four moduli for $N=2$ and two otherwise. 

Combining all contributions one finally obtains the total quantum 
anomaly on an $\IZ_N$ \I\ plane of a $\IZ_N$ orbifold
$$
\eqalign{
{\cal A}({\rm quantum})&={2\over3}\Tr_Q F^4-{1\over6}\tr R^2 \Tr_Q F^2\cr
&\quad+{1\over360}\biggl({\rm dim}(Q)-122 T(1)
-2{\rm Re} T(e^{2\pi i/N})\biggr)\tr R^4\cr
&\qquad+{1\over288}\left({\rm dim}(Q)+22T(1)-2{\rm Re} T(e^{2\pi i/N})\right)
(\tr R^2)^2\,.}
\eqn\QuantumAnomaly
$$

\noindent
{\it Inflow  contributions}:
they arise from gauge variance of the
11d SUGRA action. There is a contribution from a modified Bianchi identity
and contributions arising from Chern--Simons (CS) terms.
Explicitly [\FLOI,\KSTY]
$$
{\cal A}({\rm inflow})=
-{2g\over 3}\left({1\over 8}\tr R^4-{1\over32}(\tr R^2)^2\right)
-{g\over2}\left(\tr F_{10}^2-{1\over2}\tr R^2\right)^2\,.
\eqn\inflow
$$
Here $F_{10}$ are the M9 gauge fields and $g$ is the magnetic charge 
of the \I-plane under consideration.
The charges of all \I\ planes on one side of the interval have to  
satisfy the sum rule 
$\sum g=k$. For $\IZ_N$ orbifolds 
with $N$ prime all \I\ planes on one side are equivalent and 
the magnetic charge of each of them is easily determined once $k$ is 
known. For the $\IZ_2$ model this means that $g_1=k_1/16=-1/4=-g_2$.
For $N=4,6$ one needs to take into account the magnetic charges
of the $\IZ_2$ and (for $N=6$) $\IZ_3$ fixed planes. Their combined
charge has to be subtracted from $k$ and the remainder has to be divided 
by the number of $\IZ_N$ fixed points.

\noindent
{\it Intersection contributions}:
they arise from the electric coupling of the $\FP6$ to the three form
$C$. This coupling leads to a 7D CS term on each of the $\FP6$ planes. 
One finds [\FLOI,\KSTY]
$$
{\cal A}({\rm intersection})=\left(\tr F_{10}^2-{1\over2}\tr R^2\right)
\times\left(T(1)\tr R^2-\tr F_7^2\right)\,.
\eqn\intersection
$$
Here $F_{7}$ are the $\FP6$ gauge fields operating on \I=M9$\cap$\FP6.

Anomaly cancellation requires that  
the coefficients of $\tr R^4,\, (\tr R^2)^2,\, \tr R^2$ 
and of the term with pure gauge field dependence vanish separately.
In particular, absence of the irreducible $\tr R^4$ term 
implies that 
$$
\mathop{\rm dim}(Q\equiv Q_6+Q_7+Q_{10})\ =\ 30g\
+\left\{\vcenter{\ialign{
        \hfil ${#}$\enspace & for $N=#$\cr
        15\over 2 & 2,\cr
        121\over 9 & 3,\cr
        19 & 4,\cr
        535\over 18 & 6.\cr
        }}\right.
\eqn\AnomalyRiiii
$$
Using this, the condition ${\cal A}\equiv0$ reduces to
$$
\eqalign{
{\cal A}'\ &
\equiv\ \coeff23\Tr_Q({\cal F}^4)\ -\ \coeff16\tr(R^2)\times\Tr_Q({\cal F}^2)
        +\ (\coeff18 g+\half T(1))(\tr(R^2))^2\crr
&=\ \half g\left( \tr({\cal F}_{10D}^2)\,-\,\half\tr(R^2)\right)^2\cr
&\qquad+\ \left( \tr({\cal F}_{10D}^2)\,-\,\half\tr(R^2)\right)\times
         \left( \tr({\cal F}_{7D}^2)\,-\,T(1)\tr(R^2)\right).\cr
}\eqn\AnomalyOther
$$
For all models that we will consider we check that \AnomalyRiiii\
and \AnomalyOther\ are satisfied on each \I\ plane separately.
\foot{The relation between $\tr$ and $\Tr$ in various representations
can be found \eg\ in App.~C of [\KSTY].}

In addition to the $\IZ_2$ model with gauge group 
$(E_7\times SU(2))\times SO(16)$ which has accompanied us through this 
section, we also considered a $\IZ_3$, a $\IZ_4$ and a $\IZ_6$ model 
in [\KSTY]. We will reconsider these models in view of the 
${\rm HW}\leftrightarrow\,$\tI\ duality in section~7. 
Here we collect some basic data. Further details 
can be found in [\KSTY] 

\noindent
$\underline{\IZ_3-\hbox{orbifold with gauge group }
(E_6\times SU(3))\times SU(9)}$ 

\noindent
shift vector: $\delta=(-{2\over3},{1\over3},{1\over3},0,0,0,0,0;
{5\over6},{1\over6},{1\over6},{1\over6},
{1\over6},{1\over6},{1\over6},{1\over6})$

\noindent
untwisted matter: two moduli $\oplus~ \bf{(27,3;1)}~\oplus~ \bf{(1,1;84)}$

\noindent
twisted matter: $9\times\bf{(1,3;9)}$

\noindent
$\tilde v_{E_6}=-{3\over2},\,\tilde v_{SU(3)}=-{3\over2}+9;\,
\tilde v_{SU(9)}=+{3\over2}$

\noindent
$k_1=-3=-k_2~\Longrightarrow~ g_{\I_1}=-g_{\I_2}=-{1\over3}$

\noindent
$G_1^{\rm local}=E_6^{\rm 10D}\times SU(3)^{\rm diag},\,
G_2^{\rm local}=SU(9)^{\rm 10D}\times SU(3)^{\rm 7D}$

\noindent
\settabs 4\columns
\+ $Q_{10}^{(1)}={1\over9}[\bf{(27,3)}-\bf{(78,1)}-\bf{(1,8)}],$&&
$Q_6^{(1)}=\emptyset$,&
$Q_7^{(1)}=\phantom{-}{1\over2}{\bf 8}$\cr
\+ $Q_{10}^{(2)}={1\over9}[\bf{84}-\bf{80}]$,&&
$Q_6^{(2)}=\bf{(9;3)}$,&
$Q_7^{(2)}=-{1\over2}\bf{8}$\cr

\vskip.5cm

\noindent
$\underline{\IZ_4-\hbox{orbifold with gauge group }
(SO(10)\times SU(4))\times (SU(8)\times SU(2))}$ 

\noindent
shift vector: $\delta=(-{3\over4},{1\over4},{1\over4},{1\over4},0,0,0,0;
-{7\over8},{1\over8},{1\over8},{1\over8},
{1\over8},{1\over8},{1\over8},{1\over8})$

\noindent
untwisted matter: two moduli $\oplus~ \bf{(16,4;1,1)}~\oplus~ \bf{(1,1;28,2)}$

\noindent
twisted matter:$\,4\times\!\left[{1\over2}\bf{(1\!,6;1\!,2)}
\oplus\bf{(1\!,4;8\!,1)}\right]\Big|_{\IZ_4}\!\!
\oplus~ 6\times\left[{1\over2}\bf{(1\!,6;1\!,2)}
\oplus{1\over2}\bf{(10\!,1;1\!,2)}\right]
\Big|_{\IZ_2}$ 

\noindent
$\tilde v_{SO(10)}=0,\,\tilde v_{SU(4)}=0+4;\,
\tilde v_{SU(8)}=0,\,\tilde v_{SU(2)}=0+6$

\noindent
$k_1=k_2=0~\Longrightarrow~ g_{\I_1}={1\over4}(0-6\times{1\over4})
=-{3\over8}=-g_{\I_2}$~~(for $\IZ_4$ planes) 

\noindent
$G_1^{\rm local}=SO(10)^{\rm 10D}\times SU(4)^{\rm diag},\,
G_2^{\rm local}=[SU(8)\times SU(2)]^{\rm 10D}\times SU(4)^{\rm 7D}$

\noindent
local spectrum at the $\IZ_4$ intersection planes:
$$
\eqalign{Q_{10}^{(1)}&=-{5\over32}[({\bf 45,1})+({\bf1,15})]
+{3\over32}({\bf 10,6})+{1\over16}({\bf 16,4})\qquad\qquad\qquad\qquad\cr
Q_6^{(1)}&=\phantom{-}\emptyset\,,\qquad\qquad
Q_7^{(1)}={1\over2}{\bf 15}\cr
Q_{10}^{(2)}&=-{5\over32}[({\bf 63,1})+({\bf 1,3})]
+{3\over32}({\bf70,1})+{1\over16}({\bf28,2})\cr
Q_6^{(2)}&=({\bf8,1;4})+{1\over2}({\bf1,2;6})\,,\qquad
Q_7^{(2)}=-{1\over2}{\bf 15}}
$$

\vskip.5cm

\noindent
$\underline{\IZ_6-\hbox{orbifold with gauge group }
(SU(6)\times SU(3)\times SU(2))\times SU(9)}$ 

\noindent
shift vector: $\delta=(-{5\over6},{1\over6},{1\over6},{1\over6},
{1\over6},{1\over6},0,0;
-{5\over6},{1\over6},{1\over6},{1\over6},
{1\over6},{1\over6},{1\over6},{1\over6})$

\noindent
untwisted matter: two moduli $\oplus~ \bf{(6,3;2,1)}$

\noindent
twisted matter: $\left[({\bf6,1,1;\bar{9}})
\oplus{1\over2}({\bf20,1,1;1})\right]
\Bigl|_{\IZ_6}
\oplus~4\times({\bf1,3,1;9})]\Big|_{\IZ_3}\hfill\break
\phantom{\hbox{twisted matter}}
\oplus~ 5\times\left[{1\over2}({\bf20,1,1;1})\oplus({\bf6,3,1;1})
\oplus~2\,({\bf1,1,2;1})\right]\Big|_{\IZ_2}$ 

\noindent
$\tilde v_{SU(6)}=1+1,\,\tilde v_{SU(3)}=1+4,\,\tilde v_{SU(2)}=1;\,
\tilde v_{SU(9)}=-1$

\noindent
$k_1=-k_2=2~\Longrightarrow~ g_{\I_1}=-g_{\I_2}=-{5\over12}$~~
(for $\IZ_6$ planes) 

\noindent
$G_1^{\rm local}=[SU(3)\times SU(2)]^{\rm 10D}\times SU(6)^{\rm diag},\,~
G_2^{\rm local}=SU(9)^{\rm 10D}\times SU(6)^{\rm 7D}$

\noindent
local spectrum at the $\IZ_6$ intersection planes:
$$
\eqalign{Q_{10}^{(1)}&=-{35\over144}[({\bf35,1,1})+({\bf1,8,1})
+({\bf1,1,3})]\cr
&\qquad\qquad\qquad-{5\over72}({\bf6,3,2})
+{13\over72}({\bf15,\bar{3},1})
+{19\over144}({\bf20,1,2})\qquad\cr
Q_6^{(1)}&=\phantom{-}\emptyset\,,\qquad\qquad
Q_7^{(1)}={1\over2}{\bf 35}\cr
Q_{10}^{(2)}&={13\over72}\,{\bf84}-{35\over144}\,{\bf80}\cr
Q_6^{(2)}&=({\bf\bar{9};6})+{1\over2}({\bf1;20})\,,\qquad
Q_7^{(2)}=-{1\over2}{\bf 35}}
$$

%auto-ignore
% chapter3.tex
%% References for Orbifold2 paper
%% Written by Stefan Theisen

%%%%%%%%%%%%%%%%%%%%%%%%%%%%%%%%%%%%%%%%%%%%%%%%%%%%%%%%%%%
\chapter{Explaining $E_8$:  $\rm HW\,\leftrightarrow\, I'$ Duality}

Ten-dimensional string theories are connected through a web of 
perturbative and non-perturbative dualities; for a review, see \eg\ [\PolII]. 
One striking feature
is that if one studies the strong coupling limit of the type IIA theory, 
the string coupling constant gets geometrized and parameterizes the 
size of an additional, eleventh, dimension which is topologically 
a circle whose radius grows as the type IIA coupling increases.  
The massless degrees of freedom of the type IIA theory 
combine into representations of the eleven-dimensional Lorentz-group 
and their dynamics is governed by eleven-dimensional supergravity. 
The strong coupling limit of type I string theory is 
the heterotic $SO(32)$ theory  and vice versa; they are 
S-dual to each other. The type IIB theory, 
which is of no interest for the discussions in this paper, is self-dual. 
Finally, the strongly coupled heterotic $E_8\times E_8$ theory is
also an eleven-dimensional theory, with the additional dimension 
being the interval $I\simeq \IS^2/\IZ_2$. The gauge degrees of freedom are 
confined to the two ten-dimensional boundaries, one $E_8$ factor on each. 
The original arguments for this strong coupling limit are due to 
Ho\v rava--Witten [\HWI]. They are of purely kinematical nature  
and are based on the requirement of local anomaly 
cancellation on each of the two boundary planes. 
The theory in the bulk 
is straightforward --- it is simply type IIA string theory.  
The presence of boundaries has very non-trivial effects the result of which 
is $E_8$ SYM theory confined to each of its components. 
The dynamical origin of the gauge fields, which are 
not present in the eleven-dimensional supergravity theory stayed, however,
mysterious. New insight came from the duality between the HW theory and 
the \tI{} theory, which we will now review. It provides an explanation 
for the $E_8$ gauge symmetry on the boundaries and also of its 
regular subgroups some of which occur as gauge groups in 
$T^4/\IZ_N$ orbifold compactifications. 

The results in this section are not new but we thought it worthwhile 
to collect them as they are the basis of the brane constructions 
in the following sections. We will, however, be brief and qualitative 
and refer to the cited literature for further details. 

The type IIB and type IIA string theories compactified on a circle are related 
by T-duality [\DHS,\DLP] and so are their orientifolds, called 
type I and \tI{}, respectively. The latter lives on 
$R^{8,1}\times \IS^1/\IZ_2$. There is an orientifold 
eight-plane ${\bf O8}$ of charge $-8$ 
at each end of the interval. In addition, 16 D8 branes 
are required for charge neutrality. Their positions along the interval 
are \`a priori arbitrary: they are T-dual to the 16 Wilson-line moduli
of the type I theory on $\IS^1$. Generically the gauge group is 
$U(1)^{18}$ where sixteen factors live on the world-volumes
of the sixteen D8 branes and the remaining two gauge bosons are the 
one-form $A^{\rm RR}_{\mu}$ coupling to D0 charge and 
$B^{\rm NS}_{9\mu}$ which couples to winding along $x^9$.\foot{In 
this section $x^9$ is the compact coordinate along 
$\IS^1$ and $\IS^1/\IZ_2$.  
In the previous and all later sections, where we compactify 
further on $T^4/\IZ_N$, this coordinate is called $x^6$.} 
 
Clumping branes together one can engineer 
any regular subgroup of $SO(32)$: a $U(n)$ factor when
$n$ D8 branes coincide at a position away from the boundaries\foot{If 
$n_1$ branes sit at one position and $n_2$ at another, 
the gauge symmetry is $U(1)\times SU(n_1)\times SU(n_2)$, \etc.}
and a $SO(2n)$ factor when $n$ of them 
are located at a boundary. The massless vector 
bosons (and their partners under supersymmetry)
come from open strings of zero length connecting the  
different branes and also, for branes located on one of the 
${\bf O8}$ planes, the branes and their images under the 
space-time reflection which, together with world-sheet parity 
reversal, generates the orientifold group. 
Examples of such brane configurations with $n=8,7,6$ will appear in 
sections~4,5 and 7, respectively. For instance,
the $\IZ_3$ orbifold model of \S~5.1 
requires the following symmetric brane arrangement to engineer the 
perturbative gauge group 
$({\blue SO(14)}\times{\blue U(1)})^{(1)} 
\times({\blue SO(14)}\times{\blue U(1)})^{(2)}$:
$$
\pspicture[](-60,-50)(+70,40)
\psline(-50,+34)(+50,+34)
\psline(-50,+33)(-50,+35)       \rput[b](-50,+35){$0$}
\psline(-35,+33)(-35,+35)       \rput[b](-35,+35){$b$}
\psline(+50,+33)(+50,+35)       \rput[b](+50,+35){$L$}
\psline(+35,+33)(+35,+35)       \rput[b](+35,+35){$L-b$}
\rput[b](0,+35){$x^9$}
\psline{>->}(-10,+34)(+10,+34)
\psline[linecolor=lightgray,linewidth=4.5](-50,-30)(-50,+30)
\psline[linecolor=black,linestyle=dashed,linewidth=1](-51.75,-30)(-51.75,+30)
\psline[linecolor=blue](-51.0,-30)(-51.0,+30)
\psline[linecolor=blue](-50.5,-30)(-50.5,+30)
\psline[linecolor=blue](-50.0,-30)(-50.0,+30)
\psline[linecolor=blue](-49.5,-30)(-49.5,+30)
\psline[linecolor=blue](-49.0,-30)(-49.0,+30)
\psline[linecolor=blue](-48.5,-30)(-48.5,+30)
\psline[linecolor=blue](-48.0,-30)(-48.0,+30)
\psline(-52.5,+31)(-50,+33)(-47.5,+31)
\rput[b]{90}(-53.5,+15){${\bf O8}+7{\rm\blue D8}$}
\rput[b]{90}(-53.5,-15){$\blue SO(14)$}
\psline[linecolor=blue](-35,-30)(-35,+33)
\rput[t]{90}(-33,+15){$1\rm\blue D8$}
\rput[t]{90}(-33,-15){$\blue U(1)$}
\psline[linecolor=lightgray,linewidth=4.5](+50,-30)(+50,+30)
\psline[linecolor=black,linestyle=dashed,linewidth=1](+51.75,-30)(+51.75,+30)
\psline[linecolor=blue](+51.0,-30)(+51.0,+30)
\psline[linecolor=blue](+50.5,-30)(+50.5,+30)
\psline[linecolor=blue](+50.0,-30)(+50.0,+30)
\psline[linecolor=blue](+49.5,-30)(+49.5,+30)
\psline[linecolor=blue](+49.0,-30)(+49.0,+30)
\psline[linecolor=blue](+48.5,-30)(+48.5,+30)
\psline[linecolor=blue](+48.0,-30)(+48.0,+30)
\psline(+52.5,+31)(+50,+33)(+47.5,+31)
\rput[t]{90}(+53.5,+15){${\bf O8}+7{\rm\blue D8}$}
\rput[t]{90}(+53.5,-15){$\blue SO(14)$}
\psline[linecolor=blue](+35,-30)(+35,+33)
\rput[b]{90}(+33,+15){$1\rm\blue D8$}
\rput[b]{90}(+33,-15){$\blue U(1)$}
\psline(+65,-30)(+65,+30)
\psline{>->}(+65,-10)(+65,+10)
\rput[t]{90}(+66,0){$x^{1,\dots,8}$}
\psline(-52,-32)(-43,-35)(-34,-32)
\rput[tl](-60,-38){Dual to the $\rm M9_1$}
\psline(+35,-32)(+44,-35)(+53,-32)
\rput[tr](+60,-38){Dual to the $\rm M9_2$}
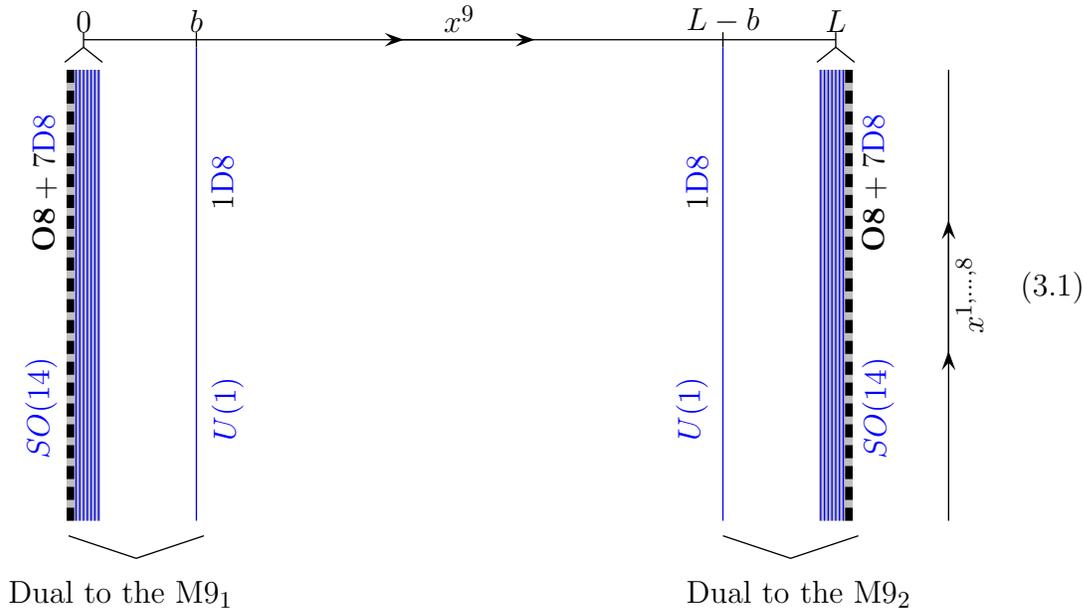
\endpspicture
\eqn\Eeight
$$
There is one feature of the \tI{} vacua that we have not yet 
addressed which is crucial for the ${\rm heterotic}\leftrightarrow$\tI{} 
duality.
In contrast to the type I dilaton, which is constant on $\IS^1$,  
the \tI{} dilaton varies across the interval. The inverse of the 
\tI{} coupling constant satisfies 
the one-dimensional Laplace equation with a source 
of unit charge at the position of each D8 brane and a source
of charge $-8$ at each end of the interval.
As a result is it a piecewise linear function whose gradient
jumps upon traversing a D8 brane. 
We will consider brane arrangements with eight D8 branes sitting 
in the vicinity of each orientifold brane, say at 
$x^9_{1,\dots,8}<b$ and $x^9_{9,\dots,16}>L-b$. 
For $b<x^9<L-b$ the dilaton is constant and if $b/L<<1$ 
it makes sense to speak about the bulk value of the dilaton and 
the \tI{} coupling constant. 

For generic D8 brane arrangements the coupling constant stays finite 
everywhere but for particular choices of their positions 
it diverges at one or both ends of the interval; 
this will play an important r\^ole below.
The dilaton is constant 
across the entire interval if and only if eight D8 branes
are located at each of the two orientifold planes, \ie\ if 
we have local charge neutrality and the (perturbative) gauge group 
$SO(16)\times SO(16)\times U(1)^2$. 

This can, however, not be the whole story for the following simple reason. 
The $\rm type~I' \dualrel{T} type~I \dualrel{S} het.[{\mit SO(32)}] 
\dualrel{T} het.[{\mit E_8\times E_8}] \dualrel{S} H$o\v{r}ava--Witten
duality chain means that we must 
be able to find \eg\ $E_8$ gauge group factors in \tI{}.
Perturbatively, neither type~I not \tI\ string theories allow exceptional
$E_n$ gauge symmetries,
hence we need a non-perturbative gauge group enhancement.  
Since the duality chain from \tI{} to het($E_8\times E_8$)  
involves an S-duality, we suspect to find $E_8$ in \tI{}  
at strong coupling. This is true, also for various subgroups 
such as $E_7\times SU(2)$ which will appear in the examples below.

A vacuum of the \tI{} theory is specified by the 
value of the dilaton $\Phi^0_{\rm I'}\equiv\Phi_{\rm I'}(x^9=0)$ 
at the orientifold plane at $x^9=0$, by the 
positions $x^9_i,\,i=1,\dots,16$ of the D8 branes and by the 
size $L$ of the interval. These parameters are mapped under the duality 
to those characterizing a vacuum of het($SO(32)$) compactified on $\IS^1$:
to the (constant) heterotic dilaton $\Phi_h$, the sixteen parameters of the 
Wilson line on $\IS^1$, 
$A=(\theta_1,\dots,\theta_{16})$ and to the radius $R_h$ of the $\IS^1$. 
The precise map of the parameter spaces follows 
by comparing the low energy effective actions
and the masses of perturbative BPS states which are related by the duality. 
This was first worked out in [\PW] and extended in 
[\KS,\DL,\MNT,\BGL,\BGS,\CV]. 
The explicit relation between the heterotic momentum and winding 
quantum numbers and the \tI{} winding and D-particle number
was established in [\BGL]. 

The \tI{} non-perturbative gauge symmetry 
enhancement, which should be mapped to the perturbative regime 
in the heterotic theory, can now be verified. 
The following qualitative discussion of a particular 
example shall illustrate this. The precise values of the parameters 
may be found in the references cited above.
Choose the D8 locations such that seven of them are 
on the ${\bf O8}$-plane at $x^9=0$, one at $x^9=b$ and, for simplicity, the 
remaining eight at $x^9=L$, \ie\ on top of the second orientifold plane. 
For this brane arrangement $\Phi_{\rm I'}$ has a constant bulk
value $\Phi_b$ for 
$b\leq x^9\leq L$ and for any such $\Phi_b$  there is a range of values  
for $b$ where the \tI{} coupling $\lambda_{\rm I'}$ 
stays small throughout the interval. 
The gauge group is $SO(14)\times U(1)\times SO(16)\times U(1)^2$.
If, however, we move the single D8 brane away from the orientifold plane 
at $x^9=0$ to a critical position 
$a(\Phi_b)$, the \tI{} coupling diverges at $x^9=0$ where the 
gauge theory on the world-volume of the D8 branes becomes strongly   
coupled. If we now use the parameter map we find that $(a,L)$  
map precisely to those values for the heterotic Wilson line and $R_h$ 
for which perturbative symmetry enhancement $SO(14)\times U(1)\to E_8$ 
occurs. \foot{Note that the position of the single D8 is frozen 
and no $U(1)$ factor is associated with it. It has combined with 
$SO(14)$ in the process of symmetry enhancement.} 
Choosing $L$ appropriately guarantees that the heterotic 
coupling is small. In the heterotic theory the additional massless 
states are BPS and carry winding numbers $\pm1$ and $\pm2$. 
They are mapped [\KS,\DL,\MNT,\BGL,\BGS] to \tI{} D-particles. 
More precisely, the heterotic states with 
winding number $\pm 1$ map to states with D-particle number $\pm 1/2$. 
Such half D-particles necessarily sit at $x^9=0$ from where they cannot 
leave because of the $\IZ_2$ orientifold symmetry. One also finds that 
they are massless precisely for $b=a$. 
The fermionic zero modes of the string connecting the 
half D-particle confined to the orientifold plane and the D8 branes 
provides the ${\bf 64}$ spinor representation  
of $SO(14)$ and the anti-D-particle the conjugate spinor representation 
${\bf 64}'$. Their $U(1)$ quantum numbers are $\pm 1/2$ respectively. 
The heterotic states with 
winding $2$ map to D-particles with particle number $1$. 
They should be viewed as
threshold bound states of two half D0 particles stuck to the orientifold
plane. Massless states only arise from the 
${\bf 14}\subset {\bf 64}\times {\bf 64}$ of $SO(14)$. 
The contribution to the massless spectrum from the ${\overline{\rm D0}}$
brane is another ${\bf 14}$. The $U(1)$ quantum numbers are $\pm 1$. 
Altogether we have thus found
those massless states which are needed for the gauge symmetry enhancement. 

In the region between $a\leq x^9\leq L $ the dilaton is constant 
and the coupling takes on its bulk value. 
Using the relations [\PW,\HWI]
$\lambda_E\sim L^{3/2}/\lambda_{\rm I'}^{1/2}$ 
and $R_E\sim\sqrt{L\lambda_{\rm I'}}$ 
where $R_E$ is the radius of the dual heterotic $E_8\times E_8$
theory measured in heterotic units $l_{\rm het}=\sqrt{\alpha'_{\rm het}}$
($L$ is measured in type I units) one sees that for $L\to\infty$
with $\lambda_{\rm I'}$ (the bulk value) fixed, $\lambda_E\to\infty$ and 
$R_E\to\infty$. Since in this limit $a/L\to 0$, 
all eight D8 branes sit at the left boundary to which the gauge degrees of
freedom are now confined. 
If we start with 
a symmetric (under $x^9\to L-x^9$) brane arrangement we get
symmetry enhancement at both ends of the interval. 
\ie{} if we place seven D8 branes on each orientifold plane,
one D8 at $x^9=a$ and one at $x^9=L-a$, we get an enhanced
$E_8\times E_8$ symmetry. 
This is the HW-theory compactified on 
$\IS^1$, \ie\ \M~theory on $R^9\times \IS^1\times(\IS^1/\IZ_2)$. 
The radius of the first $\IS^1$ is controlled by $\lambda_{\rm I'}$
via the relation $\lambda_{\rm I'}\sim R^{3/2}$.\foot{Similarly, 
starting with brane arrangements for other 
gauge groups $G^{(1)}\times G^{(2)}\subset E_8\times E_8$ 
and keeping fixed the bulk value $\lambda_{\rm I'}$, 
one finds that in the limit $L\to\infty$
all branes sit at the boundaries. Their position moduli 
have become Wilson lines on the $\IS^1$.}

In section 4 we will study in detail a $\IZ_2$ orbifold model with 
(perturbative) gauge group 
$(E_7\times SU(2))\times SO(16)$. 
The required brane configuration is as follows: 
$$
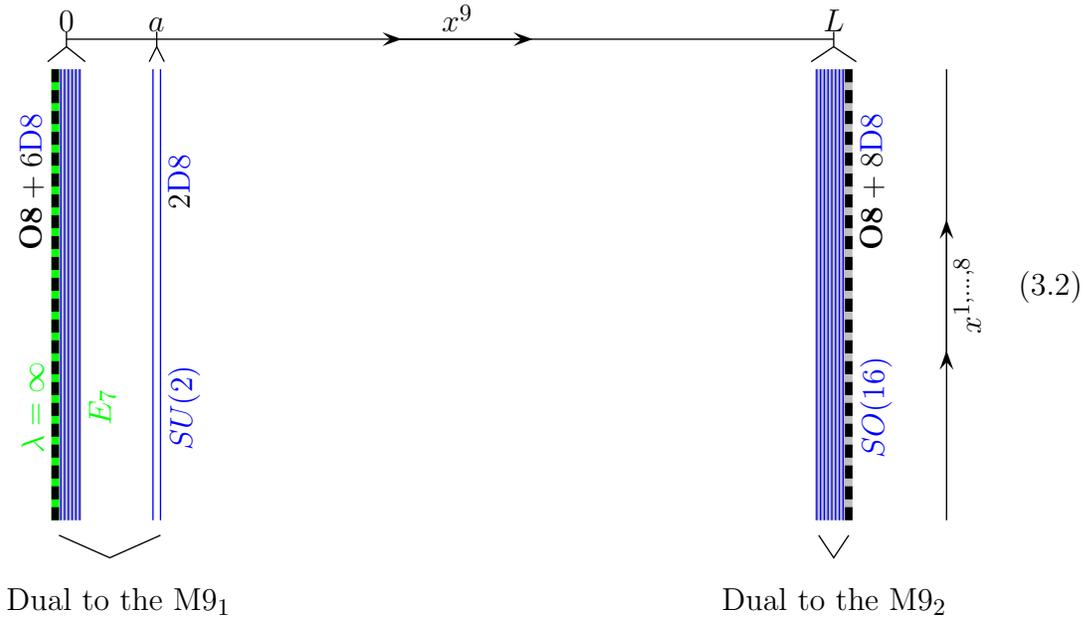
\pspicture[](-20,-43)(110,40)
\psline(-12,+34)(+90,+34)
\psline(-12,+33)(-12,+35)       \rput[b](-12,+35){$0$}
\psline(0,+33)(0,+35)           \rput[b](0,+35){$a$}
\psline(+90,+33)(+90,+35)       \rput[b](+90,+35){$L$}
\rput[b](+40,+35){$x^9$}
\psline{>->}(+30,+34)(+50,+34)
\psline[linecolor=lightgray,linewidth=4](-12,-30)(-12,+30)
\psline[linecolor=green,linewidth=1](-13.5,-30)(-13.5,+30)
\psline[linecolor=black,linestyle=dashed,linewidth=1](-13.5,-30)(-13.5,+30)
\psline[linecolor=blue](-12.75,-30)(-12.75,+30)
\psline[linecolor=blue](-12.25,-30)(-12.25,+30)
\psline[linecolor=blue](-11.75,-30)(-11.75,+30)
\psline[linecolor=blue](-11.25,-30)(-11.25,+30)
\psline[linecolor=blue](-10.75,-30)(-10.75,+30)
\psline[linecolor=blue](-10.25,-30)(-10.25,+30)
\psline(-14.5,+31)(-12,+33)(-9.5,+31)
\rput[b]{90}(-15,+15){${\bf O8}+6{\rm\blue D8}$}
\rput[t]{90}(-9,-15){$\green E_7$}
\rput[b]{90}(-15,-15){$\green\lambda=\infty$}
\psline[linecolor=blue](-0.5,-30)(-0.5,+30)
\psline[linecolor=blue](+0.5,-30)(+0.5,+30)
\psline(-1,+31)(0,+33)(+1,+31)
\rput[t]{90}(+1.5,+15){$2\rm\blue D8$}
\rput[t]{90}(+1.5,-15){$\blue SU(2)$}
\psline[linewidth=5,linecolor=lightgray](+90,-30)(+90,+30)
\psline[linecolor=black,linestyle=dashed,linewidth=1](+92,-30)(+92,+30)
\psline[linecolor=blue](+91.25,-30)(+91.25,+30)
\psline[linecolor=blue](+90.75,-30)(+90.75,+30)
\psline[linecolor=blue](+90.25,-30)(+90.25,+30)
\psline[linecolor=blue](+89.75,-30)(+89.75,+30)
\psline[linecolor=blue](+89.25,-30)(+89.25,+30)
\psline[linecolor=blue](+88.75,-30)(+88.75,+30)
\psline[linecolor=blue](+88.25,-30)(+88.25,+30)
\psline[linecolor=blue](+87.75,-30)(+87.75,+30)
\psline(+93,+31)(+90,+33)(+87,+31)
\rput[t]{90}(+93.5,+15){${\bf O8}+8{\rm\blue D8}$}
\rput[t]{90}(+93.5,-15){$\blue SO(16)$}
\psline(+105,-30)(+105,+30)
\psline{>->}(+105,-10)(+105,+10)
\rput[t]{90}(+106,0){$x^{1,\dots,8}$}
\psline(-13,-32)(-6.25,-35)(0.5,-32)
\rput[tl](-20,-39){Dual to the $\rm M9_1$}
\psline(+88,-32)(+90,-35)(+92,-32)
\rput[tr](+105,-39){Dual to the $\rm M9_2$}
\endpspicture
\eqn\Eseven
$$
The $\blue{SO(16)}$ factor is straightforward: 
place eight {\blue D8} branes at 
$x^9=L$ with $\Phi_{\rm I'}(L)$ finite. 
As long as the coupling stays finite everywhere, the branes on the 
l.h.s. of the interval give $\blue{SO(12)\times U(2)}$.
To get $\green{E_7}$ we need 
to adjust the positions of the two `outlier' {\blue D8} branes 
such that the coupling becomes infinite at $x^9=0$. The 
fact that two {\blue D8} branes must be placed at a critical 
distance $a$ means that their center-of-mass motion is frozen 
and the gauge group on their world-volume is $\blue{SU(2)}$ rather than $U(2)$.
With the help of the map between \tI{} and heterotic parameters we can 
again verify that this brane arrangement maps to the critical 
radius and Wilson line on the heterotic side where states 
with winding numbers $\pm1$ and $\pm2$ become massless. 
Again this corresponds to a half D0 brane stuck on ${\bf O8}$ whose 
fermionic zero modes provide the ${\bf 32}$ and ${\bf 32}'$ 
of $\blue{SO(12)}$ with $U(1)$ charges $\pm1/2$
and the ${\bf 1}$ component of D0-D0 and 
$\overline{\rm D0}$-$\overline{\rm D0}$ threshold bound states 
with $U(1)$ charges $\pm1$ [\BGL,\BGS]. These states, together with 
those from the 8-8 strings, provide the adjoint representation 
of $\green{E_7}$. 

The generalization to gauge groups $E_{n}\times SU(9-n)$
is straightforward. They involve the
$SO(2n-2)\times U(1)\to E_{n}$ enhancement.
For $0<n<7$ only winding states with winding number $\pm1$ become 
massless in the heterotic dual and consequently only the 
fermionic zero modes of half D0 and $\overline{\rm D0}$ branes 
contribute. They provide the two spinor representations of
$SO(2n-2)$ which in all cases are sufficient to complete the adjoint 
of $E_n$. 

At this point it seems appropriate to comment on the mixing of 
$U(1)$ factors [\BGS]. 
As already said, in addition to the 
perturbative open string gauge group there are  two $U(1)$ factors
with gauge bosons $A^{\rm RR}_\mu$ and $B^{\rm NS}_{9\mu}$ originating from 
the bulk fields of the \tI{} theory. As long as the dilaton is constant 
across the whole interval, these do not mix with the open string gauge 
group. However, as we move D8 branes into the interval, mixing sets in. 
{\it E.g.} the $U(1)$ group associated with a single D8 brane outside an 
orientifold plane with $SO(14)$ gauge group mixes with the 
two additional $U(1)$ factors where the mixing depends of the distance $a$ 
from the orientifold plane.  
The $U(1)$ which is involved in the 
$SO(14)\times U(1)\to E_8$ symmetry enhancement is the $U(1)$ 
that one gets at $a$.  
The mixing is well understood 
in the dual heterotic theory where it is caused by 
switching on Wilson lines. So, in principle, one 
should be able to push it through the duality chain to the \tI{} theory. 
However we have not attempted to done so. 

In \S~7.3. we will 
reconsider the $T^4/\IZ_3$ model of ref.~[\KSTY] where the $E_8^{(1)}$
is broken down to $E_6\times SU(3)$ and the $E_8^{(2)}$ down to 
$SU(9)$.
Given the brane engineering tools at hand the first factor is 
straightforward to obtain in 
\tI{}, but we need to introduce a new tool  
in order to explain the $SU(9)$ factor. 
Clearly, we can never get 
an $SU(9)$ group with eight D8 branes only; \cf\ however the discussion 
in [\BGS]. 
To understand how 
it can arise, we need
to elaborate further on the properties of   
the \tI{} theory at infinite coupling. 
In refs.[\CV,\Seiberg,\SM,\DKV] three different arguments are 
exploited:
(i) ${\rm heterotic}\leftrightarrow$\tI{} duality;
(ii) the world-volume theory on a $D_4$ probe in the background of 
D8 branes and \Ori\ planes in \tI; 
(iii) \M~theory compactification of Calabi-Yau threefolds.
Here we will restrict ourselves to a review of (i) since it is closest 
to the spirit of this paper. 
Consider the heterotic Wilson line $A=(0^{15}, \theta_{16})$. 
For a fixed coupling constant the moduli space of the heterotic theory  
is a strip in the $(R_h, \theta_{16})$ plane bounded by 
$1\geq\theta_{16}\geq 0$ and $R_h^2\geq 2(1- \theta_{16}^2/2)$. 
At a generic point on the strip the symmetry is $SO(30)\times U(1)^3$. 
On the boundaries 
$\theta_{16}=0$ and $\theta_{16}=1$ the gauge symmetry is enhanced to 
$SO(32)\times U(1)^2$.
Along the $R^2_h$ boundary the generic symmetry is 
$SO(30)\times SU(2)\times U(1)^2$. 
However, on the lines $\theta_{16}=0,1/2,1$ the symmetry is enhanced to
$SO(32)\times SU(2)\times U(1)$, $SO(30)\times E_2\times U(1)$ 
and $SO(34)\times U(1)$, respectively. 

In the dual \tI{} description, the heterotic Wilson line corresponds to 
having 15 D8 branes at $x^9=0$ and one at a position $0\leq x^9\leq L$
which is determined by $\theta_{16}$.
In particular, $\theta_{16}=1/2$ maps to $x^9=L$. 
Due to the invariance under $\theta_{16} \rightarrow(1-\theta_{16})$ --
they lead to identical brane configurations in the perturbative 
\tI{} theory --
the domain of the heterotic moduli space which maps to the perturbative 
\tI{} theory should be restricted to 
$R_h^2\geq 2(1-\theta_{16}^2/2)$ and 
$R_h^2\geq 2(1-{1\over2}(1-\theta_{16})^2)$.  
In particular the point with 
enhanced $SO(34)\times U(1)$ gauge symmetry where  
$R_h^2=1$ is not mapped to the perturbative \tI{} theory.
In [\SM] an extension of the \tI{} description to the
non-perturbative regime was proposed whereby a map of the complete 
heterotic moduli space to a brane configuration was achieved.   
It assumes that for $\theta_{16}=1/2$ and infinite coupling at 
$x^9=L$ an additional D8 brane can be extracted 
from the orientifold plane. For $\theta_{16}>1/2$ the original and the 
new D8 branes have left the ${\bf O8}$ plane, leaving behind an orientifold 
plane of charge $-9$ which we will henceforth refer to as an ${\bf O8}^*$
plane. 
The relative distance between these two planes is controlled by 
$R_h^2$ and they coincide for $R_h^2=2(1-\theta_{16}^2/2)$. In particular 
for $\theta_{16}=1$ and $R_h^2=1$ they have both joined the other 
15 D8 branes at $x^9=0$ and the gauge symmetry is 
enhanced to $SO(34)\times U(1)$.
In the presence of an ${\bf O^*}$ brane and the additional 
D8 brane the variation of the inverse coupling constant 
across the interval will change accordingly. 

There is no gauge symmetry associated with an ${\bf O^*}$ plane. 
In fact, if we put (8+1) D8 branes at the critical distance 
from the orientifold plane we realize the $n=0$ member of the series
$E_n\times SU(9-n)$ where $E_0$ denotes the trivial symmetry group of 
the $E_n$ series. 

We are now ready to give the dual \tI{} brane arrangement which 
reproduces the perturbative gauge group of the $T^4/\IZ_3$ 
orbifold of [\KSTY]:
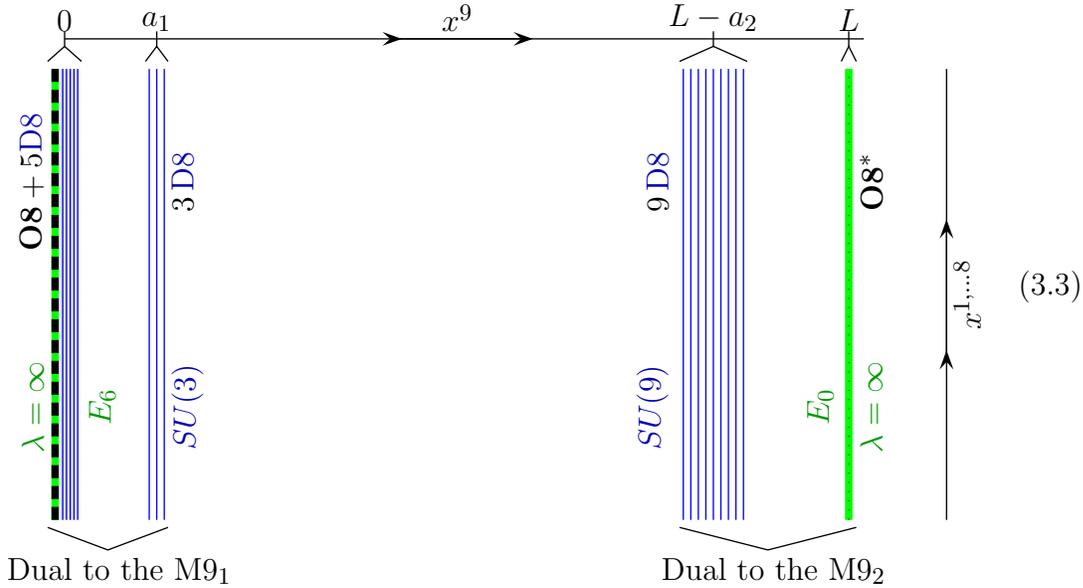
$$
\pspicture[](-20,-40)(110,40)
\psline(-12.25,+34)(+94,+34)
\rput[b](+40,+35){$x^9$}
\psline{>->}(+30,+34)(+50,+34)
\psline(-12.25,+33)(-12.25,+35)	\rput[b](-12.25,+35){$0$}
\psline(-14.5,+31)(-12.25,+33)(-10.0,+31)
\psline(0,+33)(0,+35)		\rput[b](0,+35){$a_1$}
\psline(-1.5,+31)(0,+33)(+1.5,+31)
\psline(+74,+33)(+74,+35)	\rput[b](+74,+35){$L-a_2$}
\psline(+78.5,+31)(+74,+33)(+69.5,+31)
\psline(+92,+33)(+92,+35)	\rput[b](+92,+35){$L$}
\psline(+93,+31)(+92,+33)(+91,+31)
\psline[linecolor=green,linewidth=1](-13.5,-30)(-13.5,+30)
\psline[linecolor=black,linestyle=dashed,linewidth=1](-13.5,-30)(-13.5,+30)
\psline[linecolor=palegray,linewidth=2.5](-11.75,-30)(-11.75,+30)
\psline[linecolor=blue](-12.5,-30)(-12.5,+30)
\psline[linecolor=blue](-12.0,-30)(-12.0,+30)
\psline[linecolor=blue](-11.5,-30)(-11.5,+30)
\psline[linecolor=blue](-11.0,-30)(-11.0,+30)
\psline[linecolor=blue](-10.5,-30)(-10.5,+30)
\rput[b]{90}(-15,+15){${\bf O8}+5{\rm\darkblue D8}$}
\rput[t]{90}(-9,-15){$\darkgreen E_6$}
\rput[b]{90}(-15,-15){$\darkgreen\lambda=\infty$}
\psline[linecolor=blue](-1.0,-30)(-1.0,+30)
\psline[linecolor=blue](-0.0,-30)(-0.0,+30)
\psline[linecolor=blue](+1.0,-30)(+1.0,+30)
\rput[t]{90}(+2,+15){$3\,\rm\darkblue D8$}
\rput[t]{90}(+2,-15){$\darkblue SU(3)$}
\psline[linecolor=blue](+78,-30)(+78,+30)
\psline[linecolor=blue](+77,-30)(+77,+30)
\psline[linecolor=blue](+76,-30)(+76,+30)
\psline[linecolor=blue](+75,-30)(+75,+30)
\psline[linecolor=blue](+74,-30)(+74,+30)
\psline[linecolor=blue](+73,-30)(+73,+30)
\psline[linecolor=blue](+72,-30)(+72,+30)
\psline[linecolor=blue](+71,-30)(+71,+30)
\psline[linecolor=blue](+70,-30)(+70,+30)
\rput[b]{90}(+68,+15){$9\,\rm\darkblue D8$}
\rput[b]{90}(+68,-15){$\darkblue SU(9)$}
\psline[linewidth=1,linecolor=green](+92,-30)(+92,+30)
\psline[linecolor=black,linestyle=dotted](+92,-30)(+92,+30)
\rput[t]{90}(+93.5,+15){$\bf O8^*$}
\rput[t]{90}(+93.5,-15){$\darkgreen\lambda=\infty$}
\rput[b]{90}(+90,-15){$\darkgreen E_0$}
\psline(-14.5,-31)(-6.5,-34)(+1.5,-31)
\rput[t](-5,-35){Dual to the $\rm M9_1$}
\psline(+69.5,-31)(+81.25,-34)(+93,-31)
\rput[t](+82,-35){Dual to the $\rm M9_2$}
\psline(+105,-30)(+105,+30)
\psline{>->}(+105,-10)(+105,+10)
\rput[t]{90}(+106,0){$x^{1,\dots 8}$}
\endpspicture
\eqn\ZiiiBranesPlanes
$$
Thanks to the $-9$ charge of this plane, we put nine rather than eight 
coincident {\darkblue D8} branes at the critical location $L-a_2$
where they carry an ${\darkblue SU(9)}$ SYM on their world-volume.

%auto-ignore
% chapter4.tex
%% Chapter 4 of Orbifold2 paper
%% Written by Vadim Kaplunovsky

%%%%%%%%%%%%%%%%%%%%%%%%%%%%%%%%%%%%%%%%%%%%%%%%%%%%%%%%%%%
\chapter{Brane Duals of the HW Orbifolds}
All massless charged particles in the {\sl untwisted} sector of
a heterotic orbifold are made out of 10D $E_8\times E_8$ SYM fields.
The dynamical origin of these fields in the Ho\v rava--Witten \M~theory
is rather \M ysterious, but in the previous section we saw an explanation
of this \M ystery in terms of the dual \tI\ superstring theory.
In this section, we address  HW \M ysteries of the {\sl twisted}
sectors of heterotic orbifolds; again, our explanation involves
$\rm HW\leftrightarrow type~I'$ duality.
\S4.1 below relates the $\FP6$ orbifold fixed planes
in the 11D bulk
of the HW brane world to the D6 branes of the superstring theory;
the end-of-the-world M9 branes are addressed in \S4.2 and the
$\FP6/\rm M9$ intersections in \S4.3--4.

%%%%%%%%%%%%%%%%%%
\section{Orbifold Planes, D6 branes and Taub--NUTs}
Massless states in  twisted sectors of a heterotic orbifold
live in the immediate vicinity of the appropriate fixed plane and don't
care for the overall geometry of the $T^4/\IZ_N$ space.
As far as these states are concerned, we may replace the toroidal orbifold
with the non-compact space $\IC^2/\IZ_N$ or any other space with a similar
orbifold singularity;
for supersymmetry's sake, this replacement space should have $SU(2)$
holonomy, but there are no other restrictions. 
For simplicity, we would like a non-compact replacement space with
a simple flat asymptotics;
in order to make contact with the \tI\ theory
 discussed in the previous section,
we choose the $\IR^3\times \IS^1$ flat asymptotics 
[\WittenTN] instead of the $\IR^4$.

The $SU(2)$ holonomy and the $\IR^3\times\IS^1$  asymptotics immediately
lead us to the multi--Taub--NUT
geometry of $N$ Kaluza--Klein magnetic monopoles,
$$
ds^2\ =\ V(\bx)d\bx^2\ +\ {\bigl(dy\,-\,{\bf A}(x)d\bx\bigr)^2\over V(\bx)}\,,
\eqn\MTN
$$
where $y\equiv y+ 2\pi R$ is a periodic coordinate of some radius $R$,
$$
\nabla\times{\bf A}(\bx)\ =\ \nabla V(\bx)\qquad {\rm and}\quad
V(\bx)\ =\ 1\ +\ \sum_{i=1}^N{(R/2)\over|\bx-\bx_i|}\,.
\eqno\eq
$$
The multi--Taub--NUT geometry is smooth when all the monopoles are located
at distinct points $\bx_i$ in the 3D space but develops a $\IC^2/\IZ_k$
orbifold singularity when $k$ monopoles become coincident.
In particular, when all $N$ monopoles sit at the same point, the multi--Taub--NUT
geometry is an orbifold of the simple Taub--NUT space,
$$
\TN_N\ =\ \TN_1/\IZ_N\qquad {\rm for}\quad
\IZ_N:\,(\bx,y)\mapsto(\bx,y+\coeff{2\pi R}{N}).
\eqn\MTNZN
$$
In the large radius limit $R\to\infty$ the local curvature of the Taub--NUT
becomes negligible and the multi--Taub--NUT geometry approximates the flat
orbifold $\TN_N\approx\IC^2/\IZ_N$.

The main idea of this section is to replace the $T^4/\IZ_N$ orbifold with
the $\TN_N$ geometry and then to make the Kaluza--Klein radius $R$
small rather than large;
this is legitimate because the massless twisted spectrum of the orbifold is
chiral and hence independent of continuous parameters such as~$R$.
In the 11D bulk of the HW brane world, 
the Kaluza--Klein compactification of the \M~theory on a circle is dual
to the type~IIA superstring theory in 10 flat spacetime dimensions;
for small $R$, the superstring is weakly coupled.
In the type~IIA context, each KK monopole of the multi--Taub--NUT geometry~\MTN\
becomes a D6~brane [Townsend] located at 
$(x^7,x^8,x^9)=\bx_i$ and spanning $x^0,\ldots,x^6$.
$N$ coincident monopoles of the singular $\TN_N$ space become $N$
coincident D6 branes;
the open strings beginning and ending on these branes 
 give rise to a $U(N)$ SYM theory in the D6 world volume.

{}From the dual \M~theory point of view, an open string connecting two distinct
D6 branes (corresponding to an off-diagonal element of the $U(N)$ matrix)
is an M2 membrane or anti-membrane wrapped around a 2--cycle of the multi--Taub--NUT
geometry whose area (and hence the particle's mass) vanishes when the two
KK monopoles coincide in space.
The diagonal matrix elements --- the open strings beginning and ending on the same
D6~brane --- give rise to the moduli multiplets associated with locations of the
corresponding D6~branes; from the \M\ point of view, they are the location
moduli~$\bx_i$ of the individual KK monopoles.
Among these moduli, the positions of the KK monopoles
{\sl relative to each other} are moduli of the
resolutions of the $\IC/\IZ_N$ orbifold singularity, but the overall
center-of-mass motion of the singularity is an artifact of the non-compactness
of the multi--Taub--NUT geometry.
This center-of-mass
motion --- responsible for the abelian $U(1)$ factor of the $U(N)$ gauge group
--- has nothing to do with the twisted sector of the orbifold.
Indeed, in the compact $T^4/\IZ_N$ theory, the moduli responsible for the
center-of-mass motions of complete un-resolved singularities belong to the
un-twisted sector of the orbifold.  

Therefore, on the type~IIA side of the duality, the $U(1)\subset U(N)$
associated with the center-of-mass motion of the whole stack of $N$ coincident
D6~branes is an artifact of replacing the $T^4/\IZ_N$ geometry with $\TN_N$
and has nothing to do with the twisted sector of the orbifold.
In our subsequent analysis of the twisted sectors and their brane duals,
we shall disregard such abelian factors of the D6 gauge groups and focus
on the non-abelian $SU(N)$ SYM fields.

%%%%%%%%%%%%%%%%%%%%%%%%%%%%%%%%%%%%%%%%%%%
\section{Taub--NUTs and Branes at the End of the World}
In the complete Ho\v rava--Witten context --- including both the 11D bulk
and the two end-of-the-world boundary M9~branes --- the KK reduction
leads to  \tI\ rather than type~IIA superstring (\cf\ section~3) and the D6
branes dual to the \FP6\ orbifold planes span the finite dimension of the
\tI~theory.
Let us dub this finite dimension $x^6$ while $(x^0,\ldots,x^5)$ denote
the 6D Minkowski space.
In the HW theory,  $(x^7,\ldots,x^{10})$ are coordinates of the $T^4/\IZ_N$
orbifold or its $\TN_N$ replacement;
in the small-radius limit of the multi--Taub--NUT geometry, we lose the $x^{10}=y$
coordinate to the KK reduction and the monopoles become D6 branes located at
$(x^7,x^8,x^9)=\bx_i=\bf 0$.
This naturally raises {\sl The Question:
\it\darkblue ``What happens to the D6~branes when they
reach the ends of the world at $x^6=0$ and $x^6=L$?}

Before we answer this question however, we must first clarify what happens
at $x^6=0,L$ away from the D6~branes.
Let us start by considering the effect of orbifolding on the end-of-the-world M9~branes
of the Ho\v rava--Witten theory.
For the flat toroidal orbifold $T^4/\IZ_N$, at generic points of either
M9 brane one has {\sl locally} unbroken $E_8$ gauge symmetry and 16 supercharges.
Only at the $\I=\rm M9\cap\FP6$ intersections 
with the fixed planes of the orbifold action
there is a {\sl local} effect: The gauge symmetry 
is broken down to the commutant of
the $\alpha:\IZ_N\mapsto E_8$ twist and SUSY down to 8 supercharges.
The multi--Taub--NUT space is not flat 
(apart from the singularity) but it flattens
out at large distances from the coincident monopoles, hence for $|\bx|\gg R$
we have effectively unbroken local $E_8$ symmetry and all 16 supercharges.
Or rather, the {\sl local} symmetry on the M9 brane is unbroken $E_8$ --- but
the KK reduction of the $x^{10}$ coordinate 
introduces Wilson lines into the picture.
Hence, the effective theory on the 9D boundary brane of the 10D brane world has
a reduced gauge symmetry.

In a generic KK compactification on a circle, 
the Wilson lines would be quite arbitrary,
but in the $\TN_N$ compactification~\MTNZN\ 
the asymptotic $S\I^1$ circle at $\bx\to\infty$
is topologically equivalent to the noncontractable 
loop around the $\IZ_N$ orbifold
singularity at $\bx=\bf0$.
As a stand-in for the $T^4/\IZ_N$ orbifold, the $\TN_N$ compactification
should have $F_{\mu\nu}=0$ outside the singularity itself,
hence topologically equivalent loops have equivalent Wilson lines.
Furthermore, the non-trivial $\IZ_N$ Wilson line around the singularity is
precisely the %$\alpha:\IZ_N\mapsto E_8$
action of the orbifolding symmetry on the $E_8$ gauge fields;
again, we should copy this action from the 
particular $T^4/\IZ_N$ heterotic orbifold
model under consideration.
The bottom line of this argument is that the Wilson line around the KK circle
should be precisely the $\alpha:\IZ_N\mapsto E_8$ 
twist of the heterotic orbifold;
this is not an inherent constraint of the 
Ho\v rava--Witten theory, but any other
choice would spoil the equivalence between the 
$T^4/\IZ_N$ and the $\TN_N$ twisted sectors.

In the dual \tI\ superstring theory [\PW] the KK Wilson 
lines manifest themselves as
brane arrangements within the $\Ori+8\darkblue\rm D8$ boundary 
stack at each end of $x^6$,
\cf\ section~3.
{}From the \tI\ point of view, such arrangements have absolutely no relation to the
D6~branes at $\bx=\bf0$ which are dual to the orbifold fixed plane.
However, in order to make use of $\rm HW\leftrightarrow I'$ duality in the orbifold
context, we must {\sl engineer} the specific brane arrangement in which the
gauge group of the 9D SYM living on each boundary is precisely the commutant of the
$\IZ_N$ action in the appropriate $E_8$ in the specific heterotic/HW orbifold 
under consideration.
For example, for the $T^4/\IZ_2$ orbifold in which $E_8^{(1)}$ 
is broken to $E_7\times SU_2$
and $E_8^{(2)}$ is broken to $SO_{16}$ we should engineer the branes to yield
$E_7\times SU_2$ near $x^6=0$ and $SO_{16}$ near $x^6=L$.
As discussed in section~3, this means the following brane layout:
$$
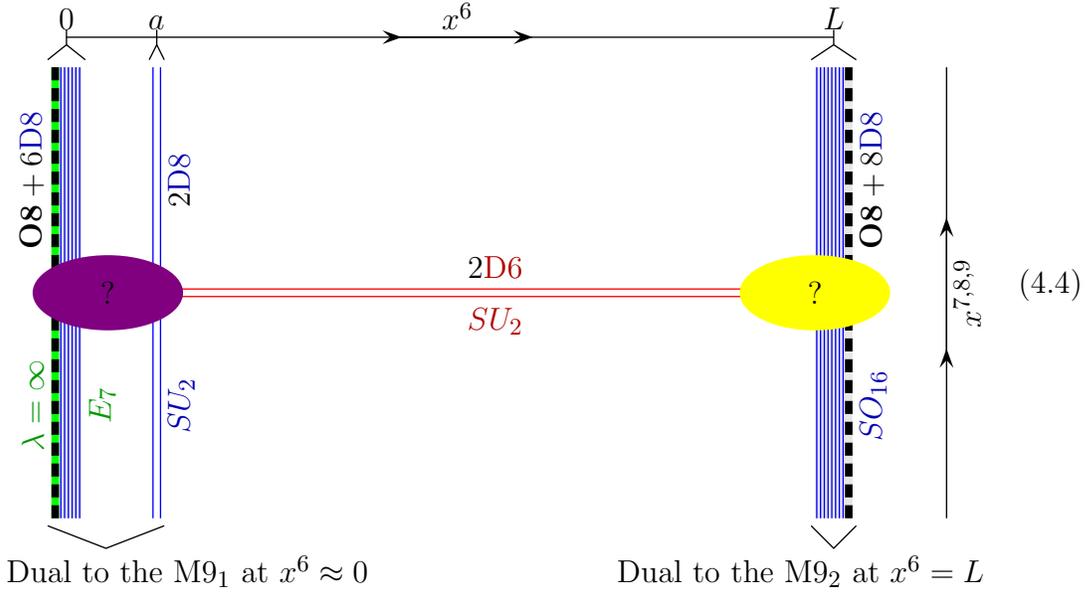
\pspicture[](-20,-40)(110,40)
\psline(-12,+34)(+90,+34)
\psline(-12,+33)(-12,+35)       \rput[b](-12,+35){$0$}
\psline(0,+33)(0,+35)           \rput[b](0,+35){$a$}
\psline(+90,+33)(+90,+35)       \rput[b](+90,+35){$L$}
\rput[b](+40,+35){$x^6$}
\psline{>->}(+30,+34)(+50,+34)
\psline[linecolor=palegray,linewidth=4](-12,-30)(-12,+30)
\psline[linecolor=green,linewidth=1](-13.5,-30)(-13.5,+30)
\psline[linecolor=black,linestyle=dashed,linewidth=1](-13.5,-30)(-13.5,+30)
\psline[linecolor=blue](-12.75,-30)(-12.75,+30)
\psline[linecolor=blue](-12.25,-30)(-12.25,+30)
\psline[linecolor=blue](-11.75,-30)(-11.75,+30)
\psline[linecolor=blue](-11.25,-30)(-11.25,+30)
\psline[linecolor=blue](-10.75,-30)(-10.75,+30)
\psline[linecolor=blue](-10.25,-30)(-10.25,+30)
\psline(-14.5,+31)(-12,+33)(-9.5,+31)
\rput[b]{90}(-15,+15){${\bf O8}+6{\rm\darkblue D8}$}
\rput[t]{90}(-9,-15){$\darkgreen E_7$}
\rput[b]{90}(-15,-15){$\darkgreen\lambda=\infty$}
\psline[linecolor=blue](-0.5,-30)(-0.5,+30)
\psline[linecolor=blue](+0.5,-30)(+0.5,+30)
\psline(-1,+31)(0,+33)(+1,+31)
\rput[t]{90}(+1.5,+15){$2\rm\darkblue D8$}
\rput[t]{90}(+1.5,-15){$\darkblue SU_2$}
\psline[linewidth=5,linecolor=palegray](+90,-30)(+90,+30)
\psline[linecolor=black,linestyle=dashed,linewidth=1](+92,-30)(+92,+30)
\psline[linecolor=blue](+91.25,-30)(+91.25,+30)
\psline[linecolor=blue](+90.75,-30)(+90.75,+30)
\psline[linecolor=blue](+90.25,-30)(+90.25,+30)
\psline[linecolor=blue](+89.75,-30)(+89.75,+30)
\psline[linecolor=blue](+89.25,-30)(+89.25,+30)
\psline[linecolor=blue](+88.75,-30)(+88.75,+30)
\psline[linecolor=blue](+88.25,-30)(+88.25,+30)
\psline[linecolor=blue](+87.75,-30)(+87.75,+30)
\psline(+93,+31)(+90,+33)(+87,+31)
\rput[t]{90}(+93.5,+15){${\bf O8}+8{\rm\darkblue D8}$}
\rput[t]{90}(+93.5,-15){$\darkblue SO_{16}$}
\psline[linecolor=red](-0.5,-0.5)(+92,-0.5)
\psline[linecolor=red](+0.5,+0.5)(+92,+0.5)
\rput[b](45,+2){$2\darkred\rm D6$}
\rput[t](45,-2){$\darkred SU_2$}
\psellipse*[linecolor=purple](-6.5,0)(10,5)
\psellipse*[linecolor=yellow](87.5,0)(10,5)
\rput(-6.5,0){?}
\rput(87.5,0){?}
\psline(-14.5,-31)(-6.75,-34)(+1,-31)
\rput[tl](-20,-35){Dual to the $\rm M9_1$ at $x^6\approx0$}
\psline(+87,-31)(+90,-34)(+93,-31)
\rput[tr](+110,-35){Dual to the $\rm M9_2$ at $x^6=L$}
\psline(+105,-30)(+105,+30)
\psline{>->}(+105,-10)(+105,+10)
\rput[t]{90}(+106,0){$x^{7,8,9}$}
\endpspicture
\eqn\ZtwoDeight
$$
where the location $a$ of the two `outlier' D8 branes at the left end of the world
is tuned such that the $x^6$--dependent string coupling~$\darkgreen\lambda$ diverges at
$x^6=0$, hence the gauge symmetry enhancement from $\darkblue SO(12)\times U(2)$
to ${\darkgreen E_7}\times{\darkblue SU(2)}$

%%%%%%%%%%%%%%%%%%%%%%%%%%%%%%%%%%%%%%%%%%%%%%%%%%%%%%%%%%%%%%%%%%%
\section{Ends of D6 Branes: The O8 Terminus}
The two elliptic blots on figure~\ZtwoDeight\ denote the two \M ysterious \I\
regions of the HW orbifold where the $\FP6$ fixed plane intersects the end-of-the-world M9s.
In the dual \tI/D6 picture, these regions contain brane junctions amenable to
string-theoretic analysis --- which will finally reveal the physical origin
of the boundary conditions for the 7D fields discovered in ref.~[\KSTY].
Let us start with the yellow junction on the right side of fig~\ZtwoDeight\ where all
8 D8 branes coincide with the \Ori~orientifold plane at $x^6=L$.

More generally, consider a junction of $N$ D6 branes terminating on an \Ori\
orientifold plane accompanied by $k$ D8 branes [\SOSP].
The orientifold plane acts like a mirror; combining the branes on figure~\ZtwoDeight\
with their reflections under $x^6\to 2L-x^6$, we have a stack of $2k$ coincident
D8 branes at $x^6=L$ crossed at the right angle with a stack of $N$ coincident
D6 branes at $\bx=\bf0$.
$$
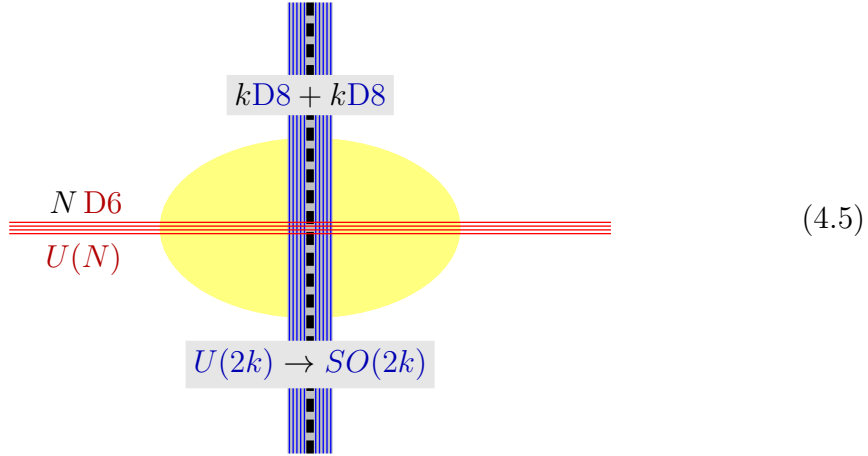
\pspicture[](-40,-30)(+40,+30)
\psellipse*[linecolor=paleyellow](0,0)(20,12)
\psline[linecolor=lightgray,linewidth=6](0,-30)(0,+30)
\psline[linecolor=black,linestyle=dashed,linewidth=1](0,-30)(0,+30)
\psline[linecolor=blue](-0.75,-30)(-0.75,+30)
\psline[linecolor=blue](-1.25,-30)(-1.25,+30)
\psline[linecolor=blue](-1.75,-30)(-1.75,+30)
\psline[linecolor=blue](-2.25,-30)(-2.25,+30)
\psline[linecolor=blue](-2.75,-30)(-2.75,+30)
\psline[linecolor=blue](+0.75,-30)(+0.75,+30)
\psline[linecolor=blue](+1.25,-30)(+1.25,+30)
\psline[linecolor=blue](+1.75,-30)(+1.75,+30)
\psline[linecolor=blue](+2.25,-30)(+2.25,+30)
\psline[linecolor=blue](+2.75,-30)(+2.75,+30)
\psline[linecolor=red](-40,+0.75)(+40,+0.75)
\psline[linecolor=red](-40,+0.25)(+40,+0.25)
\psline[linecolor=red](-40,-0.25)(+40,-0.25)
\psline[linecolor=red](-40,-0.75)(+40,-0.75)
\rput[br](-25,+2){$N\darkred\rm\,D6$}
\rput[tr](-25,-2){$\darkred U(N)$}
\rput[b](0,+15){\psframebox*[fillcolor=palegray]{$k{\rm\darkblue D8}+k{\rm\darkblue D8}$}}
\rput[t](0,-15){\psframebox*[fillcolor=palegray]{${\darkblue U(2k)}\to{\darkblue SO(2k)}$}}
\endpspicture
\eqn\YellowJunction
$$
Before the orientifold projection, the open strings connecting these D branes
produce the following massless particles:
$\darkblue U(2k)$ gauge bosons and their 9D, 16--SUSY superpartners from the {\darkblue 88} strings;
$\darkred U(N)$ gauge bosons and their 7D, 16--SUSY superpartners from the {\darkred 66} strings;
6D, 8--SUSY hypermultiplets in the bi-fundamental $({\bf\darkred N},{\bf\darkblue 2k})$
representation of the gauge group from the $\darkred6\darkblue8$ strings.
Note that only 8 out of 32 supercharges of the type~II superstring survive
at the D8--D6 brane junction.

Locally near $x^6\approx L$, the \Ori{} orientifold projection $\Omega$
reverses $L-x^6$, transposes the Chan--Patton indices of open strings
and breaks half of the supercharges.
For the 7D massless modes of the {\darkred 66} open strings, the three effects are independent,
thus $\Omega=\Omega_1\Omega_2\Omega_3$ where:
\item\bullet
$\Omega_1=\pm1$ is the parity of the spatial wave function,
$\psi(x^6-L)=\Omega_1\psi(L-x^6)$.
In $x^6\le L$ terms, this translates into the boundary conditions for the wave
functions --- and hence for the corresponding fields --- at the $x^6=L$ end of the
world: {\sl Neumann BC for the $\Omega_1$--positive fields and Dirichlet BC
for the $\Omega_1$--negative fields.}
\item\bullet
$\Omega_2$ transposes the D6 Chan--Patton indices;
in $N\times N$ matrix terms (for the two D6 indices of a $\darkred 66$ string),
the symmetric matrices are $\Omega_2$--positive and the antisymmetric matrices
are $\Omega_2$--negative.
\item\bullet
$\Omega_3$ is the 6D chirality of the 8--SUSY supermultiplet;
for the $\bf O8^-$ orientifold,
the vector multiplets are $\Omega_3$ positive and the hypermultiplets are $\Omega_3$ negative.

\par\noindent
The states surviving the projection have $\Omega=+1$ and hence $\Omega_1=\Omega_2\Omega_3$.
Thus,
\item\star
7D fields with Neumann (free) boundary conditions at $x^6=L$ comprise
(6D) vector multiplets for the symmetric $N\times N$ matrices and
hypermultiplets for the antisymmetric matrices.
\item\star
The fields with Dirichlet boundary conditions comprise  vector multiplets
for the antisymmetric matrices
and  hypermultiplets for the symmetric matrices.
\par\noindent
Note that the local gauge symmetry at $x^6=L$ must make sense group theoretically,
hence the symmetric $N\times N$ matrices must form a closed Lie algebra;
such an algebra is called symplectic and denoted $\darkred Sp(N/2)$;
it exists for even $N$ only.
Consequently, {\sl the number $N$ of {\darkred D6} branes terminating at the same
generic point on an $\bf O8^-$ orientifold plane must be even.}

In the  $\darkred Sp(N/2)$ terms, the multiplet structure of the 7D fields
includes one symmetric tensor multiplet~$\symrep$, one irreducible antisymmetric
tensor multiplet $\tilde\asymrep=\asymrep-{\bf1}$, and one singlet
--- which is responsible for the center-of-mass motion of the
{\darkred D6} branes and irrelevant for the HW orbifold problem.
Thus, as far as the 7D $\darkred SU(N)$ SYM fields living on a $\IZ_N$ $\FP6$ plane of a
HW orbifold are concerned, the \tI/D6 dual theory provides the following boundary
conditions:
\pointbegin
Locally, at $x^6=L$, the 7D gauge group is partially broken from $\darkred SU(N)$ down to
$\darkred Sp(N/2)$;
\point
(6D) vector multiplets in the adjoint $\darkred\symrep$ of the $\darkred Sp(N/2)$
and hypermultiplets
in the $\darkred\tilde\asymrep$ have Neumann boundary conditions at $x^6=L$;
\point
the remaining vector multiplets in $\darkred\tilde\asymrep$ and hypermultiplets
in the $\darkred\symrep$ have Dirichlet boundary conditions at $x^6=L$.

The action of the orientifold projection on the
$\darkblue 88$ and $\darkred 6\darkblue8$ open string states is less complicated.
The $\darkblue 88$ open strings are precisely as in the \tI\ superstring without the D6 branes:
The $\bf O8^-$ orientifold projection breaks the 9D gauge group down to $\darkblue SO(2k)$
but all 16 supercharges remain unbroken away from $\bx=0$.
At the junction, there are only eight supercharges and the massless modes of the
$\darkred6\darkblue8$ open strings form 6D hypermultiplets.
The orientifold projection removes precisely one half of each hypermultiplet,
leaving us with half-hypermultiplets in the bi-fundamental
$({\bf\darkred N},{\bf\darkblue 2k})$ representation of the net
${\darkred Sp(N/2)}\times {\darkblue SO(2k)}$ gauge symmetry at the brane junction;
`fortunately', this representation is pseudo-real so it allows half-hypermultiplets.
\foot{Obviously, this pseudoreality is not an accidental `good fortune' but a consistency
    constraint on the orbifold projection $\Omega$.
    This is precisely the constraint which requires $\Omega$ to select opposite
    (anti) symmetrizations for the D8 and D6 Chan--Patton indices, hence the net
    gauge group is either ${\darkblue SO(2k)}\times{\darkred Sp(N/2)}$ (the $\bf O8^-$
    projection) or ${\darkblue Sp(k)}\times{\darkred SO(N)}$ (the $\bf O8^+$ projection)
    but never ${\darkblue Sp(k)}\times{\darkred Sp(N/2)}$ or ${\darkblue SO(2k)}\times{\darkred SO(N)}$.}

Finally, let us return to the specific example of $N=2$, $k=8$ dual to the $\darkblue SO(16)$
side of the $\IZ_2$ HW heterotic orbifold.
Because $Sp(1)=SU(2)$, the  7D $\darkred SU(2)$ gauge group remains completely
unbroken at the $x^6=L$ terminus;
all $\bf\darkred 3$ vector multiplets satisfy Neumann boundary conditions and all $\bf\darkred 3$
hypermultiplets satisfy Dirichlet boundary conditions.
Furthermore, there are `twisted' massless fields localized at the junction, namely
half-hypermultiplets in the $({\bf\darkred2},{\bf\darkblue16})$ representation of the net
locally visible gauge symmetry ${\darkred SU(2)}\times{\darkblue SO(16)}$.
$$
\pspicture[.49](+60,-30)(94,30)
\psline[linewidth=5,linecolor=palegray](+90,-30)(+90,+30)
\psline[linecolor=black,linestyle=dashed,linewidth=1](+92,-30)(+92,+30)
\psline[linecolor=blue](+91.25,-30)(+91.25,+30)
\psline[linecolor=blue](+90.75,-30)(+90.75,+30)
\psline[linecolor=blue](+90.25,-30)(+90.25,+30)
\psline[linecolor=blue](+89.75,-30)(+89.75,+30)
\psline[linecolor=blue](+89.25,-30)(+89.25,+30)
\psline[linecolor=blue](+88.75,-30)(+88.75,+30)
\psline[linecolor=blue](+88.25,-30)(+88.25,+30)
\psline[linecolor=blue](+87.75,-30)(+87.75,+30)
\rput[b]{90}(+86.5,+15){${\bf O8}+8{\rm\darkblue D8}$}
\rput[b]{90}(+86.5,-15){$\darkblue SO_{16}$}
\psline[linecolor=red](+60,-0.5)(+92,-0.5)
\psline[linecolor=red](+60,+0.5)(+92,+0.5)
\rput[b](70,+2){$2\darkred\rm D6$}
\rput[t](70,-2){$\darkred SU_2$}
\psline[linecolor=yellow,linewidth=2](87.5,0)(92.5,0)
\endpspicture
\left\{
\eqalign{
        G_{\rm local}\ &=\ {\darkred SU(2)}\times{\darkblue SO(16)},\cr
        {\rm 7D}\,V\ &=\ {\darkred\bf 3},\cr
        {\rm 7D}\,H\ &=\ 0,\cr
        {\rm 6D}\,H\ &=\ \half({\bf\darkred2},{\bf\darkblue16}).\cr
        }
\right.
\eqn\YellowEndZtwo
$$
Note that these are precisely the boundary conditions and the local fields
we found in ref.~[\KSTY] to occur at the $\I_2$~intersection plane of the dual HW orbifold.
In the HW theory, this combination of boundary conditions and local fields was a solution
of several kinematic constraints but its dynamical origin remained a \M ystery;
this particular \M ystery is now solved in terms of the dual \tI/D6 superstring model.

%%%%%%%%%%%%%%%%%%%%%%%%%%%%%%%%%%%%%%%%%%%%%%%%%%
\section{D6 Branes Terminating on D8 branes and the Diagonal Gauge Groups.}
Now consider the other terminus of the D6 branes at the
${\darkgreen E_7}\times {\darkblue SU_2}$ side of the $N=2$ model, \cf\ the  purple ellipse in
fig~\ZtwoDeight.
In the HW theory, the $\FP6$ plane terminates at the $\rm M9_1$ at $x^6=0$,
but in the dual \tI/D6 theory this $\rm M9_1$ becomes the whole stack of $\Ori+8\darkblue\rm D8$
branes spanning $0\le x^6\le a$.
Hence, in the \tI/D6 model we have a choice of the allowed left termini for
each of the two {\darkred D6} branes:
\pointbegin
A D6 brane may cross (without termination) the two `outlier' D8 branes and
terminate  on the orientifold plane or  one of the  six D8 branes at $x^6=0$.
$$
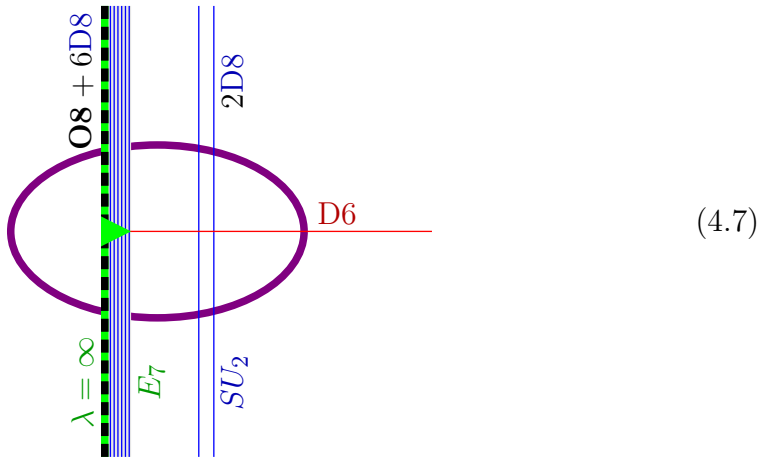
\pspicture[](-20,-30)(+30,+30)
\psellipse[linecolor=purple,linewidth=1](-6.5,0)(20,12)
\psline[linecolor=palegray,linewidth=4](-12,-30)(-12,+30)
\psline[linecolor=green,linewidth=1](-13.5,-30)(-13.5,+30)
\psline[linecolor=black,linestyle=dashed,linewidth=1](-13.5,-30)(-13.5,+30)
\psline[linecolor=blue](-12.75,-30)(-12.75,+30)
\psline[linecolor=blue](-12.25,-30)(-12.25,+30)
\psline[linecolor=blue](-11.75,-30)(-11.75,+30)
\psline[linecolor=blue](-11.25,-30)(-11.25,+30)
\psline[linecolor=blue](-10.75,-30)(-10.75,+30)
\psline[linecolor=blue](-10.25,-30)(-10.25,+30)
\rput[b]{90}(-15,+20){${\bf O8}+6{\rm\darkblue D8}$}
\rput[t]{90}(-9,-20){$\darkgreen E_7$}
\rput[b]{90}(-15,-20){$\darkgreen\lambda=\infty$}
\psline[linecolor=blue](-1,-30)(-1,+30)
\psline[linecolor=blue](+1,-30)(+1,+30)
\rput[t]{90}(+2,+20){$2\rm\darkblue D8$}
\rput[t]{90}(+2,-20){$\darkblue SU_2$}
\psline[linecolor=red](-12,0)(+30,0)
\rput[br](+20,+1){\darkred D6}
\pspolygon*[linecolor=green](-14,+2)(-10,0)(-14,-2)
\endpspicture
\eqn\GreenJunction
$$
Superficially, the terminus at $x^6=0$ is  similar to terminus at $x^6=L$
discussed in the previous subsection, 
but the divergence of the string coupling at $x^6=0$
demands a non-perturbative re-analysis of the resulting boundary conditions
and local 6D fields.
As of this writing, the physics of such  $\darkgreen\lambda=\infty$ terminal junctions
remains somewhat mysterious, but the net effect may be inferred from 
duality considerations;
we shall return to this issue in section~7.

\point %2
Alternatively, a D6 brane may terminate on one of the two outlier D8 branes
at $x^6=a$ without extending all the way to the true `end of the world' at $x^6=0$.
$$
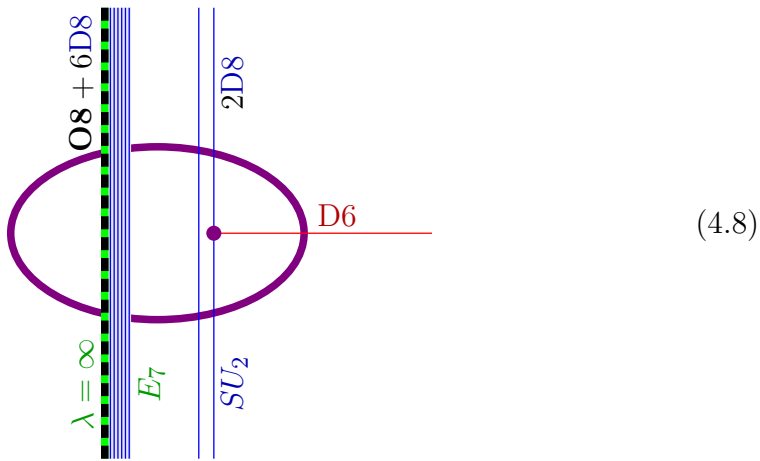
\pspicture[](-20,-30)(+30,+30)
\psellipse[linecolor=purple,linewidth=1](-6.5,0)(20,12)
\psline[linecolor=palegray,linewidth=4](-12,-30)(-12,+30)
\psline[linecolor=green,linewidth=1](-13.5,-30)(-13.5,+30)
\psline[linecolor=black,linestyle=dashed,linewidth=1](-13.5,-30)(-13.5,+30)
\psline[linecolor=blue](-12.75,-30)(-12.75,+30)
\psline[linecolor=blue](-12.25,-30)(-12.25,+30)
\psline[linecolor=blue](-11.75,-30)(-11.75,+30)
\psline[linecolor=blue](-11.25,-30)(-11.25,+30)
\psline[linecolor=blue](-10.75,-30)(-10.75,+30)
\psline[linecolor=blue](-10.25,-30)(-10.25,+30)
\rput[b]{90}(-15,+20){${\bf O8}+6{\rm\darkblue D8}$}
\rput[t]{90}(-9,-20){$\darkgreen E_7$}
\rput[b]{90}(-15,-20){$\darkgreen\lambda=\infty$}
\psline[linecolor=blue](-1,-30)(-1,+30)
\psline[linecolor=blue](+1,-30)(+1,+30)
\rput[t]{90}(+2,+20){$2\rm\darkblue D8$}
\rput[t]{90}(+2,-20){$\darkblue SU_2$}
\psline[linecolor=red](+0.5,0)(+30,0)
\rput[br](+20,+1){\darkred D6}
\pscircle*[linecolor=purple](+1,0){1}
\endpspicture
\eqn\PurpleJunction
$$
It turns out that both D6 branes of the \tI/D6 model dual to the
$T^4/\IZ_2$ heterotic orbifold should terminate in this manner at $x^6=a$.

Indeed, for the heterotic orbifold with $E_8\times E_8$ broken to
$\bigl[{\darkgreen E_7}\times {\purple SU(2)}\bigr]\times {\darkblue SO(16)}$,
all massless twisted states are $\darkgreen E_7$ singlets.
In terms of the dual \tI/D6 model, this implies spatial segregation between the
{\darkred D6} branes and the $\darkgreen E_7$ gauge fields living at $x^6=0$ --- in other words,
terminating the {\darkred D6} branes at $x^6=a>0$.
Furthermore, we need to explain the \M ystery of the locking boundary conditions
\HWLockingBC\ for the $\purple SU(2)$ fields of the HW orbifold; in the dual
\tI/D6 terms, these boundary conditions become
$$
A_\mu^{\darkred\rm 7D}(x^6=a)\ =\ A_\mu^{\rm\darkblue 9D}(\bx={\bf0}) .
\eqn\LockingBC
$$
We shall see momentarily that such gauge field locking follows from each of the
two  {\darkred D6} branes at $\bx=\bf0$ terminating on a separate outlier {\darkblue D8}
 brane at $x^6=a$ according to following figure:
$$
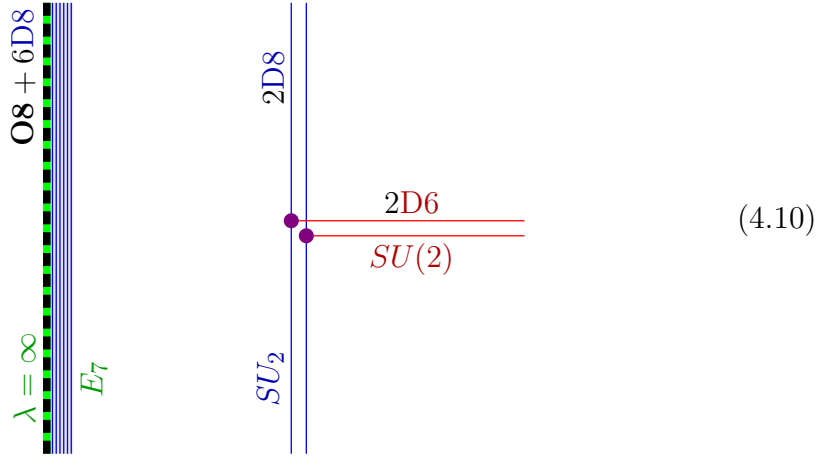
\pspicture[](-20,-30)(+50,+30)
\psline[linecolor=palegray,linewidth=4](-12,-30)(-12,+30)
\psline[linecolor=green,linewidth=1](-13.5,-30)(-13.5,+30)
\psline[linecolor=black,linestyle=dashed,linewidth=1](-13.5,-30)(-13.5,+30)
\psline[linecolor=blue](-12.75,-30)(-12.75,+30)
\psline[linecolor=blue](-12.25,-30)(-12.25,+30)
\psline[linecolor=blue](-11.75,-30)(-11.75,+30)
\psline[linecolor=blue](-11.25,-30)(-11.25,+30)
\psline[linecolor=blue](-10.75,-30)(-10.75,+30)
\psline[linecolor=blue](-10.25,-30)(-10.25,+30)
\rput[b]{90}(-15,+20){${\bf O8}+6{\rm\darkblue D8}$}
\rput[t]{90}(-9,-20){$\darkgreen E_7$}
\rput[b]{90}(-15,-20){$\darkgreen\lambda=\infty$}
\psline[linecolor=blue](+19,-30)(+19,+30)
\psline[linecolor=blue](+21,-30)(+21,+30)
\rput[b]{90}(+18,+20){$2\rm\darkblue D8$}
\rput[b]{90}(+18,-20){$\darkblue SU_2$}
\psline[linecolor=red](+19,+1)(+50,+1)
\psline[linecolor=red](+21,-1)(+50,-1)
\rput[b](+35,+2){$2\rm\darkred D6$}
\rput[t](+35,-2){$\darkred SU(2)$}
\pscircle*[linecolor=purple](+21,-1){1}
\pscircle*[linecolor=purple](+19,+1){1}
\endpspicture
\eqn\PurpleJunctionZtwo
$$

Proper understanding of the {\darkred D6}--{\darkblue D8} brane junctions
(as opposed to simple brane crossings)
involves {\it inter alia} the mechanical tension of the D--branes.
A {\darkred D6} brane pulls on the {\darkblue D8} brane on which it ends and bends it
out of planarity;
consequently, each of the two outlier {\darkblue D8} branes on figure~\PurpleJunctionZtwo\
is located at
$$
x^6\ = a\ +\ {\alpha'\over|\bx|}
\eqn\CoulombBrane
$$
instead of simply $x^6\equiv a$.
At the junction itself ($\bx=\bf0$) the {\darkblue D8} branes are singular
and the quantum string effects become important.

The simplest way to understand these effects is via T-duality.
Let us compactify one of the transverse coordinates of the {\ blue D8} branes,
\eg~$x^7$ on a large circle of radius $R_7$, then T-dualize $x^7\to\tilde x^7$.
This duality turns the {\darkblue D8} branes into {\darkblue D7} branes spanning $x^0,\ldots,x^5$
and $x^8,x^9$ with transverse coordinates $x^6$ and $\tilde x^7$.
Consequently, the co-dimension of the junction in the brane reduces from 3 down to 2,
hence the bending of the brane [\Wittenbend] by the sideways pulling
force becomes logarithmic,
$$
x^6\ =\ a\ + {\alpha'\over R_7}\,\log{\alpha'/R_7\over\sqrt{(x^8)^2+(x^9)^2}}
\eqn\LogarithmicBend
$$
instead of the Coulomb shape~\CoulombBrane.
The bend {\darkblue D7} brane preserves 8 out of 32 supercharges of
the (T-dual) type~IIB superstring;
to make SUSY manifest, it is convenient to introduce complex coordinates
$$
(x^6-a)\,+\,i\tilde x^7\ =\ \tilde R_7\times w,\qquad x^8\,+\,ix^9\ =\ \tilde R_7\times u,
\eqn\ComplexUW
$$
where $\tilde R_7=\alpha'/R_7$ is the radius of the T-dual $\tilde x^7$ circle;
note that $w$ is a cylindrical coordinate, $w\equiv w+2\pi i$.
In terms of these complex coordinates, the {\darkblue D7} branes span a holomorphic curve
$$
w\ =\ \log{1\over u}\,.
\eqn\DsevenLocus
$$

The {\darkred D6} branes at $\bx=\bf0$ are mapped by the T-duality onto $\rm\darkred D7'$ branes
spanning the $\tilde x^7$ coordinate in addition to the $x^0,\ldots,x^6$;
the $x^8,x^9$ coordinates remain transverse;
in terms of the complex coordinates~\ComplexUW, the $\rm\darkred D7'$ branes are located at
$u\equiv 0~\forall w$.

Now consider a single {\darkred D6} brane terminating on a single {\darkblue D8} brane.
The T-dual of this picture is a junction between a $\rm\darkred D7'$ and a {\darkblue D7}
brane.
Because of the {\darkblue D7} brane bending~\DsevenLocus,
 this junction is located somewhat to the right
of $x^6=a$, \ie\ at $\Re w>0$.
Re-writing eq.~\DsevenLocus\ as
$$
u\ =\ e^{-w} ,
\eqn\DsevenCurve
$$
we see that for $\Re w>0$ the {\darkblue D7} brane rapidly asymptotes to $u\equiv0$.
Consequently, the {\darkblue D7} brane {\purple\sl smoothly} connects to the $\darkred\rm D7'$
brane without any discontinuity.
In other words, {\sl the whole complex of
the {\darkblue D8} brane and the {\darkred D6} brane terminated on it is
T-dual to a single curved {\purple D7} brane spanning~\DsevenCurve.}

As a corollary, {\it the complex of two coincident {\darkred D6} branes terminated on two
coincident {\darkblue D8} branes in a one-on-one manner depicted on fig~\PurpleJunctionZtwo\ is
T-dual to a single pair of coincident smoothly curved {\purple D7} branes.}
The $\purple U(2)$ SYM generated by the $\purple 77$ open strings of the T-dual theory
has exactly one local $\purple U(2)$ gauge symmetry at every point of the {\purple D7}
world-volume.
By T-duality, this means that the \PurpleJunctionZtwo\ complex of 2 {\darkred D6} and 2
{\darkblue D8} branes has exactly one local $\purple U(2)$ gauge symmetry at every point
of the $\rm{\darkred D6}+{\darkblue D8}$ world volume, {\sl including the junction point}
$(x^6=a,\bx={\bf0})$.
Therefore, at the junction point, the $\darkred U(2)$ gauge fields living on the {\darkred D6}
world-volume and the $\darkblue U(2)$ gauge fields living on the {\darkblue D8} world-volume
must satisfy the locking boundary condition~\LockingBC.

In the \tI/D6 context, the locking boundary conditions~\LockingBC\ apply to the
${\darkred U(2)^{\rm 7D}}\times{\darkblue U(2)^{\rm 9D}}\to{\purple U(2)^{\rm diag}}$ gauge
theory involved in the brane junctions.
{}From the \M~theory point of view however, the $\darkblue U(1)$ center of the 9D $\darkblue U(2)$
is an artifact of the KK reduction of the HW theory to the \tI\ superstring and, 
likewise, the $\darkred U(1)$ center of the 7D $\darkred U(2)$ is an artifact of the 
multi--Taub--NUT geometry replacing the $T^2/\IZ_2$ orbifold.
Hence in the HW orbifold context, the \M ysterious locking boundary conditions~\HWLockingBC\
should apply to the simple ${\darkred SU(2)^{\rm 7D}}\times{\darkblue SU(2)^{\rm 10D}}
\to{\purple SU(2)^{\rm diag}}$ gauge theories only.

Next, consider the supermultiplet structure of the diagonal $\purple SU(2)$ SYM
theory.
Locally, at every point of the T-dual $\purple D7$ world volume there are
16 unbroken supersymmetries but the dimensional reduction to the effective 6D theory
preserves only 8 of the supercharges.
Consequently, the 6D vector multiplets and hypermultiplets have different wave functions
on the holomorphic curve~\DsevenCurve.
The T-dual wave functions on the {\darkblue D8}--{\darkred D6} brane junction are governed by the
boundary conditions at the junction point.
Hence, by T-duality, the hypermultiplets have different boundary conditions than the
vector multiplets;
specifically, given the Dirichlet-like locking boundary conditions for
the vector multiplets,
it follows that the hypermultiplets satisfy the free (Neumann) boundary conditions.
That is, at the junction point, there are both {\darkblue 9D} and {\darkred 7D} hypermultiplets
(each in the adjoint $\purple\bf 3$ representation of the diagonal $\purple SU(2)$
gauge group) and both are free to take whatever values they like independently of
each other.

Actually, we do not need T-duality to establish the Neumann boundary conditions for the
{\darkred 7D} hypermultiplets at the brane junction.
(The {\darkblue 9D} fields are automatically free since they cannot possibly be pinned down
at a codimension~3 junction, whatever the junction physics.)
All we need to know is that at $x^6=a$, each of the two {\darkred D6} branes is free
to move its attachment point to the {\darkblue D8} brane in the three transverse dimensions
$(x^7,x^8,x^9)=\bx$ independently of the other brane.
Note that from the 7D world-volume point of view, the transverse coordinates of the two
{\darkred D6} branes are scalars in the 7D, 16--SUSY vector multiplets in the Cartan
$\darkred U(1)^2$ subalgebra of the $\darkred U(2)$ SYM theory.
Therefore, the  freedom to move the attachment points of the {\darkred D6} branes
in the transverse directions implies free (Neumann) boundary condition for the corresponding
scalar fields.
In the 6D, 8--SUSY terms, these scalars belong to hypermultiplets,
hence thanks to SUSY and the non-abelian gauge symmetry, we must have Neumann boundary
conditions for the entire hypermultiplets in the adjoint representation of the {\darkred 7D}
gauge group.

Finally, consider the {\darkred6\darkblue8} open strings at
the {\darkred D6}--{\darkblue D8} brane junction.
In principle, such open strings may have  massless modes localized at the 6D junction
plane.
However, the T-duality shows that this does not happen.
Indeed, consider the pair of curved {\purple D7} branes dual to the junction in question.
The holomorphic curve~\DsevenCurve\ has an arbitrary scale $\tilde R_7$ which can be made
large if desired, hence the $\purple 77$ opens strings have no inherently stringy
6D massless modes trapped in the junction area.
Hence the only possible localized 6D massless fields are the {\sl normalizable} zero
modes of the 8D massless fields --- the $\purple U(2)$ SYM --- but the non-compact
curve~\DsevenCurve\ does not have any normalizable zero modes.
Altogether, the $\purple 77$ open strings of the T-dual theory do not have any localized
zero modes corresponding to  massless 6D fields, and by T-duality, the {\darkred6\darkblue8}
strings of the \tI/D6 model --- or for that matter, the {\darkred 66} or {\darkblue 88} strings ---
do not produce any massless 6D fields trapped at the junction.

To summarize, the brane junction depicted on fig.~\PurpleJunctionZtwo\ correctly reproduces
the \M ysteries at the ${\darkgreen E_7}\times {\purple SU(2)}$ terminus of the $\IZ_2$
fixed 7--plane of the HW orbifold:
The ${\darkred SU(2)^{\rm 7D}}\times {\darkblue SU(2)^{\rm 9D}}$ gauge symmetry is broken to the
diagonal $\purple SU(2)$ by the locking boundary conditions~\LockingBC\
while the $\darkgreen E_7$ gauge fields do not couple to the massless twisted states;
the hypermultiplets satisfy the Neumann boundary conditions at the junction;
and there are no massless 6D fields localized at this junction.

%%%%%%%%%%%%%%%%%%%%%%%%%%%%%%%%%%%%%%%%%%%%%%%%%%%%%%%%%%%%%%
%% end of chapter4

%auto-ignore
% chapter5.tex
%% Chapter 5 of Orbifold2 paper
%% Written by Vadim Kaplunovsky

%%%%%%%%%%%%%%%%%%%%%%%%%%%%%%%%%%%%%%%%%%%%%%%%%%%%%%%%%%%%%%%%%%%%%%%
\chapter{Brane Duals of Orbifolds with Broken 7D Gauge Symmetries}
In the previous section we focused on the simplest example of the HW heterotic orbifold,
namely the $T^4/\IZ_2$ with $E_8^{(1)}$ broken down to ${\darkgreen E_7}\times{\purple SU(2)}$
and $E_8^{(2)}$ down to $\darkblue SO(16)$; the \tI/D6 brane dual of this model has D--branes
arranged according to
$$
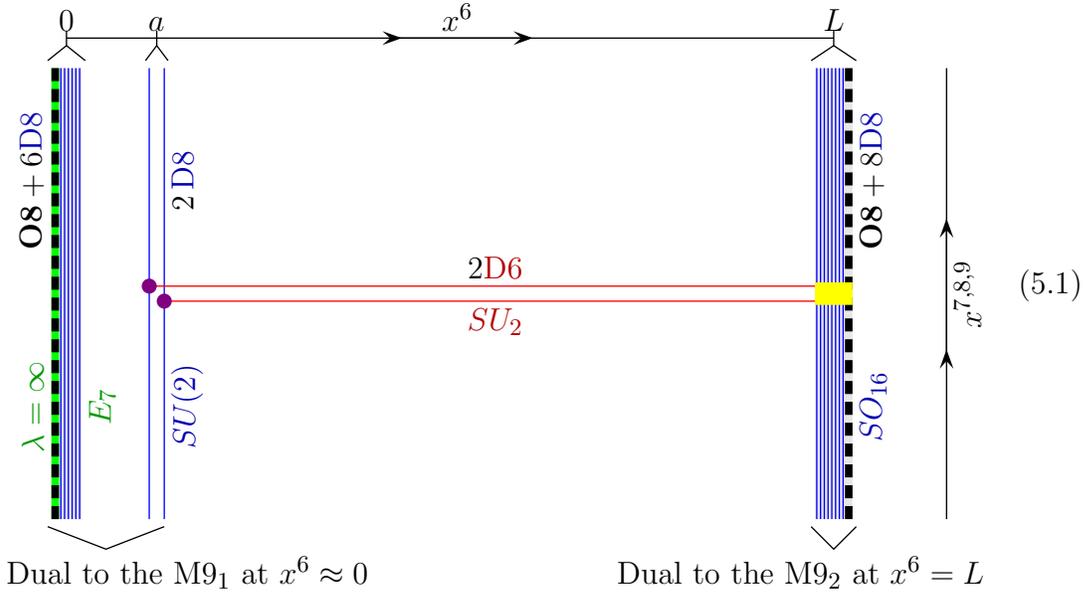
\pspicture[](-20,-40)(110,40)
\psline(-12,+34)(+90,+34)
\psline(-12,+33)(-12,+35)	\rput[b](-12,+35){$0$}
\psline(0,+33)(0,+35)		\rput[b](0,+35){$a$}
\psline(+90,+33)(+90,+35)	\rput[b](+90,+35){$L$}
\rput[b](+40,+35){$x^6$}
\psline{>->}(+30,+34)(+50,+34)
\psline[linecolor=palegray,linewidth=4](-12,-30)(-12,+30)
\psline[linecolor=green,linewidth=1](-13.5,-30)(-13.5,+30)
\psline[linecolor=black,linestyle=dashed,linewidth=1](-13.5,-30)(-13.5,+30)
\psline[linecolor=blue](-12.75,-30)(-12.75,+30)
\psline[linecolor=blue](-12.25,-30)(-12.25,+30)
\psline[linecolor=blue](-11.75,-30)(-11.75,+30)
\psline[linecolor=blue](-11.25,-30)(-11.25,+30)
\psline[linecolor=blue](-10.75,-30)(-10.75,+30)
\psline[linecolor=blue](-10.25,-30)(-10.25,+30)
\psline(-14.5,+31)(-12,+33)(-9.5,+31)
\rput[b]{90}(-15,+15){${\bf O8}+6{\rm\darkblue D8}$}
\rput[t]{90}(-9,-15){$\darkgreen E_7$}
\rput[b]{90}(-15,-15){$\darkgreen\lambda=\infty$}
\psline[linecolor=blue](-1,-30)(-1,+30)
\psline[linecolor=blue](+1,-30)(+1,+30)
\psline(-1.5,+31)(0,+33)(+1.5,+31)
\rput[t]{90}(+2,+15){$2\,\rm\darkblue D8$}
\rput[t]{90}(+2,-15){$\darkblue SU(2)$}
\psline[linewidth=5,linecolor=palegray](+90,-30)(+90,+30)
\psline[linecolor=black,linestyle=dashed,linewidth=1](+92,-30)(+92,+30)
\psline[linecolor=blue](+91.25,-30)(+91.25,+30)
\psline[linecolor=blue](+90.75,-30)(+90.75,+30)
\psline[linecolor=blue](+90.25,-30)(+90.25,+30)
\psline[linecolor=blue](+89.75,-30)(+89.75,+30)
\psline[linecolor=blue](+89.25,-30)(+89.25,+30)
\psline[linecolor=blue](+88.75,-30)(+88.75,+30)
\psline[linecolor=blue](+88.25,-30)(+88.25,+30)
\psline[linecolor=blue](+87.75,-30)(+87.75,+30)
\psline(+93,+31)(+90,+33)(+87,+31)
\rput[t]{90}(+93.5,+15){${\bf O8}+8{\rm\darkblue D8}$}
\rput[t]{90}(+93.5,-15){$\darkblue SO_{16}$}
\psline[linecolor=red](+1,-1)(+92,-1)
\psline[linecolor=red](-1,+1)(+92,+1)
\rput[b](45,+2){$2\darkred\rm D6$}
\rput[t](45,-2){$\darkred SU_2$}\psline(-14.5,-31)(-6.75,-34)(+1,-31)
\pscircle*[linecolor=purple](+1,-1){1}
\pscircle*[linecolor=purple](-1,+1){1}
\psline[linecolor=yellow,linewidth=3](+87.5,0)(92.5,0)
\rput[tl](-20,-35){Dual to the $\rm M9_1$ at $x^6\approx0$}
\psline(+87,-31)(+90,-34)(+93,-31)
\rput[tr](+110,-35){Dual to the $\rm M9_2$ at $x^6=L$}
\psline(+105,-30)(+105,+30)
\psline{>->}(+105,-10)(+105,+10)
\rput[t]{90}(+106,0){$x^{7,8,9}$}
\endpspicture
\eqn\ZtwoBranes
$$
In this section, we shall brane engineer the dual models for more complicated $\IZ_3$
and $\IZ_4$ orbifolds in which massless twisted states are charged under abelian factors
of the broken $E_8\times E_8$ gauge symmetry.
In ref.~[\KSTY] we found such abelian charges to be problematic in the
HW context as none of the 7D $\darkred SU(N)$ breaking patterns seemed to satisfy
all the kinematic constraints.
The correct solution turns out to be rather complicated or even
bizarre in purely HW terms --- but natural
and fairly simple in terms of the \tI/D6 brane engineering.

%%%%%%%%%%%%%%%%%%%%%%%%%%%%%%%%%%%%%%%%%%%%%%%%%%%%%%%%%
\section{The Symmetric $\IZ_3$ Orbifold}
Our first model is a perturbative $T^4/\IZ_3$ heterotic orbifold
with  both $E_8^{(1),(2)}$ gauge groups broken to $[SO(14)\times U(1)]^{(1),(2)}$
in identical fashion,
\foot{%
    In lattice terms, the action of the $\IZ_3$ orbifold group on each $E_8$
    corresponds to the shift vector $\delta=(\coeff23,0,0,0,0,0,0,0)$.
    }
hence the name `symmetric'.
The massless spectrum of this model comprises:
\item\bullet
Untwisted states:\brk
SUGRA + 1 tensor multiplet (the dilaton);\brk
184 vector multiplets in the adjoint representation of the $[SO(14)\times U(1)]^2$;\brk
2 moduli and 156 charged hypermultiplets,
$$
H_0\ =\ ({\bf64},+\half;{\bf1},0)\ +\ ({\bf14},-1;{\bf1},0)\
+\ ({\bf1},0;{\bf64},+\half)\ +\ ({\bf1},0;{\bf14},-1)\ +\ 2M .
\eqn\ZthreeUntw
$$
\item\bullet
Twisted states:\brk
30 charged hypermultiplets for each of the 9 $\IZ_3$ fixed points on the $T^4$,
$$
H_1\ =\ 9\Bigl[ ({\bf14},-\coeff13;{\bf1},+\coeff23)\,
+\,({\bf1},+\coeff23;{\bf14},-\coeff13)\,+\,2({\bf1},+\coeff23;{\bf1},+\coeff23)\Bigr] .
\eqn\ZthreeTw
$$

{}From the Ho\v rava--Witten point of view,
the charges~\ZthreeTw\ indicate that the abelian $U(1)\times U(1)$ gauge fields
must somehow span the $x^6$ along the $\FP6$ fixed planes, thus
$$
{\purple U(1)\times U(1)}\ =\ \diag \left[
\left({\darkblue U(1)\times U(1)}\right)_{\rm 10D}\times
\smash{\prod_{\rm fixed\atop planes}}\vphantom{\prod}
\left({\darkred U(1)\times U(1)}\right)_{\rm 7D}\right].
\eqno\eq
$$
Or rather
$$
{\purple U(1)\times U(1)}\ =\ \diag \left[
\left({\darkblue U(1)\times U(1)}\right)_{\rm 10D}\times
\smash{\prod_{\rm fixed\atop planes}}\vphantom{\prod}
\left({\darkred U(1)\times U(1)\subset SU(3)}\right)_{\rm 7D}\right] 
\eqn\SUiiiUiUi
$$
since the actual 7D gauge symmetry living on each $\IZ_3$ $\FP6$ plane is ${\darkred SU(3)}
\supset\darkred U(1)\times U(1)$.
We shall see momentarily that the symmetry breaking~\SUiiiUiUi\ follows from the
boundary conditions for the 7D and 10D gauge fields at the $\I=\FP6\cap\rm M9$ intersections
of the HW theory, but the boundary conditions are so \M ysterious it took us months
to find them.
Again, the key to this \M ystery is provided by the brane engineering.

The general setup of the \tI/D6 brane dual of the symmetric $\IZ_3$ fixed plane
is quite clear:
$$
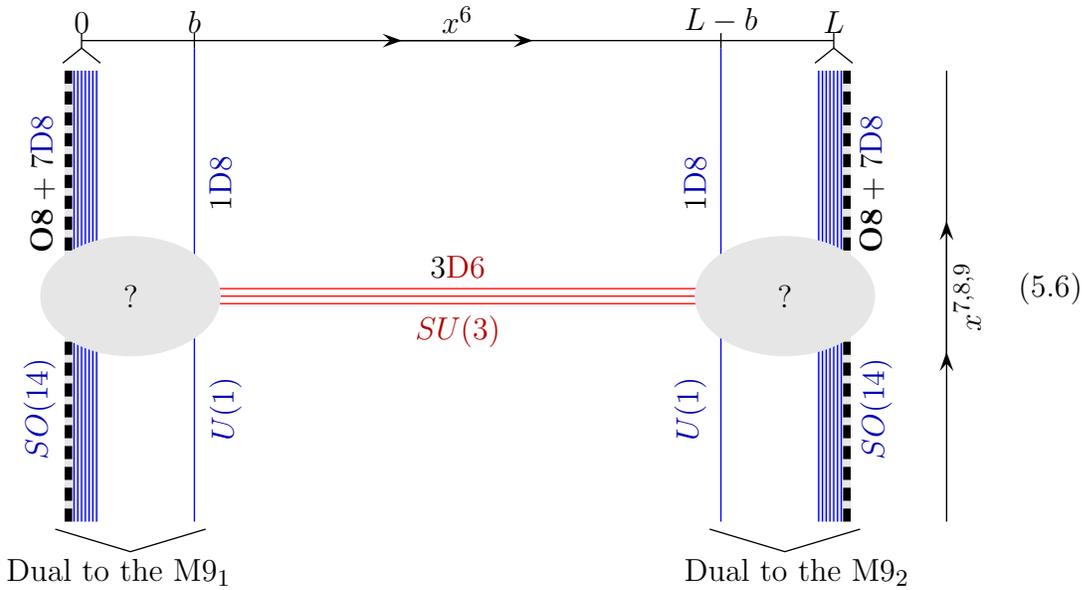
\pspicture[](-60,-40)(+70,40)
\psline(-50,+34)(+50,+34)
\psline(-50,+33)(-50,+35)	\rput[b](-50,+35){$0$}
\psline(-35,+33)(-35,+35)	\rput[b](-35,+35){$b$}
\psline(+50,+33)(+50,+35)	\rput[b](+50,+35){$L$}
\psline(+35,+33)(+35,+35)	\rput[b](+35,+35){$L-b$}
\rput[b](0,+35){$x^6$}
\psline{>->}(-10,+34)(+10,+34)
\psline[linecolor=palegray,linewidth=4.5](-50,-30)(-50,+30)
\psline[linecolor=black,linestyle=dashed,linewidth=1](-51.75,-30)(-51.75,+30)
\psline[linecolor=blue](-51.0,-30)(-51.0,+30)
\psline[linecolor=blue](-50.5,-30)(-50.5,+30)
\psline[linecolor=blue](-50.0,-30)(-50.0,+30)
\psline[linecolor=blue](-49.5,-30)(-49.5,+30)
\psline[linecolor=blue](-49.0,-30)(-49.0,+30)
\psline[linecolor=blue](-48.5,-30)(-48.5,+30)
\psline[linecolor=blue](-48.0,-30)(-48.0,+30)
\psline(-52.5,+31)(-50,+33)(-47.5,+31)
\rput[b]{90}(-53.5,+15){${\bf O8}+7{\rm\darkblue D8}$}
\rput[b]{90}(-53.5,-15){$\darkblue SO(14)$}
\psline[linecolor=blue](-35,-30)(-35,+33)
\rput[t]{90}(-33,+15){$1\rm\darkblue D8$}
\rput[t]{90}(-33,-15){$\darkblue U(1)$}
\psline[linecolor=palegray,linewidth=4.5](+50,-30)(+50,+30)
\psline[linecolor=black,linestyle=dashed,linewidth=1](+51.75,-30)(+51.75,+30)
\psline[linecolor=blue](+51.0,-30)(+51.0,+30)
\psline[linecolor=blue](+50.5,-30)(+50.5,+30)
\psline[linecolor=blue](+50.0,-30)(+50.0,+30)
\psline[linecolor=blue](+49.5,-30)(+49.5,+30)
\psline[linecolor=blue](+49.0,-30)(+49.0,+30)
\psline[linecolor=blue](+48.5,-30)(+48.5,+30)
\psline[linecolor=blue](+48.0,-30)(+48.0,+30)
\psline(+52.5,+31)(+50,+33)(+47.5,+31)
\rput[t]{90}(+53.5,+15){${\bf O8}+7{\rm\darkblue D8}$}
\rput[t]{90}(+53.5,-15){$\darkblue SO(14)$}
\psline[linecolor=blue](+35,-30)(+35,+33)
\rput[b]{90}(+33,+15){$1\rm\darkblue D8$}
\rput[b]{90}(+33,-15){$\darkblue U(1)$}
\psline[linecolor=red](-50,-1)(+50,-1)
\psline[linecolor=red](-50,0)(+50,0)
\psline[linecolor=red](-50,+1)(+50,+1)
\rput[b](0,+2.5){$3\darkred\rm D6$}
\rput[t](0,-2.5){$\darkred SU(3)$}
\psellipse*[linecolor=palegray](-43.5,0)(12,8)
\psellipse*[linecolor=palegray](+43.5,0)(12,8)
\rput(-43.5,0){?}
\rput(+43.5,0){?}
\psline(-53.5,-31)(-43.5,-34)(-33.5,-31)
\rput[tl](-60,-35){Dual to the $\rm M9_1$}
\psline(+53.5,-31)(+43.5,-34)(+33.5,-31)
\rput[tr](+60,-35){Dual to the $\rm M9_2$}
\psline(+65,-30)(+65,+30)
\psline{>->}(+65,-10)(+65,+10)
\rput[t]{90}(+66,0){$x^{7,8,9}$}
\endpspicture
\eqn\ZthreeBranes
$$
At each end of the $x^6$, the distance $b$ between the \Ori\ orientifold plane and the
outlier {\darkblue D8} brane is less than critical, hence finite string coupling $\lambda$
at the orientifold plane and the 9D gauge symmetry is $\darkblue SO(14)\times U(1)$
rather than $\darkgreen E_8$, \cf\ section~3.
The 7D $\darkred SU(3)$ gauge symmetry follows from three coincident {\darkred D6} branes
at $\bx=\bf0$.
The only non-obvious features of the brane model --- denoted by the gray areas of
 fig.~\ZthreeBranes\ --- are the terminal regions of the {\darkred D6} branes at the
two ends of the $x^6$.

The two terminal regions are related by the $E_8^{(1)}\leftrightarrow E_8^{(2)}$
symmetry of the model, so let us focus on the left terminal.
The existence of twisted hypermultiplets in the vector $\bf 14$ representation of
the $\darkblue SO(14)_1$ group living at $x^6=0$ indicates that at least some of the
{\darkred D6} branes must reach all the way to the orientifold plane.
On the other hand, we saw in section 4.3 that the net number of  {\darkred D6} branes
terminating on an \Ori{} orientifold plane must be even.
Altogether, we have 3 {\darkred D6} branes, which means that 2 of them should terminate
on the orientifold plane at $x^6=0$ while the third {\darkred D6} terminates on the
outlier {\darkblue D8} brane at $x^6=b$:
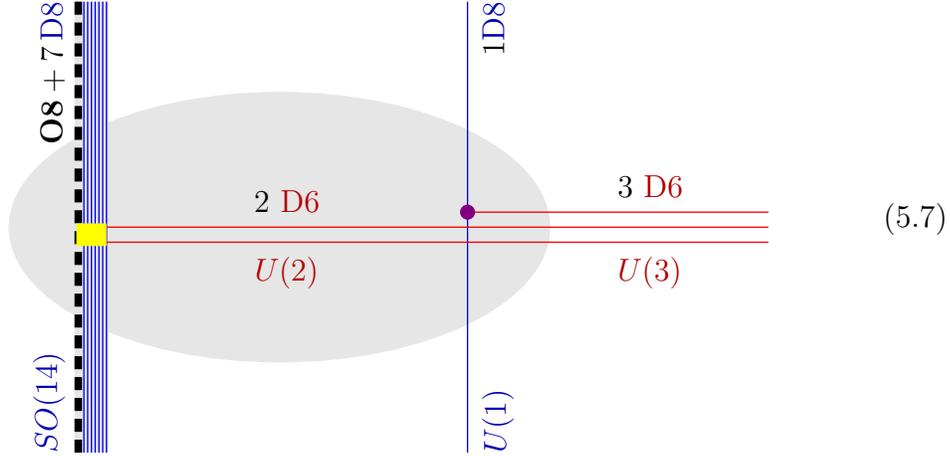
$$
\pspicture[](-60,-30)(+40,+30)
\psellipse*[linecolor=palegray](-25,0)(36,18)
\psline[linecolor=palegray,linewidth=4.5](-50,-30)(-50,+30)
\psline[linecolor=black,linestyle=dashed,linewidth=1](-51.75,-30)(-51.75,+30)
\psline[linecolor=blue](-51.0,-30)(-51.0,+30)
\psline[linecolor=blue](-50.5,-30)(-50.5,+30)
\psline[linecolor=blue](-50.0,-30)(-50.0,+30)
\psline[linecolor=blue](-49.5,-30)(-49.5,+30)
\psline[linecolor=blue](-49.0,-30)(-49.0,+30)
\psline[linecolor=blue](-48.5,-30)(-48.5,+30)
\psline[linecolor=blue](-48.0,-30)(-48.0,+30)
\rput[br]{90}(-53.5,+30){${\bf O8}+7\,{\rm\darkblue D8}$}
\rput[bl]{90}(-53.5,-30){$\darkblue SO(14)$}
\psline[linecolor=blue](0,-30)(0,+30)
\rput[tr]{90}(2,+30){$1\rm\darkblue D8$}
\rput[tl]{90}(2,-30){$\darkblue U(1)$}
\psline[linecolor=red](0,+2)(+40,+2)
\psline[linecolor=red](-50,0)(+40,+0)
\psline[linecolor=red](-50,-2)(+40,-2)
\rput[b](-24,+2){2 \darkred D6}
\rput[t](-24,-4){$\darkred U(2)$}
\rput[bl](+20,+4){3 \darkred D6}
\rput[tl](+20,-4){$\darkred U(3)$}
\pscircle*[linecolor=purple](0,+2){1}
\psline[linecolor=yellow,linewidth=3](-52,-1)(-48,-1)
\endpspicture
\eqn\ZthreeLeft
$$

String theory of the {\darkred D6}--{\darkblue D8} brane junction at $x^6=b$ follows from T-duality
along the lines of \S~4.4.
The outlier {\darkblue D8} brane and the {\darkred D6} brane which terminates on it are
together T-dual to a single curved $\rm\purple D7$ brane at $u=e^{-w}$
(in the coordinates of eq.~\ComplexUW) while the other two {\darkred D6} branes
are T-dual to flat $\rm\darkred D7'$ branes at $u\equiv0$:
$$
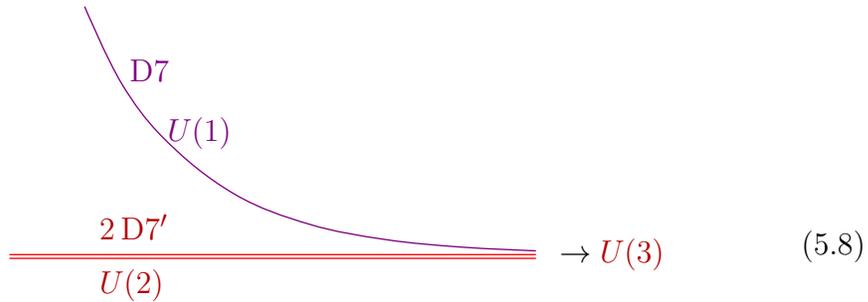
\pspicture[](-10,-10)(70,33)
\psline[linecolor=red](-10,0)(60,0)
\psline[linecolor=red](-10,-0.5)(60,-0.5)
\rput[bl](2,2){$\darkred\rm 2\, D7'$}
\rput[tl](2,-2){$\darkred U(2)$}
\pscurve[linecolor=purple](0,33)(5,22.62)(10,16)(20,8)(30,4)(40,2)(50,1)(60,0.5)
\rput[bl](6,23){\purple D7}
\rput[bl](11,14){$\purple U(1)$}
\rput[Bl](62,-1){${}\to\darkred U(3)$}
\endpspicture
\eqn\ZthreeTdual
$$
The curved {\purple D7} brane carries a $\purple U(1)$ gauge theory;
it is T-dual to the 7D $\darkred U(1)$ for $x^6\ge b$ locked onto a 9D $\darkblue U(1)$
via boundary condition~\LockingBC\ at $x^6=b$.
At the same time, the two flat $\darkred\rm D7'$ branes carry a $\darkred U(2)$ SYM
whose T-dual is obviously a 7D $\darkred U(2)$ SYM which continues from $x^6\ge b$
to $x^6<b$ without anything happening to it at the junction point $x^6=b$.

The new element here are the $\purple7\darkred7'$ open strings whose length
asymptotes to zero for $\Re w\to+\infty$.
The lowest-energy modes of these strings give rise to $\darkred U(3)/[U(1)\times U(2)]$
SYM fields which are massless in the $\Re w\to+\infty$ limit but become massive
for finite $\Re w$ and super-heavy for $\Re w\to-\infty$.
{}From the 8D field theory point of view, we have a SYM with $\darkred U(3)$ gauge
symmetry {\sl spontaneously} broken to $\darkred U(2)\times U(1)$ by an $x$-dependent
VEV of the adjoint scalar field.
In the unitary gauge,
$$
\vev\Phi\ =\ \diag (1,0,0)\times u(w)\
=\ \diag (1,0,0)\times e^{-w}.
\eqn\HiggsVEV
$$
Generally, gradients of scalar VEVs break supersymmetry but holomorphic
VEVs such as~\HiggsVEV\ preserve half of the supercharges.
Consequently, we have two different 2D wave equations for the field modes corresponding
to vector  and hypermultiplets in 6D, namely
$$
\eqalign{
\left[ \nabla^2\,-\,\left|\vev\Phi\right|^2\right]\,(A_\mu,\psi_L)\ &=\
0,\crr
\left|\vev\Phi\right|^2\Bigl(\nabla{1\over\left|\vev\Phi\right|^2}\nabla\,-\,1\Bigr)\,
	(\phi,\psi_R)\ &=\ 0.\cr
}\eqno\eq
$$
In terms of the $(w,w^*)$ coordinates, the massless vector modes satisfy
$$
\left[ 4\,\partder{}{w}\partder{}{w^*}\, -\, e^{-(w+w^*)}\right] (A_\mu,\psi_L)\ =\ 0
\eqno\eq
$$
while the massless hyper modes satisfy
$$
\left[ 4\,\partder{}{w}\partder{}{w^*}\, +\, 2\left(\partder{}{w}+\partder{}{w^*}\right)\
-\, e^{-(w+w^*)}\right] (\phi,\psi_R)\ =\ 0;
\eqno\eq
$$
furthermore, physical wavefunctions should not blow up for ${\Re w\to\pm\infty}$.
These conditions uniquely specify the 2D wavefunctions:
$$
(A_\mu,\psi_L)\ \propto \int_1^\infty\!{dt\over\sqrt{t^2-1}}
	\times\exp\left( -t e^{-\Re w}\right)\
\becomes{\Re w\to+\infty}\ \Re w\ +\ {\rm a\ small\choose constant}
\eqn\VectorSolution
$$
for the massless 6D vector multiplets and
$$
(\phi,\psi_R)\ \propto \int_1^\infty\!{dt\over\sqrt{t^2-1}}
	\times te^{-\Re w}\,\exp\left( -t e^{-\Re w}\right)\
\becomes{\Re w\to+\infty}\ 1
\eqn\HyperSolution
$$
for the massless hypermultiplets.
Note that both wavefunctions describe (the zero energy limit of) unbound motion of a particle
in the semi-infinite $\Re w\gsim 0$ region; there are no bound states.

In T-dual terms, we have a 7D $\darkred U(3)$ SYM living on the three {\darkred D6} branes at
$x^6\ge b$.
At the junction, the $\darkred U(3)$ group is abruptly broken to its $\darkred U(1)\times U(2)$
subgroup while the remaining $\darkred U(3)/[U(1)\times U(2)]$ SYM fields satisfy
reflecting boundary conditions at $x^6=b$.
According to eqs.~\VectorSolution\ and \HyperSolution, the boundary conditions are
Dirichlet for the 8--SUSY vector multiplet fields and Neumann for the hypermultiplet fields.
Furthermore, there are no massless 6D fields localized at the junction.
Although classically the $\darkred6\darkblue8$ open strings have zero length at the junction,
they do not have massless modes because of world-sheet quantum corrections.
Indeed, the $\darkred6\darkblue8$  strings are T-dual to the $\darkred7'\purple7$ strings
giving rise to the $\darkred U(3)/[U(1)\times U(2)]$ SYM fields --- which  have
no {\sl normalizable}  zero modes localized in the junction area.

The above discussion concerns the {\darkred D6}--{\darkblue D8} junction at $x^6=b$.
There is another junction at $x^6=0$ --- denoted by the yellow rectangle in
fig.~\ZthreeLeft\ ---
where two {\darkred D6} branes reach the \Ori{} orientifold plane accompanied by
seven {\darkblue D8} branes.
As explained in \S4.3, at this second junction the $\darkred U(2)$ gauge symmetry is broken to
$\darkred Sp(1)=SU(2)$, the $\darkred\bf 3$ vector multiplets satisfy Neumann boundary conditions
at $x^6=0$ while the $\bf\darkred3$ hypermultiplets satisfy Dirichlet conditions
and the $\darkblue8\darkred6$ open strings give rise to localized 6D massless particles
forming  half-hypermultiplets in the $({\darkred\bf2},{\darkblue\bf14})$ representation of the
${\darkred SU(2)}\times{\darkblue SO(14)}$ gauge group.

The two junctions at $x^6=b$ and $x^6=0$ are distinct in the \tI/D6  theory, but in
the Ho\v rava--Witten theory $b\to0$ and the two junctions collapse into a single
\I~intersection plane.
The local physics on this plane is simply the net effect of the two junctions of the
dual \tI/D6 model translated into HW orbifold terms:
\pointbegin
The 7D $\darkred SU(3)$ gauge symmetry (we discard the $U(1)$ center of the $\darkred U(3)$
as explained in \S4.2) is broken to $\darkred SU(2)\times U(1)$.
Furthermore, the $\darkred U(1)\subset SU(3)$ and the 10D $\darkblue U(1)\subset E_8$
are broken to the diagonal $\purple U(1)$.
The net gauge symmetry on the intersection plane is therefore
$$
G^{\rm 6D}_{\rm local}\
=\ {\darkred SU(2)}\times{\purple U(1)}\times{\darkblue SO(14)}
\eqn\ZthreeG
$$
\point %2
The local  6D massless fields at the \I\
comprise a half-hypermultiplet in the $({\darkred\bf2},{\purple0},{\darkblue\bf14})$
representation of~\ZthreeG.
\point %3
The 7D SYM fields have boundary conditions depending on their
${\darkred SU(2)}\times{\purple U(1)}$ and 8--SUSY quantum numbers.
Decomposing the $\darkred SU(3)$ adjoint $\bf\darkred 8$ as $({\bf\darkred3},{\purple0})+
({\bf\darkred1},{\purple0})+({\bf\darkred2},{\purple\pm1})$ and 16--SUSY vector multiplet
as 8--SUSY $\rm vector+hyper$-multiplets, we have
$$
\def\qqq#1#2{$({\darkred\bf #1},{\purple #2})$}
\def\jjj (#1,#2) #3 #4 {\qqq{#1}{#2} & #3 & \qqq{#1}{#2} & #4 \cr }
\vcenter{\ialign{%
    \hfil # vector:\enspace & # b.c.,\hfil\qquad & \hfil # hyper:\enspace & # b.c.,\hfil\cr
    \jjj (3,0) Neumann Dirichlet
    \jjj (1,0) Locking Neumann
    \jjj (2,\pm1) Dirichlet {\omit Neumann b.c.\hfil}
    }}\qquad
\eqn\ZthreeBC
$$

Note that the boundary conditions \ZthreeBC\ are rather complicated compared to the
models presented in ref.~[\KSTY] --- where all vector multiplets at the same boundary
had similar boundary conditions and ditto for the hypermultiplets.
Furthermore, the specific gauge symmetry breaking~\ZthreeG\ following from
the conditions~\ZthreeBC\ is totally counter-intuitive from the heterotic point of view;
indeed, the unbroken $\darkred SU(2)$ factor of the local symmetry~\ZthreeG\ does not exist
in the heterotic orbifold's spectrum and the twisted states' charges~\ZthreeTw\
do not indicate any hidden $SU(2)$ symmetry either.
A bit later in this section, we shall brane engineer the removal of this unwanted $\darkred SU(2)$
from the massless spectrum of the 6D theory, but first we would like to show that its existence
as a local symmetry at the \I\ is crucial to the local anomaly cancellations.

Indeed, the locally visible charged chiral fields at the intersection plane
weighed by their contributions to the local anomaly (\cf\ section~2) comprise
$$
\def\spect#1#2#3{({\darkred\bf #1},{\purple #2},{\darkblue\bf #3})} 
\eqalign{
Q_6\ &
=\ \half\spect{2}{0}{14} ,\cr
Q_7\ &
=\ {1\over2}\left[ 2\spect{2}{+1}{1}{1}\ +\ \spect{1}{0}{1}\ -\ \spect{3}{0}{1} \right],\cr
Q_{10}\ &
=\ {1\over9}\left[ \spect{1}{+\half}{64}\ +\ \spect{1}{-1}{14}\
	-\ \spect{1}{0}{91}\ -\ \spect{1}{0}{1}\right],\cr
}\eqn\ZthreeLocalSpectrum
$$
and it's a matter of (boring) algebra to verify conditions~\AnomalyRiiii\ and \AnomalyOther\
of complete anomaly cancellation.
Specifically, the  magnetic charge $g$ vanishes for the symmetric orbifold, thus
$$
\mathop{\rm dim}(Q=Q_6+Q_7+Q_{10})=\ 0\ +\ \coeff{121}{9}
\eqno\eq
$$
which provides for cancellation of the irreducible $\tr(R^4)$ anomaly, and
$$
\!\eqalign{
{\cal A}'\ &
=\ \coeff23 \Tr_Q(F^4)\ -\ \coeff16\tr(R^2)\times\Tr_Q(F^2)\ +\ \coeff1{18}(\tr(R^2))^2\cr
&=\ 0\ +\, \left( \tr F^2_{\darkblue SO(14)}\ +\ 2F^2_{\purple U(1)}\ -\ \half\tr R^2\right)\times
    \left( \tr F^2_{\rm\darkred SU(2)}\ +\ \coeff43 F^2_{\purple U(1)}\ -\coeff19 \tr R^2\right)\cr
}\!\eqn\ZthreeOLAnomaly
$$
which lets the rest of the one-loop anomaly cancel against
the inflow and intersection anomalies.
\foot{Actually, the inflow anomaly vanishes for this model because of $g=0$.}

It is very important that the $(\tr F^2_{\darkred SU(2)})^2$ term cancels out of the one-loop
anomaly~\ZthreeOLAnomaly\ because the inflow and intersection anomalies
 cannot possibly cancel un-mixed anomalies of gauge symmetries of purely 7D origins.
Such cancellation requires both hyper- and vector multiplets with non-trivial $\darkred SU(2)$
quantum numbers to be locally visible at the intersection plane --- and of course the
vector multiplets are the 7D fields with Neumann boundary conditions which are responsible
for the $\darkred SU(2)$'s existence in the first place.
Now suppose we did not have the unbroken $\darkred SU(2)$ but only its $\darkred U(1)$ subgroup
(which eventually mixes with the $\darkblue U(1)\subset E_8^{(2)}$ at the other end of $x^6$).
In such a hypothetical model, two out of $\bf\darkred3$ vector multiplets would have Dirichlet
rather than Neumann boundary condition and the only vector multiplet visible at the
intersection plane would be the generator of the $\darkred U(1)$ --- which is neutral.
On the other hand, there would be plenty of locally present $\darkred U(1)$--charged hypermultiplets
--- and we would be stuck with the resulting $F^4_{\darkred U(1)}$ local anomaly we would have
no way of canceling.

In other words, cancellation of the $F^4_{\darkred U(1)}$ local anomaly requires local presence
of $\darkred U(1)$--charged vector multiplets --- and hence embedding the $\darkred U(1)$ into
a locally unbroken non-abelian symmetry group.
For the $\IZ_3$ model in question, local anomaly cancellation favors local spectrum~\ZthreeLocalSpectrum\
and hence boundary condition~\ZthreeBC.
Such boundary conditions are very strange from the heterotic point of view;
discovering them without the benefit of the dual \tI/D6 model would have been rather difficult.

Our next  task is therefore  to brane engineer the correct twisted spectrum~\ZthreeTw\ of
the heterotic orbifold; in particular, we need to break the un-observed
$\darkred SU(2)$ symmetry.
In the dual \tI/D6 model, let us shift our attention from the left terminus~\ZthreeLeft\
of the three {\darkred D6} branes to the big picture~\ZthreeBranes.
Thanks to  $E_8^{(1)}\leftrightarrow E_8^{(2)}$ symmetry of the model, the right
terminus of the {\darkred D6} branes also looks like~\ZthreeLeft\ (modulo $x^6\to L-x^6$),
but this leaves open the question whether the {\darkred D6} branes terminating on the
outlier {\darkblue D8} branes on the left and on the right are two ends of the same {\darkred D6}
brane or two distinct {\darkred D6} branes.
Thus we have two distinct brane models, namely
$$
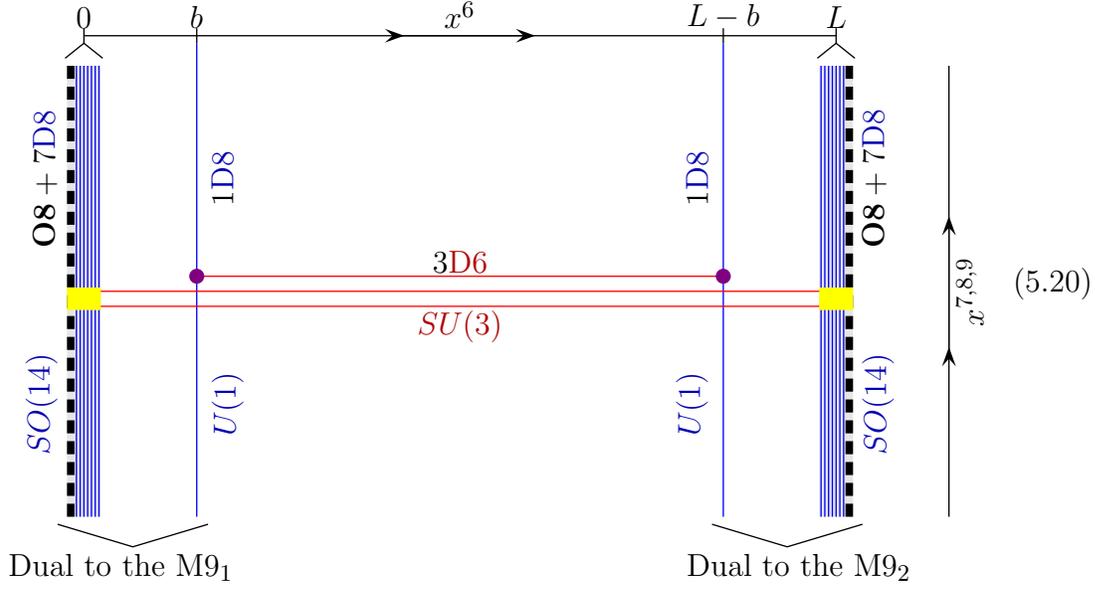
\pspicture[](-60,-40)(+70,40)
\psline(-50,+34)(+50,+34)
\psline(-50,+33)(-50,+35)	\rput[b](-50,+35){$0$}
\psline(-35,+33)(-35,+35)	\rput[b](-35,+35){$b$}
\psline(+50,+33)(+50,+35)	\rput[b](+50,+35){$L$}
\psline(+35,+33)(+35,+35)	\rput[b](+35,+35){$L-b$}
\rput[b](0,+35){$x^6$}
\psline{>->}(-10,+34)(+10,+34)
\psline[linecolor=palegray,linewidth=4.5](-50,-30)(-50,+30)
\psline[linecolor=black,linestyle=dashed,linewidth=1](-51.75,-30)(-51.75,+30)
\psline[linecolor=blue](-51.0,-30)(-51.0,+30)
\psline[linecolor=blue](-50.5,-30)(-50.5,+30)
\psline[linecolor=blue](-50.0,-30)(-50.0,+30)
\psline[linecolor=blue](-49.5,-30)(-49.5,+30)
\psline[linecolor=blue](-49.0,-30)(-49.0,+30)
\psline[linecolor=blue](-48.5,-30)(-48.5,+30)
\psline[linecolor=blue](-48.0,-30)(-48.0,+30)
\psline(-52.5,+31)(-50,+33)(-47.5,+31)
\rput[b]{90}(-53.5,+15){${\bf O8}+7{\rm\darkblue D8}$}
\rput[b]{90}(-53.5,-15){$\darkblue SO(14)$}
\psline[linecolor=blue](-35,-30)(-35,+33)
\rput[t]{90}(-33,+15){$1\rm\darkblue D8$}
\rput[t]{90}(-33,-15){$\darkblue U(1)$}
\psline[linecolor=palegray,linewidth=4.5](+50,-30)(+50,+30)
\psline[linecolor=black,linestyle=dashed,linewidth=1](+51.75,-30)(+51.75,+30)
\psline[linecolor=blue](+51.0,-30)(+51.0,+30)
\psline[linecolor=blue](+50.5,-30)(+50.5,+30)
\psline[linecolor=blue](+50.0,-30)(+50.0,+30)
\psline[linecolor=blue](+49.5,-30)(+49.5,+30)
\psline[linecolor=blue](+49.0,-30)(+49.0,+30)
\psline[linecolor=blue](+48.5,-30)(+48.5,+30)
\psline[linecolor=blue](+48.0,-30)(+48.0,+30)
\psline(+52.5,+31)(+50,+33)(+47.5,+31)
\rput[t]{90}(+53.5,+15){${\bf O8}+7{\rm\darkblue D8}$}
\rput[t]{90}(+53.5,-15){$\darkblue SO(14)$}
\psline[linecolor=blue](+35,-30)(+35,+33)
\rput[b]{90}(+33,+15){$1\rm\darkblue D8$}
\rput[b]{90}(+33,-15){$\darkblue U(1)$}
\psline[linecolor=red](-52,-2)(+52,-2)
\psline[linecolor=red](-52,0)(+52,0)
\psline[linecolor=red](-35,+2)(+35,+2)
\pscircle*[linecolor=purple](-35,+2){1}
\pscircle*[linecolor=purple](+35,+2){1}
\psline[linecolor=yellow,linewidth=3](-52.25,-1)(-47.75,-1)
\psline[linecolor=yellow,linewidth=3](+52.25,-1)(+47.75,-1)
\rput[b](0,+2.5){$3\darkred\rm D6$}
\rput[t](0,-2.5){$\darkred SU(3)$}
\psline(-53.5,-31)(-43.5,-34)(-33.5,-31)
\rput[tl](-60,-35){Dual to the $\rm M9_1$}
\psline(+53.5,-31)(+43.5,-34)(+33.5,-31)
\rput[tr](+60,-35){Dual to the $\rm M9_2$}
\psline(+65,-30)(+65,+30)
\psline{>->}(+65,-10)(+65,+10)
\rput[t]{90}(+66,0){$x^{7,8,9}$}
\endpspicture
\eqn\ZthreeBad
$$
\par\noindent
and
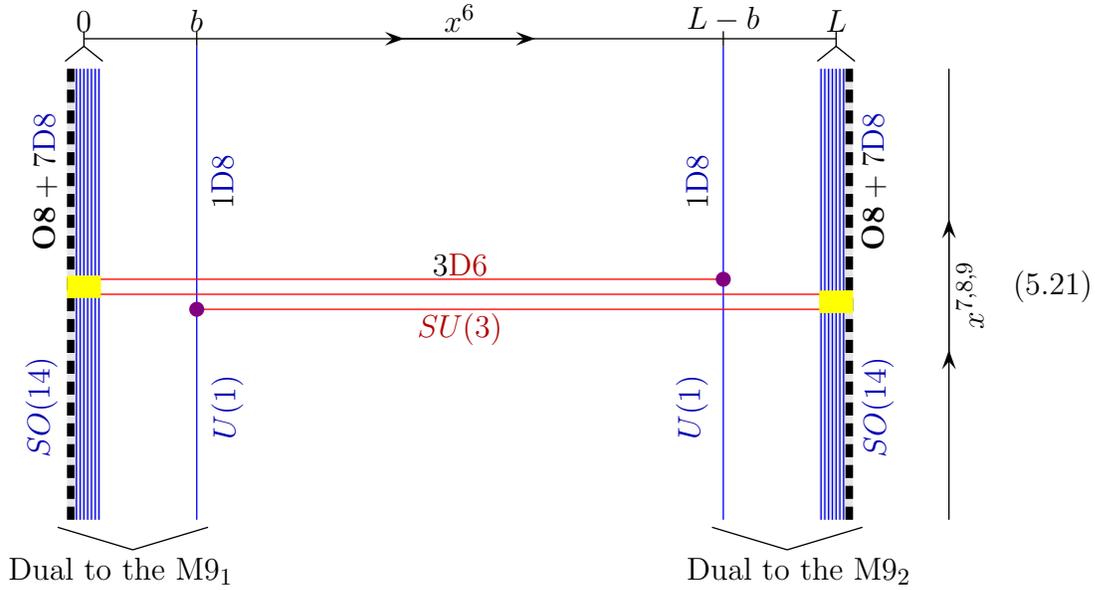
$$
\pspicture[](-60,-40)(+70,40)
\psline(-50,+34)(+50,+34)
\psline(-50,+33)(-50,+35)	\rput[b](-50,+35){$0$}
\psline(-35,+33)(-35,+35)	\rput[b](-35,+35){$b$}
\psline(+50,+33)(+50,+35)	\rput[b](+50,+35){$L$}
\psline(+35,+33)(+35,+35)	\rput[b](+35,+35){$L-b$}
\rput[b](0,+35){$x^6$}
\psline{>->}(-10,+34)(+10,+34)
\psline[linecolor=palegray,linewidth=4.5](-50,-30)(-50,+30)
\psline[linecolor=black,linestyle=dashed,linewidth=1](-51.75,-30)(-51.75,+30)
\psline[linecolor=blue](-51.0,-30)(-51.0,+30)
\psline[linecolor=blue](-50.5,-30)(-50.5,+30)
\psline[linecolor=blue](-50.0,-30)(-50.0,+30)
\psline[linecolor=blue](-49.5,-30)(-49.5,+30)
\psline[linecolor=blue](-49.0,-30)(-49.0,+30)
\psline[linecolor=blue](-48.5,-30)(-48.5,+30)
\psline[linecolor=blue](-48.0,-30)(-48.0,+30)
\psline(-52.5,+31)(-50,+33)(-47.5,+31)
\rput[b]{90}(-53.5,+15){${\bf O8}+7{\rm\darkblue D8}$}
\rput[b]{90}(-53.5,-15){$\darkblue SO(14)$}
\psline[linecolor=blue](-35,-30)(-35,+33)
\rput[t]{90}(-33,+15){$1\rm\darkblue D8$}
\rput[t]{90}(-33,-15){$\darkblue U(1)$}
\psline[linecolor=palegray,linewidth=4.5](+50,-30)(+50,+30)
\psline[linecolor=black,linestyle=dashed,linewidth=1](+51.75,-30)(+51.75,+30)
\psline[linecolor=blue](+51.0,-30)(+51.0,+30)
\psline[linecolor=blue](+50.5,-30)(+50.5,+30)
\psline[linecolor=blue](+50.0,-30)(+50.0,+30)
\psline[linecolor=blue](+49.5,-30)(+49.5,+30)
\psline[linecolor=blue](+49.0,-30)(+49.0,+30)
\psline[linecolor=blue](+48.5,-30)(+48.5,+30)
\psline[linecolor=blue](+48.0,-30)(+48.0,+30)
\psline(+52.5,+31)(+50,+33)(+47.5,+31)
\rput[t]{90}(+53.5,+15){${\bf O8}+7{\rm\darkblue D8}$}
\rput[t]{90}(+53.5,-15){$\darkblue SO(14)$}
\psline[linecolor=blue](+35,-30)(+35,+33)
\rput[b]{90}(+33,+15){$1\rm\darkblue D8$}
\rput[b]{90}(+33,-15){$\darkblue U(1)$}
\psline[linecolor=red](-52,+2)(+35,+2)
\psline[linecolor=red](-52,0)(+52,0)
\psline[linecolor=red](-35,-2)(+52,-2)
\pscircle*[linecolor=purple](-35,-2){1}
\pscircle*[linecolor=purple](+35,+2){1}
\psline[linecolor=yellow,linewidth=3](-52.25,+1)(-47.75,+1)
\psline[linecolor=yellow,linewidth=3](+52.25,-1)(+47.75,-1)
\rput[b](0,+2.5){$3\darkred\rm D6$}
\rput[t](0,-2.5){$\darkred SU(3)$}
\psline(-53.5,-31)(-43.5,-34)(-33.5,-31)
\rput[tl](-60,-35){Dual to the $\rm M9_1$}
\psline(+53.5,-31)(+43.5,-34)(+33.5,-31)
\rput[tr](+60,-35){Dual to the $\rm M9_2$}
\psline(+65,-30)(+65,+30)
\psline{>->}(+65,-10)(+65,+10)
\rput[t]{90}(+66,0){$x^{7,8,9}$}
\endpspicture
\eqn\ZthreeGood
$$
In each model, the 7D $\darkred SU(3)$ gauge symmetry is broken down to an
$\darkred SU(2)\times U(1)$ subgroup at both purple junctions ($x^6=b$ and $x^6=(L-b)$),
but there is one crucial difference:
In the first model~\ZthreeBad, the two junctions preserve {\it the same} $\darkred SU(2)\times
U(1)\subset SU(3)$ at both ends, hence the same $\bf\darkred 3$ 7D gauge fields have
Neumann boundary conditions at both ends of $x^6$ and therefore zero modes.
Consequently, the 6D effective theory contains massless $\darkred SU(2)$ gauge fields
of purely 7D origin.
In the  dual heterotic terms, this means a {\it non-perturbative}
6D $\darkred SU(2)$ SYM at each fixed plane of the $T^4/\IZ_3$
orbifold in addition to the perturbative gauge fields in the untwisted sector.
This is a very interesting heterotic string model in its own right, but unfortunately
it is quite different from the perturbative $T^4/\IZ_3$ orbifold we started from.

By contrast, in the second model~\ZthreeGood\ the two purple junctions preserve
two {\sl different} $\darkred SU(2)\times U(1)$ subgroups of the $\darkred SU(3)$.
\foot{%
    To be precise, the two subgroups are equivalent via an $SU(3)$ isomorphism $W\neq0$.
    In gauge invariant terms, $W=\mathop{\hbox{\rm P exp}}\Bigl(\int_b^{L-b}\!A_6 dx^6\Bigr)
    \times{}$a non-trivial element of the Weyl group of the $SU(3)$.
    }
Because of this mis-alignment, the 3 vector fields with Neumann boundary
conditions at $x^6=b$ have Dirichlet or locking boundary conditions at $x^6=(L-b)$
and vice verse, hence no zero modes and no purely non-perturbative vector multiplets
in the twisted sector of the heterotic orbifold.
The abelian $\darkred U(1)\times U(1)\subset SU(3)$ vector fields which mix with the
9D abelian fields according to eq.~\SUiiiUiUi\ belong to the overlap of the
two surviving subgroups of the $\darkred SU(3)$ at each junction,
$$
\left[ {\darkred SU(2)\times U(1)}\right]_1\,\cap\,\left[ {\darkred SU(2)\times U(1)}\right]_2\
=\ {\darkred U(1)\times U(1)}.
\eqn\SUiiiOverlap
$$

To keep track of the $\darkred U(1)\times U(1)$ charges of various 7D fields we need
two orthogonal commuting generators of the $\darkred SU(3)$.
The $SU(3)\to[SU(2)\times U(1)]_1\to U(1)\times U(1)$ chain of symmetry breaking
suggests
$$
T_1\ =\ \diag (+\half,-\half,0)\qquad{\rm and}\quad
Y_1\ =\ \diag (+\coeff13,+\coeff13,-\coeff23)
\eqn\FirstPair
$$
while the $SU(3)\to[SU(2)\times U(1)]_2\to U(1)\times U(1)$ chain suggests
$$
T_2\ =\ \diag (0,+\half,-\half)\qquad{\rm and}\quad
Y_2\ =\ \diag (-\coeff23,+\coeff13,+\coeff13);
\eqn\SecondPair
$$
the two sets of charges are related to each other according to
$$
\left\{\eqalign{
    T_1\ &=\ -\half T_2\ - \coeff34 Y_2\cr
    Y_1\ &=\ +T_2\ -\ \half Y_2\cr
    }\right\}\
\Leftrightarrow\
\left\{\eqalign{
    T_2\ &=\ -\half T_1\ + \coeff34 Y_1\cr
    Y_2\ &=\ -T_1\ -\ \half Y_1\cr
    }\right\} .
\eqn\ZthreeConversion
$$
For completeness sake, the table below lists both sets of charges as well as boundary
conditions for the all the 7D SYM fields.
$$
\vcenter{\ialign{
    \hfil\vrule width 0pt height 15pt depth 5pt \vrule #&
    \enspace\hfil $(#)$\hfil\enspace & \vrule #\hfil &
    \enspace\hfil $(#)$\hfil\enspace & \vrule #\hfil &
    \enspace\hfil\frenchspacing(#\unskip)\hfil\enspace &\vrule #\hfil &
    \enspace\hfil\frenchspacing(#\unskip)\hfil\enspace &\vrule #\hfil \cr
\noalign{\hrule}
& \multispan3 \hfil Charges\hfil && \multispan3 \hfil Boundary Conditions\hfil &\cr
\noalign{\hrule}
& T_1,Y_1 && T_2,Y_2 && \omit\hfil 8--SUSY vector\hfil && \omit\hfil hyper\hfil &\cr
\noalign{\hrule}
& \pm1,0 && \mp\half,\mp1 && Neumann,Dirichlet && Dirichlet,Neumann &\cr
\noalign{\hrule}
& \pm\half,\pm1 && \pm\half,\mp1 && Dirichlet,Dirichlet && Neumann,Neumann &\cr
\noalign{\hrule}
& \mp\half,\pm1 && \pm1,0 && Dirichlet,Neumann && Neumann,Dirichlet &\cr
\noalign{\hrule}
& 0,0 && 0,0 && locking,Neumann && Neumann,Dirichlet &\cr
\noalign{\hrule}
& 0,0 && 0,0 && Neumann,locking && Dirichlet,Neumann &\cr
\noalign{\hrule}
}}
\eqn\ZthreeTable
$$
As promised, none of the vector multiplets have Neumann--Neumann boundary conditions.
On the other hand, two hypermultiplets with similar charges
\foot{%
    For a hypermultiplet, the overall sign of all its charges is a matter of convention.
    Hence, two hypermultiplets with exactly opposite charges are equivalent to
    two hypermultiplets with identical charges.
    }
have Neumann--Neumann conditions and hence zero modes.
In the dual heterotic terms, these two zero modes manifest themselves as twisted
hypermultiplets charged with respect to $\purple U(1)\times U(1)$
(thanks to the 10D/7D abelian field mixing~\SUiiiUiUi) but singlets with respect to
$\darkblue SO(14)\times SO(14)$.
And indeed the twisted spectrum~\ZthreeTw\ of the perturbative heterotic orbifold
contains precisely two such singlets per fixed point of the $T^4/\IZ_3$.

Altogether, the hypermultiplets localized at $\bx=\bf0$ in the brane model~\ZthreeGood\
comprise one $\bf\darkblue14$ of the ${\darkblue SO(14)}_1$ at $x^6=0$, one $\bf\darkblue14$ of the
${\darkblue SO(14)}_2$ at $x^6=L$, and two singlets from zero modes spanning the whole $x^6$.
\item\bullet
The $\darkblue({\bf14},{\bf1})$ fields have abelian charges $(T_1=\half,Y_1=0)$
(from $\half({\darkred2},{\purple0})$ of the $[{\darkred SU(2)}\times {\purple U(1)}]_1$ local symmetry
at $x^6=0$) and hence $(T_2=-\coeff14,Y_2=-\coeff12)$.
\item\bullet
Likewise, the $\darkblue({\bf1},{\bf14})$ fields have abelian charges $(T_2=\half,Y_2=0)$ and hence
$(T_1=-\coeff14,Y_1=+\coeff12)$.
\item\bullet
Finally, the two $\darkblue({\bf1},{\bf1})$ singlet fields have $(T_1=+\half,Y_1=+1)$ and
$(T_2=+\half,Y_2=-1)$.
\par\noindent
Comparing this spectrum to the heterotic twisted spectrum~\ZthreeTw\ we see full agreement,
provided we identify the abelian $\purple U(1)\times U(1)$ charges according to
$$
\eqalign{
C_1^{\rm heterotic}\ &
=\ C^{\rm HW}_{U(1)\subset E_8^{(1)}}\
    + \sum_{\rm fixed\atop planes}\left[ Y_1\,-\,\coeff23 T_1\ =\ \coeff43 T_2\right] ,\cr
C_2^{\rm heterotic}\ &
=\ C^{\rm HW}_{U(1)\subset E_8^{(2)}}\ 
    + \sum_{\rm fixed\atop planes}\left[ -Y_2\,-\,\coeff23 T_2\ =\ \coeff43 T_1\right] .\cr
}\eqn\ZthreeCharges
$$
Weirdly, the observed abelian symmetries~\SUiiiUiUi\ of the heterotic $\IZ_3$ orbifold
mix the 10D abelian charge living on the left end of the world with the 7D charge
that happens to be part of an unbroken non-abelian symmetry ${\darkred SU(2)}_2$ at the right
end of the world and vice verse.
In section~7 we shall see  similar coincidences happening for other orbifold models;
alas, we have no explanation for this phenomenon.
Instead, we have an independent confirmation of the charges~\ZthreeCharges\
through the gauge couplings of the model.

Specifically, the $(\coeff43 T_2,\coeff43 T_1)$ 7D abelian charges at each fixed plane come with
gauge couplings
$$
\eqalign{
{1\over g^2_{7D}[{\darkred U(1)\times U(1)}]}\ &
=\ {1\over g^2[{\darkred SU(3)}]}\times\pmatrix{
	\tr(\coeff43 T_2)^2 & \tr((\coeff43 T_2)(\coeff43 T_1)) \cr
	\tr((\coeff43 T_2)(\coeff43 T_1)) & \tr(\coeff43 T_1)^2 \cr
	}^{\footsymbolgen}\crr
&=\ {1\over g^2[{\darkred SU(3)}]}\times\pmatrix{
	 \coeff{16}{9} & -\coeff89 \cr -\coeff 89 & \coeff{16}{9} \cr } ,\cr
}\eqno\eq
$$
\vfootnote{\footsymbol}{Note normalization $\tr\equiv2\Tr_{\bf 3}\,$.}
hence the observed abelian charges~\ZthreeCharges\ have
$$
{1\over g^2[{\purple U(1)\times U(1)}]}\
=\ {2\over g^2[{\darkblue E_8\times E_8}]}\ +\ {9\over g^2[{\darkred SU(3)}]}\times
\pmatrix{\coeff{16}{9} & -\coeff89 \cr -\coeff 89 & \coeff{16}{9} \cr } 
\eqno\eq
$$
where the factor 9 in the second term stems from 9 fixed planes of the $T^4/\IZ_3$ orbifold.
(The factor 2 in the first term comes from the normalization of the $\darkblue U(1)\subset E_8$
charges, ${1\over30}\Tr_{\bf 248}(C_{U(1)\subset E_8})^2=2$.)
The two $E_8$ couplings are equal for the symmetric $\IZ_3$ orbifold (12 instantons in each $E_8$
hence $k_1=k_2=0$), thus in terms of the coefficients $v,\tilde v$ in eq.~\GaugeCouplings\
the abelian charges have
$$
v[U(1)\times U(1)]\ =\ \pmatrix{ 2 & 0\cr 0 & 2\cr},\qquad
\tilde v[U(1)\times U(1)]\ =\ \pmatrix{ 16 & -8 \cr -8 & 16 \cr } .
\eqn\ZthreeAbelianVtV
$$
On the other hand, for any perturbative heterotic model in 6D, the coefficients $v,\tilde v$ follow
from the model's massless spectrum via factorization of the net 6D anomaly polynomial according
to eq.~\NetAnomalyFactorization.
Evaluating and factorizing the  anomaly polynomial for the symmetric $T^4/\IZ_3$ orbifold model
is a straightforward albeit tedious exercise which eventually yields
$$
\eqalign{
v[SO(14)_1]\ =\ v[SO(14)_2]\ =\ 1,\qquad &
v[U(1)\times U(1)]\ =\ {\Tenpoint\pmatrix{2 & 0 \cr 0 & 2 \cr }} ,\cr
\tilde v[SO(14)_1]\ =\ \tilde v[SO(14)_2]\ =\ 0,\qquad &
\tilde v[U(1)\times U(1)]\ =\ {\Tenpoint\pmatrix{16 & -8 \cr -8 & 16 \cr }} ,\cr
}\eqn\ZthreeVtV
$$
in full agreement with eq.~\ZthreeAbelianVtV\ based on the specific charges~\ZthreeCharges.

To summarize, we have a HW description of the symmetric $T^4/\IZ_3$ heterotic orbifold which
passes all the consistency conditions, the twisted spectrum, the gauge couplings, the local
anomalies, the works.
Furthermore,
this description is based on the \tI/D6 brane model~\ZthreeGood\ which explains all the
\M ysterious phenomena at the \I~intersection planes.
Indeed, the fairly straightforward logic of  brane engineering leads directly to the correct
HW solution --- which looks weird but works well.

%%%%%%%%%%%%%%%%%%%%%%%%%%%%%%%%%%%%%%%%%%%%%%%%%%%%%%%%%%%%%%%%%%%%%%%%%%%%%%%%%
%
\section{A $\IZ_4$ Model}

Our next model is a $T^4/\IZ_4$ heterotic orbifold in which the first $E_8^{(1)}$ is broken to
$SO(12)\times SU(2)\times U(1)$ (lattice shift vector $\delta_1=(-\coeff34,\coeff14,0,0,0,0,0,0)$)
while the second $E_8^{(2)}$ is broken to $SO(16)$ (shift vector $\delta_2=(1,0,0,0,0,0,0,0)$).
In terms of the unbroken subgroups, the $\alpha_1:\IZ_4\mapsto E_8^{(1)}$ twist acts according to
$$
\def\qqq(#1,#2,#3){({\bf #1},{\bf #2},#3)}
\eqalign{
\alpha_1({\bf248})\ =\ {}&
+[\qqq(66,1,0)+\qqq(1,3,0)+(1,1,0)]\ -\ [\qqq(32',2,0)+\qqq(1,1,\pm2)]\cr
&+\ i\,[\qqq(32,1,+1)+\qqq(12,2,-1)]\ -\ i\,[\qqq(32,1,-1)+\qqq(12,2,+1)].\cr
}\eqn\ZfourSOxiiUii
$$
while the $\alpha_2:\IZ_4\mapsto E_8^{(2)}$ twist acts as
$$
\alpha_2({\bf 248}={\bf120}+{\bf128})\ =\ +({\bf120})\ -\ ({\bf128}) .
\eqno\eq
$$
Note  $(\alpha_2)^2=1$,
hence all massless hypermultiplets in the untwisted and the doubly-twisted sectors
are singlets with respect to the $SO(16)\subset E_8^{2)}$.
Altogether, the massless spectrum of this model comprises:
\item\bullet
In the untwisted sector:
$\rm SUGRA + 1$~tensor multiplet (the dilaton);
190 vector multiplets in the adjoint representation of the
$$
G\ =\ [SO(12)\times SU(2)\times U(1)]\times SO(16);
\eqn\ZfourG
$$
2 moduli and 56 charged hypermultiplets,
$$
H_0\ =\ ({\bf12},{\bf2},+1;{\bf1})\ +\ ({\bf32},{\bf1},-1;{\bf1})\ +\ 2M.
\eqn\ZfourUntw
$$
\item\bullet
In the singly-twisted sector:
32 charged hypermultiplets for each of the 4 $\IZ_4$ fixed points on the $T^4$,
$$
H_1\ =\ 4({\bf1},{\bf2},-\half;{\bf16}).
\eqn\ZfourTwOne
$$
\item\bullet
In the doubly-twisted sector:
$$
H_2\ =\ 6\times\half({\bf32},{\bf1},0;{\bf1})\ +\ 10\times\half({\bf12},{\bf2},0;{\bf1})\
+\ 32({\bf1},{\bf1},+1;{\bf1}).
\eqn\ZfourTwTwo
$$
\par\noindent
In terms of individual $\FP6$ fixed planes, the orbifold has 6 $\IZ_2$ fixed planes where
the $E_8^{(1)}$ is broken to $[E_7\supset SO(12)\times SU(2)]\times[SU(2)\supset U(1)]$,
the $E_8^{(2)}$ remains unbroken and the twisted hypermultiplets comprise $\half({\bf56},{\bf1};{\bf1})
+2({\bf1},{\bf2};{\bf1})$,
plus 4 $\IZ_4$ fixed planes each having gauge symmetry~\ZfourG\ and twisted hypermultiplets
$$
H_{\rm tw}\ 
=\ ({\bf1},{\bf2},-\half;{\bf16})\ +\ \half({\bf12},{\bf2},0;{\bf1})\ +\ 2({\bf1},{\bf1},+1;{\bf1}).
\eqn\ZfourHtw
$$
\par

In the HW picture, the twisted states' charges~\ZfourHtw\ require the $SU(2)\times U(1)$ gauge fields
to span the $x^6$ dimension between the end-of-the-world M9 branes along the $\FP6$ fixed planes, thus
$$
{\purple SU(2)\times U(1)}_{\rm net}\ =\ \diag \left[
\left({\darkblue SU(2)\times U(1)}\right)_{\rm 10D}\times
\smash{\prod_{\IZ_4\,\rm fixed\atop planes}}\vphantom{\prod}
\left({\darkred SU(2)\times U(1)}\right)_{\rm 7D}\right].
\eqn\ZfourGaugeMixing
$$
(The 7D gauge fields on the $\IZ_2$ fixed planes do not participate in this mixing, \cf\
discussion of the $[E_7\times SU(2)]\times E_8^{(2)}$ model in ref.~[\KSTY].)
The actual 7D gauge symmetry on a $\IZ_4$ fixed plane is of course $\darkred SU(4)$;
somehow, we need to break it down to a non-maximal $\darkred SU(2)\times U(1)$ subgroup, then
impose locking boundary conditions on the 7/10~D $\purple SU(2)\times U(1)$ fields at
the $\I_1$ intersection.

Again, the solution comes courtesy of the dual \tI/D6 brane model:
$$
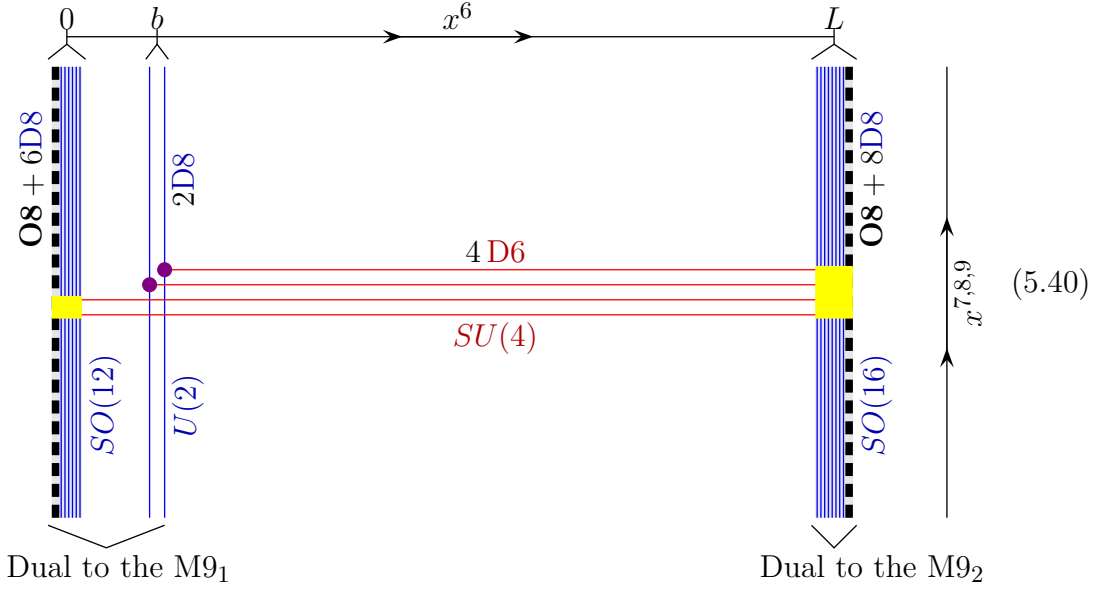
\pspicture[](-20,-40)(110,40)
\psline(-12,+34)(+90,+34)
\psline(-12,+33)(-12,+35)	\rput[b](-12,+35){$0$}
\psline(0,+33)(0,+35)		\rput[b](0,+35){$b$}
\psline(+90,+33)(+90,+35)	\rput[b](+90,+35){$L$}
\rput[b](+40,+35){$x^6$}
\psline{>->}(+30,+34)(+50,+34)
\psline[linecolor=palegray,linewidth=4](-12,-30)(-12,+30)
\psline[linecolor=black,linestyle=dashed,linewidth=1](-13.5,-30)(-13.5,+30)
\psline[linecolor=blue](-12.75,-30)(-12.75,+30)
\psline[linecolor=blue](-12.25,-30)(-12.25,+30)
\psline[linecolor=blue](-11.75,-30)(-11.75,+30)
\psline[linecolor=blue](-11.25,-30)(-11.25,+30)
\psline[linecolor=blue](-10.75,-30)(-10.75,+30)
\psline[linecolor=blue](-10.25,-30)(-10.25,+30)
\psline(-14.5,+31)(-12,+33)(-9.5,+31)
\rput[b]{90}(-15,+15){${\bf O8}+6{\rm\darkblue D8}$}
\rput[t]{90}(-9,-15){$\darkblue SO(12)$}
\psline[linecolor=blue](-1,-30)(-1,+30)
\psline[linecolor=blue](+1,-30)(+1,+30)
\psline(-1.5,+31)(0,+33)(+1.5,+31)
\rput[t]{90}(+2,+15){$2\rm\darkblue D8$}
\rput[t]{90}(+2,-15){$\darkblue U(2)$}
\psline[linewidth=5,linecolor=palegray](+90,-30)(+90,+30)
\psline[linecolor=black,linestyle=dashed,linewidth=1](+92,-30)(+92,+30)
\psline[linecolor=blue](+91.25,-30)(+91.25,+30)
\psline[linecolor=blue](+90.75,-30)(+90.75,+30)
\psline[linecolor=blue](+90.25,-30)(+90.25,+30)
\psline[linecolor=blue](+89.75,-30)(+89.75,+30)
\psline[linecolor=blue](+89.25,-30)(+89.25,+30)
\psline[linecolor=blue](+88.75,-30)(+88.75,+30)
\psline[linecolor=blue](+88.25,-30)(+88.25,+30)
\psline[linecolor=blue](+87.75,-30)(+87.75,+30)
\psline(+93,+31)(+90,+33)(+87,+31)
\rput[t]{90}(+93.5,+15){${\bf O8}+8{\rm\darkblue D8}$}
\rput[t]{90}(+93.5,-15){$\darkblue SO(16)$}
\psline[linecolor=red](+1,+3)(+92,+3)
\psline[linecolor=red](-1,+1)(+92,+1)
\psline[linecolor=red](-13,-1)(+92,-1)
\psline[linecolor=red](-13,-3)(+92,-3)
\rput[b](45,+4){$4\,\darkred\rm D6$}
\rput[t](45,-4){$\darkred SU(4)$}
\pscircle*[linecolor=purple](+1,+3){1}
\pscircle*[linecolor=purple](-1,+1){1}
\psline[linecolor=yellow,linewidth=7](+87.5,0)(92.5,0)
\psline[linecolor=yellow,linewidth=3](-14,-2)(-10,-2)
\psline(-14.5,-31)(-6.75,-34)(+1,-31)
\rput[tl](-20,-35){Dual to the $\rm M9_1$}
\psline(+87,-31)(+90,-34)(+93,-31)
\rput[tr](+110,-35){Dual to the $\rm M9_2$}
\psline(+105,-30)(+105,+30)
\psline{>->}(+105,-10)(+105,+10)
\rput[t]{90}(+106,0){$x^{7,8,9}$}
\endpspicture
\eqn\ZfourBranes
$$
At the right terminus of the four coincident {\darkred D6} branes dual to a $\IZ_4$ fixed plane,
all four {\darkred D6} branes end on the \Ori{} orientifold at $x^6=L$.
Junctions of this type were discussed in \S4.3,
so let us simply quote the results for the case at hand ($N=4$, $k=8$):
\item\bullet
The $\darkred SU(4)$ gauge symmetry on the {\darkred D6} world volume is broken at $x^6=L$ down to
$\darkred Sp(2)\subset SU(4)$.
\item\bullet
The 7D SYM fields form the adjoint $\darkred\bf15$ representation of $\darkred SU(4)$.
In terms of the $\darkred Sp(2)\subset SU(4)$,
${\darkred\bf15}={\darkred\bf10}(\symrep)+{\darkred\bf5}(\tilde\asymrep)$
and the boundary conditions for the corresponding fields are as follows:
$$
\vcenter{\ialign{
	&\hfil\darkred\bf #\unskip\enspace & #\unskip:\hfil & \enspace #\hfil\qquad\cr
	10 & vector & Neumann, & 10 & hyper & Dirichlet,\cr
	5 & vector & Dirichlet, & 5 & hyper & Neumann.\cr
	}}
\eqn\ZfourRightBC
$$
\item\bullet
The $\darkred6\darkblue8$ open strings produce 6D hypermultiplets localized at the junction.
Their ${\darkred Sp(2)}\times{\darkblue SO(16)}$ quantum numbers are
$$
{\rm 6D}\,H\ =\ \half({\darkred\bf4},{\darkblue\bf16}).
\eqn\ZfourRightH
$$

Almost at the other end of the world, at $x^6=b$ we have two of the {\darkred D6} branes terminating
at the two outlier {\darkblue D8} branes while the other two {\darkred D6} branes continue toward
the orientifold plane at $x^6=0$.
For the sake of
${\darkred SU(2)}_{\rm 9D}\times{\darkblue SU(2)}_{\rm 7D}\to{\purple SU(2)}_{\rm diag}$
gauge symmetry mixing we let each of the first two {\darkred D6} branes terminate on
a separate {\darkblue D8}.
Altogether, the junction at $x^6=b$ is T-dual to
$$
\pspicture[](-10,-10)(70,33)
\psline[linecolor=red](-10,0)(60,0)
\psline[linecolor=red](-10,-0.5)(60,-0.5)
\rput[bl](2,2){$\darkred\rm 2\, D7'$}
\rput[tl](2,-2){$\darkred U(2)_2$}
\pscurve[linecolor=purple](0,33)(5,22.62)(10,16)(20,8)(30,4)(40,2)(50,1)(60,0.5)
\pscurve[linecolor=purple](0.5,33.5)(5.5,23.12)(10.5,16.5)(20.5,8.5)(30.5,4.5)(40.5,2.5)(50,1.5)(60,1)
\rput[bl](7,23){\purple2 D7}
\rput[bl](12,15){$\purple U(2)_1$}
\rput[Bl](62,-1){${}\to\darkred U(4)$}
\endpspicture
\eqn\ZfourTdual
$$
in the same manner as a similar junction in \S5.1 is T-dual to~\ZthreeTdual.
The physical consequences are also similar:
There is a ${\purple U(2)}_1$ SYM  living on the two curved {\purple D7} branes,
another ${\darkred U(2)}_2$ SYM  living on the two flat $\darkred\rm D7'$ branes,
plus $\darkred U(4)/[U(2)_1\times U(2)_2]$ SYM fields which are asymptotically massless
for $x^6\gg b$ but become heavy and decouple for $x^6\lsim b$.
It T-dual terms, we have a 7D SYM whose $\darkred U(4)$ gauge symmetry (at $x^6\ge b$) is abruptly
broken down to $\darkred U(2)_1\times U(2)_2$ by boundary conditions at $x^6=b$.
Specifically, the coset $\darkred U(4)/[U(2)_1\times U(2)_2]$ SYM fields satisfy Dirichlet
boundary conditions for the 8--SUSY vector multiplet components and Neumann for the hypermultiplet
components.
Also, in spite of zero length of the $\darkred6\darkblue8$ open strings at the junction, there are
no 6D massless fields localized at $x^6=b$.
Furthermore, the $\darkred U(2)_1$ vector fields satisfy locking boundary conditions which break
${\darkred U(2)}_1^{\rm 7D}\times{\darkblue U(2)}^{\rm 9D}\to{\purple U(2)}^{\rm diag}$
while the corresponding hypermultiplets satisfy Neumann boundary conditions.

On the other hand, the $\darkred U(2)_2$ SYM fields do not have any boundary conditions at $x^6=b$
and continue unmolested toward the ultimate boundary at $x^6=0$.
The physics at this boundary follows from two {\darkred D6} branes terminating
on an \Ori{} orientifold plane, \cf~\S4.3:
The $\darkred U(2)$ gauge symmetry is broken to $\darkred Sp(1)=SU(2)$, 
the $\bf\darkred3$ vector multiplets
satisfy Neumann boundary conditions while the $\bf\darkred3$ hypermultiplets satisfy Dirichlet condition,
and the $\darkred6\darkblue8$ open strings give rise to a localized 6D massless half-hypermultiplet 
in the $({\darkred\bf2},{\darkblue\bf12})$ representation of the locally visible
${\darkred SU(2)}\times{\darkblue SO(12)}$ gauge group.

In the HW picture of the \tI/D6 model~\ZfourBranes, the two separate brane junctions at $x^6=0$
and $x^6=b$ collapse into a single $\I_1$ intersections plane.
The local physics at this plane is simply the net effect of the two junctions, modulo decoupling
of the $\darkred U(1)$ center of the 7D $\darkred U(4)$ SYM related to the center-of-mass motion of the
four {\darkred D6} branes.
Thus:
\pointbegin
The 7D gauge symmetry $\darkred SU(4)$ is broken down to $\darkred SU(2)_1\times U(1)\times SU(2)_2$
and furthermore, the $\darkred [SU(2)_1\times U(1)]\subset SU(4)$ and the 10D
$\darkblue [SU(2)\times U(1)]\subset E_8^{(1)}$ are broken to the diagonal $\purple SU(2)\times U(1)$.
The net gauge symmetry at the $\I_1$ is therefore
$$
G^{\rm 6D}_{\rm local}\ =\ {\darkblue SO(12)}^{\rm 10D}\times{\purple[SU(2)_1\times U(1)]}^{\rm diag}
\times{\darkred SU(2)_2}^{\rm 7D}.
\eqn\ZfourLeftG
$$
\point %2
The boundary conditions for the 7D SYM fields depend on their
$\purple SU(2)_1\times U(1)\darkred\times SU(2)_2$
as well as 8--SUSY quantum numbers.
Decomposing the $\darkred SU(4)$ adjoint $\darkred\bf15$ as $({\purple\bf3},{\purple0},{\darkred\bf1})+
({\purple\bf1},{\purple0},{\darkred\bf3})+({\purple\bf1},{\purple0},{\darkred\bf1})+
({\purple\bf2},{\purple\pm1},{\darkred\bf2})$, we have
$$
\def\qqq#1#2#3{$({\purple\bf #1},{\purple #2},{\darkred\bf #3})$}
\def\jjj (#1,#2,#3) #4 #5 {\qqq{#1}{#2}{#3} & #4 & \qqq{#1}{#2}{#3} & #5 \cr }
\vcenter{\ialign{%
    \hfil # vector:\enspace & # b.c.,\hfil\qquad & \hfil # hyper:\enspace & # b.c.,\hfil\cr
    \jjj (3,0,1) Locking Neumann 
    \jjj (1,0,1) Locking Neumann 
    \jjj (1,0,3) Neumann Dirichlet
    \jjj (2,\pm1,2) Dirichlet {\omit Neumann b.c.\hfil}
    }}
\eqn\ZfourBC
$$
\point %3
The local 6D massless fields at the intersection comprise a half-hypermultiplet in the
$({\darkblue\bf12},{\purple\bf1},{\purple0},{\darkred\bf2})$ representation of~\ZfourLeftG.

Note that the heterotic twisted spectrum~\ZfourHtw\ contains a similar $({\bf12},{\bf2},0)$
half-hypermultiplet, but the $({\darkblue\bf12},{\purple\bf1},{\purple0},{\darkred\bf2})$ particles
we see at the $\I_1$ intersection  are {\sl doublets of the wrong $SU(2)$!}
Hence, for duality's sake, we must somehow mix
this purely 7D $\darkred SU(2)_2$ with the already mixed
${\purple SU(2)}_1=\diag \left( {\darkred SU(2)}_1^{\rm 7D}\times{\darkblue SU(2)}^{\rm 10D}\right)$.
This second diagonalization occurs not at the $\I_1$ but at the $\I_2$ at the other end
of the world: 
the $\darkred SU(4)\to Sp(2)$ breaking  locks the
two 7D $\darkred SU(2)_{1,2}$ gauge symmetries together.
To see how this works, consider the net result of the $\darkred SU(4)$ breaking at both ends of the $\FP6$,
$$
\vcenter{\openup 1\jot
    \ialign{&\hfil\enspace$\displaystyle{{}#{}}$\enspace\hfil\cr
        SU(4) & \longrightarrow & SU(2)_1\times SU(2)_2\times U(1)\cr
        \downarrow && \downarrow \cr
        Sp(2) & \longrightarrow & ?? \cr
        }}
\eqno\eq
$$
To clarify this diagram, we identify
$$
SU(4)\ =\ SO(6),\quad
Sp(2)\ =\ SO(5),\quad
SU(2)\times SU(2)\times U(1)\ =\ SO(4)\times SO(2),
\eqno\eq
$$
and note two distinct options for the overlap
$$
SO(5)\cap[SO(4)\times SO(2)]\
=\cases{SO(3)\times SO(2)=SU(2)\times U(1)\cr {\rm or}\ SO(4)=SU(2)\times SU(2).\cr }
\eqno\eq
$$
Because the $U(1)$ is observed in the heterotic theory, we choose the first option
and consequently
$$
\vcenter{\openup 1\jot
    \ialign{&\hfil\enspace$\displaystyle{{}#{}}$\enspace\hfil\cr
        SU(4) & \longrightarrow & SU(2)_1\times SU(2)_2\times U(1),\cr
        \downarrow && \downarrow \cr
        Sp(2) & \longrightarrow & SU(2)_{1+2}\times U(1) \cr
        }}
\eqn\SUivBreaking
$$
where $SU(2)_{1+2}=\diag \left(SU(2)_1\times SU(2)_2\right)$
(\cf\ $SO(3)\subset SO(4)$).

Ultimately, the $\purple SU(2)$ observed in the heterotic theory is the common diagonal of the
$$
{\darkblue SU(2)}^{\rm 10D}\times\prod_{\IZ_4\,\rm fixed \atop planes}
	[{\darkred SU(2)_1\times SU(2)_2}]^{\rm 7D} .
\eqn\ZfourSUii
$$
At each $\FP6[\IZ_4]$ fixed plane of the HW picture, the $SU(2)$ quantum numbers of the $\half({\bf12},{\bf2})$
twisted hypermultiplet follow a weird trajectory
$$
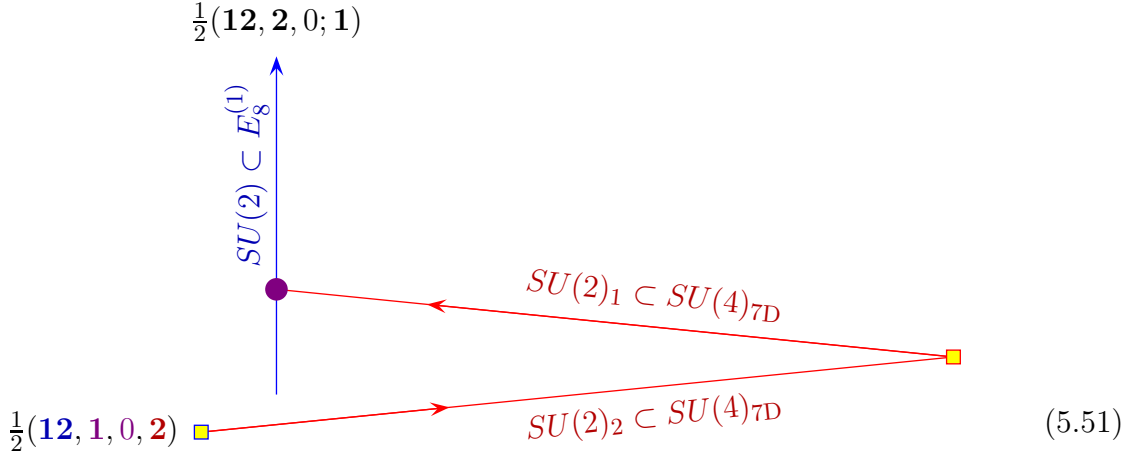
\pspicture[](-15,-10)(102,60)
\psline[linecolor=red](0,0)(100,10)(10,19)
\psline[linecolor=red]{->}(0,0)(33,3.3)
\psline[linecolor=red]{->}(100,10)(30,17)
\psline[linecolor=blue]{->}(10,5)(10,50)
\psframe[linecolor=blue,fillstyle=solid,fillcolor=yellow](-1,-1)(+1,+1)
\psframe[linecolor=red,fillstyle=solid,fillcolor=yellow](99,9)(101,11)
\pscircle*[linecolor=purple](10,19){1.5}
\rput[rB](-3,-1){$\half({\darkblue\bf12},{\bf\purple1},{\purple0},{\bf\darkred2})$}
\rput[t]{6}(60,4){$\darkred SU(2)_2\subset SU(4)_{\rm 7D}$}
\rput[b]{354}(60,16){$\darkred SU(2)_1\subset SU(4)_{\rm 7D}$}
\rput[bl]{90}(9,22){$\darkblue SU(2)\subset E_8^{(1)}$}
\rput[b](10,52){$\half({\bf12},{\bf2},0;{\bf1})$}
\endpspicture
\eqn\SUiiTrajectory
$$
from the $x^6=0$ end of the fixed plane all the way to the other end $x^6=L$ and back to
$x^6=b\approx0$.
At the end of this trajectory, the $\half({\darkblue\bf12},{\bf\purple1},{\purple0},{\bf\darkred2})$
multiplet of the local gauge symmetry~\ZfourLeftG\ indeed becomes the $\half({\bf12},{\bf2},0;{\bf1})$
multiplet of the net 6D gauge symmetry~\ZfourG.
We admit however that such a HW origin of a twisted state's quantum numbers is so bizarre,
we would have never found it without the dual \tI/D6  model~\ZfourBranes.

The HW origins of the remaining twisted states~\ZfourHtw\ are less complicated.
The $\half({\bf\darkred4},{\bf\darkblue16})$ multiplet of the
${\darkred Sp(2)}\times{\darkblue SO(16)}$
living at $\I_2$ at the right end of the fixed plane becomes $({\bf1},{\bf2},+\half;{\bf16})$
of the net 6D symmetry~\ZfourG\ after the $\darkred SP(2)$ is broken down to
$\darkred SU(2)\times U(1)$
which eventually mixes with the 10D $\darkblue SU(2)\times U(1)$ at $x^6=b\approx0$.
Indeed, in $SO(5)$ terms, the $\bf 4$ representation of the $Sp(2)$ is the spinor, hence
it decomposes as ${\bf4}=({\bf2},+\half)+({\bf2},-\half)$ with respect to the
$SO(3)\times SO(2)\subset SO(5)$;
therefore, for a hypermultiplet $\half({\bf4})=({\bf2},+\half)$.

Finally, the two charged singlets in the twisted spectrum~\ZfourHtw\ arise from the zero modes
of the 7D fields.
Indeed, let us arrange the $\bf\darkred15$ 7D SYM fields according to their
$\darkred SU(2)_{1+2}\times U(1)$
quantum numbers and note their boundary conditions at both ends of the $x^6$:
$$
\vcenter{\ialign{
    \hfil\vrule width 0pt height 15pt depth 5pt \vrule #&
    \qquad\hfil $\bigl({\bf #},{}$& ${} #\bigr)$\hfil\enspace & \vrule #\hfil &
    \enspace\hfil\frenchspacing(#\unskip)\hfil\enspace &\vrule #\hfil &
    \enspace\hfil\frenchspacing(#\unskip)\hfil\enspace &\vrule #\hfil \cr
\noalign{\hrule}
& \multispan2 \hfil Charges\hfil && \multispan3 \hfil Boundary Conditions\hfil &\cr
\noalign{\hrule}
& \multispan2 \hfil\enspace $SU(2)_{1+2}\times U(1)$\enspace\hfil &&
	\omit\hfil 8--SUSY vector\hfil && \omit\hfil hyper\hfil &\cr
\noalign{\hrule}
& 3 & 0 && locking,Neumann && Neumann,Dirichlet &\cr
\noalign{\hrule}
& 1 & 0 && locking,Neumann && Neumann,Dirichlet &\cr
\noalign{\hrule}
& 3 & 0 && Neumann,Dirichlet && Dirichlet,Neumann &\cr
\noalign{\hrule}
& 3 & \pm1 && Dirichlet,Neumann && Neumann,Dirichlet &\cr
\noalign{\hrule}
& 1 & \pm1 && Dirichlet,Dirichlet && Neumann,Neumann &\cr
\noalign{\hrule}
}}
\eqn\ZfourTable
$$
According to this table, two hypermultiplets on the last line have
Neumann boundary conditions at both ends
and hence massless (in 6D) zero modes.
The $({\bf1},\pm1)$ charges of these hypermultiplets are exactly opposite and therefore equivalent;
in the~\ZfourG\ terms, we have $2({\bf1},{\bf1},+1,{\bf1})$, \cf\ the last entry in the heterotic
twisted spectrum~\ZfourHtw.

To summarize the above discussion, the HW dual of the \tI/D6 brane model \ZfourBranes\ correctly
reproduces the twisted spectrum of the heterotic $\IZ_4$ fixed plane, albeit in a rather weird manner.
To justify this weirdness, we conclude this section by verifying the other kinematical requirements
of the HW orbifold, namely the heterotic gauge couplings and the local anomaly cancelation at both
$\I_{1,2}$ intersection planes.

We begin with the gauge couplings:
According to eq.~\ZfourSUii, the net $\purple SU(2)$ gauge theory of the orbifold involves
two copies of the $\darkred SU(2)$  embedded in the $\darkred SU(4)$ at each $\IZ_4$ fixed plane.
Consequently,
$$
{1\over g^2[{\purple SU(2)}]}\ =\ {1\over g^2[{\darkblue E_8^{(1)}}]}\
+\ {4\times2\over g^2[{\darkred SU(4)}]}
\eqno\eq
$$
or in terms of the $v,\tilde v$ coefficients,
$$
v[SU(2)]\ =\ 1,\qquad \tilde v[SU(2)]\ =\ \half k_1\ +\ 8.
\eqn\ZfourSUiiG
$$
The generator $C_{\rm 7D}$ of the $\darkred U(1)\subset SU(4)$ has eigenvalues
$(+\half,+\half,-\half,-\half)$
hence the norm $\tr(C^2_{\rm 7D})=2$ while the generator $C_{\rm 10D}$
of the $\darkblue U(1)\subset E_8^{(1)}$ is normalized to $\tr(C^2_{\rm 10D})=4$, thus
the net $\purple U(1)$ of the orbifold has
$$
v[U(1)]\ =\ 4,\qquad \tilde v[U(1)]\ =\ 2k_1\ +\ 4\times2.
\eqn\ZfourUiG
$$
Finally, the remaining $SO(12)$ and $SO(16)$ gauge factors are of purely 10D origins, therefore
$$
v[SO(12)]\ =\ v[SO(16)]\ =\ 1,\qquad
\tilde v[SO(12)]\ =\ \half k_1\,,\qquad \tilde v[SO(16)]\ =\ \half k_2\,.
\eqn\ZfourSOG
$$

Formul\ae~\ZfourSUiiG\ through \ZfourSOG\ are predictions of the HW picture of the orbifold
which is itself a prediction of the \tI/D6 dual model~\ZfourBranes.
To verify these predictions, we  calculated the net anomaly polynomial of the orbifold
and factorized it according to eq.~\NetAnomalyFactorization.
After some boring arithmetic, we arrived at
$$
\def\qqq #1 #2 #3 {v[#1] & #2 & \tilde v[#1] & #3 \cr }
\vcenter{\openup 1\jot \tabskip=0pt
    \halign{
	&\hfil$\displaystyle{#\ ={}}$ & ${#}$,\qquad\hfil\cr
	\qqq SO(12) 1 +2
	\qqq SU(2)  1 +10
	\qqq U(1)   4 +16
	\qqq SO(16) 1 -2
	}}
\eqn\ZfourCouplings
$$
which indeed agrees with eqs.~\ZfourSUiiG, \ZfourUiG\ and \ZfourSOG\ for $k_1=+4$, $k_2=-4$
(\ie, 16 instantons in the $E_8^{(1)}$ and 8 in the $E_8^{(2)}$).

Next, consider the local anomalies at the $\I_1$ intersection plane
where the local gauge symmetry is given by eq.~\ZfourLeftG.
The anomaly-weighed chiral spectrum at this plane comprises
$$
\def\qqq(#1,#2,#3,#4){({\bf\darkblue #1},{\bf\purple #2},{\purple #3},{\bf\darkred #4})}
\eqalign{
Q_6\ &=\ \half\qqq(12,1,0,2),\cr
Q_7\ &
=\ {1\over2}\Bigl[\qqq(1,3,0,1)\,+\,\qqq(1,1,0,1)\,-\,\qqq(1,1,0,3)\,+\,\qqq(1,2,\pm1,2)\Bigr],\cr
Q_{10}\ &
=\ {1\over32}\Bigl[\qqq(12,2,\pm1,1)\,+\,\qqq(32,1,\pm1,1)\Bigr]\
	+\ {3\over32}\Bigl[\qqq(32',2,0,1)\,+\,\qqq(1,1,\pm2,1)\Bigr]\cr
&\qquad-\ {5\over32}\Bigl[\qqq(66,1,0,1)\,+\,\qqq(1,3,0,1)\,+\,\qqq(1,1,0,1)\Bigr] ,\cr
}\eqn\ZfourLeftEnd
$$
while the plane's magnetic charge  is
$$
g_1[\IZ_4]\ =\ {1\over4}\Bigl(k_1\,-\,6g_1[\IZ_2]\Bigr)\
=\ {1\over4}\Bigl( 4\,-\,6\times{3\over4}\Bigr)\ =\ -{1\over8}\,,
\eqno\eq
$$
thus we can easily verify that 
$$
\mathop{\rm dim}(Q=Q_6+Q_7+Q_{10})\ =\ {61\over4}\ =\ 19\ +\ 30\times{-1\over8}
\eqno\eq
$$
and hence the $\tr(R^4)$ anomaly cancels out, \cf~eq.~\AnomalyRiiii.
The rest of the anomaly follows via straightforward if tedious evaluation of
$$
\eqalign{
{\cal A}'\ &
\equiv\ \coeff23\Tr_Q({\cal F}^4)\ -\ \coeff16\tr(R^2)\times\Tr_Q({\cal F}^2)
	+\ (\coeff18 g+\half T(1)=\coeff{1}{16})\bigl(\tr(R^2)\bigr)^2\crr
&=\ {-1\over16}\Bigl( \tr(F^2_{\darkblue SO(12)})\,+\,\tr(F^2_{\purple SU(2)})\,
		+\,4F^2_{\purple U(1)}\,-\,\half\tr(R^2)\Bigr)^2\cr
&\qquad+\ \Bigl( \tr(F^2_{\darkblue SO(12)})\,+\,\tr(F^2_{\purple SU(2)_1})\,
		+\,4F^2_{\purple U(1)}\,-\,\half\tr(R^2)\Bigr)\cr
&\qquad\qquad\qquad{}\times\Bigl( \tr(F^2_{\purple SU(2)_1})\,+\,2F^2_{\purple U(1)}\,
		+\,\tr(F^2_{\darkred SU(2)_2})\,-\,\coeff{5}{32}\tr(R^2)\Bigr),\cr
}\eqn\ZfourAnomalyL
$$
which indeed shows cancellation of the one-loop anomaly against the inflow and intersection anomalies,
\cf~eq.~\AnomalyOther.

The $SU(2)$ trajectory~\SUiiTrajectory\ goes through the $\I_1$ intersection twice, hence two
separate $SU(2)$ gauge factors in the local symmetry~\ZfourLeftG.
To see the importance of this setup anomaly-wise, suppose we had only one $SU(2)_{1+2}$ factor instead
and focus on the $\bigl(\tr(F^2_{SU(2)})\bigr)^2$ anomaly term.
In the $SU(2)_{1+2}$ terms, the $\bf\darkred 15$ 7D SYM fields comprise four triplets and three singlets,
and in our hypothetical model
 all the triplets would have to have Dirichlet or locking boundary conditions for the vector fields
in order to prevent the appearance of a second $SU(2)$ factor.
Anomaly-wise, the effect of this change is to change the sign of the third term in eq.~\ZfourLeftEnd\
for the $Q_7$ from negative to positive --- and therefore to increase the one-loop
$\Tr_Q(F^4_{SU(2)})$ by $+2\bigl(\tr(F^2_{SU(2)})\bigr)^2$.
At the same time, the $Q_6$, the $Q_{10}$ and all the terms in the inflow and intersection anomalies
are completely fixed by the heterotic data, so they must remain unchanged.
Altogether, we would have $+2\bigl(\tr(F^2_{SU(2)})\bigr)^2$ worth of un-canceled local anomaly --- which
rules out the single $SU(2)$ hypothesis.
On the other hand, the setup with two local $SU(2)$ gauge symmetries leads to complete anomaly
cancellation.
Thus, the trajectory~\SUiiTrajectory\ may look weird, but it works and nothing else seem to do the job,
so it must be right!

Finally, at the $\I_2$ intersection plane, the local gauge symmetry 
is ${\darkblue SO(16)}\times{\darkred Sp(2)}$, the anomaly-weighed chiral spectrum comprises
$$
\eqalign{
Q_6\ &=\ \half({\bf\darkblue16},{\bf\darkred4}),\cr
Q_7\ &=\ {1\over2}\Bigl[({\bf\darkblue1},{\bf\darkred5})\,-\,({\bf\darkblue1},{\bf\darkred10})\Bigr],\cr
Q_{10}\ &=\ {3\over32}\,({\bf\darkblue128},{\bf\darkred1})\
	-\ {5\over32}\,({\bf\darkblue120},{\bf\darkred1}),\cr
}\eqn\ZfourLocalSpectrumR
$$
and the magnetic charge is $g_2=-g_1=+\coeff18$.
We immediately see that eq.~\AnomalyRiiii\ is satisfied,
$$
\mathop{\rm dim}(Q=Q_6+Q_7+Q_{10})\ =\ {91\over4}\ =\ 19\ +\ 30\times{+1\over8}\,,
\eqno\eq
$$
and after a bit of arithmetic we can see that eq.~\AnomalyOther\ is satisfied as well,
$$
\eqalign{
{\cal A}'\ &
\equiv\ \coeff23\Tr_Q({\cal F}^4)\ -\ \coeff16\tr(R^2)\times\Tr_Q({\cal F}^2)
	+\ (\coeff18 g+\half T(1)=\coeff{3}{32})\bigl(\tr(R^2)\bigr)^2\crr
&=\ {+1\over16}\Bigl( \tr(F^2_{\darkblue SO(16)})\,-\,\half\tr(R^2)\Bigr)^2\cr
&\qquad+\ \Bigl( \tr(F^2_{\darkblue SO(12)})\,-\,\half\tr(R^2)\Bigr)
	\times\Bigl( \tr(F^2_{\purple Sp(4)})\,-\,\coeff{5}{32}\tr(R^2)\Bigr).\cr
}\eqn\ZfourAnomalyR
$$
Again, cancelation of the net $\tr({\cal F}^4_{\rm 7D})$ anomaly does not allow any changes
in $Q_7$ (since all the other contributions to this anomaly are fixed by the heterotic data),
which confirms that the 7D vector fields with Neumann boundary conditions should indeed comprise
the adjoint $\bf\darkred10$ multiplet of the $\darkred Sp(2)\supset SU(2)\times U(1)$,
exactly as in the dual \tI/D6 brane model. 

%%%%%%%%%%%%%%%%%%%%%%%%%%%%%%%%%%%%%%%%%%%%%%%%%%%%%%%%%%%%%%
%% end of chapter5

%auto-ignore
% chapter6.tex
%% Chapter 6 of Orbifold2 paper
%% Written by Vadim Kaplunovsky

%%%%%%%%%%%%%%%%%%%%%%%%%%%%%%%%%%%%%%%%%%%%%%%%%%%%%%%%%%%%%%%%%%%%%%
\chapter{NS5 Half--Branes at the End of the World}

In this section, we add another tool to our brane engineering toolkit, namely NS5 branes serving
as terminals of several coincident D6 branes.
Or rather NS5 {\it half-branes}, stuck on the \Ori{} orientifold planes and unable to move
in the $x^6$ direction.
Such half-branes are explained in some detail in
refs.~[\HanZaf,\BruKar]; the following couple of pages
give a brief summary of relevant phenomena.

An NS5 half-brane terminus of $N$ D6 branes 
results from  \Ori{} orientifold projection of the following picture:
$$
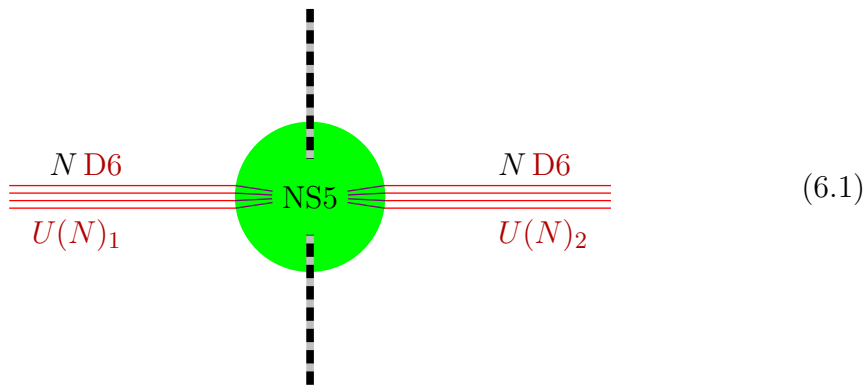
\pspicture[](-40,-25)(+40,+25)
\psline[linecolor=red](-40,+1.5)(+40,+1.5)
\psline[linecolor=red](-40,+0.5)(+40,+0.5)
\psline[linecolor=red](-40,-0.5)(+40,-0.5)
\psline[linecolor=red](-40,-1.5)(+40,-1.5)
\rput[br](-25,+3){$N\darkred\rm\,D6$}
\rput[tr](-25,-3){$\darkred U(N)_1$}
\rput[bl](+25,+3){$N\darkred\rm\,D6$}
\rput[tl](+25,-3){$\darkred U(N)_2$}
\pscircle*[linecolor=green](0,0){10}
\psline[linecolor=purple](-9.9,-1.5)(+9.9,+1.5)
\psline[linecolor=purple](-9.9,+1.5)(+9.9,-1.5)
\psline[linecolor=purple](-10,+0.5)(+10,-0.5)
\psline[linecolor=purple](-10,-0.5)(+10,+0.5)
\psline[linecolor=lightgray,linewidth=1](0,-25)(0,+25)
\psline[linecolor=black,linestyle=dashed,linewidth=1](0,-25)(0,+25)
\cput*[fillcolor=green,linecolor=green](0,0){NS5}
\endpspicture
\eqn\NSbrane
$$
In the middle of this picture we have an NS5 brane --- a supersymmetric co-dimension 4
soliton of the metric, dilaton and $B_{\mu\nu}$ fields of the type~IIA superstring.
The metric is asymptotically flat for $r\to\infty$ but develops an infinite $S^3\times R_+$ throat
at $r\to0$; deep down the throat, the string coupling $\lambda=e^\varphi$
increases and eventually becomes strong.
The D6 branes approaching the NS5 brane from the right (\ie, $\bx=\bf0$, $x^6\to+0$) plunge down the
throat and eventually suffer some kind of a `meltdown' in the strong coupling region.
In the local metric (string frame), the $x^6\to+0$ dimension of these D6 branes is infinite and there
is no terminus, but the continuously rising string coupling causes reflection of the 7D particles
living on the D6 world-volume back to $x^6\to+\infty$.
Hence, from the low-energy / long-distance point of view, the 6D branes {\sl appear to terminate}
on the NS5 brane where the 7D $\darkred U(N)$ SYM fields have reflecting boundary conditions:
Neumann for the 8--SUSY vector multiplet components and Dirichlet for the hypermultiplet components.

On the left side of the picture~\NSbrane\ we have $N$ more D6 branes at $x^6<0$;
they also plunge down the NS5 throat for $x^6\to-0$, but the two sets of D6 branes remain at finite
distance from each other all the way down.
Therefore, the $\darkred U(N)$ SYM fields at $x^6<0$ suffer reflecting boundary conditions at
$x^6=-0$ without any locking onto the similar SYM fields at $x^6=+0$, hence {\sl locally at
$x^6=0$} we have a double gauge symmetry $\darkred U(N)_L\times U(N)_R$.
Or rather $\darkred SU(N)_L\times SU(N)_R\times U(1)_{L-R}$ whereas the other $\darkred U(1)_{L+R}$
is Higgsed out by quantum corrections (the Fayet--Iliopoulos term and its superpartners);
the hypermultiplet `eaten up' by this Higgs effect corresponds to moving the NS5 brane in the
$x^{7,8,9,10}$ directions separately from the D6 branes.
On the other hand, moving the NS5 brane in the $x^6$ direction corresponds to the scalar in the
8--SUSY tensor multiplet, which remains in the massless spectrum of the configuration.

Finally, we have  open strings connecting the two sets of the D6 branes in the throat region.
Such strings have short $O(\sqrt{\alpha'})$ but non-zero length; nevertheless they have zero
modes giving rise to 6D massless hypermultiplets localized at $x^6=0$.
Naturally, the gauge quantum numbers of these particles are $({\bf\darkred N},{\bf\darkred N})$.

The \Ori{} orientifold projection identifies the two halves of the picture~\NSbrane\ as mirror
images of each other:  The physical part of the NS5 brane is only a half-brane and there is only
one independent set of $N$ D6 branes.
The 7D SYM fields surviving the projection comprise the diagonal $\darkred SU(N)_{L+R}$ while
the $\darkred U(1)$ fields are projected out altogether.
The tensor multiplet is also projected out;
consequently, the NS5 half-brane cannot move in the $x^6$ direction any longer and remains forever
stuck at the orientifold plane.
Finally, the $({\bf\darkred N},{\bf\darkred N})$ multiplet of localized 6D fields splits into
a symmetric $\darkred\symrep$ and an antisymmetric $\darkred\asymrep$ multiplets of
the diagonal $\darkred SU(N)$ gauge symmetry.
The two multiplets have opposite signs with respect to the orientifold projection $\Omega$:
The $\darkred\symrep$ is $\Omega$--negative while the $\darkred\asymrep$ is $\Omega$--positive,
thus only the antisymmetric $\darkred\asymrep$ multiplet survives the projection.

Ultimately, from the low-energy, $x^6>0$ point of view, the NS5 half-brane serves as a new kind
of a terminal junction between the {\darkred D6} branes and the \Ori{} orientifold plane,
$$
\left\{ \eqalign{
    G^{\rm local}\ &=\ {\darkred SU(N)},\cr
    {\rm 7D}\,V\ &=\ {\darkred\rm Adj.},\cr
    {\rm 7D}\,H\ &=\ 0,\cr
    {\rm 7D}\,H\ &=\ {\darkred\asymrep}\,,\cr
    }\right\}
{\darkgreen\half\,\rm NS5}\
\pspicture*[0.475](-0.5,-25)(+40,+25)
\psline[linecolor=palegray,linewidth=1](0,-25)(0,+25)
\psline[linecolor=black,linestyle=dashed,linewidth=1](0,-25)(0,+25)
\psline[linecolor=red](0,+1.5)(+40,+1.5)
\psline[linecolor=red](0,+0.5)(+40,+0.5)
\psline[linecolor=red](0,-0.5)(+40,-0.5)
\psline[linecolor=red](0,-1.5)(+40,-1.5)
\rput[bl](20,+3){$N$ \darkred D6}
\rput[tl](20,-3){$\darkred SU(N)$}
\pscircle*[linecolor=green](0,0){3}
\endpspicture
\eqn\NShalf
$$
Unlike the junctions discussed in \S4.3 (\cf\ fig.~\YellowEndZtwo), the $\rm\darkgreen\half NS5$
junction~\NShalf\ provides Neumann boundary conditions at $x^6=0$ for all the 7D $\darkred SU(N)$
8--SUSY vector multiplets while all the 7D hypermultiplets satisfy Dirichlet conditions.
\foot{As usual, we disregard the $\darkred U(1)$ factor of the 7D $\darkred U(N)$ as irrelevant to
    the orbifold problem.}
Consequently, the entire $\darkred SU(N)$ symmetry is visible at the junction and it is no longer
necessary for the number $N$ of the {\darkred D6} branes to be even.
Finally, the $\rm\darkgreen\half NS5$ junction supports localized 6D massless hypermultiplets
which form an antisymmetric tensor representation $\darkred\asymrep$ of the $\darkred SU(N)$.
{}From the brane engineering point of view, such $\asymrep$ hypermultiplets are characteristic of
of NS5 half-branes and do not occur at other types of brane junctions.

As an example of a $\rm\darkgreen\half NS5$ junction in a \tI/D6 brane dual of a
perturbative heterotic orbifold,
consider a $T^4/\IZ_6$ model in which $E_8^{(1)}$ is broken down to
$SU(6)\times E_3\equiv SU(6)\times SU(3)\times SU(2)$
and $E_8^{(2)}$ down to $SU(8)\times U(1)$.
In terms of the lattice shift vectors,
$\delta_1=(-\coeff56,\coeff16,\ldots,\coeff16,0,0)$ and
$\delta_2=(-\coeff{11}{12},\coeff{+1}{12},\ldots,\coeff{+1}{12})$;
in terms of the unbroken subgroups, the $\alpha_1:\IZ_6\mapsto E_8^{(1)}$ twist acts as
$$
\eqalign{
\alpha_1({\bf248})\ =\ {}&
+[({\bf35},{\bf1},{\bf1})+({\bf1},{\bf8},{\bf1})+({\bf1},{\bf1},{\bf3})]\
	-\ ({\bf20},{\bf1},{\bf2})\cr
&+\ e^{2\pi i/6}({\bf\bar6},{\bf\bar3},{\bf2})\ +\ e^{-2\pi i/6}({\bf6},{\bf3},{\bf2})\cr
&+\ e^{4\pi i/6}(\overline{\bf15},{\bf3},{\bf1})\
	+\ e^{-4\pi i/6}({\bf15},{\bf\bar3},{\bf1})\cr
}\eqn\ZsixAlphaLeft
$$
while the $\alpha_2:\IZ_6\mapsto E_8^{(2)}$ twist acts according to
$$
\eqalign{
\alpha_2({\bf248})\ =\ {}&
+[({\bf63},0)+({\bf1},0)]\ -\ ({\bf70},0)\cr
&+\ e^{2\pi i/6}[(\overline{\bf28},-1)+({\bf1},+2)]\
	 +\ e^{-2\pi i/6}[({\bf28},+1)+({\bf1},-2)]\cr
&+\ e^{4\pi i/6}(\overline{\bf28},+1)\ +\ e^{-4\pi i/6}({\bf28},-1).\cr
}\eqn\ZsixAlphaRight
$$
In the untwisted sector of the orbifold, massless particles comprise the usual
SUGRA and dilaton multiplets, 110 vector multiplets in the adjoint of
$$
G\ =\ [SU(6)\times SU(3)\times SU(2)]\,\times\,[SU(8)\times U(1)] ,
\eqn\ZsixG
$$
two moduli and 65 charged hypermultiplets,
$$
H_{\rm untw}\ =\ ({\bf\bar6},{\bf\bar3},{\bf2};{\bf1},0)\
+\ ({\bf1},{\bf1},{\bf1};{\bf1},+2)\ +\ ({\bf1},{\bf1},{\bf1};\overline{\bf28},-1)\ +\ 2M.
\eqn\ZsixHuntw
$$
The twisted sectors contain further 287 charged massless hypermultiplets;
arranging them according to specific fixed $\FP5$ planes of the model
(\ie, fixed points of the $T^4/\IZ_6$), we have:
\item\bullet
One $\IZ_6$ fixed plane carries 63 hypermultiplets,
$$
H_{\rm tw}[\IZ_6]\ =\ ({\bf6},{\bf1},{\bf1};{\bf8},+\coeff16)\
+\ ({\bf15},{\bf1},{\bf1};{\bf1},-\coeff23).
\eqn\ZsixHtwSix
$$
\item\bullet
Five $\IZ_2$ fixed planes, each carrying 16 hypermultiplets,
$$
H_{\rm tw}[\IZ_2]\ 
=\ ({\bf1},{\bf1},{\bf2};{\bf8},-\half)\
\equiv\ \half({\bf1},{\bf2};{\bf16}).
\eqn\ZsixHtwTwo
$$
The second expression on the right hand side shows the quantum numbers of these hypermultiplets
with respect to the locally surviving gauge symmetry $G_{\IZ_2}=[E_7\times SU(2)]\times SO(16)$.
Note that each of these $\IZ_2$ fixed planes works exactly like the fixed planes of the
$T^2/\IZ_2$ orbifold of section~4.
\item\bullet
Four $\IZ_3$ fixed planes, each carrying 36 hypermultiplets,
$$
\eqalign{
H_{\rm tw}[\IZ_3]\ &
=\ ({\bf15},{\bf1},{\bf1};{\bf1},-\coeff23)\ +\ ({\bf\bar6},{\bf1},{\bf2};{\bf1},-\coeff23)\cr
&\qquad+\ 2({\bf1},{\bf3},{\bf1};{\bf1},-\coeff23)\ +\ ({\bf1},{\bf3},{\bf1};{\bf1},+\coeff43)\cr
&\equiv\ ({\bf27},{\bf1};{\bf1},-\coeff23)\ +\ 2({\bf1},{\bf3};{\bf1},-\coeff23)\
    +\ ({\bf1},{\bf3};{\bf1},+\coeff43).\cr
}\eqn\ZsixHtwThree
$$
Again, the second expression on the RHS indicates the representation of the locally surviving symmetry
$G_{\IZ_3}=[E_6\times SU(3)]\times[E_7\times U(1)]$.
Naturally, there is a $T^4/\IZ_3$ model with similar fixed planes,
but for technical reasons
we do not discuss this model in the present article;
hopefully, we shall return to it in a future publication.

For our present purposes, we are interested in the HW point of view
of the $\IZ_6$ fixed plane.
In 7D, this $\FP6$ plane carries an $\darkred SU(6)$ SYM, and it is clear from the
twisted spectrum~\ZsixHtwSix\ that the entire $\darkred SU(6)$ gauge group
is involved in communicating quantum numbers between the two M9 branes.
That is, at the $\I_1$ intersection at $x^6=0$, all the 7D $\darkred SU(6)$ vector fields lock
onto the 10D $\darkblue SU(6)$ vector fields according to eq.~\HWLockingBC\ and all $\darkred\bf 35$
hypermultiplet components have Neumann boundary conditions.
All the massless twisted states~\ZsixHtwSix\ are localized at the other intersection $\I_2$
at $x^6=L$ where the 7D vector multiplets have Neumann~BC and hypermultiplets Dirichlet~BC.
Thus,
$$
\eqalignno{
\I_1 &
\left\{\eqalign{
    G^{\rm local}\ &
    =\ {\purple SU(6)}^{\rm diag}\times[{\darkblue SU(3)\times SU(2)}]^{\rm 10D} ,\cr
    {\rm 7D}\,H\ &=\ {\purple 35},\cr
    {\rm 7D}\,V\ &=\ 0,\cr
    {\rm 6D}\,H\ &=\ 0;\cr
    }\right.&
\eqname\ZsixHWleft \xdef\jj{(\chapterlabel.\number\equanumber} \cr
\noalign{\smallskip}
\I_2 &
\left\{\eqalign{
    G^{\rm local}\ &
    =\ {\darkred SU(6)}^{\rm 7D}\times[{\darkblue SU(8)\times U(1)}]^{\rm 10D} ,\cr
    {\rm 7D}\,H\ &=\ 0,\cr
    {\rm 7D}\,V\ &=\ {\darkred 35},\cr
    {\rm 6D}\,H\ &
    =\ ({\darkred\bf6};{\darkblue\bf8},{\darkblue+\coeff16})\
	+\ ({\darkred\bf15};{\darkblue\bf1},{\darkblue-\coeff23}).\cr
    }\right.&
\eqname\ZsixHWright  \cr
}$$
Clearly, this HW picture does lead to the correct twisted spectrum, and in the Appendix
we shall verify the rest of the kinematical constraints (the 6D gauge couplings and the
local anomaly cancelation at both $\I_1$ and $\I_2$ intersections),
but for now let us focus on brane engineering a dual model.
 
Brane-wise, two features of the $\I_2$ intersection
are particularly noteworthy:
First, {\sl all} of the $\bf\darkred 35$ 7D vector fields have Neumann boundary conditions
at the $\I_2$ which preserves the entire $\darkred SU(6)$ gauge symmetry.
Second, the massless hypermultiplets localized at the $\I_2$ include an antisymmetric
tensor representation $\darkred\asymrep=\bf15$ of this 7D symmetry.
Both features cry out for an {\darkgreen NS5} half-brane being present at the
junction dual to the $\I_2$, so let us engineer the following model:
$$
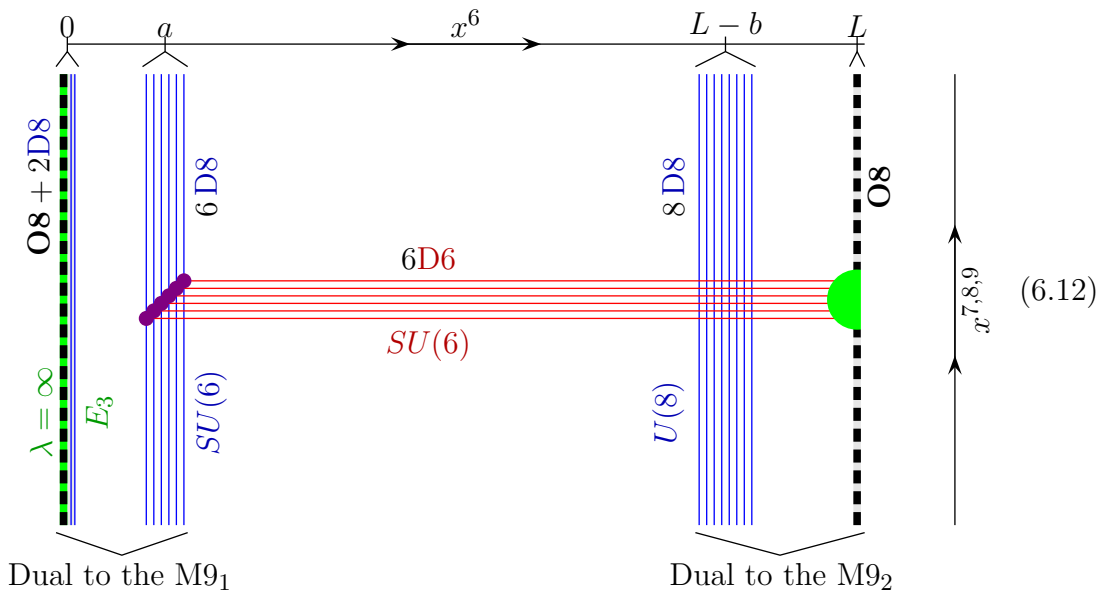
\pspicture[](-20,-40)(110,40)
\psline(-13,+34)(+93,+34)
\rput[b](+40,+35){$x^6$}
\psline{>->}(+30,+34)(+50,+34)
\psline(-13,+33)(-13,+35)	\rput[b](-13,+35){$0$}
\psline(-14.5,+31)(-13,+33)(-11.5,+31)
\psline(0,+33)(0,+35)		\rput[b](0,+35){$a$}
\psline(-3,+31)(0,+33)(+3,+31)
\psline(+74.5,+33)(+74.5,+35)	\rput[b](+74.5,+35){$L-b$}
\psline(+78.5,+31)(+74.5,+33)(+70.5,+31)
\psline(+92,+33)(+92,+35)	\rput[b](+92,+35){$L$}
\psline(+93,+31)(+92,+33)(+91,+31)
\psline[linecolor=palegray,linewidth=2](-13,-30)(-13,+30)
\psline[linecolor=green,linewidth=1](-13.5,-30)(-13.5,+30)
\psline[linecolor=black,linestyle=dashed,linewidth=1](-13.5,-30)(-13.5,+30)
\psline[linecolor=blue](-12.5,-30)(-12.5,+30)
\psline[linecolor=blue](-12.0,-30)(-12.0,+30)
\rput[b]{90}(-15,+15){${\bf O8}+2{\rm\darkblue D8}$}
\rput[t]{90}(-10.5,-15){$\darkgreen E_3$}
\rput[b]{90}(-15,-15){$\darkgreen\lambda=\infty$}
\psline[linecolor=blue](-2.5,-30)(-2.5,+30)
\psline[linecolor=blue](-1.5,-30)(-1.5,+30)
\psline[linecolor=blue](-0.5,-30)(-0.5,+30)
\psline[linecolor=blue](+0.5,-30)(+0.5,+30)
\psline[linecolor=blue](+1.5,-30)(+1.5,+30)
\psline[linecolor=blue](+2.5,-30)(+2.5,+30)
\rput[t]{90}(+4,+15){$6\,\rm\darkblue D8$}
\rput[t]{90}(+4,-15){$\darkblue SU(6)$}
\psline[linecolor=blue](+78,-30)(+78,+30)
\psline[linecolor=blue](+77,-30)(+77,+30)
\psline[linecolor=blue](+76,-30)(+76,+30)
\psline[linecolor=blue](+75,-30)(+75,+30)
\psline[linecolor=blue](+74,-30)(+74,+30)
\psline[linecolor=blue](+73,-30)(+73,+30)
\psline[linecolor=blue](+72,-30)(+72,+30)
\psline[linecolor=blue](+71,-30)(+71,+30)
\rput[b]{90}(+69,+15){$8\,\rm\darkblue D8$}
\rput[b]{90}(+69,-15){$\darkblue U(8)$}
\psline[linewidth=1,linecolor=palegray](+92,-30)(+92,+30)
\psline[linecolor=black,linestyle=dashed,linewidth=1](+92,-30)(+92,+30)
\rput[t]{90}(+93.5,+15){${\bf O8}$}
\psline[linecolor=red](-2.5,-2.5)(+93,-2.5)
\psline[linecolor=red](-1.5,-1.5)(+93,-1.5)
\psline[linecolor=red](-0.5,-0.5)(+93,-0.5)
\psline[linecolor=red](+0.5,+0.5)(+93,+0.5)
\psline[linecolor=red](+1.5,+1.5)(+93,+1.5)
\psline[linecolor=red](+2.5,+2.5)(+93,+2.5)
\rput[b](35,+4){$6\darkred\rm D6$}
\rput[t](35,-4){$\darkred SU(6)$}
\pscircle*[linecolor=purple](-2.5,-2.5){1}
\pscircle*[linecolor=purple](-1.5,-1.5){1}
\pscircle*[linecolor=purple](-0.5,-0.5){1}
\pscircle*[linecolor=purple](+0.5,+0.5){1}
\pscircle*[linecolor=purple](+1.5,+1.5){1}
\pscircle*[linecolor=purple](+2.5,+2.5){1}
\pscircle*[linecolor=green](+92,0){4}
\psframe*[linecolor=white](+92.5,-5)(+100,+5)
\psline(-14.5,-31)(-5.75,-34)(+3,-31)
\rput[t](-6,-35){Dual to the $\rm M9_1$}
\psline(+70.5,-31)(+81.75,-34)(+93,-31)
\rput[t](+82,-35){Dual to the $\rm M9_2$}
\psline(+105,-30)(+105,+30)
\psline{>->}(+105,-10)(+105,+10)
\rput[t]{90}(+106,0){$x^{7,8,9}$}
\endpspicture
\eqn\ZsixBranes
$$

On the left side of this diagram, the distance $a$ between the orientifold and the
outlier {\darkblue D8} branes is critical, hence $\darkgreen\lambda=\infty$ at $x^6=0$
and the enhancement of the perturbative 9D gauge symmetry from $\darkblue SO(4)\times U(6)$
to $\darkblue E_3\times SU(6)$.
On the right side, the distance $b$ is less then critical, hence finite $\lambda(x^6=L)$
and the 9D gauge symmetry remains $\darkblue U(8)=SU(8)\times U(1)$.
The $\FP6$ fixed plane is dual to six coincident {\darkred D6} branes
and the $\I_1$ intersection is dual to the {\darkred D6} branes terminating on the
six outlier {\darkblue D8} branes at $x^6=a$ without reaching the strongly coupled
orientifold plane.
As explained in \S4.4, junctions of this type impose locking boundary
conditions~\LockingBC\ upon the appropriate gauge fields --- in the present case
${\darkred SU(6)}^{\rm 7D}\times{\darkblue SU(6)}^{\rm 10D}\to{\purple SU(6)}^{\rm diag}$
--- without giving rise to any localized massless 6D particles.
Also, the $\darkblue E_3=SU(2)\times SU(3)$ gauge fields living at $x^6=0\neq a$
remain mere spectators at the intersection.
In other words, 
our brane model~\ZsixBranes\ correctly explains all the \M ysteries of the $\I_1$
intersection, \cf~eqs.~\ZsixHWleft.

The $\I_2$ intersection plane is dual to a combination of two distinct brane junctions
on the right side of the diagram~\ZsixBranes.
First, at $x^6=L-b$ all six {\darkred D6} branes cross eight {\darkblue D8} branes
without terminating.
At this `junction' we have zero length {\darkred6\darkblue8} open strings
which give rise to localized massless 6D hypermultiplets in
the $({\darkred\bf6},{\darkblue\bf8})$ representation of the ${\darkred SU(6)}^{\rm 7D}
\times{\darkblue SU(8)}^{\rm 9D}$ gauge symmetry.
Naturally, the 7D SYM fields suffer no boundary conditions at $x^6=L-b$ and continue
unmolested towards the second junction at
$x^6=L$ where the {\darkred D6} branes meet the {\darkgreen NS5} half-brane.
As we saw earlier in this section, the {\darkgreen NS5} half-brane
preserves the entire ${\darkred SU(6)}^{\rm 7D}$ gauge symmetry by  effectively
imposing Neumann boundary condition for all 7D vector fields and their 8--SUSY
fermionic partners.
At the same time, all 7D hypermultiplets suffer Dirichlet boundary conditions
while the open strings deep in the {\darkgreen NS5} half-brane's throat give rise
to localized hypermultiplets in the $\darkred \asymrep=\bf15$ of the $\darkred SU(6)$.

Together, the two junctions at $x^6=L$ and at $x^6=L-b$ correctly reproduce
all the localized twisted states and the boundary conditions of the $\I_2$
intersection plane of the HW picture, \cf\ eqs.~\ZsixHWright.
One \M ystery however remains unexplained, namely the $\darkblue U(1)$
charge of the $({\darkred\bf15},{\darkblue\bf1},{\darkblue -\coeff23})$
twisted states.
Naively, the 9D $\darkblue U(1)$ charge is a part of the $\darkblue U(8)$
symmetry living on the {\darkblue D8} world-volume at $x^6=L-b$ and hence
should not attach to particles originating elsewhere in $x^6$.
Since the $\darkred\bf15$ twisted states live on the {\darkgreen NS5} half-brane
at $x^6=L\neq(L-b)$, they should therefore remain $\darkblue U(1)$--neutral.

Clearly, this reasoning is too naive to be true, and indeed the abelian charges
in the \tI\ superstring theory are known to mix with each other
thanks to the $x^6$--dependent `cosmological constant'[\Romans,\PW] 
and its superpartners.
Besides the $\darkblue U(1)$ center of the $\darkblue U(8)$ on the {\darkblue D8}
world-volume, we also have the RR one-form of the \tI\ theory and the
$B^{\rm NS}_{6,\mu}$ field (which is a one-form from the 9D point of view).
Both of these vector fields live in the `bulk' of the \tI\ theory,
which puts them in touch with the {\darkgreen NS5} half-brane.
Although we do not quite understand the behavior of these fields in the throat
region of the half-brane, it stands to reason they might do something interesting
enough to couple to the $\darkred\bf15$ twisted states living there.
Consequently, the $\darkred\bf15$ twisted states acquire {\sl an} abelian charge
which we then need to identify as belonging to the $\darkblue U(1)\subset E_8^{(2)}$.

To back up this bit of wishful thinking with a mathematical argument, let us consider
the brane model~\ZsixBranes\ from a six-dimensional point of view.
Taking the $x^{7,8,9}$ coordinates to be genuinely non-compact, we turn off the 9D
gauge couplings --- which makes the corresponding symmetries global rather than
local.
The symmetries originating from {\sl finite} stretches of {\darkred D6} branes
keep finite 6D gauge couplings and hence remain local --- provided of course that
they do not lock onto global symmetries of 9D origins.
Since we are now interested in the \M ysteries at the right side of the diagram~\ZsixBranes\
rather than the locking happening at the left side, let us replace the whole left side
with some kind of a {\darkred D6} terminal which does not break or lock the $\darkred SU(6)$
symmetry, \eg, a free-floating {\darkgreen NS5} brane.
In other words, consider the following model: 
$$
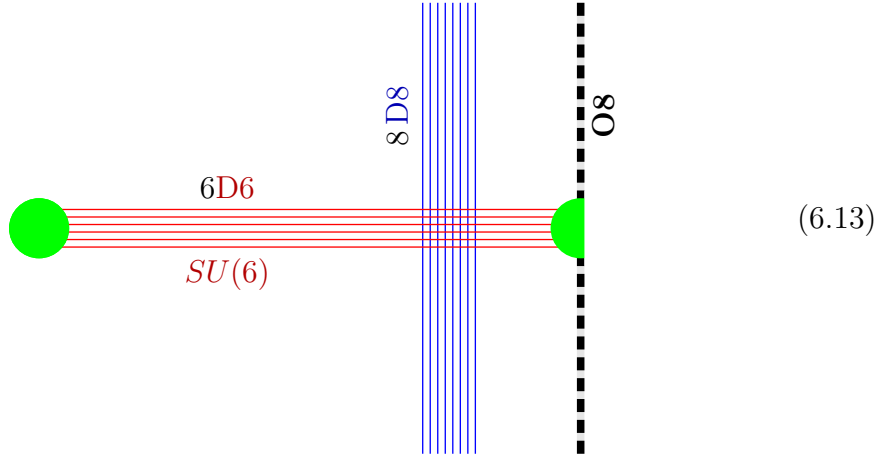
\pspicture[](+15,-40)(+100,+40)
\psline[linecolor=blue](+78,-30)(+78,+30)
\psline[linecolor=blue](+77,-30)(+77,+30)
\psline[linecolor=blue](+76,-30)(+76,+30)
\psline[linecolor=blue](+75,-30)(+75,+30)
\psline[linecolor=blue](+74,-30)(+74,+30)
\psline[linecolor=blue](+73,-30)(+73,+30)
\psline[linecolor=blue](+72,-30)(+72,+30)
\psline[linecolor=blue](+71,-30)(+71,+30)
\rput[b]{90}(+69,+15){$8\,\rm\darkblue D8$}
\psline[linewidth=1,linecolor=palegray](+92,-30)(+92,+30)
\psline[linecolor=black,linestyle=dashed,linewidth=1](+92,-30)(+92,+30)
\rput[t]{90}(+93.5,+15){${\bf O8}$}
\pscircle*[linecolor=green](+20,0){4}
\psline[linecolor=red](+20,-2.5)(+93,-2.5)
\psline[linecolor=red](+20,-1.5)(+93,-1.5)
\psline[linecolor=red](+20,-0.5)(+93,-0.5)
\psline[linecolor=red](+20,+0.5)(+93,+0.5)
\psline[linecolor=red](+20,+1.5)(+93,+1.5)
\psline[linecolor=red](+20,+2.5)(+93,+2.5)
\rput[b](45,+4){$6\darkred\rm D6$}
\rput[t](45,-4){$\darkred SU(6)$}
\pscircle*[linecolor=green](+20,0){4}
\pscircle*[linecolor=green](+92,0){4}
\psframe*[linecolor=white](+92.5,-5)(+100,+5)
\endpspicture
\eqn\NewModel
$$
{}From the 6D point of view, this new model describes an $\darkred SU(6)$ gauge theory
coupled to hypermultiplets in the $({\darkred\bf15})+8({\darkred\bf6})$ representation
of the gauge group.
\foot{%
    There is also a tensor multiplet arising from the freely floating
    {\darkgreen NS5} brane,  but it's existence does not affect the following argument.
    }
Classically, the {\sl flavor} symmetry of this model is $U(1)_{15}\times U(8)_{6}=
U(1)\times U(1)\times SU(8)$, but in the quantum theory one combination of the
abelian flavor symmetries is destroyed by the color anomaly.
The surviving anomaly-free combination is determined by the cubic index of the
color group,
$$
\Tr_{8({\bf\darkred6})}(F^3)\ =\ 8\Tr_{(\bf\darkred6)}(F^3)\
=\ 4\Tr_{(\bf\darkred15)}(F^3),
\eqno\eq
$$
hence the charge of the $\bf\darkred15$ hypermultiplet should be
exactly $-4$ times the charge of the $\bf\darkred6$.
In other words, we have
$$
\eqalign{
G\ &=\ {\darkred SU(6)}^{\rm color}\times[{\darkblue SU(8)\times U(1)}]^{\rm flavor} ,\cr
H\ &=\ ({\darkred\bf6};{\darkblue\bf8},{\darkblue+1})\
	+\ ({\darkred\bf15};{\darkblue\bf1},{\darkblue-4}) ,\cr
}\eqn\TensorCharge
$$
modulo an overall rescaling of the abelian charge.

When this picture is translated back into the \tI\ language, the flavor symmetry
$\darkblue SU(8)\times U(1)$ becomes a 9D gauge symmetry
which we would like to identify as {\sl the} $\darkblue SU(8)\times U(1)\subset E_8^{(1)}$.
Consequently, the quantum numbers of the twisted states localized at the junctions dual to
the HW~$\I_2$ should be exactly as in eq.~\TensorCharge\
--- and indeed these are precisely the quantum numbers of  the twisted states in
eq.~\ZsixHWright\ (modulo rescaling of the abelian charge by a factor $\coeff16$).

The bottom line of this exercise is to show that the brane model does somehow
provides the $\darkred\bf15$ twisted states with a correct $\darkblue U(1)$ charge.
Unfortunately, the provenance of this charge from the \tI\ point of view
remains an unsolved \M ystery.

%%%%%%%%%%%%%%%%%%%%%%%%%%%%%%%%%%%%%%%%%%%%%%%%%%%%%%%%%%%%%%%%%%%%%%%%%%%%%%%%%%%%
% End of chapter 6

%auto-ignore
% chapter7.tex
%% Chapter 7 of Orbifold2 paper
%% Written by Vadim Kaplunovsky

%%%%%%%%%%%%%%%%%%%%%%%%%%%%%%%%%%%%%%%%%%%%%%%%%%%%%%%%%%%%%%%%%%%%%%
\chapter{Junctions at Infinite String Coupling}
The {\darkgreen NS5} half-branes have strong string coupling
regions  hiding deep in their throats.
Other brane models have $\darkgreen\lambda\to\infty$ divergence at the
\Ori~orientifold planes, in full view of the 9D gauge symmetry ---
and in fact instrumental for brane engineering this symmetry
in the first place.
In this section, we consider brane models where such 
$\darkgreen\lambda=\infty$ orientifold planes are in direct contact
with the {\darkred D6} branes dual to fixed planes of the HW orbifolds.

Unfortunately, our knowledge of string theory is insufficient
to directly describe the physics of such strong-coupling brane junctions.
Instead, we put the $\rm HW\leftrightarrow I'$ duality machinery in
reverse gear and use the HW data to predict what {\sl should} happen
at the $\darkgreen\lambda=\infty$ junctions and leave the question of
{\sl how} it actually happens for future research.
Specifically, we consider two junction types, one involving a
$\darkgreen\lambda=\infty$ $\bf O8^-$ plane of D-brane charge~$-8$
and the other an $\bf O8^*$ plane [\SM] of charge~$-9$.
To keep our predictions reliable, we develop each junction using
a simple orbifold model with a clear HW picture described in our
previous paper [\KSTY], then confirm the results using
a much more complicated model with the same junction.

%%%%%%%%%%%%%%%%%%%%%%%%%%%%%%%%%%%%%%%%%%%%%%%%%%%%%%%%%%%%%%%%%
\section{A $\IZ_4$ Model with $E_1$ Extended Symmetry.}

We begin with the $\IZ_4$ orbifold model of ref.~[\KSTY] where 
the $E_8^{(1)}$ is broken to $E_5\times SU(4)$ and the $E_8^{(2)}$
to $E_1\times SU(8)$.
Conventionally, the $E_5$ symmetry is better known as the $SO(10)$
and the $E_1$ as the $SU(2)$, but here we use the $E_n$ notation
to highlight the origin of these two gauge groups in the \tI\
picture of the model:
$$
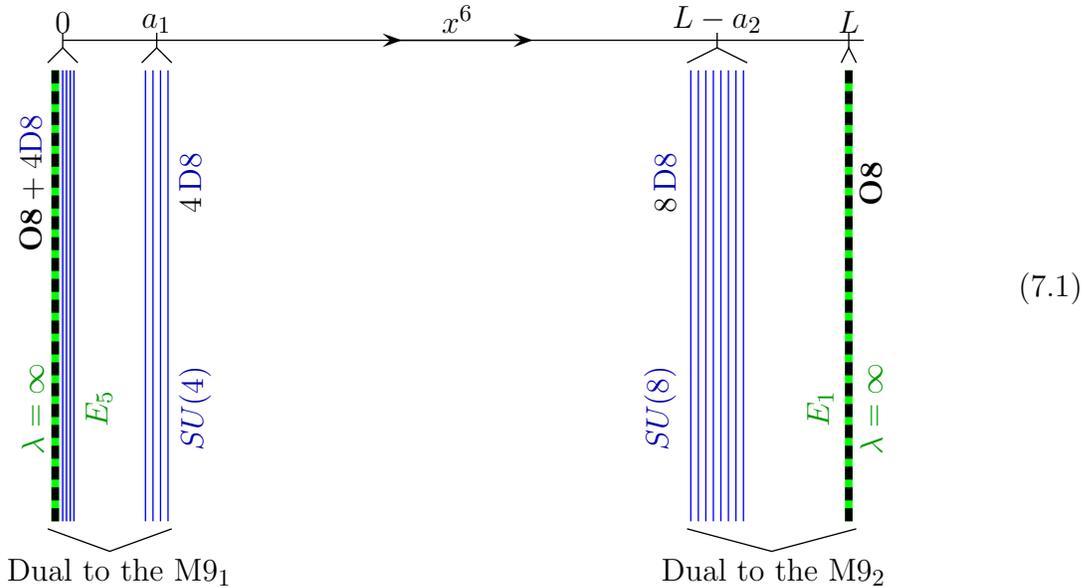
\pspicture[](-20,-40)(110,40)
\psline(-12.5,+34)(+94,+34)
\rput[b](+40,+35){$x^6$}
\psline{>->}(+30,+34)(+50,+34)
\psline(-12.5,+33)(-12.5,+35)	\rput[b](-12.5,+35){$0$}
\psline(-14.5,+31)(-12.5,+33)(-10.5,+31)
\psline(0,+33)(0,+35)		\rput[b](0,+35){$a_1$}
\psline(-2,+31)(0,+33)(+2,+31)
\psline(+74.5,+33)(+74.5,+35)	\rput[b](+74.5,+35){$L-a_2$}
\psline(+78.5,+31)(+74.5,+33)(+70.5,+31)
\psline(+92,+33)(+92,+35)	\rput[b](+92,+35){$L$}
\psline(+93,+31)(+92,+33)(+91,+31)
\psline[linecolor=palegray,linewidth=3](-12.5,-30)(-12.5,+30)
\psline[linecolor=green,linewidth=1](-13.5,-30)(-13.5,+30)
\psline[linecolor=black,linestyle=dashed,linewidth=1](-13.5,-30)(-13.5,+30)
\psline[linecolor=blue](-12.5,-30)(-12.5,+30)
\psline[linecolor=blue](-12.0,-30)(-12.0,+30)
\psline[linecolor=blue](-11.5,-30)(-11.5,+30)
\psline[linecolor=blue](-11.0,-30)(-11.0,+30)
\rput[b]{90}(-15,+15){${\bf O8}+4{\rm\darkblue D8}$}
\rput[t]{90}(-9.5,-15){$\darkgreen E_5$}
\rput[b]{90}(-15,-15){$\darkgreen\lambda=\infty$}
\psline[linecolor=blue](-1.5,-30)(-1.5,+30)
\psline[linecolor=blue](-0.5,-30)(-0.5,+30)
\psline[linecolor=blue](+0.5,-30)(+0.5,+30)
\psline[linecolor=blue](+1.5,-30)(+1.5,+30)
\rput[t]{90}(+3,+15){$4\,\rm\darkblue D8$}
\rput[t]{90}(+3,-15){$\darkblue SU(4)$}
\psline[linecolor=blue](+78,-30)(+78,+30)
\psline[linecolor=blue](+77,-30)(+77,+30)
\psline[linecolor=blue](+76,-30)(+76,+30)
\psline[linecolor=blue](+75,-30)(+75,+30)
\psline[linecolor=blue](+74,-30)(+74,+30)
\psline[linecolor=blue](+73,-30)(+73,+30)
\psline[linecolor=blue](+72,-30)(+72,+30)
\psline[linecolor=blue](+71,-30)(+71,+30)
\rput[b]{90}(+69,+15){$8\,\rm\darkblue D8$}
\rput[b]{90}(+69,-15){$\darkblue SU(8)$}
\psline[linewidth=1,linecolor=green](+92,-30)(+92,+30)
\psline[linecolor=black,linestyle=dashed,linewidth=1](+92,-30)(+92,+30)
\rput[t]{90}(+93.5,+15){${\bf O8}$}
\rput[t]{90}(+93.5,-15){$\darkgreen\lambda=\infty$}
\rput[b]{90}(+90,-15){$\darkgreen E_1$}
\psline(-14.5,-31)(-6.25,-34)(+2,-31)
\rput[t](-5,-35){Dual to the $\rm M9_1$}
\psline(+70.5,-31)(+81.75,-34)(+93,-31)
\rput[t](+82,-35){Dual to the $\rm M9_2$}
\endpspicture
\eqn\ZivDeight
$$
On both sides of this diagram, the outlier {\darkblue D8} branes are at critical
distances from the \Ori\ planes, hence divergent $\darkgreen\lambda\to\infty$
both at $x^6=0$ and at $x^6=L$ and therefore enhancement of the 9D gauge symmetry
${\darkblue SO(8)\times U(4)}\to{\darkgreen E_5}\times{\darkblue SU(4)}$
on the left side and
${\darkblue SO(0)\times U(8)}\to{\darkgreen E_1}\times{\darkblue SU(8)}$
on the right side.

In the HW picture of this model (\cf\ section~2), each $\IZ_4$ fixed
plane of the orbifold carries an $\darkred SU(4)$ which locks upon the
10D $\darkblue SU(4)$ at the left intersection~$\I_1$.
At the right intersection~$\I_2$, the ${\darkred SU(4)}^{\rm 7D}$
gauge fields have free (Neumann) boundary conditions, and that's where
the twisted states live.
Altogether, we have
$$
\eqalignno{
\I_1 &
\left\{\eqalign{
    G^{\rm local}\ &
    =\ {\purple SU(4)}^{\rm diag}\times[{\darkgreen SO(10)}]^{\rm 10D} ,\cr
    {\rm 7D}\,H\ &=\ {\purple 15},\cr
    {\rm 7D}\,V\ &=\ 0,\cr
    {\rm 6D}\,H\ &=\ 0;\cr
    }\right.&
\eqname\ZivHWleft \xdef\jj{(\chapterlabel.\number\equanumber} \cr
\noalign{\smallskip}
\I_2 &
\left\{\eqalign{
    G^{\rm local}\ &
    =\ {\darkred SU(4)}^{\rm 7D}\times
	[{\darkblue SU(8)}\times {\darkgreen SU(2)}]^{\rm 10D} ,\cr
    {\rm 7D}\,H\ &=\ 0,\cr
    {\rm 7D}\,V\ &=\ {\darkred 15},\cr
    {\rm 6D}\,H\ &
    =\ ({\darkred\bf4};{\darkblue\bf8},{\darkgreen\bf1})\
	+\ \half({\darkred\bf6};{\darkblue\bf1},{\darkgreen2}).\cr
    }\right.&
\eqname\ZivHWright  \cr
}$$

In brane engineering, the $\I_1$ intersection is obviously dual to a junction where
the four {\darkred D6} branes (dual to the $\FP6$) terminate on the
four outlier {\darkblue D8} branes at $x^6=a_1$, \cf~\S4.4.
Engineering the $\I_2$ intersection is less obvious, but the fact that both
the $({\darkred\bf4};{\darkblue\bf8},{\darkgreen\bf1})$ and the
$\half ({\darkred\bf6};{\darkblue\bf1},{\darkgreen2})$ twisted states
live there evidently requires the {\darkred D6} branes to cross the
eight outlier {\darkblue D8} branes at $x^6=L-a_2$, reach all the way
to the $\darkgreen\lambda=\infty$ orientifold plane at $x^6=L$ and
terminate there somehow.
Thus, 
$$
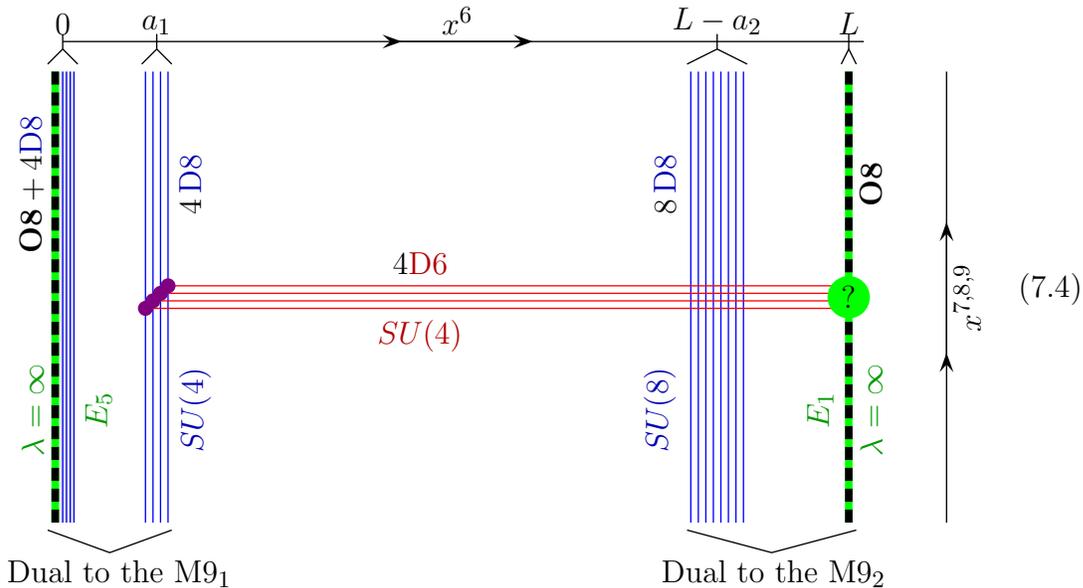
\pspicture[](-20,-40)(110,40)
\psline(-12.5,+34)(+94,+34)
\rput[b](+40,+35){$x^6$}
\psline{>->}(+30,+34)(+50,+34)
\psline(-12.5,+33)(-12.5,+35)	\rput[b](-12.5,+35){$0$}
\psline(-14.5,+31)(-12.5,+33)(-10.5,+31)
\psline(0,+33)(0,+35)		\rput[b](0,+35){$a_1$}
\psline(-2,+31)(0,+33)(+2,+31)
\psline(+74.5,+33)(+74.5,+35)	\rput[b](+74.5,+35){$L-a_2$}
\psline(+78.5,+31)(+74.5,+33)(+70.5,+31)
\psline(+92,+33)(+92,+35)	\rput[b](+92,+35){$L$}
\psline(+93,+31)(+92,+33)(+91,+31)
\psline[linecolor=palegray,linewidth=3](-12.5,-30)(-12.5,+30)
\psline[linecolor=green,linewidth=1](-13.5,-30)(-13.5,+30)
\psline[linecolor=black,linestyle=dashed,linewidth=1](-13.5,-30)(-13.5,+30)
\psline[linecolor=blue](-12.5,-30)(-12.5,+30)
\psline[linecolor=blue](-12.0,-30)(-12.0,+30)
\psline[linecolor=blue](-11.5,-30)(-11.5,+30)
\psline[linecolor=blue](-11.0,-30)(-11.0,+30)
\rput[b]{90}(-15,+15){${\bf O8}+4{\rm\darkblue D8}$}
\rput[t]{90}(-9.5,-15){$\darkgreen E_5$}
\rput[b]{90}(-15,-15){$\darkgreen\lambda=\infty$}
\psline[linecolor=blue](-1.5,-30)(-1.5,+30)
\psline[linecolor=blue](-0.5,-30)(-0.5,+30)
\psline[linecolor=blue](+0.5,-30)(+0.5,+30)
\psline[linecolor=blue](+1.5,-30)(+1.5,+30)
\rput[t]{90}(+3,+15){$4\,\rm\darkblue D8$}
\rput[t]{90}(+3,-15){$\darkblue SU(4)$}
\psline[linecolor=blue](+78,-30)(+78,+30)
\psline[linecolor=blue](+77,-30)(+77,+30)
\psline[linecolor=blue](+76,-30)(+76,+30)
\psline[linecolor=blue](+75,-30)(+75,+30)
\psline[linecolor=blue](+74,-30)(+74,+30)
\psline[linecolor=blue](+73,-30)(+73,+30)
\psline[linecolor=blue](+72,-30)(+72,+30)
\psline[linecolor=blue](+71,-30)(+71,+30)
\rput[b]{90}(+69,+15){$8\,\rm\darkblue D8$}
\rput[b]{90}(+69,-15){$\darkblue SU(8)$}
\psline[linewidth=1,linecolor=green](+92,-30)(+92,+30)
\psline[linecolor=black,linestyle=dashed,linewidth=1](+92,-30)(+92,+30)
\rput[t]{90}(+93.5,+15){${\bf O8}$}
\rput[t]{90}(+93.5,-15){$\darkgreen\lambda=\infty$}
\rput[b]{90}(+90,-15){$\darkgreen E_1$}
\psline[linecolor=red](-1.5,-1.5)(+92,-1.5)
\psline[linecolor=red](-0.5,-0.5)(+92,-0.5)
\psline[linecolor=red](+0.5,+0.5)(+92,+0.5)
\psline[linecolor=red](+1.5,+1.5)(+92,+1.5)
\rput[b](+35,+3){4\darkred D6}
\rput[t](+35,-3){$\darkred SU(4)$}
\pscircle*[linecolor=purple](-1.5,-1.5){1}
\pscircle*[linecolor=purple](-0.5,-0.5){1}
\pscircle*[linecolor=purple](+0.5,+0.5){1}
\pscircle*[linecolor=purple](+1.5,+1.5){1}
\cput*[fillcolor=green](+92,0){?}
\psline(-14.5,-31)(-6.25,-34)(+2,-31)
\rput[t](-5,-35){Dual to the $\rm M9_1$}
\psline(+70.5,-31)(+81.75,-34)(+93,-31)
\rput[t](+82,-35){Dual to the $\rm M9_2$}
\psline(+105,-30)(+105,+30)
\psline{>->}(+105,-10)(+105,+10)
\rput[t]{90}(+106,0){$x^{7,8,9}$}
\endpspicture
\eqn\ZivBranes
$$
the $\I_2$ splits into a simple {\darkred D6}--{\darkblue D8}
brane crossing at $x^6=L-a_2$ --- which clearly gives rise to the 
$({\darkred\bf4};{\darkblue\bf8},{\darkgreen\bf1})$
twisted hypermultiplets but does nothing to the 7D SYM fields themselves
--- plus a mysterious $\darkgreen\lambda=\infty$ terminus which
is supposed to fulfill the rest of eqs.~\ZivHWright.
In other words, for the sake of the $\rm HW\leftrightarrow I'$ duality,
we need this terminus to produce:
(1) Neumann boundary conditions for all {\darkred\bf 15} 7D vector fields,
(2) Dirichlet boundary conditions for all 7D hypermultiplets, and
(3) localized 6D massless half-hypermultiplets in the $({\darkred\bf6},
{\darkgreen\bf2})$ representations of the ${\darkred SU(4)}\times{\darkgreen SU(2)}$,
$$
\pspicture[0.475](60,-30)(96,+30)
\psline[linewidth=1,linecolor=green](92,-30)(92,+30)
\psline[linecolor=black,linestyle=dashed,linewidth=1](92,-30)(92,+30)
\rput[b]{90}(90,+17){\Ori\ @ $\darkgreen\lambda=\infty$}
\rput[b]{90}(90.5,-17){$\darkgreen E_1=SU(2)$}
\psline[linecolor=red](60,-1.5)(+92,-1.5)
\psline[linecolor=red](60,-0.5)(+92,-0.5)
\psline[linecolor=red](60,+0.5)(+92,+0.5)
\psline[linecolor=red](60,+1.5)(+92,+1.5)
\rput[b](75,+3){4\darkred D6}
\rput[t](75,-3){$\darkred SU(4)$}
\cput*[fillcolor=green](92,0){?}
\endpspicture
\left\{ \eqalign{
G^{\rm local}\ &=\ {\darkred SU(4)}\times{\darkgreen SU(2)} ,\cr
{\rm 7D}\,V\ &=\ {\darkred 15},\cr
{\rm 7D}\,H\ &=\ 0,\cr
{\rm 6D}\,H\ &=\ \half({\darkred\bf6},{\darkgreen\bf2}) .\cr
}\right.
\eqn\StrongJunction
$$

Although we do not have a complete theory of this junction, we do have a
conjecture based on a `detuned' version of the model.
That is, in the multi--Taub--NUT picture of the HW $\FP6$ fixed plane,
let us detune the Wilson line from a $\IZ_4$ twist breaking $E_8^{(2)}\to
SU(8)\times SU(2)$ to a more generic $U(1)$ twist which commutes with the
$SU(8)$ subgroup but not with the $SU(2)$.
Please note that while such a detuning breaks the duality with the
$T^4/\IZ_4$ heterotic orbifold model, it is a perfectly legitimate
deformation of the  multi--Taub--NUT configuration of the HW theory
in its own right.
In the \tI~language, this detuning corresponds to bringing the {\darkblue D8}
branes closer to the orientifold, $a_2\to b<a_2$ and consequently
avoiding the string coupling divergence at $x^6=L$ and the gauge symmetry
enhancement from $\darkblue U(1)$ to $\darkgreen E_1=SU(2)$.

For the detuned model, we want a terminal junction which works
as similarly to the $\darkgreen\lambda=\infty$ junction~\StrongJunction\
as mathematically possible.
That is, we want the same boundary conditions
for the 7D SYM fields as well as localized 6D hypermultiplets in
a $\darkred{\bf6}=\asymrep$ representation of the $\darkred SU(4)$;
in lieu of the half-doublet of the $\darkgreen SU(2)$ we broke
down to  the $\darkblue U(1)$,
the local states should simply have a non-zero $\darkblue U(1)$ charge.
Luckily, we already know how to engineer such a junction ---
we need to put an {\darkgreen NS5} half-brane at \Ori~plane and let the
four {\darkred D6} branes terminate on the $\darkgreen\half\,\rm NS5$,
\cf~\S6.
Indeed, according to eqs.~\NShalf\ this type of a terminus leads to
precisely the boundary conditions and the local 6D states we need
at $x^6=L$, \cf\ eqs.~\StrongJunction.

Similarly to the $\IZ_6$ model of section~6, we do not understand the
string theoretical origin of the $\darkblue U(1)$ charge of the 
$\darkred{\bf6}=\asymrep$ twisted states, but we can work it out
in terms of the anomaly-free flavor symmetry of the appropriate
6D theory.
Specifically, we build a brane model along the lines of fig.~\NewModel\
but use four {\darkred D6} branes instead of six, which
gives us a 6D $\darkred SU(4)$ gauge theory with 
$8({\darkred\bf4})+({\darkred\bf6})$ hypermultiplet matter.
Because the cubic index of the $({\darkred\bf6})$ representation of
the $\darkred SU(4)$ vanishes (it's a real representation), the
anomaly-free abelian flavor symmetry of this model acts on the
$({\darkred\bf6})$ fields only and leaves the $8({\darkred\bf4})$
fields neutral.
Translating this result back into the \tI\ language, we see that
the $({\darkred\bf6})$ states living at $x^6=0$
 have a 9D $\darkblue U(1)$ charge but the $({\darkred\bf4},{\darkblue\bf8})$
states living at $x^6=L-b$ remain neutral.
This is very important for the eventual 9D symmetry enhancement
${\darkblue U(1)}\to{\darkgreen E_1=SU(2)}$ because the
$({\darkred\bf4},{\darkblue\bf8})$ states are $\darkgreen SU(2)$ singlets.

In light of the above argument, we would like to conclude that the \M ysterious
junction~\StrongJunction\ is simply the $\lambda(L)\to\infty$ limit
of a $\darkgreen\half\,\rm NS5$ junction with four {\darkred D6} branes.
Unfortunately, this is not a well defined limit because the {\darkgreen NS5}
half-brane's tension is proportional to the $\lambda^{-1}(L)$
while its geometric size as a soliton is proportional to the
$\lambda^{+1}(L)$.
Indeed, in the strong coupling limit, the {\darkgreen NS5} half-brane
is best described as a magnetic monopole [\HanZaf] of the
$\darkgreen SU(2)$ SYM living on the orientifold plane.
This $\darkgreen SU(2)$ is spontaneously broken down to $\darkblue U(1)$
by the adjoint Higgs ${\rm VEV}\propto\lambda^{-1}(L)$, hence magnetic
monopoles.
Unfortunately, when the Higgs VEV vanishes and the non-abelian gauge symmetry
is restored, the monopoles become zero-tension infinite-size notional entities
rather than physical objects located at some particular places in~9D.
In particular, in the $\lambda(L)\to\infty$ limit we cannot affix such a
monopole / {\darkgreen NS5} half-brane to the {\darkred D6}--\Ori\ junction
at $\bx=0$, at least not by any 9D means at our disposal.

Therefore, we {\it\darkblue conjecture} that somehow
{\it\darkblue the {\darkred D6} branes pin
down the {\darkgreen NS5} half-brane to the junction and prevent it from
bloating to infinite size despite ${\lambda(L)=\infty}$}.
This conjecture is not based on any brane dynamics we know;
instead, we are driven to it by the logic of $\rm heterotic\leftrightarrow HW
\leftrightarrow type~I'$ duality in the orbifold context.
It would be very interesting to find out how the conjectured pinning down
of the {\darkgreen NS5} half-brane actually works --- or even to verify that
it indeed works --- but it's clearly a subject of future research.

Finally, to complete the duality, we need two more conjectures.
First, {\it the $\darkred\asymrep$ hypermultiplets} made of open $\darkred66$
strings in the throat of such a pinned-down {\darkgreen NS5} half-brane
{\it are half-doublets of the 9D gauge symmetry $\darkgreen E_1=SU(2)$.}
Again, we do not know the \tI\ origin of such $\darkgreen E_1$ quantum numbers,
we simply infer them from the $\rm heterotic\to HW\to I'$ duality chain.
Furthermore, we note that half-doublets of an $SU(2)$ symmetry are allowed only
for hypermultiplets in a {\sl real} representation of all other symmetries.
Consequently, the $\darkred\asymrep$ representation of the 7D $\darkred SU(N)$
symmetry must be real, which happens only for the $\darkred N=4$.
Therefore, we {\it\darkblue conjecture} than 
{\it\darkblue it takes precisely four {\darkred D6} branes to pin down a
{\darkgreen NS5} half-brane to a $\darkgreen\lambda=\infty$ junction.}

%%%%%%%%%%%%%%%%%%%%%%%%%%%%%%%%%%%%%%%%%%%%%%%%%%%%%%%%%%%%%%%%%%%%%%%%%%%
\section{A $\IZ_6$ Model with $E_1$ Symmetry.}
The conjectures we made would be better for a proof or at least for another
example of a similar junction~\StrongJunction.
Let us therefore consider a $T^4/\IZ_6$ heterotic orbifold in which
the $E_8^{(1)}$ is broken down to $SO(12)\times SU(2)\times U(1)$
(lattice shift vector $\delta_1=(-\coeff56,\coeff16,0,\ldots,0)$)
and the $E_8^{(2)}$ down to $SU(6)\times SU(2)\times SU(2)\times U(1)$
(shift vector $\delta_2=(-\coeff34,\coeff14,\coeff1{12},\cdots,\coeff1{12})$).
In terms of the unbroken subgroups, the $E_8$--breaking twists
act according to
$$
\def\qqq(#1,#2,#3){({\bf #1},{\bf #2},#3)}
\def\qqqq(#1,#2,#3,#4){({\bf #1},{\bf #2},{\bf #3},#4)}
\eqalign{
\alpha_1({\bf248})\ ={}\ &
{+\bigl[}\qqq(66,1,0)+\qqq(1,3,0)+\qqq(1,1,0)\bigr]\cr
&+ e^{+2\pi i/6}\,\qqq(12,2,-1)\ +\ e^{-2\pi i/6}\,\qqq(12,2,+1)\cr
&+ e^{+4\pi i/6}\bigl[\qqq(32,1,+1)+\qqq(1,1,-2)\bigr]\cr
&+ e^{-4\pi i/6}\bigl[\qqq(32,1,-1)+\qqq(1,1,+2)\bigr]\cr
&-\ \qqq(32',2,0),\crr
\alpha_2({\bf248})\ ={}\ &
{+\bigl[}\qqqq(35,1,1,0)+\qqqq(1,3,1,0)+\qqqq(1,1,3,0)+\qqqq(1,1,1,0)\bigr]\cr
&+ e^{+2\pi i/6}\bigl[\qqqq(\overline{15},1,2,+1)+\qqqq(6,2,1,-2)\bigr]\cr
&+ e^{-2\pi i/6}\bigl[\qqqq(15,1,2,-1)+\qqqq(\bar6,2,1,+2)\bigr]\cr
&+ e^{+4\pi i/6}\bigl[\qqqq(15,1,1,+2)+\qqqq(\bar6,2,2,-1)\bigr]\cr
&+ e^{-4\pi i/6}\bigl[\qqqq(\overline{15},1,1,-2)+\qqqq(6,2,2,+1)\bigr]\cr
&-\ \bigl[\qqqq(20,2,1,0)+\qqqq(1,1,2,\pm3)\bigr].\cr
}\eqn\MonsterAlphas
$$
The untwisted sector of the orbifold's spectrum comprises  SUGRA and dilaton
multiplets, 112 vector multiplets in the adjoint representation of 
$$
G\ =\ \bigl[ SO(12)\times SU(2)_A\times U(1)_1\bigr]\times
\bigl[SU(6)\times SU(2)_B\times SU(2)_C\times U(1)_2]\,,
\eqn\MonsterG
$$
two moduli and 66 charged hypermultiplets,
$$
\def\qqq(#1,#2,#3;#4,#5,#6,#7){({\bf #1},{\bf #2},#3;\,{\bf #4},{\bf #5},{\bf #6},#7)}
H_0\ =\ \qqq(12,2,-1;1,1,1,0)\ +\ \qqq(1,1,0;\overline{15},1,2,+1)\
+\ \qqq(1,1,0;6,2,1,-2)\ +\ 2M.
\eqno\eq
$$
Organizing the twisted sectors according to the fixed planes of the orbifold,
we have five $\IZ_2$ fixed planes (16 hypermultiplets per plane),
four $\IZ_3$ fixed planes (36 hypermultiplets per plane), and one $\IZ_6$
fixed plane carrying 60 hypermultiplets with rather complicated quantum numbers:
$$
\def\qqq(#1,#2,#3;#4,#5,#6,#7){({\bf #1},{\bf #2},#3;\,{\bf #4},{\bf #5},{\bf #6},#7)}
\eqalign{
H_{\rm tw}[\IZ_6]\ &
=\ \qqq(1,2,+\coeff23;6,1,1,-1)\ +\ \qqq(1,1,-\coeff23;6,2,1,0)\cr
&\qquad+\ \half\qqq(1,2,0;1,2,2,0)\ +\ \qqq(1,1,+\coeff43;1,1,2,-1)\cr
&\qquad+\ \qqq(12,1,-\coeff13;1,2,1,+1)\cr
&\qquad+\ 2\qqq(1,2,+\coeff23;1,2,1,+1)\ +\ 2\qqq(1,1,+\coeff23;1,1,1,-2).
}\eqn\MonsterTW
$$

At a first glance, these quantum numbers look too complicated for
any HW picture we might be able to write down.
Fortunately, brane engineering comes to rescue,
so let us consider the following monsterpiece: 
$$
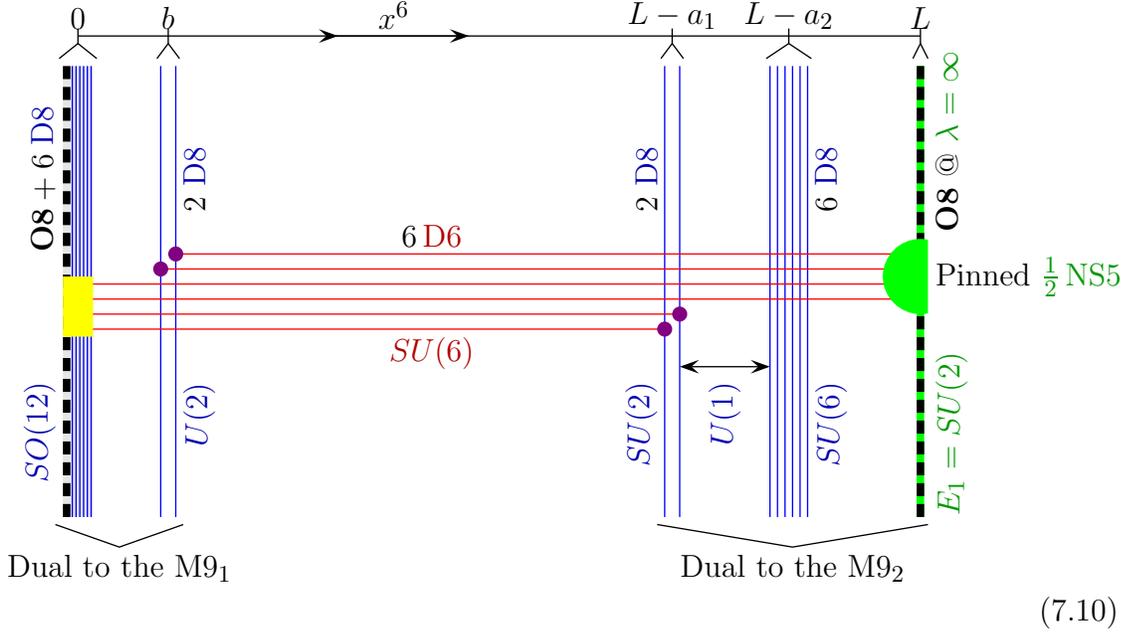
\pspicture[](-20,-40)(125,40)
\psline(-12,+34)(+100,+34)
\psline(-12,+33)(-12,+35)	\rput[b](-12,+35){$0$}
\psline(-14.5,+31)(-12,+33)(-9.5,+31)
\psline(0,+33)(0,+35)		\rput[b](0,+35){$b$}
\psline(-1.5,+31)(0,+33)(+1.5,+31)
\psline(+67,+33)(+67,+35)	\rput[b](+67,+35){$L-a_1$}
\psline(+65.5,+31)(+67,+33)(+68.5,+31)
\psline(+82.5,+33)(+82.5,+35)	\rput[b](+82.5,+35){$L-a_2$}
\psline(+79.5,+31)(+82.5,+33)(+85.5,+31)
\psline(+100,+33)(+100,+35)	\rput[b](+100,+35){$L$}
\psline(+99,+31)(+100,+33)(+101,+31)
\rput[b](+30,+35){$x^6$}
\psline{>->}(+20,+34)(+40,+34)
\psline[linecolor=palegray,linewidth=4](-12,-30)(-12,+30)
\psline[linecolor=black,linestyle=dashed,linewidth=1](-13.5,-30)(-13.5,+30)
\psline[linecolor=blue](-12.75,-30)(-12.75,+30)
\psline[linecolor=blue](-12.25,-30)(-12.25,+30)
\psline[linecolor=blue](-11.75,-30)(-11.75,+30)
\psline[linecolor=blue](-11.25,-30)(-11.25,+30)
\psline[linecolor=blue](-10.75,-30)(-10.75,+30)
\psline[linecolor=blue](-10.25,-30)(-10.25,+30)
\rput[b]{90}(-15,+15){${\bf O8}+6\;{\rm\darkblue D8}$}
\rput[br]{90}(-15,-12){$\darkblue SO(12)$}
\psline[linecolor=blue](-1,-30)(-1,+30)
\psline[linecolor=blue](+1,-30)(+1,+30)
\rput[t]{90}(+2,+15){2 \darkblue D8}
\rput[tr]{90}(+2,-12){$\darkblue U(2)$}
\psline[linecolor=blue](+85,-30)(+85,+30)
\psline[linecolor=blue](+84,-30)(+84,+30)
\psline[linecolor=blue](+83,-30)(+83,+30)
\psline[linecolor=blue](+82,-30)(+82,+30)
\psline[linecolor=blue](+81,-30)(+81,+30)
\psline[linecolor=blue](+80,-30)(+80,+30)
\rput[t]{90}(+86,+15){6 \darkblue D8}
\rput[tr]{90}(+86,-12){$\darkblue SU(6)$}
\psline[linecolor=blue](+68,-30)(+68,+30)
\psline[linecolor=blue](+66,-30)(+66,+30)
\rput[b]{90}(+65,+15){2 \darkblue D8}
\rput[br]{90}(+65,-12){$\darkblue SU(2)$}
\psline{<->}(+68,-10)(+80,-10)
\rput[r]{90}(+74,-12){$\darkblue U(1)$}
\psline[linecolor=green,linewidth=1](+100,-30)(+100,+30)
\psline[linecolor=black,linestyle=dashed,linewidth=1](+100,-30)(+100,+30)
\rput[tl]{90}(102,+8){\Ori\ @ $\darkgreen\lambda=\infty$}
\rput[tr]{90}(102,-8){$\darkgreen E_1=SU(2)$}
\psline[linecolor=red](+1,+5)(+100,+5)
\psline[linecolor=red](-1,+3)(+100,+3)
\psline[linecolor=red](-13,+1)(+100,+1)
\psline[linecolor=red](-13,-1)(+100,-1)
\psline[linecolor=red](-13,-3)(+68,-3)
\psline[linecolor=red](-13,-5)(+66,-5)
\rput[b](35,+6){$6\,\darkred\rm D6$}
\rput[t](35,-6){$\darkred SU(6)$}
\pscircle*[linecolor=purple](+1,+5){1}
\pscircle*[linecolor=purple](-1,+3){1}
\pscircle*[linecolor=purple](+68,-3){1}
\pscircle*[linecolor=purple](+66,-5){1}
\psframe*[linecolor=yellow](-14,-6)(-10,+2)
\pscircle*[linecolor=green](+100,+2){5}
\psframe*[linecolor=white](+101,-3)(+106,+7)
\rput[l](+102,+2){Pinned \darkgreen $\half\,$NS5}
\psline(-15,-31)(-6.5,-34)(+2,-31)
\rput[t](-6.5,-35){Dual to the $\rm M9_1$}
\psline(+65,-31)(+83,-34)(+101,-31)
\rput[t](+83,-35){Dual to the $\rm M9_2$}
\endpspicture
\eqn\MonsterBranes
$$
On the left side of this brane diagram,
the two outlier {\darkblue D8} branes are at less-than-critical distance
$b<a_c$ from the orientifold plane, hence finite $\lambda(0)$ and the
9D gauge symmetry is purely classical $\darkblue SO(12)\times U(2)$.
Of the six {\darkred D6} branes dual to the $\IZ_6$ $\FP6$ plane,
two have their left termini on the outlier {\darkblue D8} branes at $x^6=b$.
Brane junctions of this type were discussed in detail in section~5;
applying the same general rules to the junction at hand, we find the
${\darkred SU(6)}^{\rm 7D}$ gauge symmetry broken down to
$\darkred SU(2)_1\times SU(4)\times U(1)$ and furthermore
$[{\darkred SU(2)_1\times U(1)}]^{\rm 7D}\times
[{\darkblue SU(2)_A\times U(1)_1}]^{\rm 9D}
\to[{\purple SU(2)_A\times U(1)_1}]^{\rm diag}$.
The other four {\darkred D6} branes end on the \Ori~plane at $x^6=0$
where the 7D gauge symmetry is further broken ${\darkred SU(4)}
\to{\darkred Sp(2)}$ and the $\darkred6\darkblue8$ open strings give rise
to localized 6D massless half-hypermultiplets in the bi-fundamental
representation of the ${\darkred Sp(2)}\times{\darkblue SO(12)}$.

According to the $\rm I'\leftrightarrow HW$ duality, the two brane junctions
at $x^6=0$ and $x^6=b$ are together dual to the $\I_1$ intersection plane.
Totalling their combined effect, we have
$$
\def\qqq(#1,#2,#3,#4){({\darkblue\bf #1},{\bf\purple #2},{\purple #3},{\darkred\bf #4})}
\I_1\left\{
\eqalign{
G^{\rm local}\ &
=\ {\darkblue SO(12)}^{\rm 10D}\times\bigl[{\purple SU(2)_A\times U(1)_1}\bigr]^{\rm diag}
	\times{\darkred Sp(2)}^{\rm 7D} ,\cr
{\rm 7D}\,H\ &
=\ \qqq(1,3,0,1)\,+\,\qqq(1,1,0,1)\,+\,2\qqq(1,2,+1,4)\,+\,\qqq(1,1,0,5),\cr
{\rm 7D}\,V\ &=\ \qqq(1,1,0,10),\cr
{\rm 6D}\,H\ &=\ \half\qqq(12,1,0,4).\cr
    }\right.
\eqn\MonsterHWleft
\xdef\jj{(\chapterlabel.\number\equanumber}
$$

On the right side of the brane diagram~\MonsterBranes\ we have a critical
combination of the distances $a_1$ and $a_2$, hence $\darkgreen\lambda(L)=\infty$
and the 9D gauge symmetry enhancement from $\darkblue U(2)\times U(6)$ to
${\darkblue S(U(2)\times U(6))}\times{\darkgreen E_1}\equiv
{\darkblue SU(6)\times SU(2)_B\times U(1)_2}\times{\darkgreen SU(2)_C}$.
For the {\darkred D6} branes, the right side offers three junctions:
First, at $(L-a_1)$ two of the six {\darkred D6} branes terminate on the
{\darkblue D8} branes.
Consequently, the ${\darkred SU(6)}^{\rm 7D}$ gauge symmetry breaks down to
$\darkred SU(4)\times SU(2)_3\times U(1)$ and furthermore
$[{\darkred SU(2)_3\times U(1)}]^{\rm 7D}\times
[{\darkblue SU(2)_B\times U(1)_2}]^{\rm 9D}
\to[{\purple SU(2)_B\times U(1)_2}]^{\rm diag}$.
Second, at $(L-a_2)$ there is a {\darkred D6}/{\darkblue D8} brane crossing
which produces localized 6D hypermultiplets in the bi-fundamental of the
${\darkred SU(4)}\times{\darkblue SU(6)}$.
Finally, at $x^6=L$ the four {\darkred D6} branes pin down an {\darkgreen NS5}
half-brane to the $\darkgreen\lambda=\infty$ orientifold plane.
We presume this junction works exactly as conjectured in the previous section,
\cf\ eqs.~\StrongJunction, thus unbroken $\darkred SU(4)$ and 
localized half-hypermultiplets with $({\darkred\bf6},{\darkgreen2})$ quantum numbers.

Together, the three junctions are dual to the $\I_2$ intersection plane of the
HW picture.
Their net effect amounts to
$$
\def\qqq(#1,#2,#3,#4,#5){({\darkred\bf #1},{\bf\purple #2},{\purple #3},%
	{\darkblue\bf #4},{\darkgreen\bf #5})}
\I_2\left\{
\eqalign{
G^{\rm local}\ &
=\ {\darkred SU(4)}^{\rm 7D}\times\bigl[{\purple SU(2)_C\times U(1)_2}\bigr]^{\rm diag}
	\times\bigl[{\darkblue SU(6)}\times{\darkgreen SU(2)_C}\bigr]^{\rm 10D} ,\cr
{\rm 7D}\,H\ &
=\ \qqq(1,3,0,1,1)\,+\,\qqq(1,1,0,1,1)\,+\,2\qqq(4,2,\beta,1,1),\cr
{\rm 7D}\,V\ &=\ \qqq(15,1,0,1,1),\cr
{\rm 6D}\,H\ &=\ \qqq(4,1,\gamma,6,1)\,+\,\half\qqq(6,1,0,1,2).\cr
    }\right.
\eqn\MonsterHWright
\xdef\MonsterHW{\jj--\number\equanumber)}
$$
The abelian charges  $\beta$ and $\gamma$ in these formul\ae\ depend on the
precise manner of the ${\darkred U(1)}^{\rm 7D}\times{\darkblue  U(1)_2}^{\rm 9D}
\to{\purple  U(1)_2}^{\rm diag}$ diagonalization.
The simplest way to determine these charges is via anomaly considerations;
in the Appendix we show that all local anomalies at the $\I_2$ cancel out
provided $\beta=+\coeff32$ and $\gamma=-\half$.

In order to make sense out of the local quantum numbers in eqs.~\MonsterHW\
we need to combine the ${\darkred SU(6)}^{\rm 7D}$ breaking effects at
both ends of the $x^6$.
Diagrammatically,
$$
\vcenter{
\tabskip=0.5em plus 0.5em
\openup 1\jot
\halign{
&\hfil$\displaystyle{{}#{}}$\hfil\cr
SU(6) & \to & SU(4)_{12}\times SU(2)_3 \times U(1) \cr
\downarrow && \downarrow \cr
SU(2)_1\times SU(4)_{23}\times U(1) & \to &
    SU(2)_1\times SU(2)_2\times SU(2)_3\times U(1)^2\cr
\downarrow && \downarrow \cr
SU(2)_1\times Sp(2)\times U(1) & \to &
    SU(2)_1\times SU(2)_{2+3}\times U(1)^2\cr
}}\eqn\MonsterBreaking
$$
where the upper block follows from the left and right termini of the six
{\darkred D6} branes in fig.~\MonsterBranes\ matching each other in
in three distinct pairs while the lower block accounts for the
orientifold projection at $x^6=0$, \cf\ eq.~\SUivBreaking.
The abelian charges may be identified via either chain of symmetry breaking,
hence we may use
$$
\eqalign{
{\rm either}\ &
\left\{\eqalign{
    X_1\ &=\ \diag(0,0,\coeff{+1}2,\coeff{+1}2,\coeff{-1}2,\coeff{-1}2),\cr
    Y_1\ &=\ \diag(\coeff{+2}3,\coeff{+2}3,
		\coeff{-1}3,\coeff{-1}3,\coeff{-1}3,\coeff{-1}3),\cr
    }\right.\crr
{\rm or}\ &
\left\{\eqalign{
    X_2\ &=\ \diag(\coeff{+1}2,\coeff{+1}2,\coeff{-1}2,\coeff{-1}2,0,0),\cr
    Y_2\ &=\ \diag(\coeff{+1}3,\coeff{+1}3,
		\coeff{+1}3,\coeff{+1}3,\coeff{-2}3,\coeff{-2}3);\cr
    }\right.\cr
}\eqn\MonsterCharges
$$
the two sets of charges are related according to
$$
\left\{ \eqalign{
	X_1\ &=\ \coeff34 Y_2\ -\ \half X_2\cr
	Y_1\ &=\ \half Y_2\ +\ \phantom{\half} X_2\cr
	}\right\}\
\longleftrightarrow\
\left\{ \eqalign{
	X_2\ &=\ \coeff34 Y_1\ -\ \half X_1\cr
	Y_2\ &=\ \half Y_1\ +\ \phantom{\half} X_1\cr
	}\right\} .
\eqno\eq
$$

Locally at the $\I_2$ intersection, the ${\purple U(1)}_2^{\rm diag}$ charge
is $\beta Y_2$ while $X_2$ is one of the $\darkred SU(4)$ generators.
Consequently, the 
$({\darkred\bf4},{\purple\bf1},{\purple\gamma},{\darkblue\bf6},
{\darkgreen\bf1})$ hypermultiplets localized at the~$\I_2$
have $Y_2=(\gamma/\beta)=-\coeff13$ and $X_2=\pm\half$.
Specifically, we have doublets of the $\darkred SU(2)_1$ with $X_2=+\half$
and doublets of the $\darkred SU(2)_2$ with $X_2=-\half$.
Following 7D/10D symmetry mixings at both ends of the world,
the quantum numbers of these states evolve according to the 
following trajectories
$$
\pspicture[0.5](-2,-10)(115,60)
\psline[linecolor=red](100,50)(10,50)
\psline[linecolor=red](100,50)(0,40)(90,31)
\psline[linecolor=red]{->}(100,50)(30,50)
\psline[linecolor=red]{->}(100,50)(30,43)
\psline[linecolor=red]{->}(0,40)(70,33)
\psline[linecolor=blue]{->}(90,45)(90,-5)
\psline[linecolor=blue]{->}(10,55)(10,-5)
\psframe[linecolor=red,fillstyle=solid,fillcolor=blue](99,49)(101,51)
\psframe[linecolor=red,fillstyle=solid,fillcolor=yellow](-1,39)(+1,41)
\pscircle*[linecolor=purple](90,31){1.5}
\pscircle*[linecolor=purple](10,50){1.5}
\rput[l](103,50){$({\darkred\bf4},{\purple\bf1},{\purple-\half},
	{\darkblue\bf6},{\darkgreen\bf1})$}
\rput[b]{0}(60,51){$\darkred SU(2)_1\subset SU(6)_{\rm 7D}$}
\rput[t]{6}(60,45){$\darkred SU(2)_2\subset SU(6)_{\rm 7D}$}
\rput[t]{354}(60,32){$\darkred SU(2)_3\subset SU(6)_{\rm 7D}$}
\rput[tr]{90}(91.5,28){$\darkblue SU(2)_B\subset E_8^{(2)}$}
\rput[br]{90}(9,28){$\darkblue SU(2)_A\subset E_8^{(1)}$}
\rput[t](90,-6){$({\bf1},{\bf1},-\coeff23;{\bf6},{\bf2},{\bf1},-1)$}
\rput[t](10,-6){$({\bf1},{\bf2},+\coeff23;{\bf6},{\bf1},{\bf1},0)$}
\endpspicture
\eqn\MonsterTrajectoryOne
$$
and eventually become precisely as on the first line of eq.~\MonsterTW,
provided we identify the abelian charges according to
$$
\eqalign{
C_1^{\rm heterotic}\ =\ C^{\rm HW}_{U(1)\subset E_8^{(1)}}\ &
+\ \bigl( \coeff43 X_2\,=\,Y_1-\coeff23 X_1\bigr),\cr
C_2^{\rm heterotic}\ =\ C^{\rm HW}_{U(1)\subset E_8^{(2)}}\ &
+\ \bigl( \coeff32 Y_2 -X_2\,=\,2X_1 \bigr).\cr
}\eqn\MonsterChargeID
$$
\par\noindent
Similarly, the $\half({\darkred\bf6},{\purple\bf1},{\purple0},{\darkblue\bf1},
{\darkgreen\bf2})$ states at $\I_2$ (originating from the
pinned down {\darkgreen NS5} half-brane)
have $Y_2=0$ while the $\darkred{\bf6}\in SU(4)$ splits
into a bi-doublet of $\darkred SU(2)_1\times SU(2)_2$ with $X_2=0$
and two singlets with $X_2=\pm1$.
Again, these quantum numbers evolve according to
$$
\pspicture[0.5](-2,-10)(115,60)
\psline[linecolor=red](100,50)(10,50)
\psline[linecolor=red](100,50)(0,40)(90,31)
\psline[linecolor=red]{->}(100,50)(30,50)
\psline[linecolor=red]{->}(100,50)(30,43)
\psline[linecolor=red]{->}(0,40)(70,33)
\psline[linecolor=blue]{->}(90,45)(90,-5)
\psline[linecolor=blue]{->}(10,55)(10,-5)
\psframe[linecolor=red,fillstyle=solid,fillcolor=green](99,49)(101,51)
\psframe[linecolor=red,fillstyle=solid,fillcolor=yellow](-1,39)(+1,41)
\pscircle*[linecolor=purple](90,31){1.5}
\pscircle*[linecolor=purple](10,50){1.5}
\rput[l](103,50){$\half({\darkred\bf6},{\purple\bf1},{\purple0},
	{\darkblue\bf1},{\darkgreen\bf2})$}
\rput[b]{0}(60,51){$\darkred SU(2)_1\subset SU(6)_{\rm 7D}$}
\rput[t]{6}(60,45){$\darkred SU(2)_2\subset SU(6)_{\rm 7D}$}
\rput[t]{354}(60,32){$\darkred SU(2)_3\subset SU(6)_{\rm 7D}$}
\rput[tr]{90}(91.5,28){$\darkblue SU(2)_B\subset E_8^{(2)}$}
\rput[br]{90}(9,28){$\darkblue SU(2)_A\subset E_8^{(1)}$}
\rput[t](90,-6){$\half({\bf1},{\bf2},0;{\bf1},{\bf2},{\bf2},0)$}
\rput[t](10,-6){$({\bf1},{\bf1},+\coeff43;{\bf1},{\bf1},{\bf2},-1)$}
\endpspicture
\eqn\MonsterTrajectoryTwo
$$
and eventually become precisely as on the second line of eq.~\MonsterTW.

Next, consider the $\I_1$ intersection plane and the 
$\half({\darkblue\bf12},{\purple\bf1},{\purple0},{\darkred\bf4})$
hypermultiplets which live there.
In terms of the unbroken symmetries of the model, these states have
$Y_1=0$ ($Y_1$ being the local $\purple U(1)_1$ abelian charge) and
$X_1=\half$ (which follows from $X_1\in\darkred Sp(2)$);
they are also doublets of the $\darkred SU(2)_{2+3}$, which translates
into the heterotic language as $SU(2)_B\subset E_8^{(2)}$.
Consequently, we have
$({\bf12},{\bf1},-\coeff13;{\bf1},{\bf2},{\bf1},+1)$,
exactly as on the third line of eq.~\MonsterTW.

The remaining twisted states  arise as
zero modes of 7D hyper fields with Neumann boundary conditions at both ends.
The following table lists  the ${\darkred SU(6)}^{\rm 7D}$ SYM fields
according to their $\darkred SU(2)_1\times SU(2)_{2+3}\times U(1)^2$
and 8--SUSY quantum numbers and shows the
 boundary conditions at $\I_{1,2}$ according to eqs.~\MonsterHW:
$$
\def\qqq #1,#2{({\bf #1},{\bf#2})}
\vcenter{\Tenpoint
    \ialign{
	\hfil\vrule width 0pt height 12pt depth 4pt \vrule #&
	\enspace\hfil $\qqq #$\hfil\enspace & \vrule #\hfil &
	\enspace\hfil $(#)$\hfil\enspace & \vrule #\hfil &
	\enspace\hfil $(#)$\hfil\enspace & \vrule #\hfil &
	\enspace\hfil\frenchspacing(#\unskip)\hfil\enspace &\vrule #\hfil &
	\enspace\hfil\frenchspacing(#\unskip)\hfil\enspace &\vrule #\hfil \cr
	\noalign{\hrule}
	& \multispan5 \hfil Charges\hfil &&
	    \multispan3 \hfil Boundary Conditions\hfil &\cr
	\noalign{\hrule}
	&\omit\hfil $\,SU(2)_1\times SU(2)_{2+3}\,$\hfil && X_1,Y_1 && X_2,Y_2 &&
	    \omit\hfil 8--SUSY vector\hfil && \omit\hfil hyper\hfil &\cr
	\noalign{\hrule}
	& 3,1 && 0,0 && 0,0 && locking,Neumann && Neumann,Dirichlet &\cr
	\noalign{\hrule}
	& 1,1 && 0,0 && 0,0 && locking,Neumann && Neumann,Dirichlet &\cr
	\noalign{\hrule}
	& 1,3 && 0,0 && 0,0 && Neumann,locking && Dirichlet,Neumann &\cr
	\noalign{\hrule}
	& 1,1 && 0,0 && 0,0 && Neumann,locking && Dirichlet,Neumann &\cr
	\noalign{\hrule}
	& 1,3 && \pm1,0 && \mp\half,\pm1 && Neumann, Dirichlet &&
		Dirichlet, Neumann &\cr
	\noalign{\hrule}
	& 1,3 && 0,0 && 0,0 && Dirichlet,Neumann && Neumann,Dirichlet &\cr
	\noalign{\hrule}
	& 2,2 && \pm\half,\mp1 && \mp1,0 && Dirichlet,Neumann &&
		Neumann, Dirichlet &\cr
	\noalign{\hrule}
	& 1,1 && \pm1,0 && \mp\half,\pm1 && Dirichlet, Dirichlet &&
		Neumann, Neumann &\cr
	\noalign{\hrule}
	& 2,2 && \pm\half,\pm1 && \pm\half,\pm1 && Dirichlet, Dirichlet &&
		Neumann, Neumann &\cr
	\noalign{\hrule\smallskip}
	}
    }
\eqn\MonsterTable
$$
On the last two lines of this table we indeed find  hypermultiplets
with Neumann BC at both ends and hence zero modes.
Translating their abelian charges into the heterotic language according to
eqs.~\MonsterChargeID\ and identifying the two $\darkred SU(2)_1\times SU(2)_{2+3}$
as $SU(2)_A\times SU(2)_B$, we find precisely the twisted states
on the last line of eq.~\MonsterTW.

This completes our verification of the $\rm heterotic\leftrightarrow HW
\leftrightarrow I'$ duality as far as the massless spectrum of the model is concerned.
In the Appendix we verify the remaining duality constraints due to 6D gauge
couplings and local anomaly cancellation.
The bottom line is, in spite of formidable complexity of this model,
the duality works like \M agic!

Among other things, this \M agic involves a junction where four {\darkred D6}
branes pin down an {\darkgreen NS5} half-brane on an
$\Ori^{\darkgreen\lambda=\infty}$ orientifold plane.
To maintain the duality, this junction must work precisely according to
eqs.~\StrongJunction; this gives us strong `experimental' evidence in favor
of the conjectures we made in the previous section. 
In particular, we confirm that the
${\darkgreen\half\rm NS}@\Ori^{\darkgreen\lambda=\infty}$
junction involves precisely $N=4$ {\darkred D6} branes:
Although our model~\MonsterBranes\ has six {\darkred D6} branes in the
middle of the $x^6$ dimension, only four of them reach the {\darkgreen NS5}
half-brane while the other two terminate elsewhere!

%%%%%%%%%%%%%%%%%%%%%%%%%%%%%%%%%%%%%%%%%%%%%%%%%%%%%%%%%%%%%%%%%%%%%%%%%%%
\section{Junctions at $\Ori^*$ Planes: A $\IZ_3$ Example.}
All the \tI/D6 brane models we considered thus far were either within or at the
limit of the classical moduli space of the \tI\ superstring.
That is, even in the models where the string coupling diverged at the orientifold
planes, we could `detune' such divergence and keep $\darkgreen\lambda$ finite
after an infinitesimal change of Wilson lines around the multi--Taub--NUT
configuration of the HW theory.
In this section, we go beyond the classical limits and encounter an
{\sl excited} orientifold plane [\SM] $\Ori^*$ which has D-brane
charge $-9$
rather than $-8$ and {\sl requires} $\darkgreen\lambda=\infty$ for its
very existence.

Consider the the $T^4/\IZ_3$ model of ref.~[\KSTY] where the $E_8^{(1)}$
is broken down to $E_6\times SU(3)$ and the $E_8^{(2)}$ down to $E_0\times SU(9)
\equiv SU(9)$.
The $E_0$ factor here is trivial as a symmetry group, but in the \tI\ terms
it denotes a very non-trivial $\Ori^*$ plane at $x^6=L$
(see section~3 for discussion):
$$
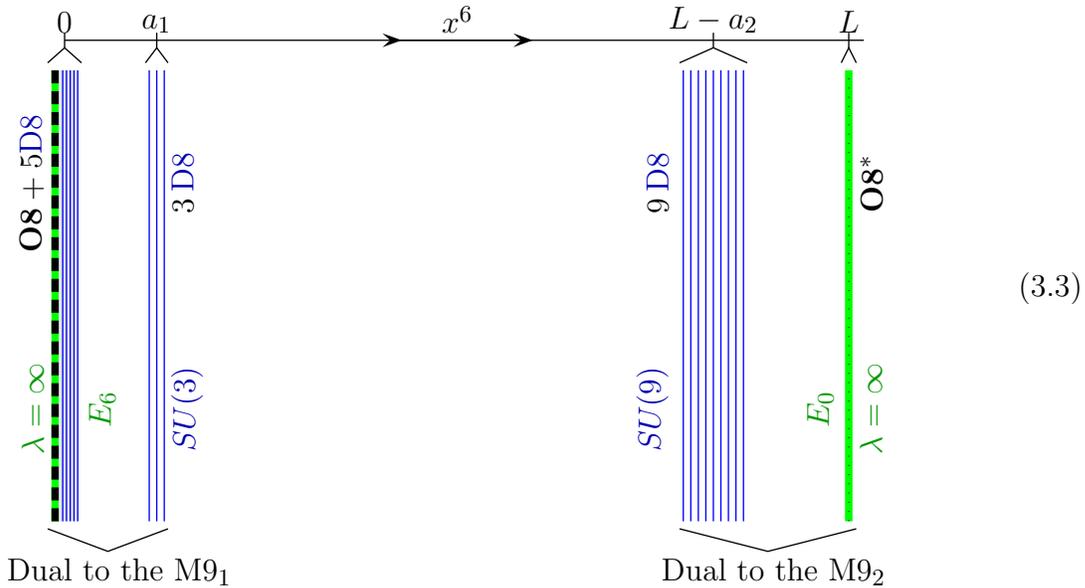
\pspicture[](-20,-40)(110,40)
\psline(-12.25,+34)(+94,+34)
\rput[b](+40,+35){$x^6$}
\psline{>->}(+30,+34)(+50,+34)
\psline(-12.25,+33)(-12.25,+35)	\rput[b](-12.25,+35){$0$}
\psline(-14.5,+31)(-12.25,+33)(-10.0,+31)
\psline(0,+33)(0,+35)		\rput[b](0,+35){$a_1$}
\psline(-1.5,+31)(0,+33)(+1.5,+31)
\psline(+74,+33)(+74,+35)	\rput[b](+74,+35){$L-a_2$}
\psline(+78.5,+31)(+74,+33)(+69.5,+31)
\psline(+92,+33)(+92,+35)	\rput[b](+92,+35){$L$}
\psline(+93,+31)(+92,+33)(+91,+31)
\psline[linecolor=green,linewidth=1](-13.5,-30)(-13.5,+30)
\psline[linecolor=black,linestyle=dashed,linewidth=1](-13.5,-30)(-13.5,+30)
\psline[linecolor=palegray,linewidth=2.5](-11.75,-30)(-11.75,+30)
\psline[linecolor=blue](-12.5,-30)(-12.5,+30)
\psline[linecolor=blue](-12.0,-30)(-12.0,+30)
\psline[linecolor=blue](-11.5,-30)(-11.5,+30)
\psline[linecolor=blue](-11.0,-30)(-11.0,+30)
\psline[linecolor=blue](-10.5,-30)(-10.5,+30)
\rput[b]{90}(-15,+15){${\bf O8}+5{\rm\darkblue D8}$}
\rput[t]{90}(-9,-15){$\darkgreen E_6$}
\rput[b]{90}(-15,-15){$\darkgreen\lambda=\infty$}
\psline[linecolor=blue](-1.0,-30)(-1.0,+30)
\psline[linecolor=blue](-0.0,-30)(-0.0,+30)
\psline[linecolor=blue](+1.0,-30)(+1.0,+30)
\rput[t]{90}(+2,+15){$3\,\rm\darkblue D8$}
\rput[t]{90}(+2,-15){$\darkblue SU(3)$}
\psline[linecolor=blue](+78,-30)(+78,+30)
\psline[linecolor=blue](+77,-30)(+77,+30)
\psline[linecolor=blue](+76,-30)(+76,+30)
\psline[linecolor=blue](+75,-30)(+75,+30)
\psline[linecolor=blue](+74,-30)(+74,+30)
\psline[linecolor=blue](+73,-30)(+73,+30)
\psline[linecolor=blue](+72,-30)(+72,+30)
\psline[linecolor=blue](+71,-30)(+71,+30)
\psline[linecolor=blue](+70,-30)(+70,+30)
\rput[b]{90}(+68,+15){$9\,\rm\darkblue D8$}
\rput[b]{90}(+68,-15){$\darkblue SU(9)$}
\psline[linewidth=1,linecolor=green](+92,-30)(+92,+30)
\psline[linecolor=black,linestyle=dotted](+92,-30)(+92,+30)
\rput[t]{90}(+93.5,+15){$\bf O8^*$}
\rput[t]{90}(+93.5,-15){$\darkgreen\lambda=\infty$}
\rput[b]{90}(+90,-15){$\darkgreen E_0$}
\psline(-14.5,-31)(-6.5,-34)(+1.5,-31)
\rput[t](-5,-35){Dual to the $\rm M9_1$}
\psline(+69.5,-31)(+81.25,-34)(+93,-31)
\rput[t](+82,-35){Dual to the $\rm M9_2$}
\endpspicture
\eqno\ZiiiBranesPlanes
$$
Thanks to the $-9$ charge of this plane, we put nine rather than eight 
coincident {\darkblue D8} branes at the critical location $L-a_2$
where they carry an ${\darkblue SU(9)}$ SYM on their world-volume.
The $\Ori^*$ plane itself does not carry any 9D fields, but 
manifests itself via non-trivial junctions with the {\darkred D6} branes.
On the left side of diagram \ZiiiBranesPlanes,
we have a more conventional $\bf O8^-$
plane accompanied by 5 {\darkblue D8} branes while 3 more {\darkblue D8} branes
are at critical distance $a_1$ away.
Due to this criticality, $\darkgreen\lambda(0)=\infty$ and the classical
$\darkblue SO(10)\times U(3)$ gauge symmetry is enhanced to the
${\darkgreen E_6}\times{\darkblue SU(3)}$.

In the HW picture of the $T^4/\IZ_3$ orbifold, each $\FP6$ fixed plane
carries an $\darkred SU(3)$, which locks upon the
10D $\darkblue SU(3)\subset E_1^{(1)}$ at the left intersection~$\I_1$;
the twisted $({\bf1},{\bf3},{\bf9})$ states live at the
right intersection~$\I_2$.
Altogether, we have
$$
\eqalignno{
\I_1 &
\left\{\eqalign{
    G^{\rm local}\ &
    =\ {\purple SU(3)}^{\rm diag}\times[{\darkgreen E_6}]^{\rm 10D} ,\cr
    {\rm 7D}\,H\ &=\ {\purple 8},\cr
    {\rm 7D}\,V\ &=\ 0,\cr
    {\rm 6D}\,H\ &=\ 0;\cr
    }\right.&
\eqname\ZiiiHWleft \xdef\jj{(\chapterlabel.\number\equanumber} \cr
\noalign{\smallskip}
\I_2 &
\left\{\eqalign{
    G^{\rm local}\ &
    =\ {\darkred SU(3)}^{\rm 7D}\times
	{\darkblue SU(9)}^{\rm 10D} ,\cr
    {\rm 7D}\,H\ &=\ 0,\cr
    {\rm 7D}\,V\ &=\ {\darkred 8},\cr
    {\rm 6D}\,H\ &
    =\ ({\darkred\bf3},{\darkblue\bf9}).\cr
    }\right.&
\eqname\ZiiiHWright  \cr
}$$
Brane-wise, the left intersection~$\I_1$ is evidently dual to the \S4.4 type
of a junction where the three {\darkred D6} branes (dual to the $\IZ_3$ $\FP6$)
end on the three outlier {\darkred D8} branes at $x^6=a_1$.
On the right side of the brane picture, the existence of the
$({\darkred\bf3},{\darkblue\bf9})$ localized states as well as absence of
$\darkred SU(3)$ locking clearly calls for the {\darkred D6} branes crossing
the {\darkblue D8} branes at $x^6=L-a_2$ without termination.
This leaves only one option for their eventual right terminus --- right
on the $\bf O8^*$ plane at $x^6=L$, \cf\ the following diagram:
$$
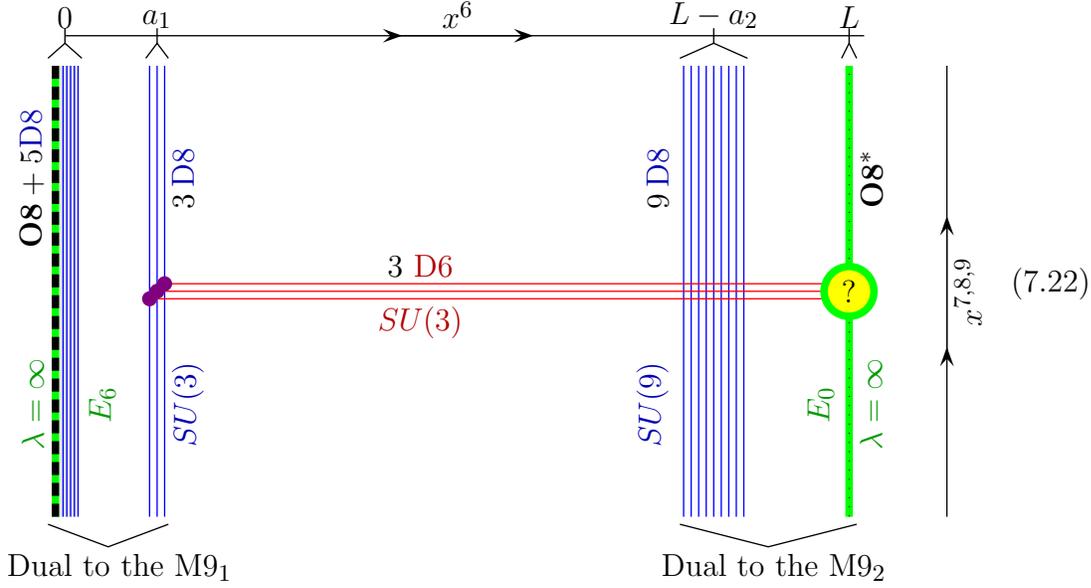
\pspicture[](-20,-40)(110,40)
\psline(-12.25,+34)(+94,+34)
\rput[b](+40,+35){$x^6$}
\psline{>->}(+30,+34)(+50,+34)
\psline(-12.25,+33)(-12.25,+35)	\rput[b](-12.25,+35){$0$}
\psline(-14.5,+31)(-12.25,+33)(-10.0,+31)
\psline(0,+33)(0,+35)		\rput[b](0,+35){$a_1$}
\psline(-1.5,+31)(0,+33)(+1.5,+31)
\psline(+74,+33)(+74,+35)	\rput[b](+74,+35){$L-a_2$}
\psline(+78.5,+31)(+74,+33)(+69.5,+31)
\psline(+92,+33)(+92,+35)	\rput[b](+92,+35){$L$}
\psline(+93,+31)(+92,+33)(+91,+31)
\psline[linecolor=green,linewidth=1](-13.5,-30)(-13.5,+30)
\psline[linecolor=black,linestyle=dashed,linewidth=1](-13.5,-30)(-13.5,+30)
\psline[linecolor=palegray,linewidth=2.5](-11.75,-30)(-11.75,+30)
\psline[linecolor=blue](-12.5,-30)(-12.5,+30)
\psline[linecolor=blue](-12.0,-30)(-12.0,+30)
\psline[linecolor=blue](-11.5,-30)(-11.5,+30)
\psline[linecolor=blue](-11.0,-30)(-11.0,+30)
\psline[linecolor=blue](-10.5,-30)(-10.5,+30)
\rput[b]{90}(-15,+15){${\bf O8}+5{\rm\darkblue D8}$}
\rput[t]{90}(-9,-15){$\darkgreen E_6$}
\rput[b]{90}(-15,-15){$\darkgreen\lambda=\infty$}
\psline[linecolor=blue](-1.0,-30)(-1.0,+30)
\psline[linecolor=blue](-0.0,-30)(-0.0,+30)
\psline[linecolor=blue](+1.0,-30)(+1.0,+30)
\rput[t]{90}(+2,+15){$3\,\rm\darkblue D8$}
\rput[t]{90}(+2,-15){$\darkblue SU(3)$}
\psline[linecolor=blue](+78,-30)(+78,+30)
\psline[linecolor=blue](+77,-30)(+77,+30)
\psline[linecolor=blue](+76,-30)(+76,+30)
\psline[linecolor=blue](+75,-30)(+75,+30)
\psline[linecolor=blue](+74,-30)(+74,+30)
\psline[linecolor=blue](+73,-30)(+73,+30)
\psline[linecolor=blue](+72,-30)(+72,+30)
\psline[linecolor=blue](+71,-30)(+71,+30)
\psline[linecolor=blue](+70,-30)(+70,+30)
\rput[b]{90}(+68,+15){$9\,\rm\darkblue D8$}
\rput[b]{90}(+68,-15){$\darkblue SU(9)$}
\psline[linewidth=1,linecolor=green](+92,-30)(+92,+30)
\psline[linecolor=black,linestyle=dotted](+92,-30)(+92,+30)
\rput[t]{90}(+93.5,+15){$\bf O8^*$}
\rput[t]{90}(+93.5,-15){$\darkgreen\lambda=\infty$}
\rput[b]{90}(+90,-15){$\darkgreen E_0$}
\psline[linecolor=red](-1,-1)(+92,-1)
\psline[linecolor=red](0,0)(+92,0)
\psline[linecolor=red](+1,+1)(+92,+1)
\rput[b](+35,+2){3 \darkred D6}
\rput[t](+35,-2){$\darkred SU(3)$}
\pscircle*[linecolor=purple](-1,-1){1}
\pscircle*[linecolor=purple](0,0){1}
\pscircle*[linecolor=purple](+1,+1){1}
\cput[fillcolor=yellow,fillstyle=solid,linecolor=green,linewidth=1](+92,0){?}
\psline(-14.5,-31)(-6.5,-34)(+1.5,-31)
\rput[t](-5,-35){Dual to the $\rm M9_1$}
\psline(+69.5,-31)(+81.25,-34)(+93,-31)
\rput[t](+82,-35){Dual to the $\rm M9_2$}
\psline(+105,-30)(+105,+30)
\psline{>->}(+105,-10)(+105,+10)
\rput[t]{90}(+106,0){$x^{7,8,9}$}
\endpspicture
\eqn\ZiiiBranes
$$

\par\noindent
The yellow circle with a `?' here denotes a ${\rm\darkred D6}/{\bf O8^*}$ junction
whose string theory is beyond our present knowledge.
Instead, we may use $\rm HW\leftrightarrow I'$ duality to argue that this junction
--- whatever it is --- must accomplish the $\I_2$ intersection's job
(\cf\ eqs.~\ZiiiHWright) which isn't accomplished by the
the {\darkred D6}/{\darkblue D8} brane crossing at $x^6=(L-a_2)$.
Consequently, for duality's sake, {\it\darkblue we conjecture} that
the ${\rm\darkred D6}/{\bf O8^*}$ junction somehow produces:
(1)~Neumann boundary conditions for all {\darkred\bf8} 7D vector fields,
(2)~Dirichlet boundary conditions for the 7D hypermultiplet fields, and
(3)~no localized 6D massless particles,
$$
\pspicture[0.475](60,-30)(97,+30)
\psline[linewidth=1,linecolor=green](92,-30)(92,+30)
\psline[linecolor=black,linestyle=dotted](92,-30)(92,+30)
\rput[b]{90}(90,+17){$\bf O8^*$}
\rput[b]{90}(90.5,-17){$\darkgreen E_0$}
\psline[linecolor=red](60,-1)(+92,-1)
\psline[linecolor=red](60,0)(+92,0)
\psline[linecolor=red](60,+1)(+92,+1)
\rput[b](75,+2){3 \darkred D6}
\rput[t](75,-2){$\darkred SU(3)$}
\cput[fillcolor=yellow,fillstyle=solid,linecolor=green,linewidth=1](92,0){?}
\endpspicture
\left\{ \eqalign{
G^{\rm local}\ &=\ {\darkred SU(3)}\times{\darkgreen E_0} ,\cr
{\rm 7D}\,V\ &=\ {\darkred 8},\cr
{\rm 7D}\,H\ &=\ 0,\cr
{\rm 6D}\,H\ &=\ 0.\cr
}\right.
\eqn\UltraJunction
$$
Note that the survival of the whole ${\darkred SU(3)}^{\rm 7D}$ gauge symmetry
at this junction is quite different from the orientifold projection
$\darkred SU(N)\to Sp(N/2)$ at ordinary {\darkred D6}/\Ori\ junctions,
\cf~\S4.3.
Consequently, while the ordinary \Ori~planes require an even number
$N$ of {\darkred D6} branes to terminate at the same point $\bx$,
the $\bf O8^*$ plane is evidently quite happy with an odd {\darkred D6}
number~$N=3$.

%%%%%%%%%%%%%%%%%%%%%%%%%%%%%%%%%%%%%%%%%%%%%%%%%%%%%%%%%%%%%%%%%%%%%%%%%%%%%%%%%%
\section{A $\IZ_4$ Example of an $\bf O8^*$ Junction.}
Again, to affirm the conjectures we made about the ${\darkred\rm D6}/{\bf O8^*}$
junction~\UltraJunction\ we present another orbifold model with a similar junction.
In heterotic terms, the model is an $T^4/\IZ_4$ orbifold with $E_8^{(1)}$
broken down to $SO(12)\times SU(2)\times U(1)$ (\cf\ eq.~\ZfourSOxiiUii)
and $E_8^{(2)}$ down to $SU(8)\times U(1)$;
in lattice terms, $\delta_2=(\coeff58,\coeff18,\ldots,\coeff18)$ while
in terms of the surviving gauge symmetry,
$$
\eqalign{
\alpha_2({\bf248})\ &
=\ +\bigl[ ({\bf63},0) + ({\bf1},0)\bigr]\
	-\ \bigl[ ({\bf28},+1) + (\overline{\bf28},-1) \bigr]\cr
&\qquad+\ i\bigl[ (\overline{\bf56},+\half) + ({\bf8},-\coeff32) \bigr]\
	-\ i\bigl[ ({\bf56},-\half) + ({\bf\bar8},+\coeff32) \bigr] .\cr
}\eqn\ZIVSUviiiUi
$$
The untwisted sector of this model comprises the usual SUGRA and dilaton multiplets,
134 vector multiplets in the adjoint of
$$
G\ =\ \bigl[SO(12)\times SU(2)\times U(1)\bigr]\,
\times\,\bigl[SU(8)\times U(1)\bigr] ,
\eqn\ZIVG
$$
2 moduli and 120 charged hypermultiplets,
$$
\def\qqq(#1,#2,#3,#4,#5){({\bf #1},{\bf #2}, #3; {\bf #4}, #5 )}
H_0\ =\ \qqq(32,1,+1,1,0)\, +\, \qqq(12,2,-1,1,0)\,
+\, \qqq(1,1,0,\overline{56},+\half)\, +\, \qqq(1,1,0,8,-\coeff32)\, +\, 2M.
\eqno\eq
$$
Arranging the twisted sector according to the $\FP5$ fixed planes, we have
6 $\IZ_2$ planes carrying 16 hypermultiplets per plane,
$$
H_{\rm tw}[\IZ_2]\ =\ ({\bf1},{\bf1},\pm1;{\bf8},+\half),
\eqno\eq
$$
and 4 $\IZ_4$ planes carrying 40 hypermultiplets per plane,
$$
\def\qqq(#1,#2,#3,#4,#5){({\bf #1},{\bf #2}, #3; {\bf #4}, #5 )}
\eqalign{
H_{\rm tw}[\IZ_4]\  =\ \qqq(12,1,-\half,1,+1)\ &
+\ \qqq(1,2,+\half,8,-\half)\ +\ \qqq(1,1,-1,8,+\half)\cr
&+\ 2\qqq(1,2,+\half,1,+1) .\cr
}\eqn\ZIVHtw
$$

The brane dual of such a $\IZ_4$ fixed plane is engineered as follows:
$$
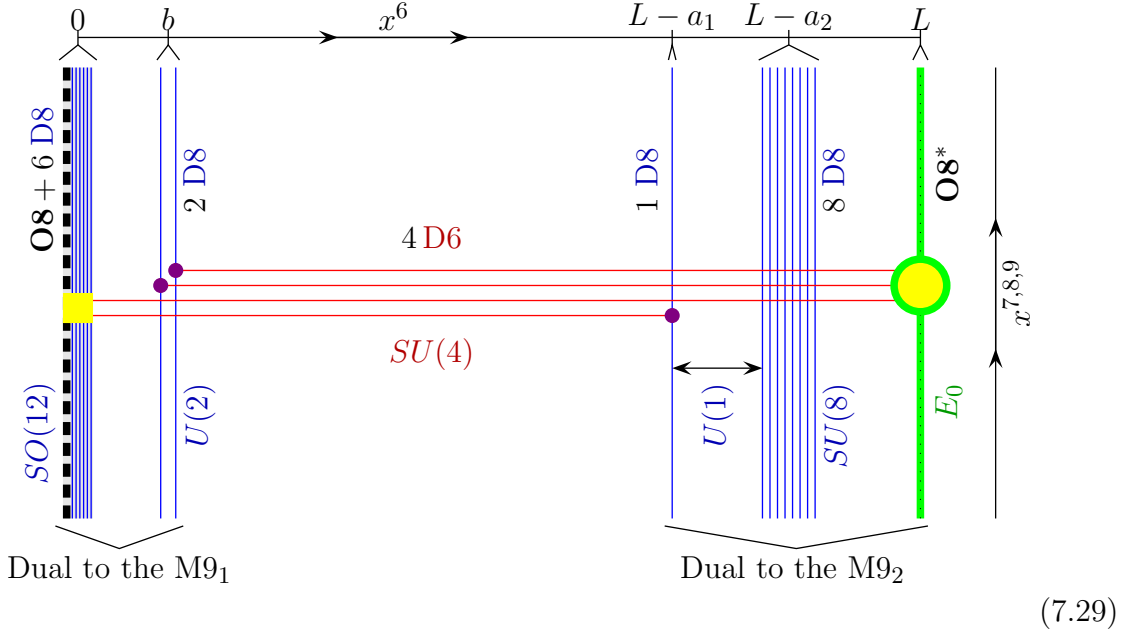
\pspicture[](-20,-40)(125,40)
\psline(-12,+34)(+100,+34)
\psline(-12,+33)(-12,+35)	\rput[b](-12,+35){$0$}
\psline(-14.5,+31)(-12,+33)(-9.5,+31)
\psline(0,+33)(0,+35)		\rput[b](0,+35){$b$}
\psline(-1.5,+31)(0,+33)(+1.5,+31)
\psline(+67,+33)(+67,+35)	\rput[b](+67,+35){$L-a_1$}
\psline(+66.5,+31)(+67,+33)(+67.5,+31)
\psline(+82.5,+33)(+82.5,+35)	\rput[b](+82.5,+35){$L-a_2$}
\psline(+78.5,+31)(+82.5,+33)(+86.5,+31)
\psline(+100,+33)(+100,+35)	\rput[b](+100,+35){$L$}
\psline(+99,+31)(+100,+33)(+101,+31)
\rput[b](+30,+35){$x^6$}
\psline{>->}(+20,+34)(+40,+34)
\psline[linecolor=palegray,linewidth=4](-12,-30)(-12,+30)
\psline[linecolor=black,linestyle=dashed,linewidth=1](-13.5,-30)(-13.5,+30)
\psline[linecolor=blue](-12.75,-30)(-12.75,+30)
\psline[linecolor=blue](-12.25,-30)(-12.25,+30)
\psline[linecolor=blue](-11.75,-30)(-11.75,+30)
\psline[linecolor=blue](-11.25,-30)(-11.25,+30)
\psline[linecolor=blue](-10.75,-30)(-10.75,+30)
\psline[linecolor=blue](-10.25,-30)(-10.25,+30)
\rput[b]{90}(-15,+15){${\bf O8}+6\;{\rm\darkblue D8}$}
\rput[br]{90}(-15,-12){$\darkblue SO(12)$}
\psline[linecolor=blue](-1,-30)(-1,+30)
\psline[linecolor=blue](+1,-30)(+1,+30)
\rput[t]{90}(+2,+15){2 \darkblue D8}
\rput[tr]{90}(+2,-12){$\darkblue U(2)$}
\psline[linecolor=blue](+86,-30)(+86,+30)
\psline[linecolor=blue](+85,-30)(+85,+30)
\psline[linecolor=blue](+84,-30)(+84,+30)
\psline[linecolor=blue](+83,-30)(+83,+30)
\psline[linecolor=blue](+82,-30)(+82,+30)
\psline[linecolor=blue](+81,-30)(+81,+30)
\psline[linecolor=blue](+80,-30)(+80,+30)
\psline[linecolor=blue](+79,-30)(+79,+30)
\rput[t]{90}(+87,+15){8 \darkblue D8}
\rput[tr]{90}(+87,-12){$\darkblue SU(8)$}
\psline[linecolor=blue](+67,-30)(+67,+30)
\rput[b]{90}(+65,+15){1 \darkblue D8}
\psline{<->}(+67,-10)(+79,-10)
\rput[r]{90}(+73,-12){$\darkblue U(1)$}
\psline[linecolor=green,linewidth=1](+100,-30)(+100,+30)
\psline[linecolor=black,linestyle=dotted](+100,-30)(+100,+30)
\rput[tl]{90}(102,+12){$\bf O8^*$}
\rput[tr]{90}(102,-12){$\darkgreen E_0$}
\psline[linecolor=red](+1,+3)(+100,+3)
\psline[linecolor=red](-1,+1)(+100,+1)
\psline[linecolor=red](-13,-1)(+100,-1)
\psline[linecolor=red](-13,-3)(+67,-3)
\rput[b](35,+6){$4\,\darkred\rm D6$}
\rput[t](35,-6){$\darkred SU(4)$}
\pscircle*[linecolor=purple](+1,+3){1}
\pscircle*[linecolor=purple](-1,+1){1}
\pscircle*[linecolor=purple](+67,-3){1}
\psframe*[linecolor=yellow](-14,-4)(-10,0)
\pscircle[linecolor=green,linewidth=1,fillcolor=yellow,fillstyle=solid](+100,+1){4}
\psline(-15,-31)(-6.5,-34)(+2,-31)
\rput[t](-6.5,-35){Dual to the $\rm M9_1$}
\psline(+66,-31)(+83.5,-34)(+101,-31)
\rput[t](+83,-35){Dual to the $\rm M9_2$}
\psline(+110,-30)(+110,+30)
\psline{>->}(+110,-10)(+110,+10)
\rput[t]{90}(+111,0){$x^{7,8,9}$}
\endpspicture
\eqn\ZIVBranes
$$
The left half of this diagram is similar to that of fig~\ZfourBranes\
and works in exactly the same way:
First, at $x^6=b$ the ${\darkred SU(4)}^{\rm 7D}$ gauge symmetry breaks down to
$\darkred SU(2)_1\times SU(2)_2\times U(1)$ and the $\darkred  SU(2)_2\times U(1)$
gauge fields lock onto the 9D $\darkblue SU(2)\times U(1)$ fields.
Second, at $x^6=0$ the {\darkred6\darkblue8} opens strings produce localized
half-hypermultiplets in the bi-fundamental representation of the
${\darkred SU(2)_1}\times{\darkblue SO(12)}$.
Together, the two junctions are dual to the $\I_1$ intersection of the HW picture
whose net effect is therefore
$$
\def\qqq(#1,#2,#3,#4){({\darkblue\bf #1},{\purple\bf #2},{\purple #3},{\darkred\bf #4})}
\I_1\,\left\{
\eqalign{
G^{\rm local}\ &
=\ {\darkblue SO(12)}^{\rm 10D} \times
    \bigl[{\purple SU(2)_1 \times U(1)}\bigr]^{\rm diag}\times
    {\darkred SU(2)_2}^{\rm 7D} ,\cr
{\rm 7D}\,V\ &=\ \qqq(1,1,0,3) ,\cr
{\rm 7D}\,H\ &=\ \qqq(1,3,0,1)\,+\,\qqq(1,1,0,1)\,+\,\qqq(1,2,\pm1,2) ,\cr
{\rm 6D}\,H\ &=\ \half\qqq(12,1,0,2) .\cr
}\right.
\eqn\ZIVHWleft
\xdef\jj{(\chapterlabel.\number\equanumber}
$$

The right half of the diagram~\ZIVBranes\ is more complicated.
The $SU(8)\times U(1)$ subgroup of $E_8$ described in eq.~\ZIVSUviiiUi\
is actually $S\bigl(U(8)\times U(1)\bigr)\times E_0$, hence the $\bf O8^*$
plane at $x^6=L$ and nine {\darkblue D8} branes, eight at $(L-a_2)$
and one further away at $(L-a_1)$.
\foot{%
    There are two in-equivalent $SU(8)\times U(1)$ subgroups of $E_8$
    distinguished by the respective adjoint decompositions,
    \cf\ eq.~\ZsixAlphaRight\ {\it v.} eq.~\ZIVSUviiiUi.
    Brane-wise, the first alternative is depicted on fig.~\ZsixBranes\
    and the second on fig.~\ZIVBranes.
    }
Therefore, the $\I_2$ intersection of the HW picture is dual to three distinct
brane junctions:
First, at $(L-a_1)$ one of the {\darkred D6} branes ends on the outlier
{\darkblue D8} brane.
Consequently, the ${\darkred SU(4)}^{\rm 7D}$ gauge symmetry breaks to
$\darkred SU(3)\times U(1)$ and furthermore
${\darkred U(1)}^{\rm 7D}\times{\darkblue U(1)}^{\rm 9D}\to{\purple U(1)}^{\rm diag}$.
Second, there is a {\darkred D6}/{\darkblue D8} brane crossing at $(L-a_2)$
which yields localized hypermultiplets in the bi-fundamental representation
of the ${\darkred SU(3)}\times{\darkblue SU(8)}$.
Finally, three {\darkred D6} branes terminate on $\bf O8^*$ at $x^6=L$;
we presume this junction to work exactly as in the previous model,
\cf\ eqs.~\UltraJunction.

In HW terms, the net effect of the three junctions is as follows:
$$
\def\qqq(#1,#2,#3){({\darkblue\bf #1},{\purple #2},{\darkred\bf #3})}
\I_2\,\left\{
\eqalign{
G^{\rm local}\ &
=\ {\darkblue SU(8)}^{\rm 10D} \times {\purple U(1)}^{\rm diag}\times
    {\darkred SU(3)}^{\rm 7D} ,\cr
{\rm 7D}\,V\ &=\ \qqq(1,0,8) ,\cr
{\rm 7D}\,H\ &=\ \qqq(1,0,1)\ +\ 2\qqq(1,\beta,3) ,\cr
{\rm 6D}\,H\ &=\ \qqq(8,\gamma,3) ,\cr
}\right.
\eqn\ZIVHWright
\xdef\ZIVHW{\jj--\number\equanumber)}
$$
where the abelian charges $\beta$ and $\gamma$ are non-zero but their exact
values depend on the details of the $\purple U(1)$ gauge fields locking.
As in \S7.2, we use anomaly considerations to determine
$\beta=+\coeff23$, $\gamma=-\coeff16$, \cf\  calculation in the Appendix.

To verify that eqs.~\ZIVHWleft\ and \ZIVHWright\ correctly describe the
HW picture of the heterotic model,
we need to combine the ${\darkred SU(4)}^{\rm 7D}$ breaking effects at both
ends of the $x^6$,
$$
\vcenter{
    \tabskip=0.5em plus 0.5em
    \openup 1\jot
    \halign{
	&\hfil$\displaystyle{{}#{}}$\hfil\cr
	SU(4) & \to & SU(3)\times U(1) \cr
	\downarrow && \downarrow \cr
	SU(2)_1\times U(1)\times SU(2)_2 & \to & SU(2)_1\times U(1)\times U(1).\cr
	}
    }
\eqn\ZIVSUiv
$$
Note that the surviving $SU(2)$ subgroup is the $SU(2)_1$ (which locks
onto the $SU(2)\subset E_8^{(1)}$ at the~$\I_1$) rather than the $SU(2)_2$
(which acts freely at the~$\I_1$);
this identification follows from matching the left termini of each of the
four {\darkred D6} branes in fig.~\ZIVBranes\ with the appropriate right
termini.
The abelian charges may be identified via either chain of symmetry breaking,
hence we may use
$$
\vcenter{\openup 2\jot
    \ialign{
	&\hfil #\unskip\quad &
	\hfil $\displaystyle{#}$\ &
	$\displaystyle{{}=\ #}$\quad\hfil\cr
	either & X & \diag(+\half,+\half,-\half,-\half) &
	and & T & \diag(0,0,+\half,-\half) \cr
	or & Z & \diag(+\coeff14,+\coeff14,+\coeff14,-\coeff34) &
	and & Y & \diag(+\coeff13,+\coeff13,-\coeff23,0) ;\cr
	}
    }
\eqn\ZIVCharges
$$
the two sets of charges are related according to
$$
\left\{ \eqalign{
	X\ &=\ \coeff23 Z\ +\ \phantom{\half} Y\cr
	T\ &=\ \coeff23 Z\ -\ \half Y\cr
	}\right\}\
\longleftrightarrow\
\left\{ \eqalign{
	Z\ &=\ \half X\ +\ \phantom{\half} T\cr
	Y\ &=\ \coeff23 X\ -\ \coeff23 T\cr
	}\right\} .
\eqno\eq
$$
Locally at the $\I_1$ intersection, the manifest abelian charge is $X$
while $T$ is a generator of the $\darkred SU(2)_2$.
Therefore, the $\darkblue 12$ hypermultiplets localized at the $\I_1$
have $(X=0,T=\half)$ and hence $(Z=\half,Y=-\coeff13)$.
Similarly, locally at the $\I_2$ the manifest abelian charge is  $\beta Z$
while $Y$  is a generator of the $\darkred SU(3)$.
Consequently, the $({\darkblue 8},{\darkred 3})$ hypermultiplets living at the $\I_2$
have $Z=(\gamma/\beta)=-\coeff18)$ and split into
$\darkred SU(2)_1$ doublets with $Y=\coeff13\Rightarrow(X=\coeff14,T=-\coeff14)$
 and $\darkred SU(2)_1$ singlets with
$Y=-\coeff23\Rightarrow(X=-\coeff34,T=+\coeff14)$.

Thanks to the ${\darkred SU(2)}_1^{\rm 7D}\times{\darkblue SU(2)}^{\rm 10D}\to
{\purple SU(2)}^{\rm diag}$ symmetry locking, the $\darkred SU(2)_1$ quantum numbers
of the twisted states  appear in the heterotic picture as belonging to
the $SU(2)\subset E_8^{(1)}$.
Consequently, the non-abelian quantum numbers of the twisted states on the
first line of eq.~\ZIVHtw\ precisely correspond to the localized 6D states
of the intersection planes $\I_1+\I_2$ of the HW picture.
The abelian quantum numbers have a similar correspondence provided we identify
$$
\eqalign{
C_1^{\rm heterotic}\ =\ C^{\rm HW}_{U(1)\subset E_8^{(1)}}\ &
+\sum_{\IZ_4\rm\ fixed\atop planes}\left( X\,-\,T\ =\ \coeff32 Y \right) ,\cr
C_2^{\rm heterotic}\ =\ C^{\rm HW}_{U(1)\subset E_8^{(2)}}\ &
+\sum_{\IZ_4\rm\ fixed\atop planes}\left( \coeff43 Z\,-\, Y\ =\ 2T\right) .\cr
}\eqn\ZIVHetCharges
$$
As in models of \S5.1 and \S7.2, the 7D abelian charge which mixes with the
10D abelian charge at the left end of the world happen to be a part of an unbroken
nonabelian symmetry at the right end of the world and vice verse;
we do not know why.

Finally, let us check for 7D fields
with Neumann boundary conditions at both ends of the $x^6$.
Organizing the 7D SYM fields according to their $\darkred SU(2)_1
\times U(1)\times U(1)$ and 8--SUSY quantum numbers, we tabulate their
respective boundary conditions at $\I_1$ and at $\I_2$ as follows:
$$
\vcenter{\ialign{
    \hfil\vrule width 0pt height 15pt depth 5pt \vrule #&
    \enspace\hfil $({\bf #})$\hfil\enspace & \vrule #\hfil &
    \enspace\hfil $(#)$\hfil\enspace & \vrule #\hfil &
    \enspace\hfil $(#)$\hfil\enspace & \vrule #\hfil &
    \enspace\hfil\frenchspacing(#\unskip)\hfil\enspace &\vrule #\hfil &
    \enspace\hfil\frenchspacing(#\unskip)\hfil\enspace &\vrule #\hfil \cr
\noalign{\hrule}
& \multispan5 \hfil Charges\hfil && \multispan3 \hfil Boundary Conditions\hfil &\cr
\noalign{\hrule}
&\omit\hfil $\,SU(2)\,$\hfil && X,T && Z,Y &&
\omit\hfil 8--SUSY vector\hfil && \omit\hfil hyper\hfil &\cr
\noalign{\hrule}
& 3 && 0,0 && 0,0 && locking,Neumann && Neumann,Dirichlet &\cr
\noalign{\hrule}
& 1 && 0,0 && 0,0 && locking,Neumann && Neumann,Dirichlet &\cr
\noalign{\hrule}
& 2 && \pm1,\mp\half && 0,\pm1 && Dirichlet,Neumann && Neumann,Dirichlet &\cr
\noalign{\hrule}
& 1 && 0,0 && 0,0 && Neumann,locking && Dirichlet,Neumann &\cr
\noalign{\hrule}
& 1 && 0,\pm1 && \pm1,\mp\coeff23 && Neumann,Dirichlet && Dirichlet,Neumann &\cr
\noalign{\hrule}
& 2 && \pm1,\pm\half && \pm1,\pm\coeff13 && Dirichlet,Dirichlet && Neumann,Neumann &\cr
\noalign{\hrule}
}}
\eqn\ZIVTable
$$
On the last line of this table, we indeed find hypermultiplets
with Neumann--Neumann BC and hence zero modes.
In heterotic terms, these zero modes manifest itself as
twisted states which are $SU(2)$ doublets
with $C_1=\half$, $C_2=1$, --- \cf\ the second line of the heterotic twisted
spectrum~\ZIVHtw.

The bottom line of the above discussion is that the HW picture we deduced from the
brane model~\ZIVBranes\ yields the correct spectrum from the heterotic point of view.
In the Appendix we verify the other kinematic constraints due to 6D gauge couplings
and local anomaly cancellation.
The conclusion is that our HW picture is correct and therefore,
{\sl the $\sl HW\leftrightarrow I'$ duality we used to derive eqs.~\ZIVHW\ is correct.}
In particular, {\it our analysis of various brane junctions of fig.~\ZIVBranes\
was correct,  including the $\bf O8^*$ junction~\UltraJunction\ at $x^6=L$.}

We conclude this section with three simple observations.
First, we cannot eliminate the ${\bf O8^*}/{\darkred\rm D6}$ brane junction from the
brane model~\ZIVBranes\ without a major disruption of the model's spectrum.
Indeed, for the sake of $3({\bf8})$ twisted states, three {\darkred D6} branes
have to cross the eight {\darkblue D8} branes and hence terminate on the $\bf O8^*$ plane
simply because they don't have any other place to end.
Second, unless the ${\bf O8^*}/{\darkred\rm D6}$ junction works exactly
as advertised in eqs.~\UltraJunction, the twisted spectrum of the brane model
does not match that of the heterotic orbifold.
Indeed, our analysis~\ZIVSUiv\ of the ${\darkred SU(4)}^{\rm 7D}$ symmetry breaking
depends on the unbroken $\darkred SU(3)$ --- and hence Neumann BC for the vector
fields --- at $x^6=L$.
Likewise, the rules of Dirichlet BC for all the 7D hypermultiplets and no local
states at the $\bf O8^*$ junction are important for avoiding extra twisted
states not present in the heterotic spectrum.

Finally,  assuming eqs.~\UltraJunction\ for the ${\bf O8^*}/{\darkred\rm D6}$
junction at $x^6=L$, we have the $\rm heterotic\leftrightarrow HW
\leftrightarrow I'$ duality working like \M agic.
Naturally, in light of these observation we come to the evident conclusion
that eqs.~\UltraJunction\ must hold true, string only knows how.

%%%%%%%%%%%%%%%%%%%%%%%%%%%%%%%%%%%%%%%%%%%%%%%%%%%%%%%%%%%%%%%%%%%%%%%%%%%%%%%%%
\section{An \Ori* Junction with Six D6 Branes.}
Both our previous examples of ${\bf O8^*}/{\darkred\rm D6}$ junctions
had $N=3$ {\darkred D6} branes terminating at the same point of the $\bf O8^*$
plane.
Despite diligently searching for other examples involving $N=2$, 4 or 5
{\darkred D6} branes, we did not find any.
We suspect such junctions may be forbidden, although that remains to be
confirmed via more extensive model building.
We do however have an example with $N=6$ branes, so the rule for
the {\darkred D6} branes ending on an ${\bf O8^*}$ plane seems to be
$N\equiv0$ modulo~3 (we wonder why).

Our example is the good old $T^4/\IZ_6$ model of ref.~[\KSTY] in which
the $E_8^{(1)}$ is broken down to $SU(6)\times E_3\equiv SU(6)\times SU(3)\times SU(2)$
(\cf\ eq.~\ZsixAlphaLeft) while the $E_8^{(2)}$ is broken down to the $SU(9)\times E_0
\equiv SU(9)$ similarly to the $T^4/\IZ_3$ model of \S7.3
(and indeed the $\alpha_2:\IZ_6\mapsto E_8^{(2)}$ twist satisfies $\alpha_2^3=1$).
The HW picture of the $\IZ_6$ fixed plane is similar to the model of section~6:
The 7D $\darkred SU(6)$ gauge fields lock on the  10D $\darkblue SU(6)$ gauge fields
at the left intersection~$\I_1$ while the twisted states are localized at the
right intersection~$\I_2$, thus
$$
\eqalignno{
\I_1 &
\left\{\eqalign{
    G^{\rm local}\ &
    =\ {\purple SU(6)}^{\rm diag}\times[{\darkblue SU(3)\times SU(2)}]^{\rm 10D} ,\cr
    {\rm 7D}\,H\ &=\ {\purple 35},\cr
    {\rm 7D}\,V\ &=\ 0,\cr
    {\rm 6D}\,H\ &=\ 0;\cr
    }\right.&
\eqname\ZviHWleft \xdef\jj{(\chapterlabel.\number\equanumber} \cr
\noalign{\smallskip}
\I_2 &
\left\{\eqalign{
    G^{\rm local}\ &
    =\ {\darkred SU(6)}^{\rm 7D}\times {\darkblue SU(9)}^{\rm 10D} ,\cr
    {\rm 7D}\,H\ &=\ 0,\cr
    {\rm 7D}\,V\ &=\ {\darkred 35},\cr
    {\rm 6D}\,H\ &
    =\ ({\darkred\bf6},{\darkblue\bf9})\
	+\ \half({\darkred\bf20},{\darkblue\bf1}).\cr
    }\right.&
\eqname\ZviHWright  \cr
}$$

The brane dual of this HW fixed plane is quite straightforward:
$$
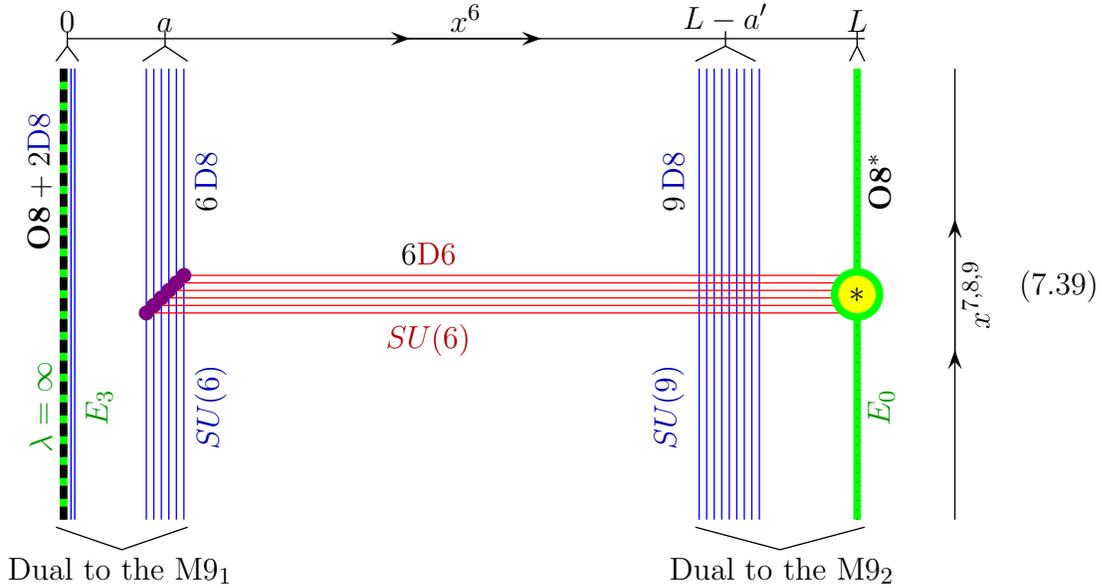
\pspicture[](-20,-40)(110,40)
\psline(-13,+34)(+93,+34)
\rput[b](+40,+35){$x^6$}
\psline{>->}(+30,+34)(+50,+34)
\psline(-13,+33)(-13,+35)	\rput[b](-13,+35){$0$}
\psline(-14.5,+31)(-13,+33)(-11.5,+31)
\psline(0,+33)(0,+35)		\rput[b](0,+35){$a$}
\psline(-3,+31)(0,+33)(+3,+31)
\psline(+74.5,+33)(+74.5,+35)	\rput[b](+74.5,+35){$L-a'$}
\psline(+78.5,+31)(+74.5,+33)(+70.5,+31)
\psline(+92,+33)(+92,+35)	\rput[b](+92,+35){$L$}
\psline(+93,+31)(+92,+33)(+91,+31)
\psline[linecolor=palegray,linewidth=2](-13,-30)(-13,+30)
\psline[linecolor=green,linewidth=1](-13.5,-30)(-13.5,+30)
\psline[linecolor=black,linestyle=dashed,linewidth=1](-13.5,-30)(-13.5,+30)
\psline[linecolor=blue](-12.5,-30)(-12.5,+30)
\psline[linecolor=blue](-12.0,-30)(-12.0,+30)
\rput[b]{90}(-15,+15){${\bf O8}+2{\rm\darkblue D8}$}
\rput[t]{90}(-10.5,-15){$\darkgreen E_3$}
\rput[b]{90}(-15,-15){$\darkgreen\lambda=\infty$}
\psline[linecolor=blue](-2.5,-30)(-2.5,+30)
\psline[linecolor=blue](-1.5,-30)(-1.5,+30)
\psline[linecolor=blue](-0.5,-30)(-0.5,+30)
\psline[linecolor=blue](+0.5,-30)(+0.5,+30)
\psline[linecolor=blue](+1.5,-30)(+1.5,+30)
\psline[linecolor=blue](+2.5,-30)(+2.5,+30)
\rput[t]{90}(+4,+15){$6\,\rm\darkblue D8$}
\rput[t]{90}(+4,-15){$\darkblue SU(6)$}
\psline[linecolor=blue](+79,-30)(+79,+30)
\psline[linecolor=blue](+78,-30)(+78,+30)
\psline[linecolor=blue](+77,-30)(+77,+30)
\psline[linecolor=blue](+76,-30)(+76,+30)
\psline[linecolor=blue](+75,-30)(+75,+30)
\psline[linecolor=blue](+74,-30)(+74,+30)
\psline[linecolor=blue](+73,-30)(+73,+30)
\psline[linecolor=blue](+72,-30)(+72,+30)
\psline[linecolor=blue](+71,-30)(+71,+30)
\rput[b]{90}(+69,+15){$9\,\rm\darkblue D8$}
\rput[b]{90}(+69,-15){$\darkblue SU(9)$}
\psline[linewidth=1,linecolor=green](+92,-30)(+92,+30)
\psline[linecolor=black,linestyle=dotted](+92,-30)(+92,+30)
\rput[t]{90}(+93.5,+15){$\bf O8^*$}
\rput[t]{90}(+93.5,-15){$\darkgreen E_0$}
\psline[linecolor=red](-2.5,-2.5)(+93,-2.5)
\psline[linecolor=red](-1.5,-1.5)(+93,-1.5)
\psline[linecolor=red](-0.5,-0.5)(+93,-0.5)
\psline[linecolor=red](+0.5,+0.5)(+93,+0.5)
\psline[linecolor=red](+1.5,+1.5)(+93,+1.5)
\psline[linecolor=red](+2.5,+2.5)(+93,+2.5)
\rput[b](35,+4){$6\darkred\rm D6$}
\rput[t](35,-4){$\darkred SU(6)$}
\pscircle*[linecolor=purple](-2.5,-2.5){1}
\pscircle*[linecolor=purple](-1.5,-1.5){1}
\pscircle*[linecolor=purple](-0.5,-0.5){1}
\pscircle*[linecolor=purple](+0.5,+0.5){1}
\pscircle*[linecolor=purple](+1.5,+1.5){1}
\pscircle*[linecolor=purple](+2.5,+2.5){1}
\cput[linecolor=green,linewidth=1,fillcolor=yellow,fillstyle=solid](+92,0){$*$}
\psline(-14.5,-31)(-5.75,-34)(+3,-31)
\rput[t](-6,-35){Dual to the $\rm M9_1$}
\psline(+70.5,-31)(+81.75,-34)(+93,-31)
\rput[t](+82,-35){Dual to the $\rm M9_2$}
\psline(+105,-30)(+105,+30)
\psline{>->}(+105,-10)(+105,+10)
\rput[t]{90}(+106,0){$x^{7,8,9}$}
\endpspicture
\eqn\ZviBranes
$$
The left side of this diagram is similar to that of fig.~\ZsixBranes\ due
to similarity of the respective HW $\I_1$ intersection planes of the two models.
On the right side, for the sake of the $({\darkred\bf6},{\darkred\bf9})$
bi-fundamentals localized at the $\I_2$, we have a {\darkred D6}/{\darkblue D8}
brane crossing at $x^6=(L-a')$.
Since the {\darkred D6} branes do not terminate at this crossing, they
have to continue all the way to the $\bf O8^*$ plane.
Consequently, the $\rm HW\leftrightarrow I'$ duality tells us that the
${\darkred\rm D6}/{\bf O8^*}$ junction must somehow produce:
(1) Neumann BC for all the ${\darkred SU(6)}^{\rm 7D}$ gauge fields,
(2) Dirichlet BC for all $\darkred\bf 35$ 7D hypermultiplets, and
(3) {\sl local 6D half-hypermultiplets} in the $\darkred\thrindex$
representation of the $\darkred SU(6)$,
$$
\pspicture[0.475](60,-30)(97,+30)
\psline[linewidth=1,linecolor=green](92,-30)(92,+30)
\psline[linecolor=black,linestyle=dotted](92,-30)(92,+30)
\rput[b]{90}(90,+17){$\bf O8^*$}
\rput[b]{90}(90.5,-17){$\darkgreen E_0$}
\psline[linecolor=red](60,-2.5)(+92,-2.5)
\psline[linecolor=red](60,-1.5)(+92,-1.5)
\psline[linecolor=red](60,-0.5)(+92,-0.5)
\psline[linecolor=red](60,+0.5)(+92,+0.5)
\psline[linecolor=red](60,+1.5)(+92,+1.5)
\psline[linecolor=red](60,+2.5)(+92,+2.5)
\rput[b](75,+3.5){6 \darkred D6}
\rput[t](75,-3.5){$\darkred SU(6)$}
\cput[fillcolor=yellow,fillstyle=solid,linecolor=green,linewidth=1](92,0){$*$}
\endpspicture
\left\{ \eqalign{
G^{\rm local}\ &=\ {\darkred SU(6)}\times{\darkgreen E_0} ,\cr
{\rm 7D}\,V\ &=\ ({\darkred\bf 35}),\cr
{\rm 7D}\,H\ &=\ 0,\cr
{\rm 6D}\,H\ &=\ \half({\darkred\bf 20}).\cr
}\right.
\eqn\UltraSixJunction
$$

The boundary conditions for the 7D fields here
are similar to eqs.~\UltraJunction;
they appear to be characteristic of the ${\darkred\rm D6}/{\bf O8^*}$ junctions
with any number $N$ of {\darkred D6} branes.
On the other hand, for $N=6$ we have local 6D hypermultiplets which we did not
have for $N=3$.
It would be very interesting to find a string-theoretical reason for this difference,
but this is clearly a subject of future research.

%%%%%%%%%%%%%%%%%%%%%%%%%%%%%%%%%%%%%%%%%%%%%%%%%%%%%%%%%%%%%%%%%%%%%%%%%%%%%%%%%

%auto-ignore
% chapter8.tex
%% Chapter 8 of Orbifold2 paper
%% version 1 --- Jacob Sonnenchein
%% version 2 --- Stefan Theisen
%% version 3 --- Vadim Kaplunovsky

%%%%%%%%%%%%%%%%%%%%%%%%%%%%%%%%%%%%%%%%%%%%%%%%%%%%%%%%%%%%%%%%%%%%%%

\chapter{Summary}
The main result of this paper is a dynamical, string-theoretical explanation
of the \M ysteries involved in the dual Ho\v rava--Witten picture of
heterotic $T^4/\IZ_N$ orbifolds and their twisted sectors.
In our previous paper~[\KSTY] we explained how  massless twisted states
become charged under both the $G^{(1)}\subset E_8^{(1)}$ living on one
end-of-the-world M9 brane of the HW theory and the $G^{(2)}\subset E_8^{(2)}$
living on the other M9 brane at the other end of eleventh dimension.
Our resolution of this apparent paradox depends on the 7D $\darkred SU(N)$ SYM
fields living on the \FP6 fixed planes (of the $\IZ_N$ orbifold action in the
11D bulk of the HW theory) and on their mixing with the 10D SYM fields living
on the M9 branes along the $\I=\FP6\cap M9$ intersection planes.
We had no mechanism for such mixing;
instead, we {\sl assumed} a complicated pattern of boundary conditions
for various 7D SYM fields --- including the locking boundary conditions~\HWLockingBC\
for the 7D and 10D gauge fields --- as needed
to explain the twisted spectrum of a heterotic orbifold,
and then subjected the resulting HW models to  stringent tests of local
anomaly cancellation and correctness of the 6D gauge couplings.
At the end of this process, we had an answer to  the {\sl kinematical}
questions of the $\rm heterotic\leftrightarrow HW$ duality for the orbifolds
but the dynamical, \M-theoretical origins of our assumed boundary
conditions and local fields remained  unexplained \M ysteries.

In this paper we explain these \M ysteries in terms of the
$\rm HW\leftrightarrow I'$ duality which maps each end-of-the-world M9 brane
onto an $\bf O8^-$ orientifold plane accompanied by 8 {\darkblue D8} branes
(or an $\bf O8^*$ plane accompanied by 9 {\darkblue D8} branes) and a $\IZ_N$ \FP6
fixed plane onto a stack of $N$ coincident {\darkred D6} branes.
The \I~intersection planes therefore become  brane junctions ---
or  combinations of several brane junctions --- and the boundary conditions
and the local fields at such junctions follow from the superstring theory.
Consequently, resolving the \M ysteries of the \I~intersections becomes
a matter of brane engineering, \ie\ arranging appropriate junctions for
the \tI/D6 dual models of  specific heterotic $T^4/\IZ_N$ orbifolds.

There are several distinct types of brane junctions, some of which we encountered
is sections 4--7 of this paper, plus a few we left out for future research.
Let us briefly review them, starting with the perturbative junctions of sections 4--6:

\pointbegin
\Ori~terminus:
An even number $N$ {\darkred D6} branes terminate on an \Ori~plane accompanied by $k$
coincident {\darkblue D8} branes (\S4.3).\brk
For $N\ge4$ the 7D gauge symmetry $\darkred SU(N)$ is broken down to $\darkred Sp(N/2)$;
the $\symrep$ gauge fields have Neumann boundary condition at the junction while
the $\tilde\asymrep$ fields have Dirichlet BC.
The junction plane supports localized massless half-hypermultiplets in the
$({\darkred\bf N},{\darkblue\bf 2k})$ bi-fundamental representation of the
${\darkred Sp(N/2)}^{\rm 7D}\times{\darkblue SO(2k)}^{\rm 10D}$.

\point %2
{\darkblue D8} terminus:
Several  {\darkred D6} branes terminate on an equal number of {\darkblue D8}
branes in a one-on-one fashion (\S4.4).\brk
This is the junction which causes 7D/10D gauge symmetry mixing via locking
boundary conditions~\LockingBC\ for the ${\darkred SU(N)}^{\rm 7D}\times
{\darkblue SU(N)}^{\rm 10D}\to{\purple SU(N)}^{\rm diag}$ gauge fields.
There are no localized 6D massless particles at this junction.

\point %3
Brane crossing:
Several  {\darkred D6} branes cross a stack of {\darkblue D8}
branes without termination (\S6).\brk
At this junction, nothing happens to the 7D SYM fields themselves, but there are
localized massless 6D hypermultiplets in the bi-fundamental representation
of the ${\darkred SU(N)}^{\rm 7D}\times{\darkblue SU(k)}^{\rm 10D}$.

\point %4
Partial termination:
In a stack of $N$ {\darkred D6} branes, $k<N$ branes terminate on
$k$ {\darkblue D8} branes while the remaining $(N-k)$ {\darkred D6} branes cross
the {\darkblue D8} branes without termination and continue to the next junction
(\S5).\brk
This junction combines 7D gauge symmetry breaking with 7D/10D symmetry mixing:
The ${\darkred SU(N)}^{\rm 7D}$ breaks to $\darkred SU(k)\times U(1)\times SU(N-k)$
and furthermore $({\darkred SU(k)\times U(1)})^{\rm 7D}\times
({\darkblue SU(k)\times U(1)})^{\rm 10D}\to ({\purple SU(k)\times U(1)})^{\rm diag}$.
Altogether, the
${\darkred SU(N)}^{\rm 7D}\times ({\darkblue SU(k)\times U(1)})^{\rm 10D}$
symmetry breaks to
$({\purple SU(k)\times U(1)})^{\rm diag}\times{\darkred SU(N-k)}^{\rm 7D}$.
Despite apparent brane crossing,
there are no localized 6D massless particles at this junction.

\point %5
{\darkgreen NS5} half-brane stuck on the \Ori~plane (\S6):\brk
Any number $N$ (even or odd) of {\darkred D6} branes may terminate on such a
half-brane without breaking the ${\darkred SU(N)}^{\rm 7D}$ gauge symmetry
--- all the 7D gauge fields have Neumann BC.
A characteristic feature of this junction is a $\darkred\asymrep$ multiplet
of localized 6D massless hypermultiplets.

\par\noindent
Physics of these five junctions
follows directly from the perturbative superstring theory.
In section~7 however, we encountered infinite-coupling terminal junctions
which cannot be described perturbatively --- and the appropriate
non-perturbative string theory is yet to be developed.
Instead, we used the  $\rm heterotic\leftrightarrow HW\leftrightarrow I'$ duality
to predict the overall features of these junctions:

\point %6
$\darkgreen E_1$ terminus:
The string coupling $\darkgreen\lambda$ diverges along an \Ori\ plane;
four {\darkred D6} branes terminate on this plane and somehow pin down
an {\darkgreen NS5} half-brane.\brk
All the 7D gauge fields have Neumann BC at this junction
and the localized 6D massless particles comprise half-hypermultiplets in the
$({\darkred\bf6},{\darkgreen\bf2})$ representation of the
locally visible gauge symmetry
${\darkred SU(4)}^{\rm 7D}\times({\darkgreen E_1=SU(2)})^{\rm 10D}$.

\point %7
$\darkgreen E_0$ terminus:
$N$ {\darkred D6} branes terminate on an $\bf O8^*$ plane.\brk
Again, all the 7D gauge fields have Neumann BC at this junction,
but the localized 6D massless spectrum depends on $N$:
Nothing for $N=3$ while for $N=6$ there are 20 half-hypermultiplets in the
$\darkred\thrindex$ representation of the $\darkred SU(6)$.
Furthermore, this junction appears to require $N\equiv0\bmod 3$,
we don't know why.

\par\noindent
Finally, there are two more junction types whose physics
we have not quite worked out:

\point %8
$\darkgreen E_{n+1}$ termini for $n=1,3,\ldots,6$:
Here the string coupling $\darkgreen\lambda$ diverges along an \Ori\ plane
accompanied by $k$ coincident {\darkblue D8} branes, hence extended
$\darkgreen E_{n+1}$ 9D gauge symmetry.
The junction is formed by several {\darkred D6} branes terminating on such a plane.\brk
Such junctions presumably gives rise to massless twisted states in non-trivial
representations of  extended gauge groups.
Unfortunately, we cannot derive the physics of these junctions from the perturbative
superstring theory while the  $\rm heterotic\leftrightarrow HW\leftrightarrow I'$ duality
analysis along the lines of section~7 is impeded by the lack of  suitable models.
Specifically, all the heterotic models with $E_{n+1}$--charged twisted states
we tried thus far have difficulties with their HW duals:
The local anomalies at the \I~intersection planes don't cancel out and sometimes
even the spectrum does not make local sense (\cf\ \eg~\S5.2 of ref.~[\KSTY]).

\point %9
Degenerate {\darkblue D8} termini in which several {\darkred D6} branes
terminate on the same {\darkblue D8} brane.\brk
For example, for any heterotic orbifold with an unbroken $E_8^{(2)}$ gauge symmetry,
all the {\darkred D6} branes of the \tI\ dual model have their right ends on the
single outlier {\darkblue D8} brane at $x^6=(L-a)$.
Physics of such junctions is governed by the perturbative superstring theory,
which predicts Dirichlet BC for the 7D gauge fields
and no localized 6D massless particles.
Unfortunately, applying this prediction to specific models runs into difficulties
with the dual HW picture, namely the local anomalies at the \I~planes fail
to cancel out.
We are presently investigating the reasons for this failure and hope to
present a resolution in our next publication.

Another open problem concerns the rules --- if any --- for pairing up
different junctions at the two ends of the {\darkred D6} branes
in \tI/D6 models.
Clearly, there are many constraints on such pairings for models based
on perturbative heterotic $T^4/\IZ_N$ orbifolds, but these constraints
have nothing to do with the \tI/D6 theory itself.
Likewise, requiring a consistent 11D HW picture imposes constraints
which are quite unnecessary in purely \tI~terms.
On the other hand, brane engineering has its own rules such as the
S~rule of MQCD [\HanWit];
perhaps this S~rule has analogues applicable in the present \tI/D6 context.

Ultimately, the most interesting open problem is to generalize our work
(both here and in ref.~[\KSTY]) from six Minkowski dimension to four,
\ie\ to Calabi--Yau orbifolds $T^6/\Gamma$ of the heterotic string
and their HW duals.
The main difficulty here lies in the \FP4 orbifold fixed planes of the \M~theory
which carry superconformal 5D theories instead of simple 7D SYM of their
\FP6 analogues.
Furthermore, unlike the \FP6 planes which are dual to stacks of {\darkred D6}
branes, the \FP4 planes do not have simple brane duals.
The hope remains however that a different duality would prove to be equally
productive; future research will tell.

%%%%%%%%%%%%%%%%%%%%%%%%%%%%%%%%%%%%%%%%%%%%%%%%%%%%%%%%%%%%%%%%%%

\ack
The authors are grateful to A.~Hanany and O.~Aharony
for many helpful discussions.
VK thanks the School of Physics and Astronomy of Tel Aviv University and 
the Theory Division of CERN for hospitality while much of the work 
presented here was done.
ST thanks A.~Font for useful discussions and the Physics Department
of Universidad Central de Venezuela for hospitality
during the writing-up stage.

%auto-ignore
% appendix.tex
%% Appendix  of Orbifold2 paper
%% Written by Vadim Kaplunovsky

\APPENDIX{A}{\break Verifying New Models}

This Appendix focuses on the HW picture of heterotic orbifolds that were not
discussed in [\KSTY] but made their first appearance in this paper.
For the two models discussed in section~5,
we have explicitly verified all the kinematic constraints
of the HW picture, namely the correct 6D massless spectrum, correct gauge couplings
and local anomaly cancellation at each $\I$ intersection plane. 
But in sections 6 and 7 we focused on the spectrum and the \tI/D6 brane engineering;
in this Appendix, we complete the discussion and verify the gauge couplings and the local
anomaly cancelation.
Or rather we outline such verification --- to save the gentle reader from utter tedium of
evaluating and factorizing various anomaly polynomials \NetAnomalyFactorization\ and
\AnomalyOther, we merely present the results of such calculations.

We begin with the \undertext{$T^4/\IZ_6$ model of section~6}.
The net observed 6D gauge symmetry of this model has five factors of diverse HW origins,
$$
\eqalign{
G^{\rm net}\ &
=\ {\purple SU(6)}\times {\darkblue SU(3)}\times {\purple SU(2)}
	\times {\darkblue SU(8)}\times {\purple U(1)},\crr
{\purple SU(6)}\ &
=\ \diag\left[ \left({\darkblue SU(6)\subset E_8^{(1)}}\right)\,
	\times\,\Bigl({\darkred SU(6)}\Bigr)_{\IZ_6\,\rm fixed\atop plane} \right] ,\cr
{\darkblue SU(3)}\ &
=\ \left({\darkblue SU(3)\subset E_8^{(1)}}\right) ,\cr
{\purple SU(2)}\ &
=\ \diag\left[ \left({\darkblue SU(2)\subset E_8^{(1)}}\right)\,
	\times\smash{\prod_{\IZ_2\ \rm fixed\atop planes}}\vphantom{\prod}
	\Bigl({\darkred SU(2)}\Bigr) \right] ,\cr
{\darkblue SU(8)}\ &
=\ \left({\darkblue SU(8)\subset E_8^{(2)}}\right) ,\cr
{\purple U(1)}\ &
=\ \diag\left[ \left({\darkblue U(1)\subset E_8^{(2)}}\right)\,
	\times\smash{\prod_{\IZ_3\ \rm fixed\atop planes}}\vphantom{\prod}
	\Bigl({\darkred U(1)\subset SU(3)}\Bigr) \right] .\cr
}\eqn\ZsixOrigins
$$
In terms of the observed gauge couplings, this list implies
$$
\eqalign{
{1\over g^2 [{\purple SU(6)}]}\ &
=\ {1\over g^2[{\darkblue E_8^{(1)}}]}\ +\ {1\over g^2[{\darkred SU(6)}_{\IZ_6}]}\,,\cr
{1\over g^2 [{\darkblue SU(3)}]}\ &
=\ {1\over g^2[{\darkblue E_8^{(1)}}]}\ +\ 0,\cr
{1\over g^2 [{\purple SU(2)}]}\ &
=\ {1\over g^2[{\darkblue E_8^{(1)}}]}\ +\ {5\over g^2[{\darkred SU(2)}_{\IZ_2}]}\,,\cr
{1\over g^2 [{\darkblue SU(8)}]}\ &
=\ {1\over g^2[{\darkblue E_8^{(2)}}]}\ +\ 0,\cr
{1\over g^2 [{\purple U(1)}]}\ &
=\ {1\over g^2[{\darkblue E_8^{(2)}}]}\ +\ {4\times(16/3)^{\footsymbolgen}
		\over g^2[{\darkred SU(3)}_{\IZ_3}]}\,,\cr
}\eqno\eq
$$
\vfootnote{\footsymbol}{%
    The factor $(16/3)$ comes from the normalization of the $\darkred U(1)\subset SU(3)$
    generator at the $\IZ_3$ fixed planes.
    This normalization --- namely $Y=\diag(-\coeff23,-\coeff23,+\coeff43)$ can be inferred
    from the abelian charges of the ${\darkblue SU(3)}^{\rm 10D}$ triplets in eq.~\ZsixHtwThree.
    Consequently, $\tr(Y^2)=2\Tr_{\bf3}(Y^2)=(16/3)$.
    }
or in terms of the $v,\tilde v$ coefficients  in eq.~\GaugeCouplings,
$$
\def\qqq #1 &#2&#3\\{v[{#1}] & #2, & \tilde v[{#1}] & #3\cr }
\vcenter{\openup 1\jot \tabskip=0pt
    \halign{
	&$\displaystyle{#\ }$\hfil & $\displaystyle{{}=\ #}$\qquad\hfil\cr
	\qqq \purple	SU(6)	& 1 & \half k_1\,+\,1 ,\\
	\qqq \darkblue	SU(3)	& 1 & \half k_1 ,\\
	\qqq \purple	SU(2)	& 1 & \half k_1\,+\, 5 ,\\
 	\qqq \darkblue	SU(8)	& 1 & \half k_2 ,\\
	\qqq \purple	U(1)	& 4 & 2k_2\,+\,4\times\coeff{16}{3} .\\
	}}
\eqn\ZsixCouplingsHW
$$
By comparison, using the heterotic orbifold's spectrum to evaluate and factorize the net 6D
anomaly polynomial~\NetAnomalyFactorization\ yields (after some boring algebra)
$$
\def\qqq #1 &#2&#3\\{v[{#1}] & #2, & \tilde v[{#1}] & #3\cr }
\vcenter{\openup 1\jot \tabskip=0pt
    \halign{
	&$\displaystyle{#\ }$\hfil & $\displaystyle{{}=\ #}$\qquad\hfil\cr
	\qqq \purple	SU(6)	& 1 & \coeff32 ,\\
	\qqq \darkblue	SU(3)	& 1 & \half  ,\\
	\qqq \purple	SU(2)	& 1 & \coeff{11}{2} ,\\
 	\qqq \darkblue	SU(8)	& 1 & -\half ,\\
	\qqq \purple	U(1)	& 4 & \coeff{58}{3},\\
	}}
\eqn\ZsixCouplingsHet
$$
in perfect consistency with the HW results~\ZsixCouplingsHW\
(assuming $k_1=+1$, $k_2=-1$ \ie, 13 instantons in the $E_8^{(1)}$ and 11 in the $E_8^{(2)}$).

Our next test concerns the local 6D anomaly at the $\I_1$ intersection plane.
In light of eqs.~\ZsixHWleft\ we have local gauge symmetry
$$
G^{\rm local}\ =\ {\purple SU(6)}\times{\darkblue SU(3)}\times{\darkblue SU(2)}
$$
and the anomaly-weighed chiral matter  comprises
$$
\def\qqq(#1,#2,#3){({\purple\bf #1},{\darkblue\bf #2},{\darkblue\bf #3})}
\eqalign{
Q_6\ &=\ 0,\cr
Q_7\ &=\ +\half\qqq(35,1,1),\cr
Q_{10}\ &
=\ \coeff{19}{144}\qqq(20,1,2)\ +\ \coeff{13}{72}\qqq(15,\bar3,1)\
	-\ \coeff{5}{72}\qqq(6,3,2)\cr
&\qquad-\ \coeff{35}{144}\Bigl[ \qqq(35,1,1)\,+\,\qqq(1,8,1)\,+\,\qqq(1,1,3)\Bigr] ;\cr
}\eqn\ZsixChiralLeft
$$
the last line here follows from the $\alpha_1:\IZ_6\mapsto E_8^{(1)}$ twist, \cf\
eq.~\ZsixAlphaLeft.
The magnetic charge of the $\I_1$ plane is
$$
g_1[\IZ_6]\ =\ k_1\ -\ 4g_1[\IZ_3]\ -\ 5g_1[\IZ_2]\ =\ -\coeff{5}{12}
\eqno\eq
$$
and it is easy to see that eq.~\AnomalyRiiii\ holds true,
$$
\dim(Q=Q_6+Q_7+Q_{10})\ =\ \coeff{155}{9}\ =\ \coeff{535}{18}\ +\ 30g .
\eqno\eq
$$
As usual, eq.~\AnomalyOther\ takes more work to verify, but it holds true as well,
$$
\eqalign{
{\cal A}'\ &
\equiv\ \coeff23\Tr_Q({\cal F}^4)\ -\ \coeff16\tr(R^2)\times\Tr_Q({\cal F}^2)\
	+\ (\coeff18 g+\half T(1)=\coeff{5}{72})\bigl(\tr(R^2)\bigr)^2\crr
&=\ -{5\over24}\left( \tr(F^2_{\purple SU(6)})\,+\,\tr(F^2_{\darkblue SU(3)})\,
	+\,\tr(F^2_{\darkblue SU(3)})\,-\,\half\tr(R^2) \right)^2\cr
&\qquad+\ \left( \tr(F^2_{\purple SU(6)})\,+\,\tr(F^2_{\darkblue SU(3)})\,
	+\,\tr(F^2_{\darkblue SU(3)})\,-\,\half\tr(R^2) \right)\cr
&\qquad\qquad\times\left( \tr(F^2_{\purple SU(6)})\,-\,\coeff{35}{144}\tr(R^2) \right) .\cr
}\eqno\eq
$$

Finally, the $\I_2$ intersection plane has magnetic charge $g_2=-g_1=+\coeff5{12}$,
local symmetry
$$
G^{\rm local}\ =\ {\darkred SU(6)}\times{\darkblue SU(8)}\times{\darkblue U(1)}
$$
and the anomaly-weighed chiral matter comprising
$$
\def\qqq(#1,#2,#3){({\darkred\bf #1},{\darkblue\bf #2},{\darkblue #3})}
\eqalign{
Q_6\ &=\ \qqq(6,8,-\coeff16)\ + \qqq(15,1,+\coeff23),\cr
Q_7\ &=\ -\half\qqq(35,1,1),\cr
Q_{10}\ &
=\ \coeff{19}{144}\qqq(1,70,0)\ +\ \coeff{13}{72}\qqq(1,28,-1)\
	-\ \coeff{5}{72}\Bigl[\qqq(1,28,+1)\,+\,\qqq(1,1,-2)\Bigr]\cr
&\qquad-\ \coeff{35}{144}\Bigl[ \qqq(1,63,0)\,+\,\qqq(1,1,0)\Bigr] .\cr
}\eqn\ZsixChiralRight
$$
(The first two lines here follow from eqs.~\ZsixHWright\ while the $Q_{10}$ follows
from eq.~\ZsixAlphaRight.)
Again, we find that eqs.~\AnomalyRiiii\ and \AnomalyOther\ hold true,
$$
\dim(Q=Q_6+Q_7+Q_{10})\ =\ \coeff{380}{9}\ =\ \coeff{535}{18}\ +\ 30g ,
\eqno\eq
$$
and 
$$
\eqalign{
{\cal A}'\ &
\equiv\ \coeff23\Tr_Q({\cal F}^4)\ -\ \coeff16\tr(R^2)\times\Tr_Q({\cal F}^2)\
	+\ (\coeff18 g+\half T(1)=\coeff{25}{144})\bigl(\tr(R^2)\bigr)^2\crr
&=\ +{5\over24}\left(\tr(F^2_{\darkblue SU(8)})\,+\,F^2_{\darkblue U(1)}\,
	-\,\half\tr(R^2) \right)^2\cr
&\qquad+\ \left(\tr(F^2_{\darkblue SU(8)})\,+\,F^2_{\darkblue U(1)}\,
	-\,\half\tr(R^2) \right)
    \times\left( \tr(F^2_{\darkred SU(6)})\,-\,\coeff{35}{144}\tr(R^2) \right) .\cr
}\eqno\eq
$$

Next, consider the \undertext{monster $T^4/\IZ_6$ model of \S7.2}.
The 6D gauge symmetry of this monster has seven factors whose HW provenance
includes various mixtures of the two M9 end-of-the-world branes and the ten \FP6
planes.
To wit,
$$
G^{\rm net}\ 
=\ {\darkblue SO(12)}\times{\purple SU(2)_A}\times{\purple U(1)_1}\times
{\darkblue SU(6)}\times{\purple SU(2)_B}\times
{\darkgreen SU(2)_C}\times{\purple U(1)_2} ,
\eqn\MonsterFactors
$$
where
$$
\eqalignno{
{\darkblue SO(12)}\ &
=\ \left( {\darkblue SO(12)\subset E_8^{(1)}} \right) , &
\eq \xdef\jj{(\chapterlabel.\number\equanumber} \cr
{\purple SU(2)_A}\ &
=\ \diag\left[
    \eqalign{
	\left( {\darkblue SU(2)\subset E_8^{(1)}} \right) &
	\times\Bigl( {\darkred SU(2)_1\subset SU(6)} \Bigr)_{\IZ_6\rm\ fixed\atop plane}\cr
	&\times\smash{\prod_{\IZ_2\rm\ fixed\atop planes}}\vphantom{\prod}
		\Bigl( {\darkred SU(2)} \Bigr)\cr
	}
    \right] ,&
\eq \cr \noalign{\vskip 15pt}
{\purple U(1)_1}\ &
=\ \diag\left[
    \eqalign{
	\left( {\darkblue U(1)\subset E_8^{(1)}} \right) &
	\times\Bigl( {\darkred U(1)_1\subset SU(6)} \Bigr)_{\IZ_6\rm\ fixed\atop plane}\cr
	&\times\smash{\prod_{\IZ_3\rm\ fixed\atop planes}}\vphantom{\prod}
		\Bigl( {\darkred U(1)\subset SU(3)} \Bigr)\cr
	}
    \right] ,&
\eq \cr \noalign{\vskip 15pt}
{\darkblue SU(6)}\ &
=\ \left( {\darkblue SU(6)\subset E_8^{(2)}} \right) ,&
\eq \cr \noalign{\vskip 10pt}
\global\count255=\interdisplaylinepenalty
\global\interdisplaylinepenalty=10000
{\purple SU(2)_B}\ &
=\ \diag\left[
	\left( {\darkblue SU(2)_B\subset E_8^{(2)}} \right)\times
	\Bigl( {\darkred SU(2)_2\times SU(2)_3\subset SU(6)}
		 \Bigr)_{\IZ_6\rm\ fixed\atop plane}
	 \right] ,\cr
&& \eq\cr
\global\interdisplaylinepenalty=\count255
{\darkgreen SU(2)_C}\ &
=\ \left( {\darkblue SU(2)_C\subset E_8^{(2)}} \right) ,&
\eq \cr \noalign{\vskip 10pt}
{\purple U(1)_2}\ &
=\ \diag\left[
	\left( {\darkblue U(1)\subset E_8^{(2)}} \right)\times
	\Bigl( {\darkred U(1)_2\subset SU(6)} \Bigr)_{\IZ_6\rm\ fixed\atop plane}
	 \right] .&
\eq \cr
}$$
Consequently, the five nonabelian factors have gauge couplings
$$
\eqalign{
{1\over g^2[{\darkblue SO(12)}]}\ &
=\ {1\over g^2[{\darkblue E_8^{(1)}}]}\,,\cr
{1\over g^2[{\purple SU(2)_A}]}\ &
=\ {1\over g^2[{\darkblue E_8^{(1)}}]}\
    +\ {1\over g^2[{\darkred SU(6)}_{\IZ_6}]}\
    +\ {5\over g^2[{\darkred SU(2)}_{\IZ_2}]}\,,\cr
{1\over g^2[{\darkblue SU(6)}]}\ &
=\ {1\over g^2[{\darkblue E_8^{(2)}}]}\,,\cr
{1\over g^2[{\purple SU(2)_B}]}\ &
=\ {1\over g^2[{\darkblue E_8^{(2)}}]}\
    +\ {2\over g^2[{\darkred SU(6)}_{\IZ_6}]}\,,\cr
{1\over g^2[{\darkgreen SU(2)_C}]}\ &
=\ {1\over g^2[{\darkblue E_8^{(2)}}]}\,,\cr
}\eqn\MonsterGNA
$$
while the two abelian factors have
$$
\eqalign{
{1\over g^2[{\purple U(1)\times U(1)}]}\ &
=\ \pmatrix{ {4\over g^2[{\darkblue E_8^{(1)}}]} & 0 \cr
	 0 & {12\over g^2[{\darkblue E_8^{(2)}}]} \cr }\
    +\ {4\over g^2[{\darkred SU(3)}_{\IZ_3}]}\pmatrix{ {16\over3} & 0 \cr 0 & 0 \cr }\crr
&\qquad+\ {1\over g^2[{\darkred SU(6)}_{\IZ_6}]}
%    \left[ \pmatrix{ \tr(C_1^2) & \tr(C_1C_2) \cr \tr(C_1C_2) & \tr(C_2^2) \cr}
	\pmatrix{ {32\over9} & -{8\over3} \cr -{8\over3} & 8 \cr } .\cr
}\eqn\MonsterGA
$$
The last matrix here follows from the normalization of the two
$\darkred U(1)\times U(1)\subset SU(6)$
generators $C_1=\coeff43 X_2$ and $C_2=2X_1$ (\cf\ eqs.~\MonsterChargeID):
$$
\eqalign{
\tr(C_1^2)\,&\equiv\, 2\Tr_{\bf6}(C_1^2)\ =\ \coeff{32}{9}\,,\cr
\tr(C_2^2)\,&\equiv\, 2\Tr_{\bf6}(C_2^2)\ =\ 8,\cr
\tr(C_1C_2)\,&\equiv\, 2\Tr_{\bf6}(C_1C_2)\ =\ -\coeff83\,.\cr
}\eqno\eq
$$
Translating the gauge couplings \MonsterGNA\ and \MonsterGA\ into the $v,\tilde v$
coefficients of eq.~\GaugeCouplings, we find
$$
\def\qqq #1 &#2&#3\\{v[{#1}] & #2, & \tilde v[{#1}] & #3\cr }
\vcenter{\openup 1\jot \tabskip=0pt
    \halign{
	&$\displaystyle{#\ }$\hfil & $\displaystyle{{}=\ #}$\quad\hfil\cr
	\qqq \darkblue	SO(12)	& 1 & \half k_1\,,\\
	\qqq \purple	SU(2)_A	& 1 & \half k_1\,+\,6 ,\\
 	\qqq \darkblue	SU(6)	& 1 & \half k_2 ,\\
	\qqq \purple	SU(2)_B	& 1 & \half k_2\,+\, 2 ,\\
 	\qqq \darkgreen	SU(2)_C	& 1 & \half k_2 ,\\
	\noalign{\smallskip}
	\qqq \purple	U(1)\times U(1)	& 
	    \pmatrix{ 4 & 0\cr 0 & 12\cr } &
	    \pmatrix{ 2k_1+{224\over9} & -{8\over3}\cr -{8\over3} & 6k_2+8\cr } .\\
	}}
\hskip -3em
\eqn\MonsterCouplingsHW
$$
By comparison, using the heterotic orbifold's spectrum to evaluate and factorize the net 6D
anomaly polynomial~\NetAnomalyFactorization\ yields (after some boring algebra)
$$
\def\qqq #1 &#2&#3\\{v[{#1}] & #2, & \tilde v[{#1}] & #3\cr }
\vcenter{\openup 1\jot \tabskip=0pt
    \halign{
	&$\displaystyle{#\ }$\hfil & $\displaystyle{{}=\ #}$\qquad\hfil\cr
	\qqq \darkblue	SO(12)	& 1 & -2,\\
	\qqq \purple	SU(2)_A	& 1 & +4,\\
 	\qqq \darkblue	SU(6)	& 1 & +2,\\
	\qqq \purple	SU(2)_B	& 1 & +4,\\
 	\qqq \darkgreen	SU(2)_C	& 1 & +2,\\
	\noalign{\smallskip}
	\qqq \purple	U(1)\times U(1)	& 
	    \pmatrix{ 4 & 0\cr 0 & 12\cr } &
	    \pmatrix{ {152\over9} & -{8\over3}\cr -{8\over3} & 32\cr } .\\
	}}
\hskip -1em
\eqn\MonsterCouplingsHet
$$
in perfect consistency with the HW results~\MonsterCouplingsHW\
(assuming $k_1=-4$, $k_2=+4$ \ie, 8 instantons in the $E_8^{(1)}$ and 16 in the $E_8^{(2)}$).

Next, consider the local anomalies at the \I~intersections of the HW picture.
According to eqs.~\MonsterHWleft\
the local gauge symmetry at the $\I_1$ intersection plane is
$$
G^{\rm local}\ =\
[{\darkblue SO(12)}]^{\rm 10D}\times[{\purple SU(2)_A\times U(1)}]^{\rm diag}
\times{\darkred Sp(2)}^{\rm 7D}
$$
and the anomaly-weighed chiral matter comprises
$$
\def\qqq(#1,#2,#3,#4){({\darkblue\bf #1},{\bf\purple #2},{\purple #3},{\darkred\bf #4})}
\eqalign{
Q_6\ &=\ \half\qqq(12,1,0,4),\cr
Q_7\ &=\ \half\qqq(1,3,0,1)\ +\ \half\qqq(1,1,0,1)\ +\ \qqq(1,2,1,4)\cr
&\qquad+\ \half\qqq(1,1,0,5)\ -\ \half\qqq(1,1,0,10),\cr
Q_{10}\ &
=\ \coeff{19}{144}\qqq(32',2,0,1)\
    +\ \coeff{13}{72}\bigl[\qqq(32,1,1,1)+\qqq(1,1,2,1)\bigr]
    -\ \coeff{5}{72}\qqq(12,2,1,1)\cr
&\qquad-\ \coeff{35}{144}\bigl[\qqq(66,1,0,1)+\qqq(1,3,0,1)+\qqq(1,1,0,1)\bigr] .\cr
}\eqno\eq
$$
(The last equation here follows from the first eq.~\MonsterAlphas.)
The magnetic charge of the $\I_1$ plane is
$$
g_1[\IZ_6]\ =\ k_1\ -\ 4g_1[\IZ_3]\ -\ 5g_1[\IZ_2]\ =\ -\coeff{1}{12} ,
\eqno\eq
$$
and it easy to check that eq.~\AnomalyRiiii\ holds true,
$$
\dim(Q=Q_6+Q_7+Q_{10})\ =\ \coeff{245}{9}\ =\ \coeff{535}{18}\ +\ 30g,
\eqno\eq
$$
hence no net $\tr(R^4)$ anomaly.
Verifying cancelation of all other anomalies takes more work, but at the end
eq.~\AnomalyOther\ holds true as well,
$$
\eqalign{
{\cal A}'\ &
\equiv\ \coeff23\Tr_Q({\cal F}^4)\ -\ \coeff16\tr(R^2)\times\Tr_Q({\cal F}^2)\
	+\ (\coeff18 g+\half T(1)=\coeff{1}{9})\bigl(\tr(R^2)\bigr)^2\crr
&=\ -{1\over24}\left(
	\tr(F^2_{\darkblue SO(12)})\,+\,\tr(F^2_{\purple SU(2)_A})\,
	+4\,F^2_{\purple U(1)}\,-\,\half\tr(R^2)
	\right)^2\crr
&\qquad+\ \left(
	\tr(F^2_{\darkblue SO(12)})\,+\,\tr(F^2_{\purple SU(2)_A})\,
	+4\,F^2_{\purple U(1)}\,-\,\half\tr(R^2)
	\right)\cr
&\qquad\qquad\times\left(
	\tr(F^2_{\purple SU(2)_A})\,+\,\coeff83 F^2_{\purple U(1)}\,
	+\,\tr(F^2_{\darkred Sp(2)})\,-\,\coeff{35}{144}\tr(R^2)
	\right) .\cr
}\eqno\eq
$$

At the other intersection plane~$\I_2$ we have (\cf\ eq.~\MonsterHWright) local symmetry
$$
G^{\rm local}\ 
=\ {\darkred SU(4)}^{\rm 7D}\times\bigl[{\purple SU(2)_C\times U(1)_2}\bigr]^{\rm diag}
	\times\bigl[{\darkblue SU(6)}\times{\darkgreen SU(2)_C}\bigr]^{\rm 10D} ,
$$
the anomaly-weighed chiral matter
$$
\def\qqq(#1,#2,#3,#4,#5){({\darkred\bf #1},{\bf\purple #2},{\purple #3},%
	{\darkblue\bf #4},{\darkgreen\bf #5})}
\hskip -0.25em
\eqalign{
Q_6\ &=\ \qqq(4,1,\gamma,6,1)\ +\ \half\qqq(6,1,0,1,2),\crr
Q_7\ &=\ \half\qqq(1,3,0,1,1)\ +\ \half\qqq(1,1,0,1,1)\ +\ \qqq(4,2,\beta,1,1)\
	-\ \half\qqq(15,1,0,1,1) ,\crr
Q_{10}\ &
=\ \coeff{19}{144}\qqq(1,2,0,20,1)\ +\ \coeff{19}{72}\qqq(1,1,+3,1,2)\cr
&\qquad+\ \coeff{13}{72}\bigl[\qqq(1,1,+2,15,1)+\qqq(1,2,+1,6,2)\bigr]\cr
&\qquad-\ \coeff{5}{72}\bigl[\qqq(1,1,-1,15,2)+\qqq(1,2,-2,6,1)\bigr]\cr
&\qquad-\ \coeff{35}{144}\bigl[\qqq(1,3,0,1,1)+\qqq(1,1,0,1,1)+\qqq(1,1,0,35,1)
	+\qqq(1,1,0,1,3)\bigr] \crr
}\hskip -1.75em
\eqn\MonsterChiralRight
$$
(the last line here follows from the second eq.~\MonsterAlphas),
and the magnetic charge of the plane is $g_2=-g_1=+\coeff{1}{12}$.
Hence, we can easily see the cancelation of the gravitational $\tr(R^4)$
anomaly according to eq.~\AnomalyRiiii,
$$
\dim(Q=Q_6+Q_7+Q_{10})\ =\ \coeff{290}{9}\ =\ \coeff{535}{18}\ +\ 30g.
\eqno\eq
$$
Next, consider the gauge anomalies involving one power of the abelian field
$F_{\purple U(1)}$ and three power of the nonabelian fields $F_{\darkred SU(4)}$
and $F_{\darkblue SU(6)}$.
For the anomaly-weighed spectrum~\MonsterChiralRight, we have
$$
\eqalign{
\Tr_Q\left(F_{\purple U(1)}F^3_{\darkred SU(4)}\right)\ &
=\  F_{\purple U(1)}\tr(F^3_{\darkred SU(4)}) \times(6\gamma+2\beta),\cr
\Tr_Q\left(F_{\purple U(1)}F^3_{\darkblue SU(6)}\right)\ &
=\ F_{\purple U(1)}\tr(F^3_{\darkblue SU(6)})\times (4\gamma+2) ,\cr
}\eqno\eq
$$
hence to assure cancelation of these dangerous anomalies we must have 
$\beta=+\coeff32$, $\gamma=-\half$.
As explained in~\S7.2, these coefficients determine respectively the normalization
of  the ${\darkred U(1)}^{\rm 7D}\times {\darkblue U(1)}^{\rm 10D}\to
{\purple U(1)}^{\rm diag}$ charge mixing at the $\I_2$ and the overall charge of the
twisted states which live there.

Given $\beta=+\coeff32$, $\gamma=-\half$, verifying cancelation of all the remaining
local anomalies at the $\I_2$ is the usual tedious exercise of evaluating and
factorizing anomaly polynomials.
At the end of this exercise, eq.~\AnomalyOther\ indeed holds true,
$$
\eqalign{
{\cal A}'\ &
\equiv\ \coeff23\Tr_Q({\cal F}^4)\ -\ \coeff16\tr(R^2)\times\Tr_Q({\cal F}^2)\
	+\ (\coeff18 g+\half T(1)=\coeff{19}{144})\bigl(\tr(R^2)\bigr)^2\crr
&=\ +{1\over24}\left(
	\tr(F^2_{\darkblue SU(6)})\,+\,\tr(F^2_{\purple SU(2)_B})\,
	+\,\tr(F^2_{\darkgreen SU(2)_C})\,+12\,F^2_{\purple U(1)}\,
	-\,\half\tr(R^2)\right)^2\crr
&\qquad+\ \left(
	\tr(F^2_{\darkblue SU(6)})\,+\,\tr(F^2_{\purple SU(2)_B})\,
	+\,\tr(F^2_{\darkgreen SU(2)_C})\,+12\,F^2_{\purple U(1)}\,
	-\,\half\tr(R^2)\right)\cr
&\qquad\qquad\times\left(
	\tr(F^2_{\purple SU(2)_B})\,+\,6 F^2_{\purple U(1)}\,
	+\,\tr(F^2_{\darkred SU(4)})\,-\,\coeff{35}{144}\tr(R^2)
	\right) .\cr
}\eqno\eq
$$

Finally, consider the \undertext{$T^4/\IZ_4$ model of \S7.4}.
This time, we have three nonabelian and two abelian gauge group factors
$$
G^{\rm net}\ = {\darkblue SO(12)}\times{\purple SU(2)}\times{\purple U(1)_1}
\times{\darkblue SU(8)}\times{\purple U(1)_2}
\eqno\eq
$$
of the following HW provenance:
$$
\eqalignno{
{\darkblue SO(12)}\ &
=\ \left( {\darkblue SO(12)\subset E_8^{(1)}} \right) , &
\eq \xdef\jj{(\chapterlabel.\number\equanumber} \cr
{\purple SU(2)}\ &
=\ \diag\left[
    \left( {\darkblue SU(2)\subset E_8^{(1)}} \right)
    \times\!\!\smash{\prod_{\IZ_4\rm\ fixed\atop planes}}\vphantom{\prod}\!\!
	\Bigl( {\darkred SU(2)\subset SU(4)}\Bigr)
    \right] ,&
\eq \cr %\noalign{\vskip 15pt}
{\purple U(1)_1}\ &
=\ \diag\left[
    \left( {\darkblue U(1)\subset E_8^{(1)}} \right)
    \times\!\!\!\smash{\prod_{\IZ_4\rm\ fixed\atop planes}}\vphantom{\prod}\!\!\!
	\Bigl( {\darkred U(1)_1\subset SU(4)} \Bigr)
    \times\!\!\!\smash{\prod_{\IZ_2\rm\ fixed\atop planes}}\vphantom{\prod}\!\!\!
	\Bigl( {\darkred U(1)\subset SU(2)} \Bigr)
    \right] \!,
    \everycr={\nobreak}\cr
&&\eq \cr %\noalign{\vskip 15pt}
{\darkblue SU(8)}\ &
=\ \left( {\darkblue SU(8)\subset E_8^{(2)}} \right) ,&
\eq \cr %\noalign{\vskip 10pt}
{\purple U(1)_2}\ &
=\ \diag\left[
    \left( {\darkblue U(1)\subset E_8^{(2)}} \right)
    \times\!\!\smash{\prod_{\IZ_4\rm\ fixed\atop planes}}\vphantom{\prod}\!\!
	\Bigl( {\darkred U(1)_2\subset SU(4)} \Bigr)
    \right] .&
\eq \cr
}$$
Accordingly, the nonabelian gauge couplings of the model are
$$
\eqalign{
{1\over g^2[{\darkblue SO(12)}]}\ &
=\ {1\over g^2[{\darkblue E_8^{(1)}}]}\,,\cr
{1\over g^2[{\purple SU(2)}]}\ &
=\ {1\over g^2[{\darkblue E_8^{(1)}}]}\
    +\ {4\over g^2[{\darkred SU(4)}_{\IZ_4}]}\,,\cr
{1\over g^2[{\darkblue SU(8)}]}\ &
=\ {1\over g^2[{\darkblue E_8^{(2)}}]}\,,\cr
}\eqn\ZIVGNA
$$
while the abelian couplings are
$$
\eqalign{
{1\over g^2[{\purple U(1)\times U(1)}]}\ &
=\ \pmatrix{ {4\over g^2[{\darkblue E_8^{(1)}}]} & 0 \cr
	 0 & {4\over g^2[{\darkblue E_8^{(2)}}]} \cr }\
    +\ {6\over g^2[{\darkred SU(2)}_{\IZ_2}]}\pmatrix{ 4 & 0 \cr 0 & 0 \cr }\crr
&\qquad+\ {4\over g^2[{\darkred SU(4)}_{\IZ_4}]}
	\pmatrix{ \hphantom{+}3 & -2 \cr -2 & \hphantom{+}4 \cr } .\cr
}\eqn\ZIVGA
$$
The last matrix here follows from the normalization of the two
$\darkred U(1)\times U(1)\subset SU(4)$
generators $C_1=\coeff32 Y$ and $C_2=2T$ (\cf\ eqs.~\ZIVHetCharges):
$$
\eqalign{
\tr(C_1^2)\,&\equiv\, 2\Tr_{\bf4}(C_1^2)\ =\ 3\,,\cr
\tr(C_2^2)\,&\equiv\, 2\Tr_{\bf4}(C_2^2)\ =\ 4,\cr
\tr(C_1C_2)\,&\equiv\, 2\Tr_{\bf4}(C_1C_2)\ =\ -2\,.\cr
}\eqno\eq
$$
Translating these couplings into the $v,\tilde v$ coefficients of eq.~\GaugeCouplings,
we find
$$
\def\qqq #1 &#2&#3\\{v[{#1}] & #2, & \tilde v[{#1}] & #3\cr }
\vcenter{\openup 1\jot \tabskip=0pt
    \halign{
	&$\displaystyle{#\ }$\hfil & $\displaystyle{{}=\ #}$\quad\hfil\cr
	\qqq \darkblue	SO(12)	& 1 & \half k_1\,,\\
	\qqq \purple	SU(2)	& 1 & \half k_1\,+\,4 ,\\
 	\qqq \darkblue	SU(8)	& 1 & \half k_2 ,\\
	\noalign{\smallskip}
	\qqq \purple	U(1)\times U(1)	& 
	    \pmatrix{ 4 & 0\cr 0 & 4\cr } &
	    \pmatrix{ 2k_1+36 & -8\cr -8 & 2k_2+16\cr } .\\
	}}
\hskip -3em
\eqn\MonsterCouplingsHW
$$
By comparison, using the heterotic orbifold's spectrum to evaluate and factorize the net 6D
anomaly polynomial~\NetAnomalyFactorization\ yields (after some boring algebra)
$$
\def\qqq #1 &#2&#3\\{v[{#1}] & #2, & \tilde v[{#1}] & #3\cr }
\vcenter{\openup 1\jot \tabskip=0pt
    \halign{
	&$\displaystyle{#\ }$\hfil & $\displaystyle{{}=\ #}$\qquad\hfil\cr
	\qqq \darkblue	SO(12)	& 1 & -1,\\
	\qqq \purple	SU(2)	& 1 & +3,\\
 	\qqq \darkblue	SU(8)	& 1 & +1,\\
	\noalign{\smallskip}
	\qqq \purple	U(1)\times U(1)	& 
	    \pmatrix{ 4 & 0\cr 0 & 12\cr } &
	    \pmatrix{ 32 & -8\cr -8 & 20\cr } .\\
	}}
\hskip -1em
\eqn\MonsterCouplingsHet
$$
in perfect consistency with the HW results~\MonsterCouplingsHW\
(assuming $k_1=-2$, $k_2=+2$ \ie, 10 instantons in the $E_8^{(1)}$ and 14 in the $E_8^{(2)}$).

The local anomaly at the $\I_1$ intersection in the HW picture of this model cancels out
exactly as in the model of~\S5.2 (which has a similar $\I_1$ plane).
At the $\I_2$ intersection plane, we have local symmetry
$$
G^{\rm local}\ =\ {\darkblue SU(8)}\times{\purple U(1)}\times{\darkred SU(3)}
$$
and the anomaly-weighed chiral mater comprises
$$
\def\qqq(#1,#2,#3){({\darkblue\bf #1},{\purple #2},{\darkred\bf #3})}
\eqalign{
Q_6\ &=\ \qqq(8,\gamma,3),\cr
Q_7\ &=\ \half\qqq(1,0,1)\ +\ \qqq(1,\beta,3)\ -\ \half\qqq(1,0,8),\cr
Q_{10}\ &=\ \coeff{3}{16}\qqq(28,+1,1)\
	+\ \coeff{1}{16}\bigl[\qqq(56,-\half,1)+\qqq(8,-\coeff32,1)\bigr]\cr
&\qquad-\ \coeff{5}{32}\bigl[\qqq(63,0,1)+\qqq(1,0,1)\bigr] ,\cr
}\eqn\ZIVChiralRight
$$
\cf\ eqs.~\ZIVHWright\ and \ZIVSUviiiUi.
The magnetic charge of the $\I_2$ plane is $g_2=-g_1=+\coeff18$ and we can easily see
that eq.~\AnomalyRiiii\ holds true
$$
\dim(Q=Q_6+Q_7+Q_{10})\ =\ \coeff{91}{4}\ =\ 19\ +\ 30g
\eqno\eq
$$
and hence the irreducible gravitational anomaly $\tr(R^4)$  cancels out.

Next, consider the gauge anomalies involving one power of the abelian field
$F_{\purple U(1)}$ and three power of the nonabelian fields $F_{\darkred SU(3)}$
and $F_{\darkblue SU(8)}$.
Given the anomaly-weighed spectrum~\ZIVChiralRight, we have
$$
\eqalign{
\Tr_Q\left(F_{\purple U(1)}F^3_{\darkred SU(3)}\right)\ &
=\  F_{\purple U(1)}\tr(F^3_{\darkred SU(3)}) \times(8\gamma+\beta),\cr
\Tr_Q\left(F_{\purple U(1)}F^3_{\darkblue SU(8)}\right)\ &
=\ F_{\purple U(1)}\tr(F^3_{\darkblue SU(8)})\times (3\gamma+\half) ,\cr
}\eqno\eq
$$
hence to assure cancelation of these dangerous anomalies we must have 
$\beta=+\coeff23$, $\gamma=-\coeff16$.
As explained in~\S7.4, these coefficients determine respectively the normalization
of  the ${\darkred U(1)}^{\rm 7D}\times {\darkblue U(1)}^{\rm 10D}\to
{\purple U(1)}^{\rm diag}$ charge mixing at the $\I_2$ and the overall charge of the
twisted states which live there.

Given $\beta=+\coeff23$, $\gamma=-\coeff16$, the remaining
local anomalies at the $\I_2$ duly cancel out according to
eq.~\AnomalyOther,
$$
\eqalign{
{\cal A}'\ &
\equiv\ \coeff23\Tr_Q({\cal F}^4)\ -\ \coeff16\tr(R^2)\times\Tr_Q({\cal F}^2)\
	+\ (\coeff18 g+\half T(1)=\coeff{3}{32})\bigl(\tr(R^2)\bigr)^2\crr
&=\ +{1\over16}\left(
	\tr(F^2_{\darkblue SU(8)})\,+\,4F^2_{\purple U(1)}\,
	-\,\half\tr(R^2)\right)^2\crr
&\qquad+\ \left(
	\tr(F^2_{\darkblue SU(8)})\,+\,4F^2_{\purple U(1)}\,
	-\,\half\tr(R^2)\right)\cr
&\qquad\qquad\times\left(
	\tr(F^2_{\darkred SU(3)})\,+\,\coeff83F^2_{\purple U(1)}\,
	-\,\coeff{5}{32}\tr(R^2)\right) .\cr
}\eqno\eq
$$

\par\nobreak\smallskip
\centerline{$***\qquad {\msym\aquamarine Q.~E.~D.}\qquad ***$}

%%%%%%%%%%%%%%%%%%%%%%%%%%%%%%%%%%%%%%%%%%%%%%%%%%%%%%%%%%%%%%%%%%%%%%%%%%%%%%%%%%%%%%%%

\refout
\bye